\documentclass[12pt]{ociamthesis}  

\usepackage{amssymb}

\title{Thomson backscattering diagnostics of nanosecond      
       electron bunches in high space charge 
                     regime}   

\author{Bruno Paroli}             
\college{Dipartimento di Fisica}  

\renewcommand{\submittedtext}{A dissertation submitted to the Department of Physics in partial
                              fulfillment of the requirements for the degree of}
\degree{Doctor of Philosophy}     
\degreedate{January 2012}         

\begin{document}

\baselineskip=18pt plus1pt

\setcounter{secnumdepth}{3}
\setcounter{tocdepth}{3}

\maketitle                  
\begin{dedication}
to Marta, Gabriel and Sara
\end{dedication}
   
\begin{acknowledgements}
Thanks to my supervisor Prof. Franca De Luca for her valuable assistance in carrying out this activity and to Prof. Roberto Pozzoli, Dr. Massimiliano Rom\'e, Dr. Giancarlo Maero, Dr. Marco Cavenago for their full participation in the development of this diagnostics. 
I am grateful for the trust you have put in me on many occasions! Thanks to Francesco Cavaliere, Attilio Gandini, Daniele Vigan\`o of the Mechanical Workshop of this Department for the constant and fundamental contribution in experimental research and to Claudio Marini and Federico Nespoli for their important theoretical and experimental studies. Thanks to Prof. Marco Bersanelli and the Cosmology Group for their support during my final year of work, and to Dr. Simone Cialdi, Daniele Cipriani and Fabio Villa for the useful discussions. 
\end{acknowledgements}
 
\begin{abstract}
The intra-beam repulsions play a significant role in determining the performances of free-electron devices when an high brilliance of the beam is required. The transversal and longitudinal spread of the beam, its energy and density are fundamental parameters in any beam experiment and different beam diagnostics are available to measure such parameters. A diagnostic method based on the Thomson backscattering of a laser beam impinging on the particle beam is proposed in this work for the study of nanosecond electron bunches in high space charge regime. This diagnostics, aimed to the measurement of density, energy and energy spread, was set-up in a Malmberg-Penning trap (generally used for the electron/ion confinment) in two different configurations designed to optimize sensitivity, spatial resolution and electron-beam coincidence in space and time. To this purpose an electron bunch (pulse time $\leq4$ ns), produced by a photocathode source, was preliminary characterized with different electrostatic diagnostics and used to test the diagnostics systems. The solutions are detailed, which were devised for both the laser and bunch injection in the vacuum chamber, space and time coincidence of electron and laser pulses, photon detection, optimization of the geometry in the laser-beam interaction. The results are then summarized with an estimate of the minimum sensitivity of the set-up.    
\end{abstract}

\begin{romanpages}          
\tableofcontents            
\listoffigures              
\end{romanpages}            

\chapter{Introduction}
Stationary and pulsed electron beams are of great interest in the scientific community and represent an important resource in industrial technology. They are produced in a wide range of densities and energies. At low energy 1 - 500 keV electron beams are commonly used in free electron devices as electromagnetic sources in GHz and THz regime and with applications in  many fields as telecommunication \cite{Levush}, accelerators \cite{Faure}, plasma physics, medicine \cite{Bogdanovitch}. For example beams with energy of some hundred of keV are used in Klystrons to obtain microwaves of power greater then $50$ MW \cite{Arnold} and up to (1 - 10 GW) \cite{Hua}, \cite{Serlin}. Free electron maser (FEM) sources are obtained at low power output with low energy beams $< 15$ keV \cite{Einat}, and to produce high power microwave $20 - 30$ MW \cite{Kaminsky}. At higher energy, pulsed beams are used  in a wide range of applications, for example to generate coherent X rays in free electron laser (FEL) \cite{Sparc} and in Thomson back-scattering  X ray sources \cite{Bacci}, \cite{Brown}. Electron beams are also largely used to generate intense ion beams by electron-beam-ion-source (EBIS). In storage rings and synchrotrons the electron cooling technique allows the intensity and brilliance of ion beams to be increased. To this aim the monochromaticity of the electron beams (tipically in the $1 - 100$ keV range) is fundamental \cite{Poth},\cite{Danared}. All these applications have in common the need to obtain beams with appropriate parameters (e.g. size, monochromaticity, intensity, brilliance). One of the main limitations in achieving such parameters is the intrabeam repulsion, i.e. space charge.

When space charge effects become dominant the internal dynamics of the beam affects the performance of free electron devices as well as of electron beam sources. For example in DC photoinjectors the space charge limitation follows the Child-Langmuir law \cite{Field}, for bunch in nanosecond \cite{Caretto} and picoseconds regimes \cite{Berger}, \cite{Shea}, \cite{Asakawa}. The beam energy is an important parameter that determinates the space charge regime. In relativistics beams space charge phenomena are usually assumed to scale with the beam energy as $1/\gamma^2$ \cite{Stupakov}, where $\gamma$ is the relativistic factor, but when high brightness is necessary these effects are of fundamental importance also in relativistic beams as well as in lower voltage devices \cite{Dowell}. For example complex space-time oscillations are extensively observed in low-voltage systems due to the formation of a virtual cathode, when the beam current is higher than the space-charge-limited current in the region between the beam source and the extraction electrode \cite{Kurkin}. At high current density the non-linear beam dynamics becomes complicated and collective effects lead to mechanism of chaotisation in both non-relativistics and relativistics electron beams \cite{Trubetskov}, \cite{Hramov}. A direct experimental characterization of these effects may be difficult in
high-energy ($\approx 100$ MeV) and ultrashort (few ps) bunched beams. By properly scaling density, current, magnetic field and spot size, similar effects may be measured on beams
with lower energy but exhibiting an almost identical transverse dynamics of the beams
used in these devices. Indeed the results are equivalent to those obtained in beams in different regimes if the parameter $\omega_p^2/\gamma\omega_c^2 \propto I/\beta\gamma B^2 r^2$ is kept constant \cite{Davidson}, where $\omega_p$ and $\omega_c$ are the
plasma and cyclotron frequencies, respectively, $I$ the beam current, $B$ the magnetic field, $r$ the beam radius and $\beta$ the usual relativistic factor. Such effects can be easily studied in low-energy beams (of some keV) with diagnostics instruments that are essential for monitoring and assessing any beam experiment. These diagnostics provide information on the state of the beam and on the progress and results of experiments performed on the beam, monitoring critical beam parameters such as current, size, energy, emittance, density, profile. A number of beam diagnostics are currently in use: 1) Faraday cups for low and high energy beams \cite{Tautfest} to measure the longitudinal charge distribution to the sub-nanosecond regime \cite{Bellato}, 2) Rogowsky coils to measure the net beam current \cite{Myers}, 3) Profile monitors to measure the transversal profile of the beam with screens \cite{Bernal} or crystals \cite{Graves}, 4) Capacitive probes to monitor the spatio-temporal position of beams \cite{Maurice}, 5) Electro-optical diagnostics as ultra-fast bunch length measurements \cite{Wilke}.

An alternative laser-based diagnostics is proposed in this work as an instrument to provide informations on density, density profile, energy and energy spread of low-energy electron beams and bunches in nanosecond regimes \cite{parolithomson}. Basically the interaction of a high energy IR laser with an electron ensemble produce scattered radiation. This interaction is classically described in the limit $h\,\nu/m_e c^2<<1$ by the Thomson scattering (where $h$ is the Planck constant, $\nu$ is the laser frequency, $m_e$ is the electron mass at rest and $c$ is the speed of light). The very low cross section of this interaction (of the order of the square of the classical electron radius) requires a high number of incident photons that are obviously provided by laser sources. The  developments of high-power lasers, optical technologies and photon-counting techniques have made possible the use of this diagnostics even in relatively low-temperature plasmas with electron density down to a few $10^{10}$ cm$^{-3}$ and the minimization of the noise is the main challenge for the detection of lower density beams. The Thomson backscattering diagnostics described in this work is part of the ELTEST/ELEBEAM projects founded by INFN and it was designed and implemented in the Malmberg-Penning trap (ELTRAP) generally aimed at the confinement of non-neutral electron plasmas with a magnetostatic and electrostatic fields. The electrostatic potentials up to $\pm 100$ V can be individually set on ten oxygen-free, high conductivity (OFHC) copper coaxial cylindrical electrodes. In such a way, an electric well can be formed  that confines the plasma in the longitudinal direction. By grounding the electrodes (open configuration) experiments on electron beams at low and very-low energies can also be performed \cite{Bettega}. In ELTRAP the beam is focused by the highly uniform magnetic field $<0.2$ T in a cylindrical drift tube of length $\approx1$ m . A nanosecond bunch with energy $1 - 20$ keV produced by a photocathode source is used to test the systems for the laser-bunch interaction in space and time. For this purpose the bunch was preliminary characterized in density, length and transversal profile with two electrostatic and an optical diagnostics. Two different set-up of the Thomson backscattering diagnostics are discussed. In the first set-up the laser beam was maintained collimated and the interaction could be moved in principle along the drift-tube. In the second set-up the laser is focused in a particular point to optimize the solid angle and the collected photons. In both cases we present the solutions for the stray-light reduction in the laser injection, photon detection, space and time coincidence of electron and laser pulses. The minimum sensitivity of the diagnostics was estimated in both set-ups measuring the noise and computing the expected signal with a theoretical estimate of the scattered photons in relativistic regime. The minimum density is generally limited by the stray-light of the high power laser that reduce the signal-to-noise ratio in the measurement of the scattered radiation . In the proposed set-up the stray-light is reduced advantageously exploiting the blue shift of the scattered radiation that is detected as close as possible along the direction of the bunch propagation.

The thesis is organized with a description of the ELTRAP apparatus (chapter 2), the bunch characterization with electrostatic and optical diagnostics (chapter 3) and the Thomson back scattering diagnostics with the related sub-system (chapter 4). Finally, chapter 5 collects the conclusions. An additional appendix A describes the reflectometry technique used for the characterization of the transmission line of the ELTRAP. In appendix B we describe extensively the production of a confined plasma in the trap by means of stochastic heating with a radio-frequency (1 - 20 MHz) for the study of electron beam-plasma interaction with applications in charged particles acceleration \cite{Hogan}, \cite{Pisin}. In this appendix we consider the possibility to produce diffused and compressed electron plasmas confined in ELTRAP, in UHV conditions and without the usual thermocathode or photocathode sources used in the past in this apparatus.

\chapter{Experimental apparatus}
\section{Introduction}
\begin{figure}
\begin{center}
\includegraphics[scale=0.8]{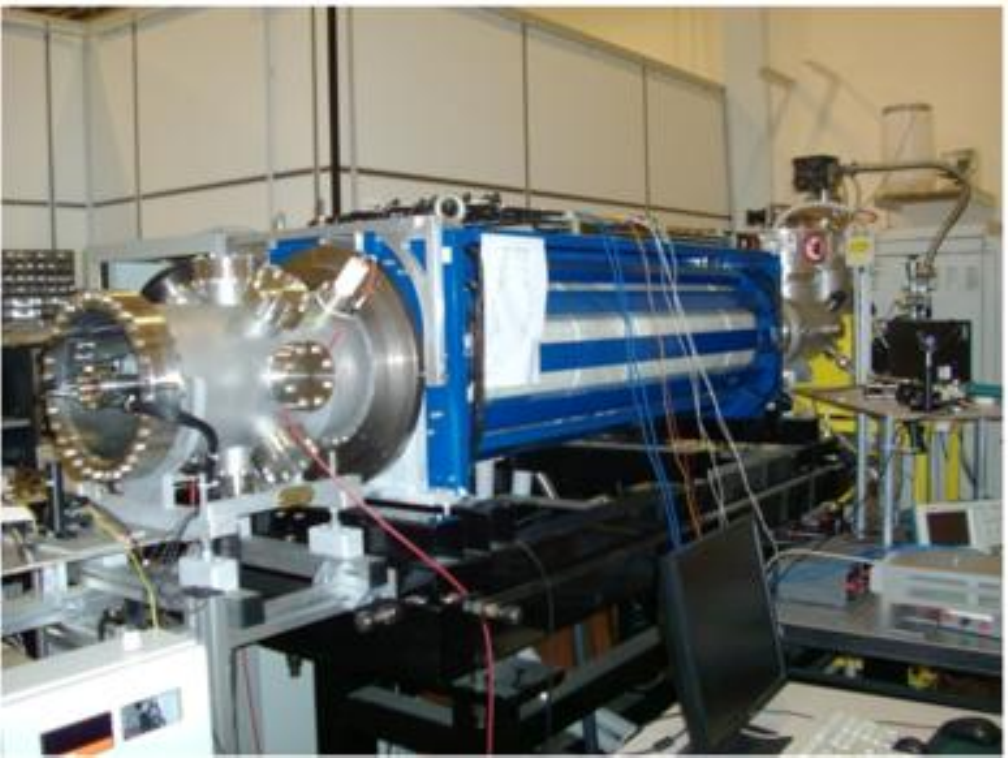}
\end{center}
\caption{\label{eltrap}Picture of the Malmberg-Penning trap Eltrap. The apparatus work at a residual gas pressure of $\approx 10^{-9}$ mbar and with a magnetic field  B $\leq 0.2$ T. The internal drift tube has a total length of $\approx 1$ m.}
\end{figure}

The experimental apparatus for the study of continuous or pulsed electron beams was implemented on a Malmberg-Penning trap ELTRAP \cite{Amoretti} (Electron Trap) working in UHV (Ultra High Vacuum) with residual gas pressure of the order of $10^{-9}$ mbar, with an uniform axial magnetic field B $\leq 0.2$ T (see fig. \ref{eltrap}) and a series of coaxial conducting cylinders that forms the drift tube for the beams, with a total length $\approx 1$ m. The UHV conditions are required to characterize the dynamics of the bunch, because the collisional time $\tau_c$ in the intrabeam scattering phenomena by electron-neutral collisions, for a pressure of  $\approx 10^{-9}$ mbar is approximatively between 320 ns and 1.40 $\mu$s (considering a beam temperature of the order of 0.1 eV and a bunch density $10^8$ - $10^9$ cm$^{-3}$). These times are an order of magnitude greater than the characteristic times of flight of the beams (30 - 100 ns). For higher pressures the collisional effects become non-negligible. The magnetic field is required to radially focus the beams. An efficient focusing requires a magnetic field of B $\geq 90$ G. The characteristic time for observing space charge effects can be estimated roughly by the plasma period $\tau_p=2\pi/(\omega_p)$ of the produced bunches $\tau_p\leq 10$ ns that fixes a minimum length of flight of the order of 1 m for bunch energies of some tens of keV. The beam source used in this experiment is a photocathode illuminated with a pulsed ($<4$ ns) UV laser. The source is located on the end of the vacuum chamber. An optical and two electrostatic diagnostics have been developed to characterize the longitudinal and transverse properties of the bunch (e.g. length, charge, transverse profile). Both electrostatics and optical diagnostics are described in details in chapter 3, while the set-up of the Thomson scattering diagnostics is described in chapter 4.

\section{Vacuum system}

\begin{figure}
\begin{center}
\includegraphics[scale=0.8]{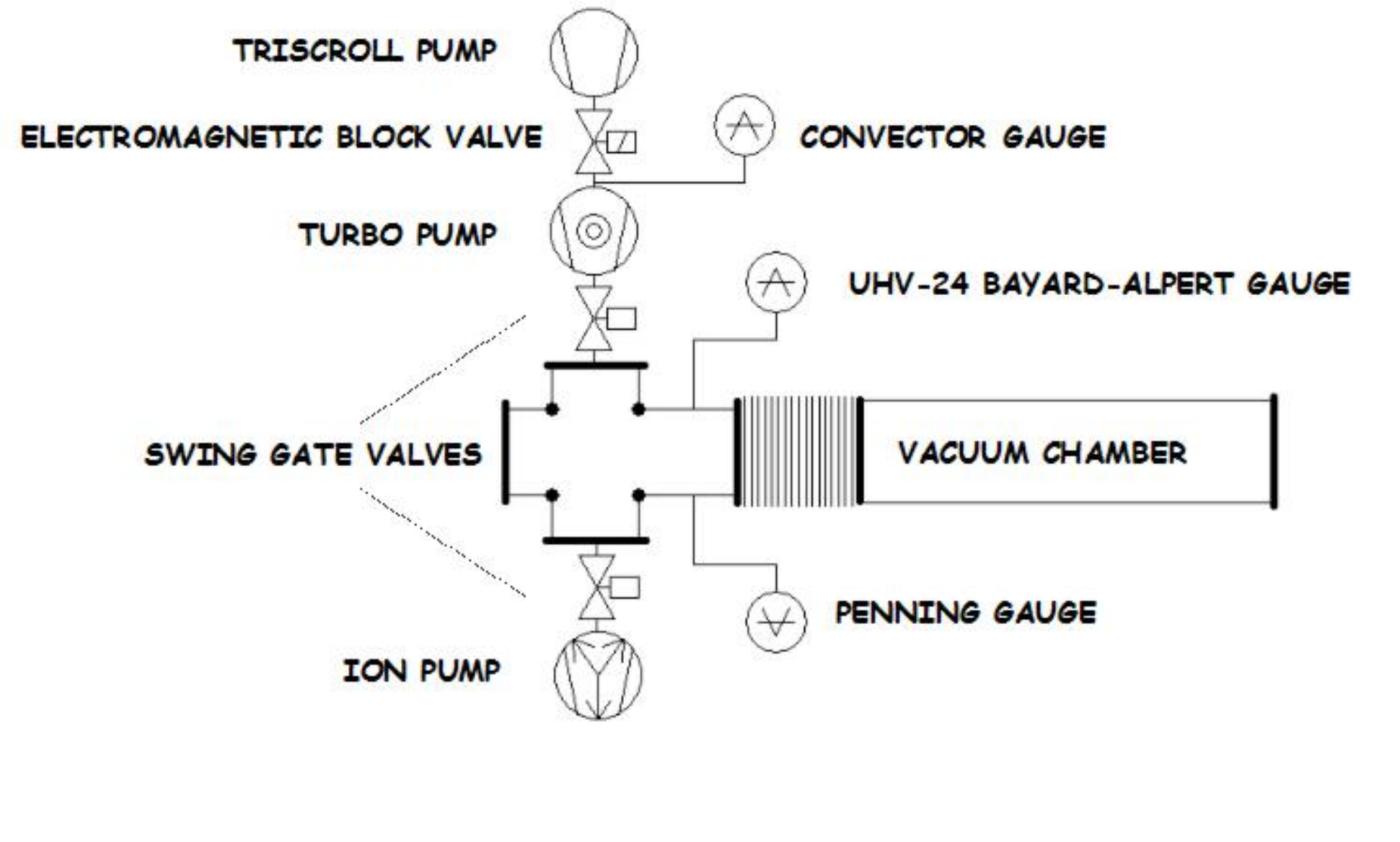}
\end{center}
\caption{\label{pumps}Schematics of the vacuum system. The working pressure of $\approx 10^{-8}$ - $10^{-9}$ mbar is obtained with three different pumps: a scroll pump, a turbo pump and an ion pump. The system is regulated by three valves and the pressure is monitored by three vacuum gauges.}
\end{figure}

The main vacuum chamber of the ELTRAP apparatus has a diameter of 25 cm and length of 1.7 m with a volume of $\approx 84$ dm$^3$. The two additional volumes of the source chamber $\approx 32$ dm$^3$ and of the detection chamber $\approx 15$ dm$^3$ must also be considered. The UHV condition of the total volume is reached with a pumping system (see fig. \ref{pumps}) composed by three different pumps working efficiently in three different regimes of pressure. A first volumetric scroll pump reduces the atmospheric pressure to about $10^{-3}$ mbar, a turbo-pump then reduces the pressure to about $10^{-7}$ mbar and an ion pump stabilizes the working pressure of the vacuum chamber at $10^{-8}$ - $10^{-9}$ mbar. The turbo and  the ionic pumps are connected in parallel by means of a pneumatic stainless steel valve controlled electrically, while the scroll pump is connected to the output of the turbo-pump by means of an electromagnetic valve. The pressure in the chamber is measured with three different vacuum gauges. A convection gauge working at higher pressure $\leq 10^{-3}$ mbar, a cold cathode Penning type gauge working at $\approx 10^{-8}$ mbar and a ionization gauge working at lower pressures ($\approx 10^{-9}$ mbar). Because the pumping speed changes for different gases, the residual gas in the chamber should be mainly composed by molecules like Hydrogen and noble gases. For example, the pumping speed of the turbopump is 280 l/s for nitrogen, 230 l/s for Helium and 210 l/s for Hydrogen (at pressure $\leq 10^{-6}$ mbar). The ion pump has internal magnets made by ferrite and the maximum stray field is $\approx6$ Gauss in the plane of the flange. Backing processes with heating bands that promote the degassing of absorbed gases from the chamber are used to reach the final pressure in a shorter time.

\section{Magnetic field}

\begin{figure}
\begin{center}
\includegraphics[scale=0.6]{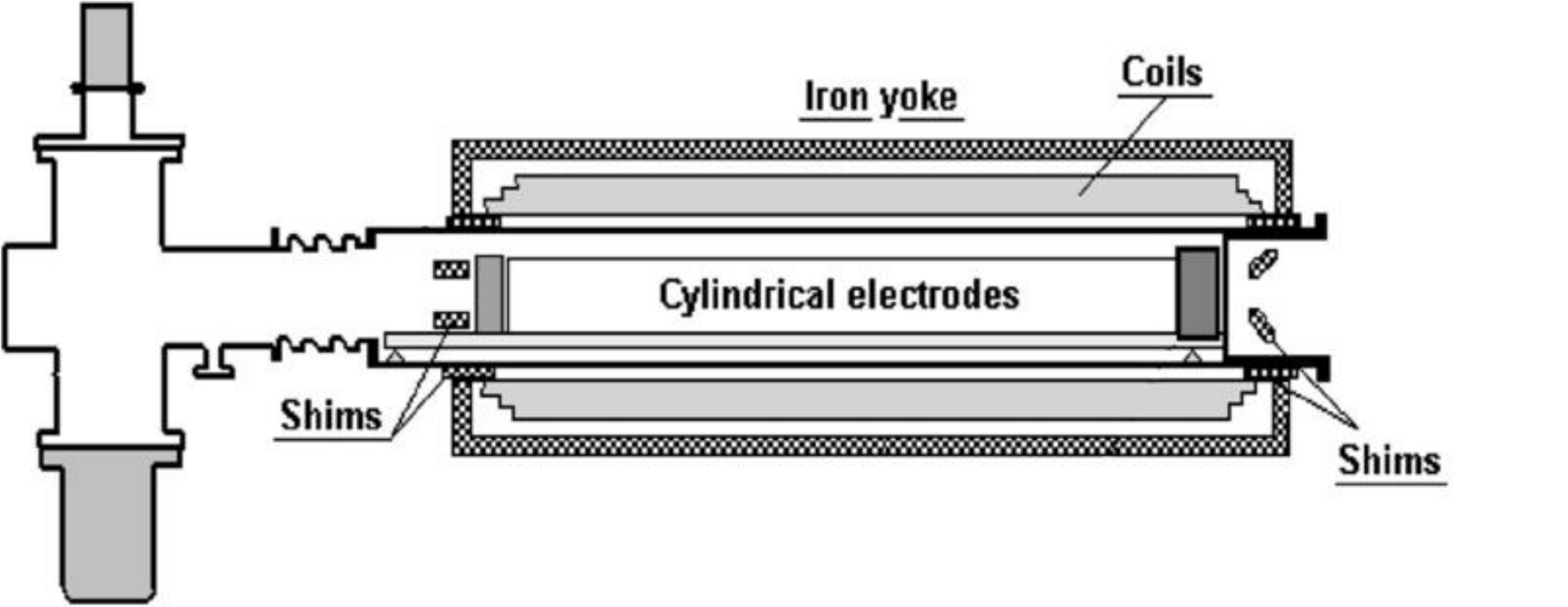}
\end{center}
\caption{\label{ironstructures}Schematics of the iron structures used to generate the magnetic field in the trap. An external iron yoke shields the stray magnetic field and some shims positioned at the coil ends concentrate the fields line to the axis. The coil is made by three windings connected in series.}
\end{figure}

\begin{figure}
\begin{center}
\includegraphics[scale=0.65]{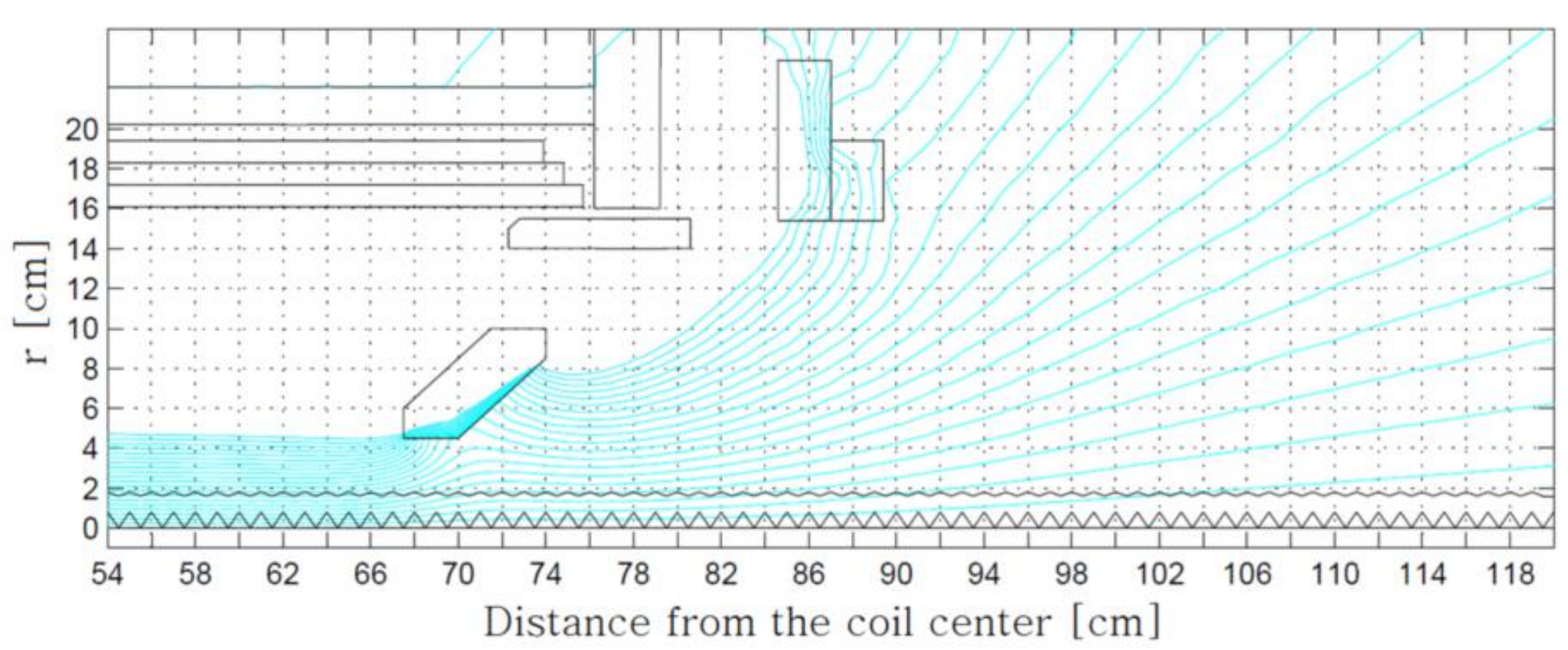}
\end{center}
\caption{\label{cavenago}Numerical analysis of the magnetic field near to the coil end. The magnetic field lines are represented in blue. The magnetic field strength decreases starting from the conical shim.}
\end{figure}

The magnetic field of Eltrap is generated by a conventional solenoid (1.5 m length, 35 cm diameter) formed by three conductors connected in series and cooled by three parallel water fluxes. A digitally controlled current generator with a current drift $\frac{dI}{dt}=10^{-5} A/h$, a maximum current of 600 A and a maximum voltage of 120 V is used as power supply of the coil. The maximum magnetic field strength obtained in the central region is 0.2 T. Two iron blocks (1 cm thickness) were inserted at the ends of the coils and conically shaped iron funnels were inserted to the greatest possible extent (see fig. \ref{ironstructures}), considering the passage of cables, the pumping needs, etc. to concentrate the field lines closely to the axis and increase the uniformity in the central region. The measured field uniformity is better than $10^{-3}$ within a distance (from the center of the magnet) of 50 cm, and within a radius of 5 cm around the axis. Four additional dipolar coils are used to correct the axial direction of the main magnetic field, the maximum deviation for higher fields being $\pm 15$ mrad at $B=0.2$ T. The external disturbances of the magnetic field i.e. the earth and the pump fields are shielded refracting the field lines in a high permittivity soft iron yoke consisting of 2 square end plates, connected by 12 return flux bars. The uniformity and the axial field direction are lost moving from the center to the end of the coil. A numerical analysis of the field (see fig. \ref{cavenago}) including the iron structures shows that the magnetic field strength on the trap axis decreases starting from $\approx 60$ cm (from the coil center) and reaches a value of $\approx 1\%$ (of the maximum field) at a distance of 113 cm. 

\section{Beam source}       
\begin{figure}
\begin{center}
\includegraphics[scale=0.8]{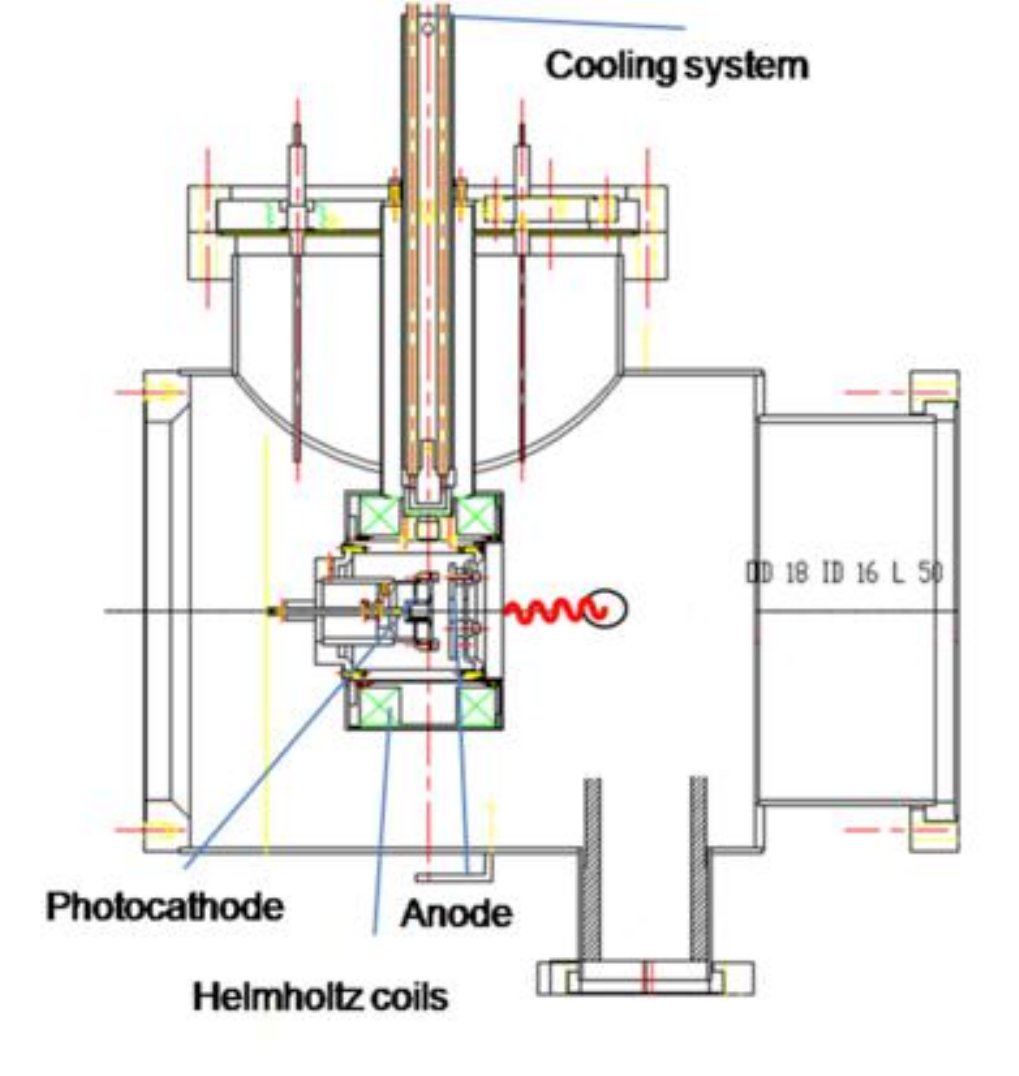}
\end{center}
\caption{\label{electronsource}Electron source used to generate nanosecond electron bunches in the experiments. An electron bunch produced by a photocathode, illuminated by a $<4$ ns UV laser pulse, is initially focalized by two Helmholtz coils and accelerated by an annular anode. A water flux provides the cooling of the coils. The output flange is connected to the front of the vacuum chamber.}
\end{figure}

\begin{figure}
\begin{center}
\includegraphics[scale=0.8]{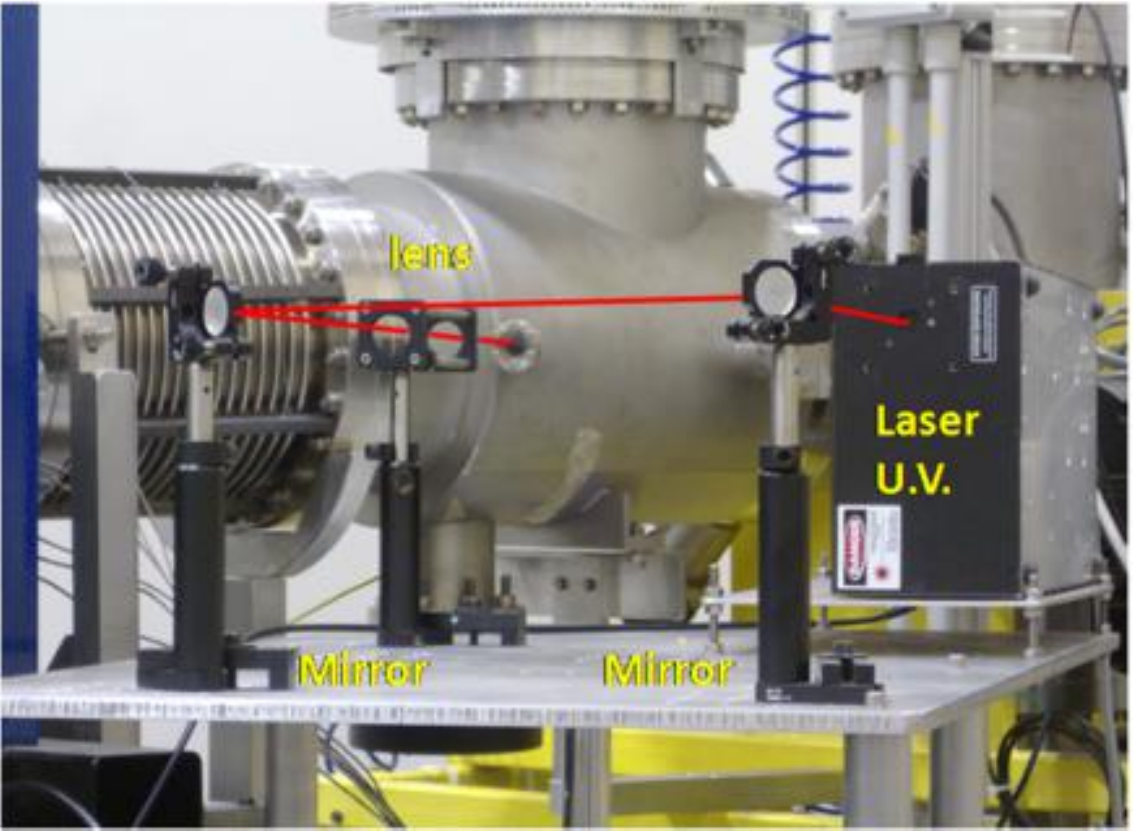}
\end{center}
\caption{\label{uvalignment}Alignment system of the UV laser beam on the photocathode. The laser trajectory is regulated adjusting two UV silica mirrors. A lens can be used to focus the laser beam on the photocathode surface. The input viewport visible in the picture is made of quartz to increase the transparency to UV radiation.}
\end{figure}

An electron source (see fig. \ref{electronsource}) is inserted at the end of the vacuum chamber and aligned to the geometrical axis of the trap. The source is a barium-tungsten dispenser photocathode mounted on a alumina body with an active area of 2.4 cm$^2$. An internal heater, supplied by a current generator in the 0 - 2 A range,  is used  to reach the working temperature needed for the surface activation (900 - 1200 $^\circ$C). The photocathode is illuminated by a pulsed ($<4$ ns) UV laser with a wavelength of 337 nm and an energy pulse $< 400$ $\mu$J. The laser is aligned to the source by means of movable UV silica mirrors (see fig. \ref{uvalignment}). The laser beam has an original size of $7\times7$ mm reduced by a circular pin-hole to 5 mm. The produced bunch is accelerated by a circular anode connected to the ground, while the photocathode is polarized at a (negative) voltage in the range 1 - 20 kV. The  initial focusing of the bunch is obtained with a local magnetic field generated by two Helmholtz coils, then the bunch enter in a more intense magnetic field where the original spot-size is radially compressed. Stationary beams from the same source are produced by thermoionic emmision heating the photocathode at higher temperature. Both emission current and transversal profile can be characterized with the electrostatics and optical diagnostics described in chapter 3. 

\section{Drift tube, electrodes and trasmission lines}  
\begin{figure}
\begin{center}
\includegraphics[scale=0.8]{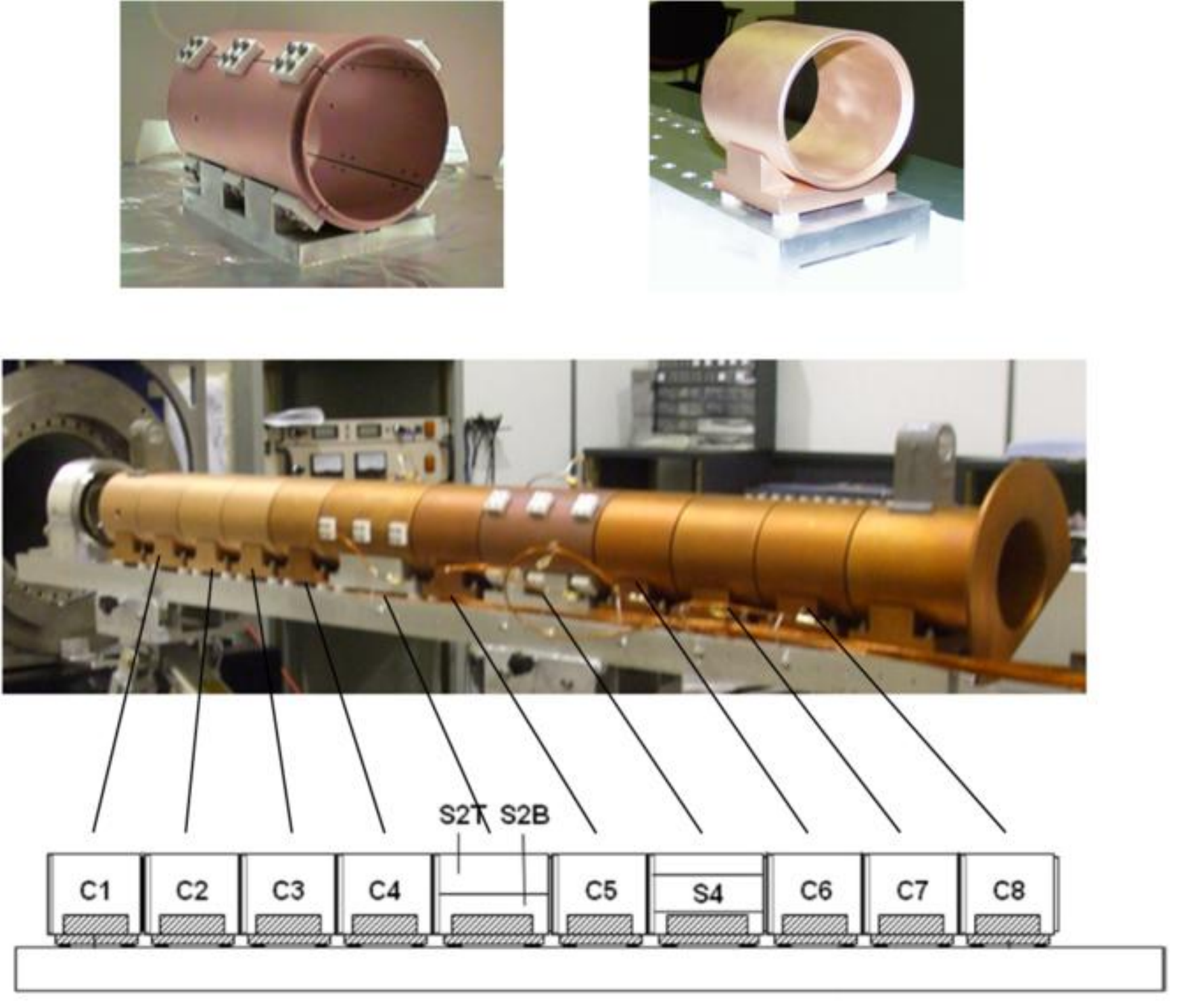}
\end{center}
\caption{\label{electrodes}Pictures of the drift tube (center) with the respectives cylindrical (top right) and sectored (top left) electrodes. The cylinders are labeled C1 - C8, S2, S4 as shown below. All cylinders are made in OFHC copper and mounted with macor insulators on an aluminum bar.}
\end{figure}

The produced beam travel in a drift tube (see fig. \ref{electrodes}) of length $\approx$ 1 m and 9 cm diameter made up by eight coaxial hollow cylinders (C1 - C8) made in OFHC copper (Oxigen Free High Conductivity) with a length of 9 cm and by two sectored cylinders (S2 - S4) of length 15 cm divided azimuthally in two and four patches, respectively. All cylinders have a copper planar base mounted on an alluminium bar with macor insulators. The area of the base is 72 cm$^2$ for C1 - C8 and 124 cm$^2$ for S2, S4, the distance from the base to the bar is $\approx$ 4.4 mm. The cylinders C1 - C8 should be characterized by the same impedance (neglecting the boundary at long distance and considering all other cylinders grounded) because each cylinder is close to the other one along the axis and with the aluminium bar on the base. This occurs also to C1 and C8 because two additional grounded cylinders are inserted to the ends of the drift tube. But the real electrical properties are quite different because the coaxial lines of C1, C3, C4, C6, C8 are brought outside by 5 not matched high voltage feedthrough while the coaxial line of C2, C5, C7 are matched externally by 3 coaxials feedthrough. Conversely the sectors of S2 and S4 are electrically different because their parasitic capacitances change due to their different azimuthal positions. All electrodes are connected with coaxial kapton insulated wires, designed for high and ultrahigh vacuum environments with an impedance of 50 $\Omega$. These trasmission lines were measured and characterized with a reflectometry technique (see appendix A) sending a pulse with a FWHM of 8 ns and receiving the reflected signal with an oscilloscope with 1 GHz bandwidth. These measurements are needed when the required bandwidth is limited by the mismatch in the cable-feedthrough transitions and by the electrode capacitances. These effects becomes significant for frequencies larger than few hundred Megahertz and are of fundamental importance in our case, to reduce the distortions in the signal produced by the fast electrostatic beam diagnostics described in the next chapter.

\chapter{Bunch characterization with electrostatic and optical diagnostics}
\section{Introduction}
In chapter 1 we pointed out the importance of space charge effects in beams and bunches for applications at low, medium and high energy in free electron devices. To characterize the basic dynamics e.g. the longitudinal spread, the times of flight and the properties of the bunch i.e. length, radius and density, a electrostatic and optical diagnostics  were developed. In the electrostatic diagnostics we measure the charge (induced or collected) and the current of the beams. The main limitation is a non negligible electrode's capacity or inductance that introduce a significant distortion on the measurement. One has to find a compromise between a higher sensitivity and a lower spatial resolution. The performances can be increased optimizing  the design \cite{Sefkow}, or with post-processing techniques. On the contrary, current or charge measurements are not affected by these limitations and currents of nA and charge of pC can be measured \cite{Rome1}. A first electrostatic destructive diagnostics based on a planar-charge collector was used; the electric signal generated by the impact of the electron bunch is distorted by the impedance mismatch between the collector and the transmission line connected to the oscilloscope. A de-convolution technique still allows to extract information about the temporal duration and the longitudinal spread of the bunch also in these conditions of mismatch \cite{Paroli1}. With a second non-destructive electrostatic diagnostics we read the signal induced by the transit of the bunch inside a cylindrical electrode of the trap. This diagnostics is limited by the spatial extension of the electrode used  for the measurement and by its capacity. Imposing a given longitudinal charge distribution of the bunch we can estimate length, spread velocity and time of  flight of the bunch \cite{Paroli2}. The transversal density profile of the bunch was characterized by an optical diagnostic based on a phosphor screen coupled with a CCD camera, information about the transversal size and charge distribution were obtained for different strength of the focusing magnetic field \cite{Rome2}. The destructive electrostatic and optical diagnostics are positioned at the end of the cylindrical stack at a distance from the source of $\approx$ 1.8 m, in a region where the magnetic field is still uniform and the phosphor screen is centered on the trap axis, while for the non destructive electrostatic diagnostics we use the cylindrical electrodes C1 - C8 or the sectored S4, S2. An additional advantage of this last diagnostics is that we can measure the properties of the same bunch in different positions during its flight. The produced electrostatic signals are acquired with a digital oscilloscope with a maximum sampling rate of 10 Giga-samples/sec. and a bandwidth of 1 GHz. The acquired data are then analyzed with deconvolution techniques.   

\begin{figure}
\begin{center}
\includegraphics[scale=0.8]{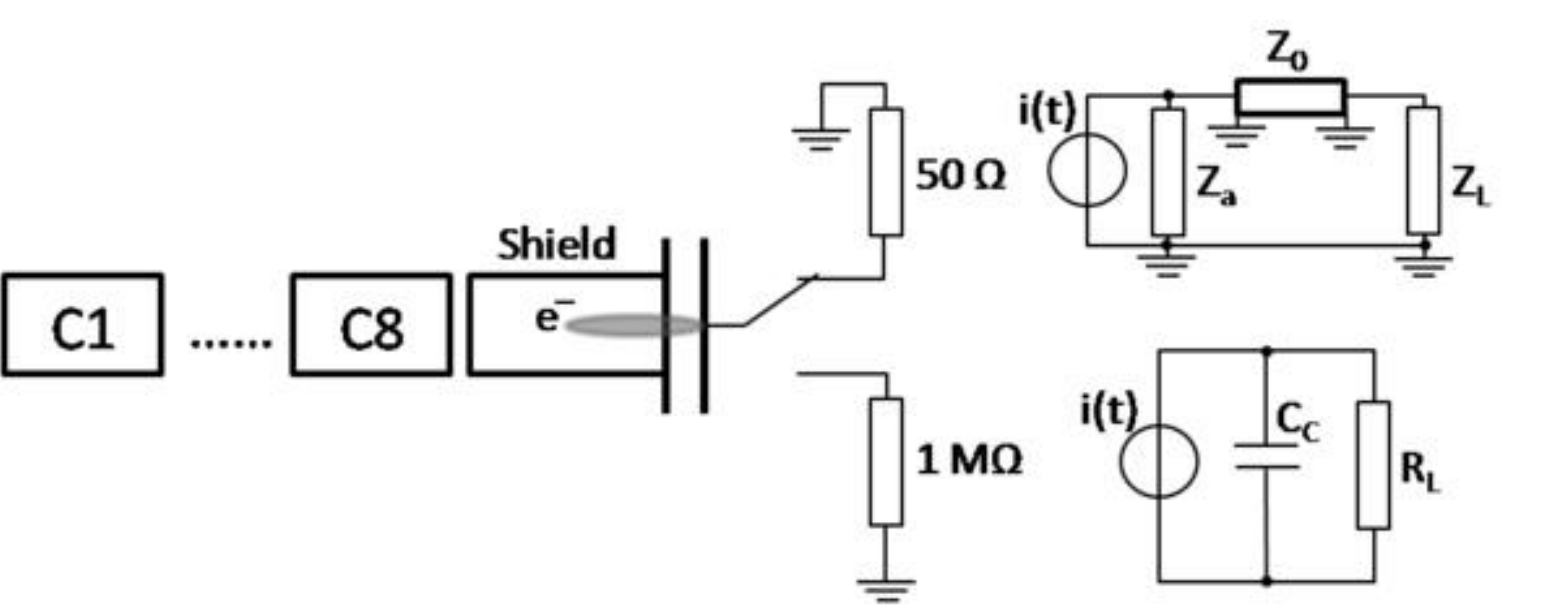}
\end{center}
\caption{\label{diagnostics}Scheme of the electrostatic diagnostics in low and high impedance mode with the equivalent circuits. The current generator $i(t)$ is the time derivative of the collected charge of the bunch across the planar charge collector. A cylindrical shield is used to reduce the image charge effects.}
\end{figure}

\section{Destructive electrostatic diagnostics}
The destructive electrostatic diagnostics is based on a planar charge collector made on a glass substrate coated with aluminium. The collector has a circular shape with a diameter of 11 cm and is shielded by a cylindrical electrode of 9 cm diameter, 15 cm length to reduce the induced image charge (fig. \ref{diagnostics}). The aluminium coating is covered by a P43 phosphor used in the optical diagnostic described later. The diagnostics is positioned at the end of the cylindrical stack and is connected to a digital oscilloscope with 1 GHz bandwidth by means a coaxial cable with an impedance of 50 $\Omega$. The measurable properties of the bunches and beams are closely related to the impedance load used for the measurement, in particular we can distinguish an high impedance  measurement, i.e. the resistive load $R_L$ has a value of 1 M$\Omega$, and a low impedance measurement with a resistive load $R_L=$ 50 $\Omega$. In the first case rapid variations of the formed signal produced by the bunch or beams are filtered by the high time constant $\tau=R_L C_c$, where $C_c$ is the capacity of the coaxial cable. Assuming a typical value of 300 pF the time constant is $\approx$ 300 $\mu$s, much larger than the characteristic time duration of the bunch. This configuration is so suitable for total charge and current measurements. In fact considering the equivalent circuit (see fig. \ref{diagnostics}) the charge is given by

\begin{equation}\label{totalcharge}
 Q(t)=V_0(t) C_c+\frac{1}{R_L} \int_{-\infty}^{t} V_0(\tau) d{\tau}
 \end{equation}

where $V_0(t)$ is the voltage measured on $R_L$. Neglecting the second term in equation (\ref{totalcharge}) we obtain $Q(t)= V_0(t) C_c$, while the instantaneous current of the beam is given by $I(t) = V_0(t)/R_L$. The low impedance measurement requires a more complicated circuital model (see figure \ref{diagnostics}). Conversely the planar charge collector has an impedance $Z_a$ not matched with the transmission line and for a rapid variation of the formed signal the distortions introduced by the mismatch are not negligible. To obtain information about the "original" signal a deconvolution method is necessary. The circuital  model of the low impedance measurement is a current generator connected to the parallel of three impedances : the antenna impedance $Z_a$,  the transmission line impedance $Z_0$ and the load impedance $Z_L$. The signal read by the oscilloscope is the voltage $V_0$ across $Z_L$ related to the detected current $i(t)$ by the transfer function
 
\begin{equation}\label{transferfunction}
 F(\omega)=V_0(\omega)/i(\omega)=\left[\frac{\cos(\omega L/v_f)}{Y_L+Y_a(\omega)}-j Z_0 \sin(\omega L/v_f)\right] 
 \end{equation}

where $i(\omega)$ is the Fourier transform of the current $i(t)$, $Y_a$, $Y_L$ are the antenna and load admittances respectively and $v_f$ is the phase velocity ($\approx 2 \cdot 10^8$ m/s) of the signal propagating in the cable of length $L$. Let us consider a bunch with a density $n(r,z,\theta)=n_0\,[1-H(r-{R_b})]\, g(z)$ i.e. with a transversal flat profile symmetric in $\theta$ and an axial profile $g(z)$. The current due to the collected bunch  for a Gaussian function $g(z)$ is $i(t)= i_0\exp(-t^2/2\sigma_t^2)$, where $\sigma_t=\Delta L/2v_b$ with $\Delta L$ the bunch length and $v_b$ the bunch velocity respectively. So the measured voltage considering equation (\ref{transferfunction}) is

\begin{equation}\label{vout}
 V_0(t)=\frac{\sigma_t i_0}{\sqrt{2\pi}}\int_{-\infty}^{+\infty} \exp\left({j\omega t-\frac{\omega^2}{2\sigma_t^2}}\right)\left[\frac{\cos(\omega L/v_f)}{Y_L+Y_a(\omega)}-j Z_0\sin(\omega L/v_f)\right] d\omega.
 \end{equation}

\begin{figure}
\centering
\subfigure[]
{\includegraphics[scale=0.55]{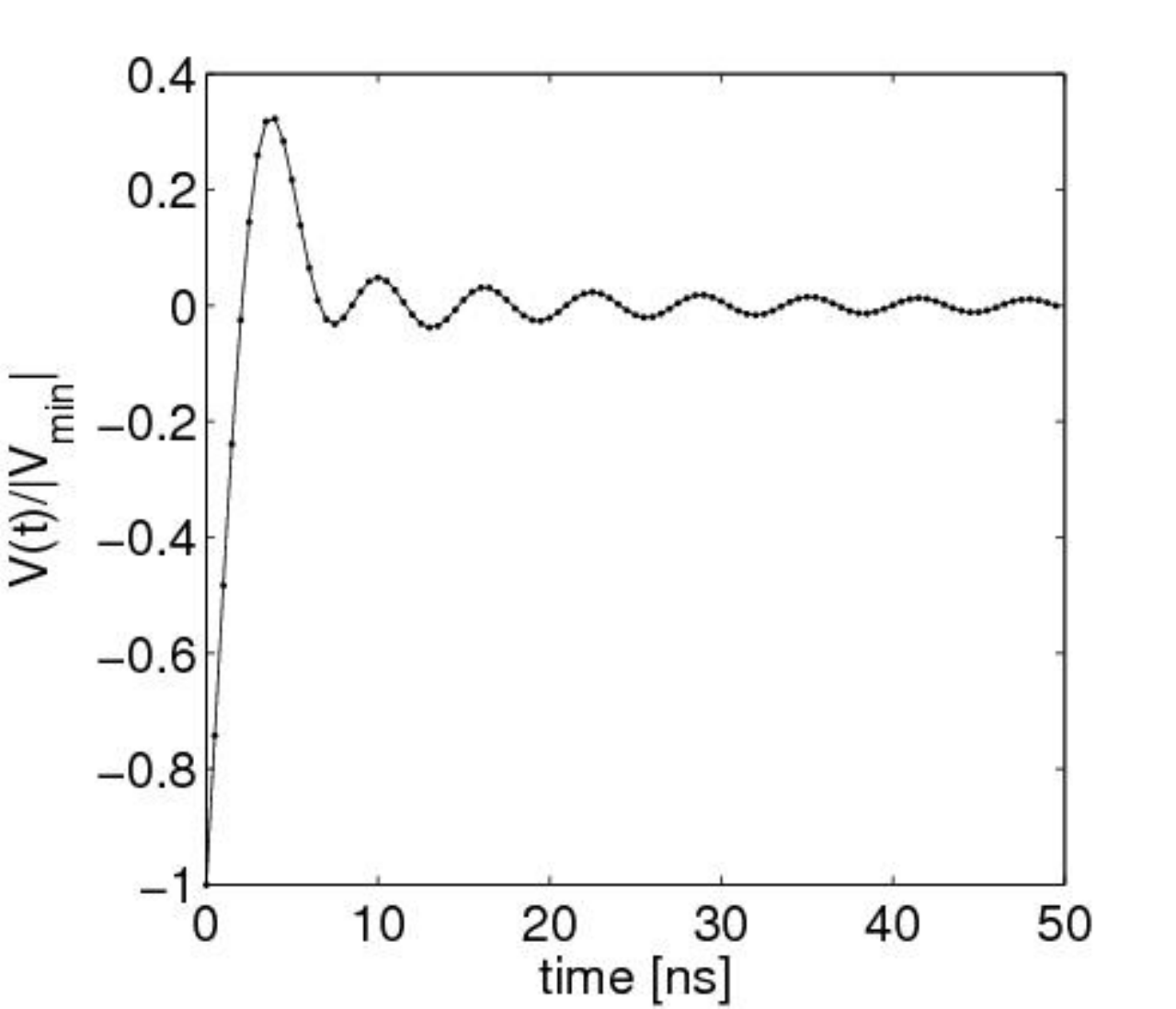}}
\hspace{5mm}
\subfigure[]
{\includegraphics[scale=0.55]{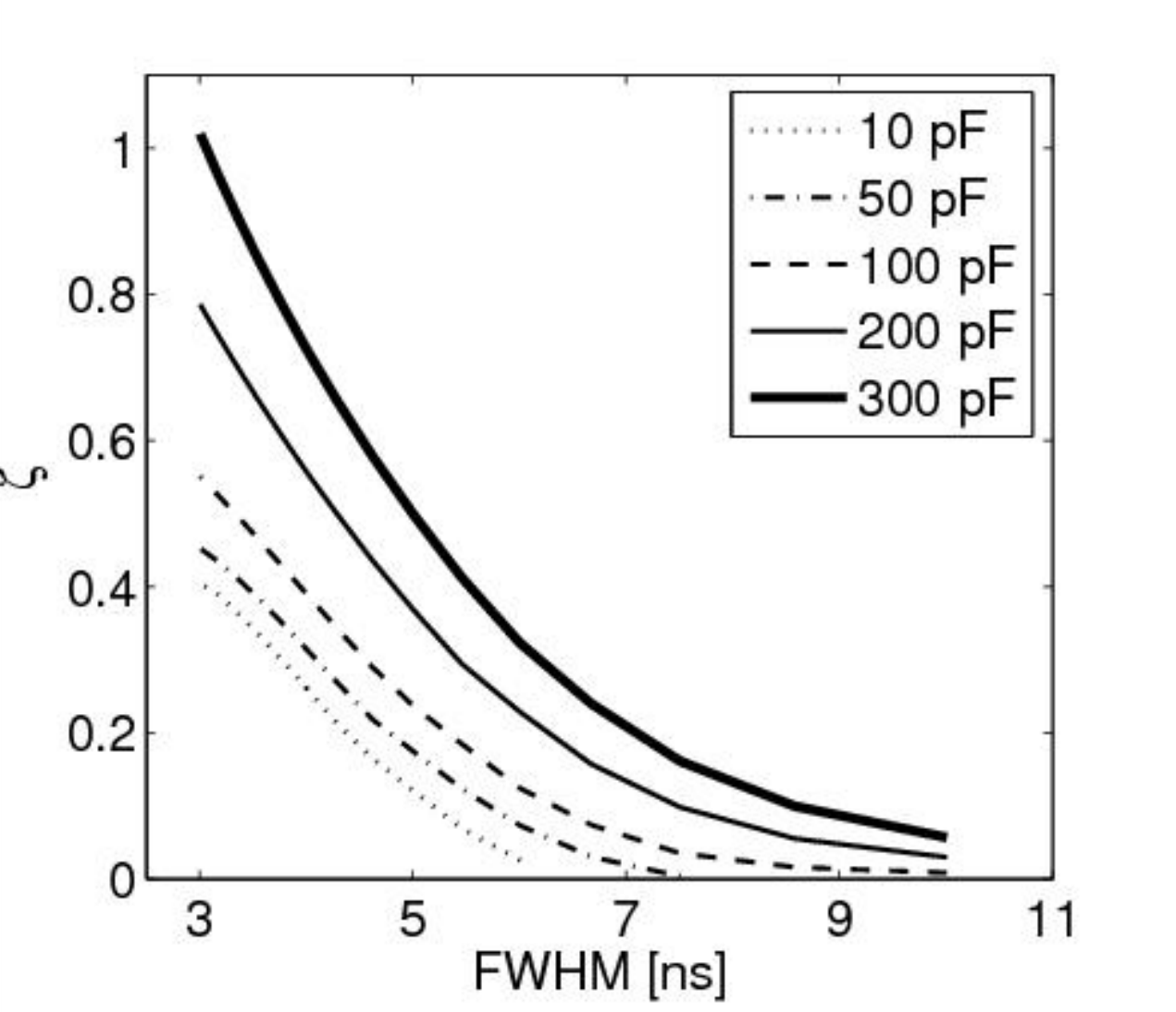}}
\caption{\label{numerical}(a) Computed output signal in the presence of a Gaussian current generator of width $\sigma_t=2.55$ ns and $C_a=300$ pF. (b) Signal shape factor $\zeta$ versus the characteristic length of the beam for different values of the antenna's capacity $C_a$ (a Gaussian axial density distribution is assumed.)}
\end{figure}

The numerical solution of equation \ref{vout} for different antenna's capacity shows an oscillatory  behavior due to the impedance mismatch in the antenna-coaxial transition figure \ref{numerical} (a). 
The ratio $\zeta \equiv V_{max}/V_{min}$  of the oscillatory output signal, where $V_{min}$ is the first minimum and $V_{max}$ is the first maximum, is in correspondence with the full width at half maximum,  defined as FWHM $\equiv 2\sigma_t\sqrt{2ln2}$, of the input Gaussian current pulse. Increasing the FWHM the ratio decreses; this is verified in a range between 3 and 10 ns. In figure \ref{numerical} (b) we report the $\zeta$ ratio for five different antenna's capacity $C_a$ (10 - 300 pF) as a function of the FWHM of the input Gaussian pulse from 3 to 10 ns. From these considerations we see that the $\zeta$ ratio is a good parameter to know the FWHM of the input Gaussian current pulse, knowing the values $V_{max}$ and $V_{min}$ of the output voltage signal. 

\subsection{Total charge and current measurements }
\begin{figure}
\begin{center}
\includegraphics[scale=0.85]{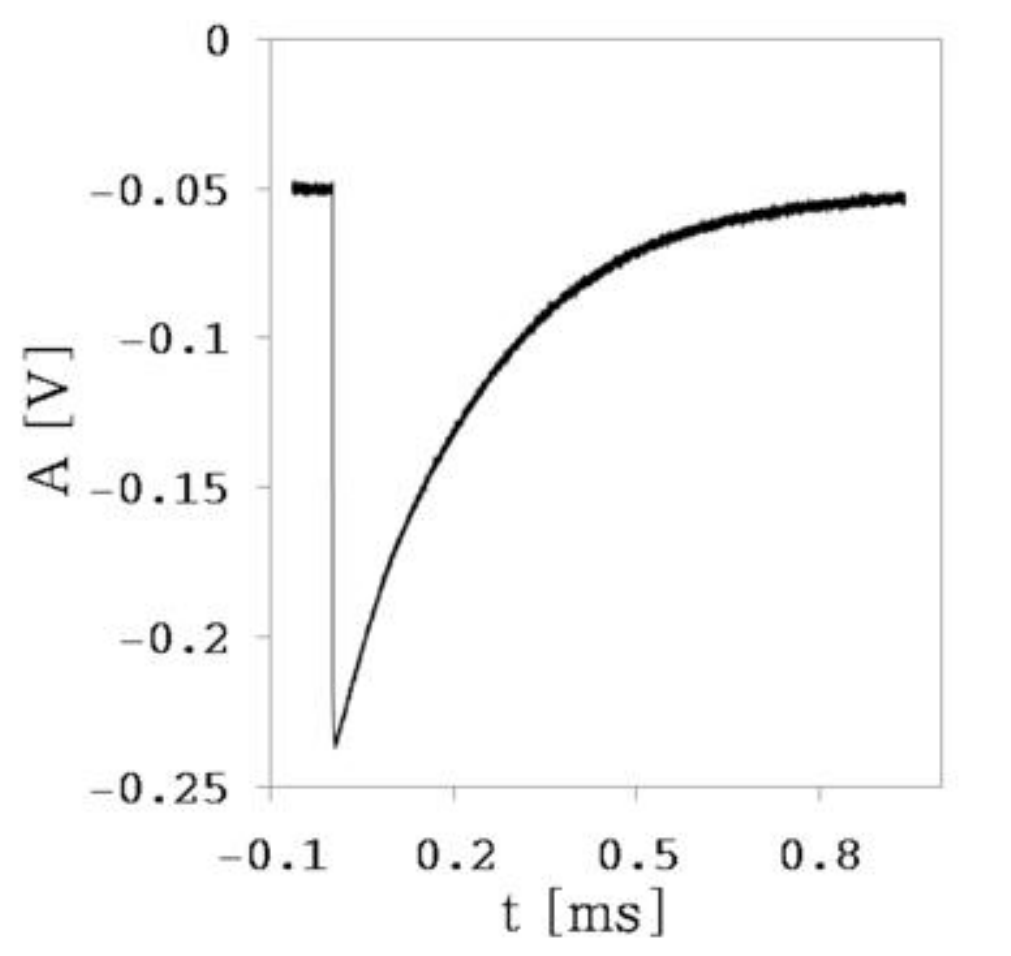}
\end{center}
\caption{\label{capacitance}Detected signal obtained with electrostatic diagnostics in high impedance mode. The capacity $C_C$ was extracted by the data fitting the exponential discharge of the RC circuit.}
\end{figure}

\begin{figure}
\centering
\subfigure[]
{\includegraphics[scale=0.7]{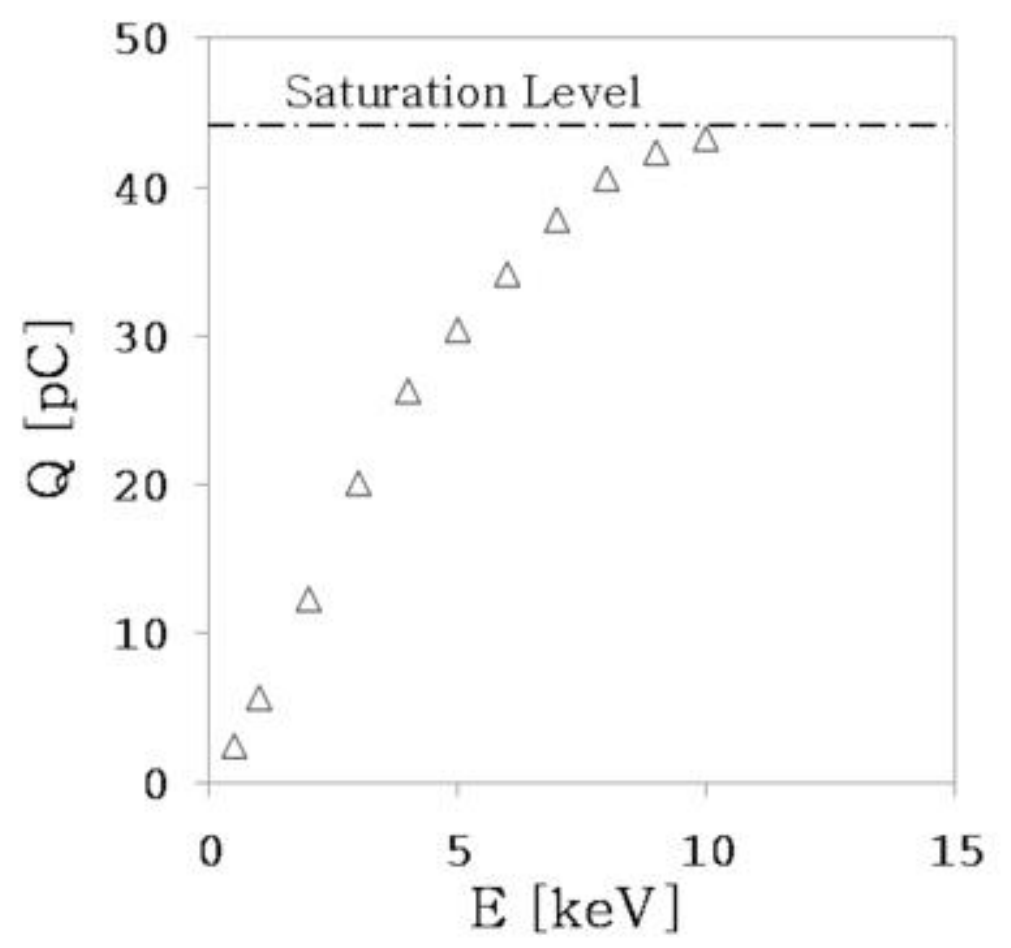}}
\hspace{5mm}
\subfigure[]
{\includegraphics[scale=0.7]{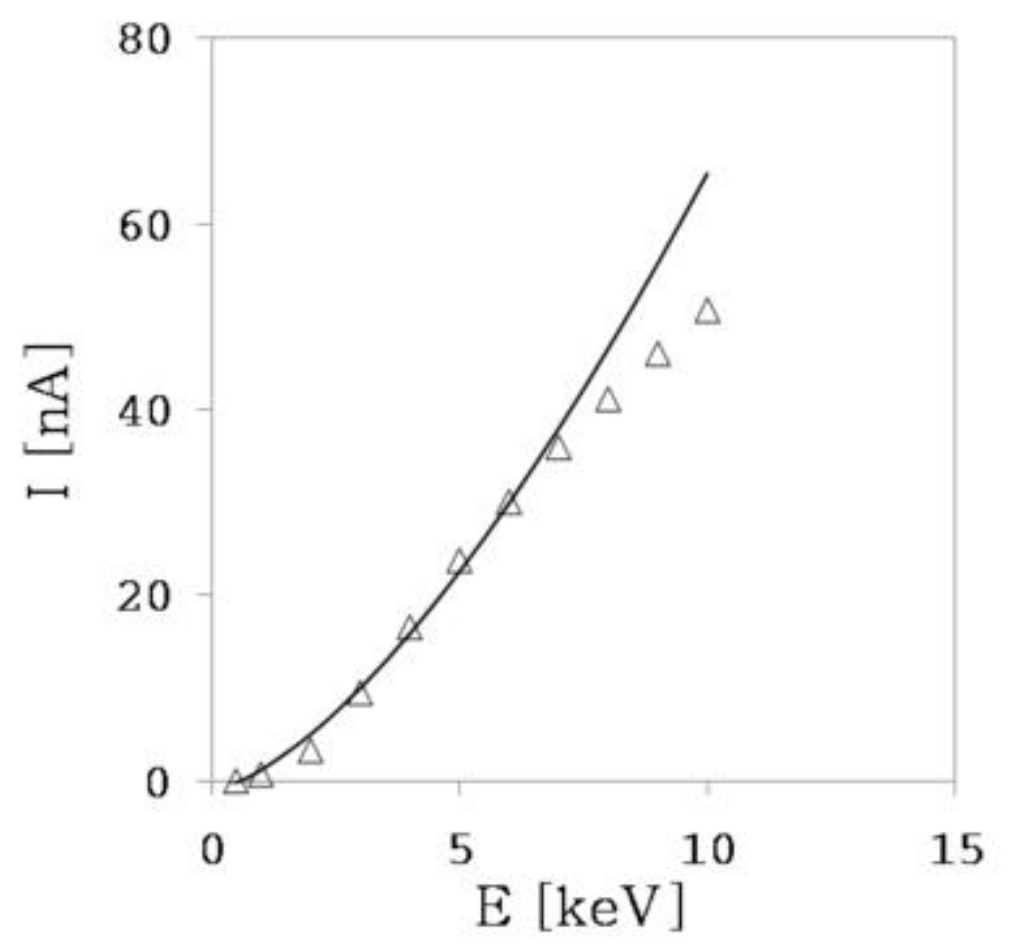}}
\caption{\label{chargemes}Charge (a) and current (b) measurements with electrostatic diagnostics in high impedance mode varying the bunch energy. The saturation level in charge emission was 43 pC and the data in (b) are fitted with the curve $I\propto V^{3/2}$ consistently with the Child-Langmuir law.}
\end{figure}

\begin{figure}
\centering
\subfigure[]
{\includegraphics[scale=0.7]{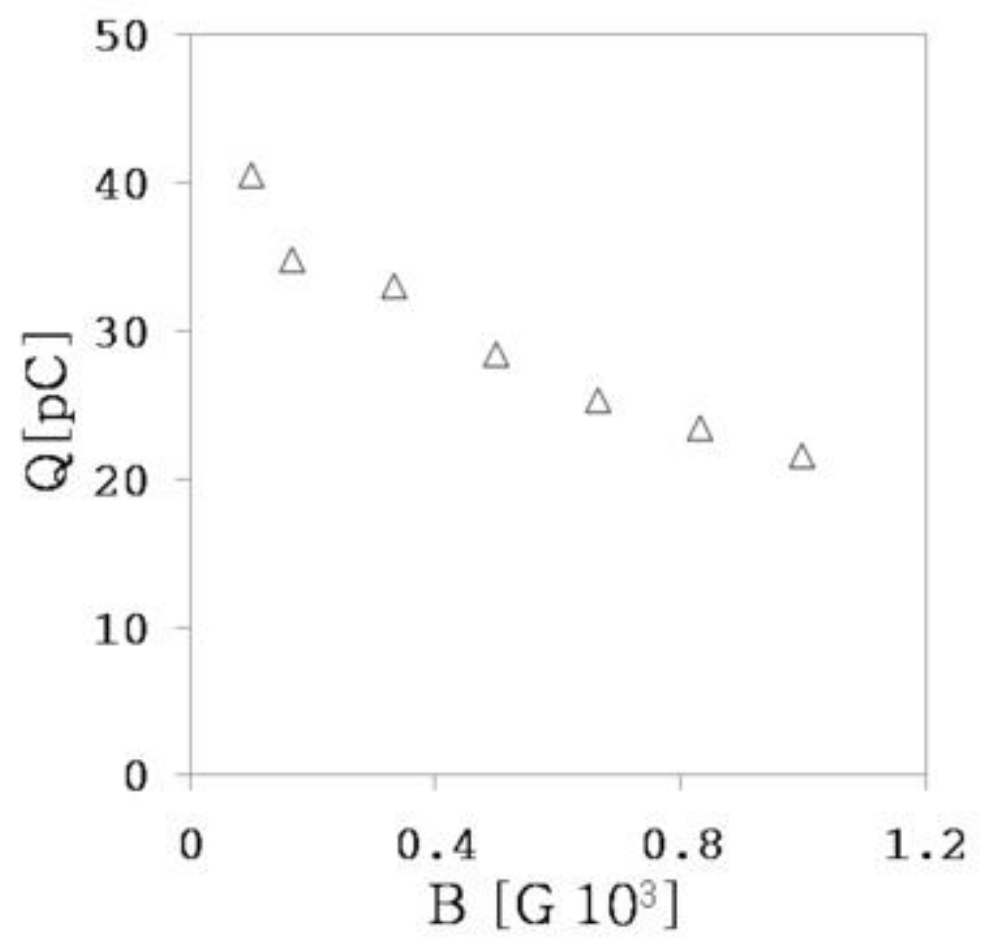}}
\hspace{5mm}
\subfigure[]
{\includegraphics[scale=0.7]{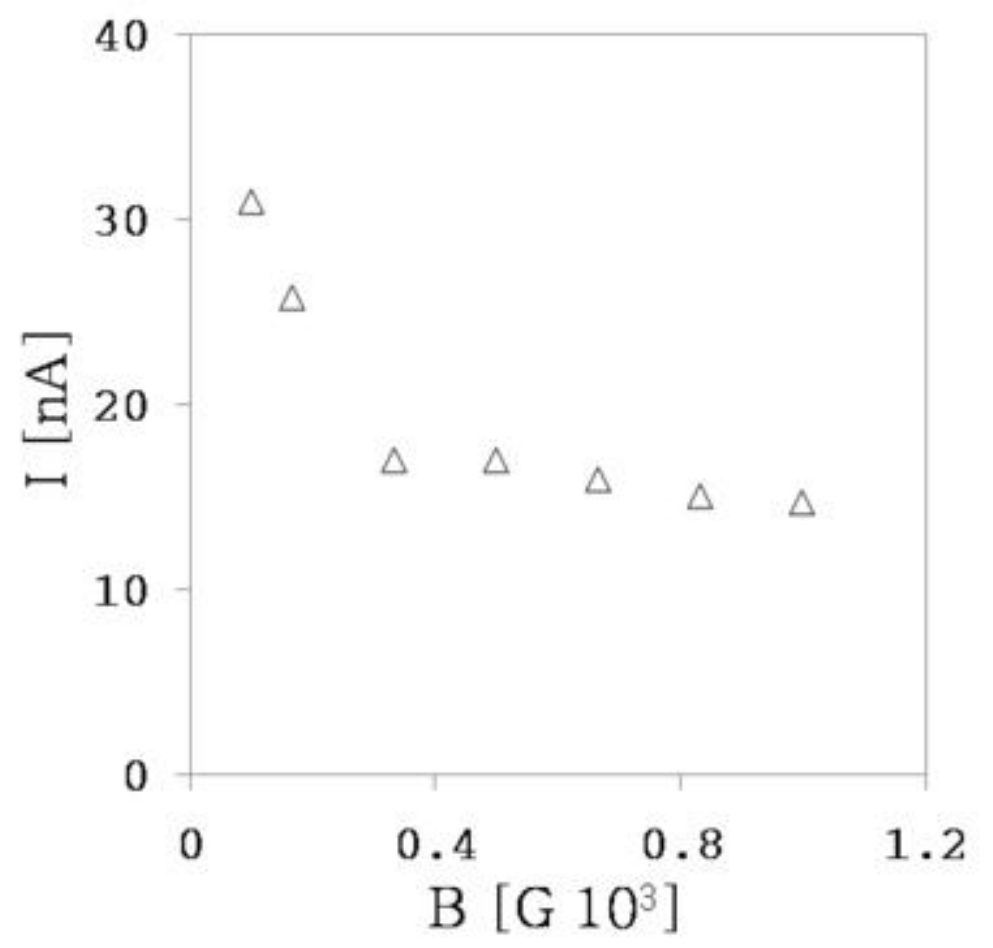}}
\caption{\label{magnmes}Charge (a) and current (b) measurements with electrostatic diagnostics in high impedance mode varying the magnetic field. Both charge and current decrease for higher values of the magnetic field. This effect is mostly due to the non-uniformity of the magnetic field from the source to the drift tube.}
\end{figure}

\begin{figure}
\centering
\subfigure[]
{\includegraphics[scale=0.7]{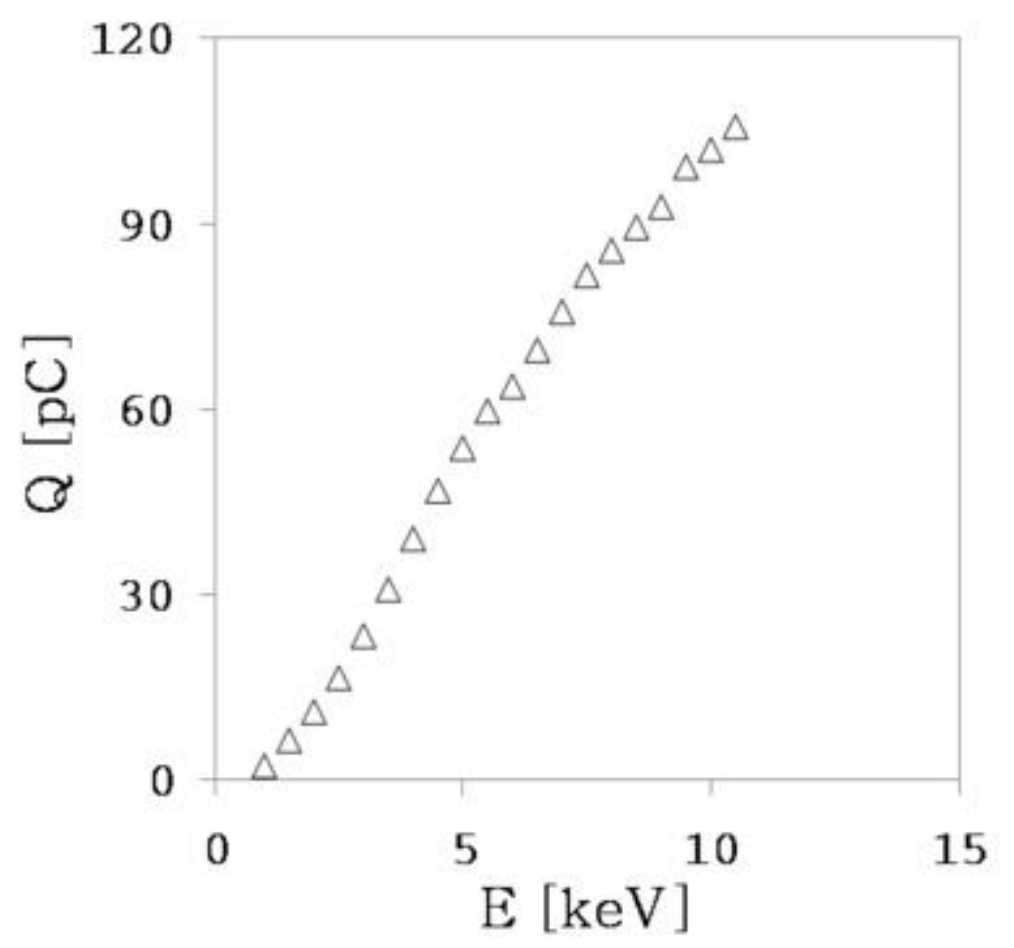}}
\hspace{5mm}
\subfigure[]
{\includegraphics[scale=0.7]{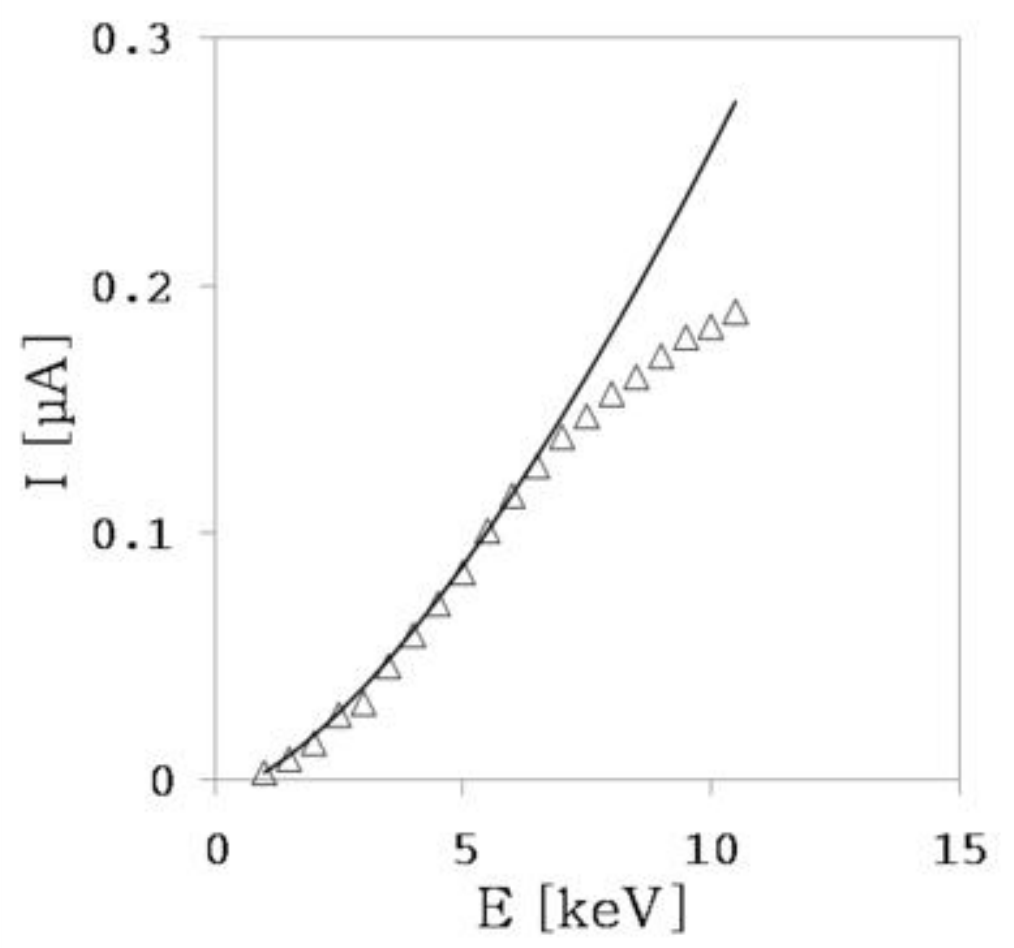}}
\caption{\label{curr1}Charge (a) and current (b) measurements with electrostatic diagnostics in high impedance mode varying the bunch energy and for a current source $I_S=1.30$ A. The measurements are obtained optimizing the laser and optics alignments.}
\end{figure}

\begin{figure}
\centering
\subfigure[]
{\includegraphics[scale=0.7]{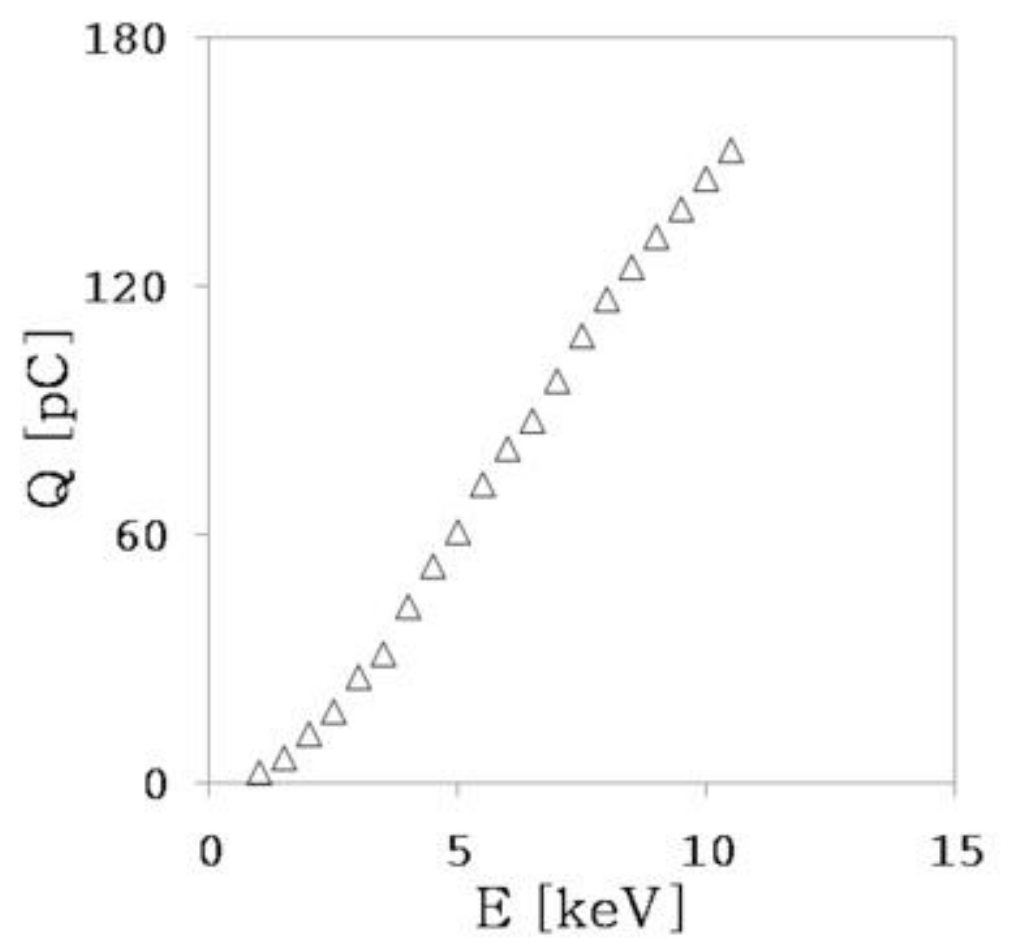}}
\hspace{5mm}
\subfigure[]
{\includegraphics[scale=0.7]{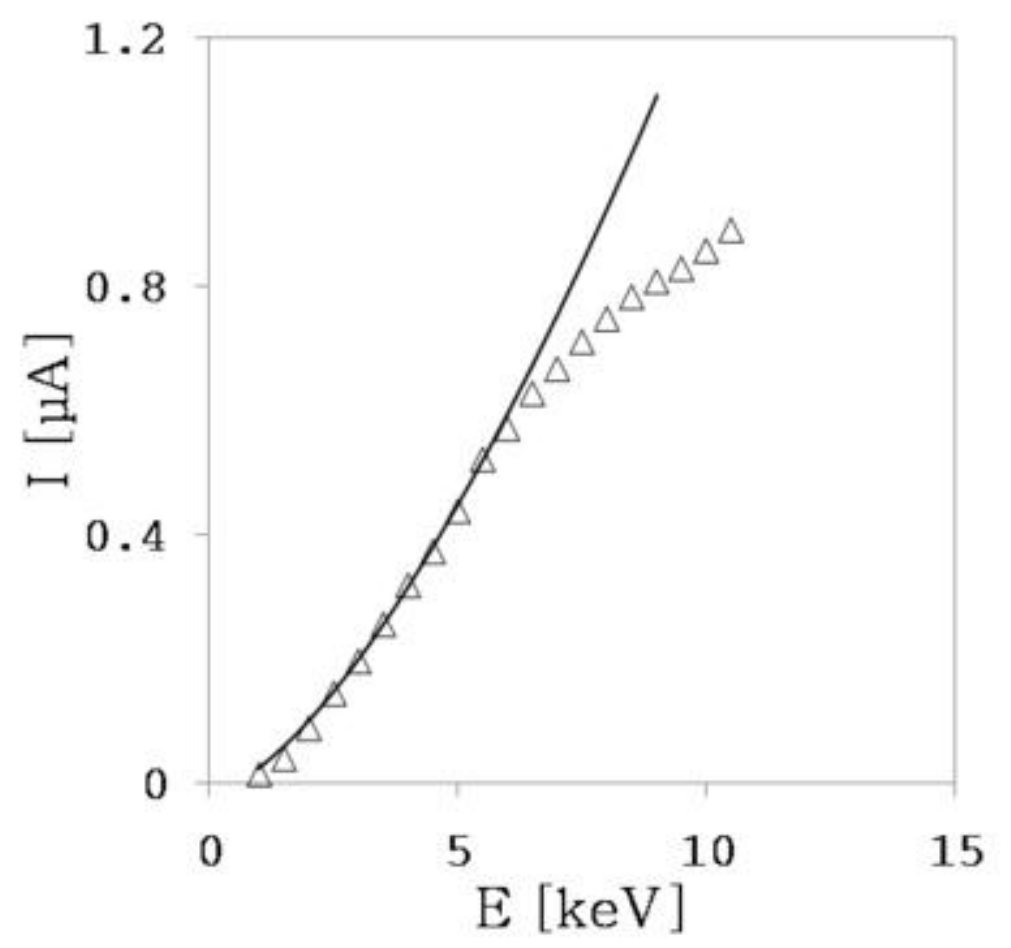}}
\caption{\label{curr2}Charge (a) and current (b) measurements with electrostatic diagnostics in high impedance mode varying the bunch energy and for a current source $I_S=1.65$ A. The measurements are obtained optimizing the laser and optics alignments.}
\end{figure}

The extracted current in pulsed or stationary beams is limited by the quantum efficiency of the source photoelectric emission expressed by:

\begin{equation}\label{quantumeff}
 q_e=\frac{Q/e}{E_L\,\eta/h\nu}
 \end{equation}

where Q is the emitted total charge, $E_L$ is the laser energy, $h$ is the Planck constant, $\nu$ is the laser frequency, $\eta$ is the total efficiency of the system, and
by space charge effects expressed in ideal conditions (two parallel infinite plates) by the Child-Langmuir law:

\begin{equation}\label{Langmuir}
 J_L=-\frac{4\epsilon_0}{9} \sqrt{\frac{2e}{m}} \frac{V^{3/2}}{d^2}
 \end{equation}

where $J_L$ is the emitted current density $\epsilon_0$ is the vacuum permettivity, V is the extraction potential, d is the cathode-anode distance, and by the saturation current density expressed by the Richardson law:

\begin{equation}\label{Richardson}
 J_R= A_0 T^2 \exp(-W/k_B T)
 \end{equation}

where $A_0=1.2\cdot 10^6 A\;m^{-2} K^{-2}$, $T$ is the cathode temperature, $W$ is the extraction potential of the metal, $k_B$ is the Boltzmann constant and $J_R$ is the current density. 
In these measurements we characterize these limitations for different values of the extraction potential 1 - 11 kV and magnetic filed 30 - 300 G in different experimental conditions and using a destructive electrostatic diagnostic in high impedance mode ($Z_L= 1$ $M\Omega$). The total charge is therefore given by equation (\ref{totalcharge}). The value of the capacitance $C_C$ is directly extracted fitting  the data of the RC discharge with the exponential $V_0[1-\exp(-t/RC)]$ (see fig. \ref{capacitance}). In figure \ref{chargemes} (a) we report the total charge measurements of bunches emitted heating the photocathode with a current of $I_s = 1.7 A$  and in a magnetic field of 330 G. The total charge reaches a saturation level of $\approx 45$ pC at higher energies, due to the complete electron emission in the photoelectric process (considering the limits imposed by the quantum efficiency and the optical and extraction efficiencies). In the same conditions we have measured the thermoionic emission current (see fig. \ref{chargemes} (b)). The data before the inflection at $\approx 7$ keV are fitted by a $J\propto V^{3/2}$ law, compatibly with the equation (\ref{Langmuir}). The fit is well overlapped with the data in the region 1 - 7 keV but then deviate from it, this is due to the maximum current density that can be extracted depending on the cathode temperature $T$ as described in equation (\ref{Richardson}). The total charge and current measurements varying the magnetic field between 30 and 300 G are reported in figure \ref{magnmes}. Note that increasing the magnetic field the total charge collected by the diagnostics is reduced, probably because the magnetic fields mismatch introduce particles lost from the source region, where the magnetic field is lower, to the trap region where the magnetic field reach the maximum uniformity and intensity. A different set of measurements were obtained optimizing the laser alignment on the source for two different values of $I_s$ and in a magnetic field of 330 G (see fig. \ref{curr1} and \ref{curr2}). The behavior remains substantially unchanged except an increase in both charge and current. The maximum value obtained was 153 pC at 10.5 keV.    

\subsection{Time width and length measurements }
\begin{figure}
\begin{center}
\includegraphics[scale=0.7]{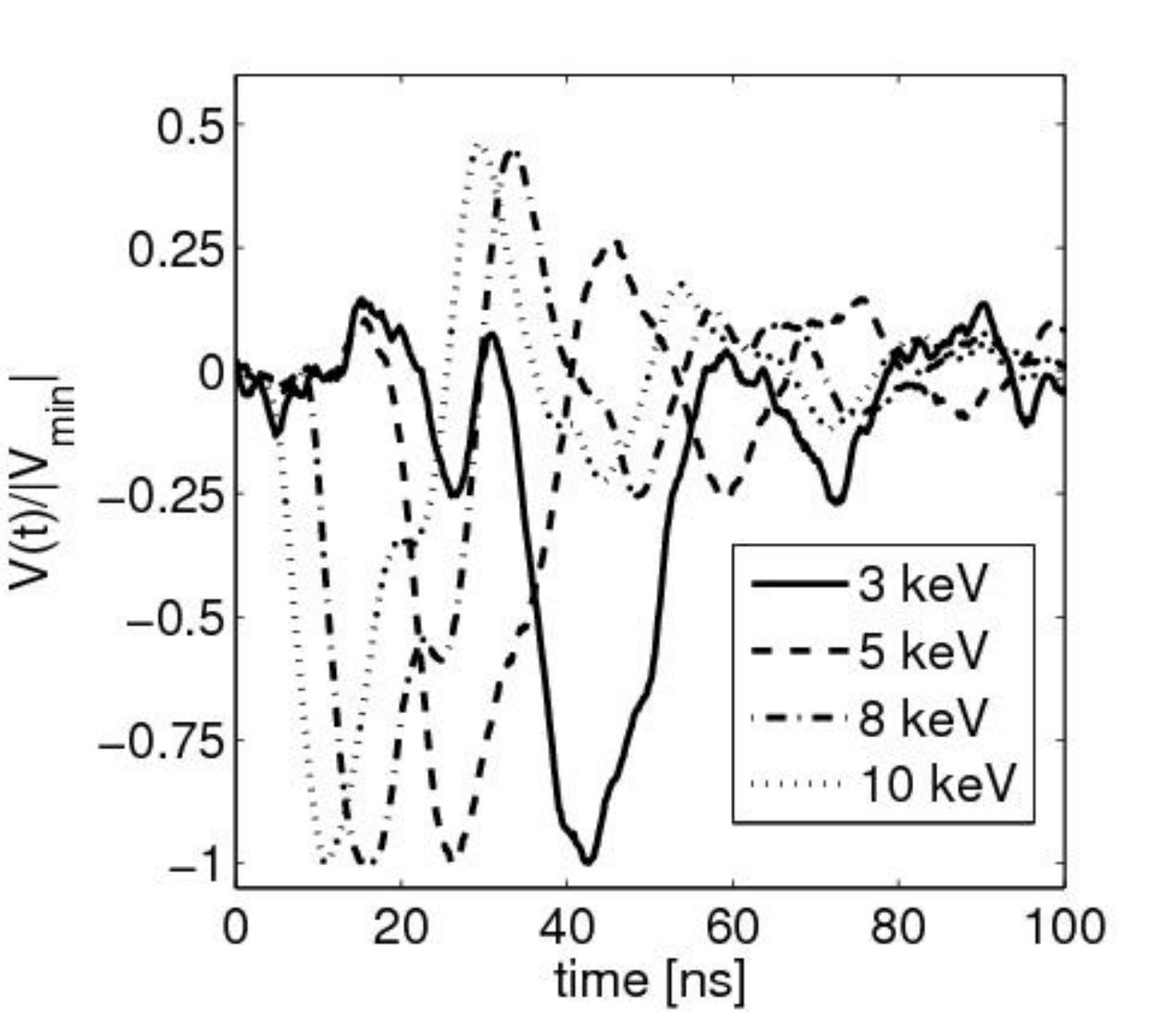}
\end{center}
\caption{\label{signals}Measured voltage signals for different beam energies. These signals have been normalized to the absolute value of the respective minima $V_{min}$.}
\end{figure}

\begin{figure}
\begin{center}
\includegraphics[scale=0.7]{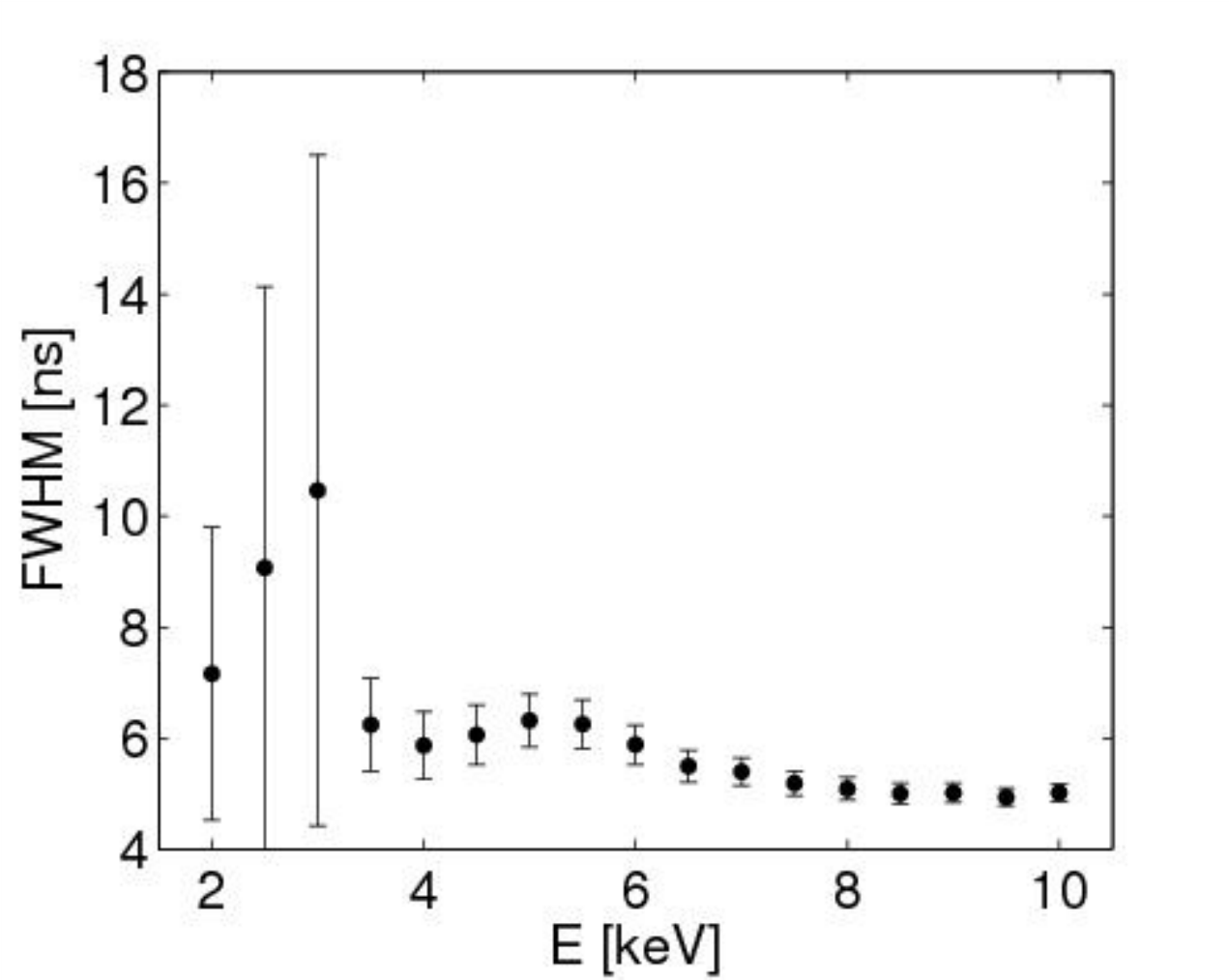}
\end{center}
\caption{\label{timewidth}Estimated time width of the Gaussian current pulse obtained from the curves of figure \ref{numerical} at different extraction energies.}
\end{figure}

\begin{figure}
\begin{center}
\includegraphics[scale=0.7]{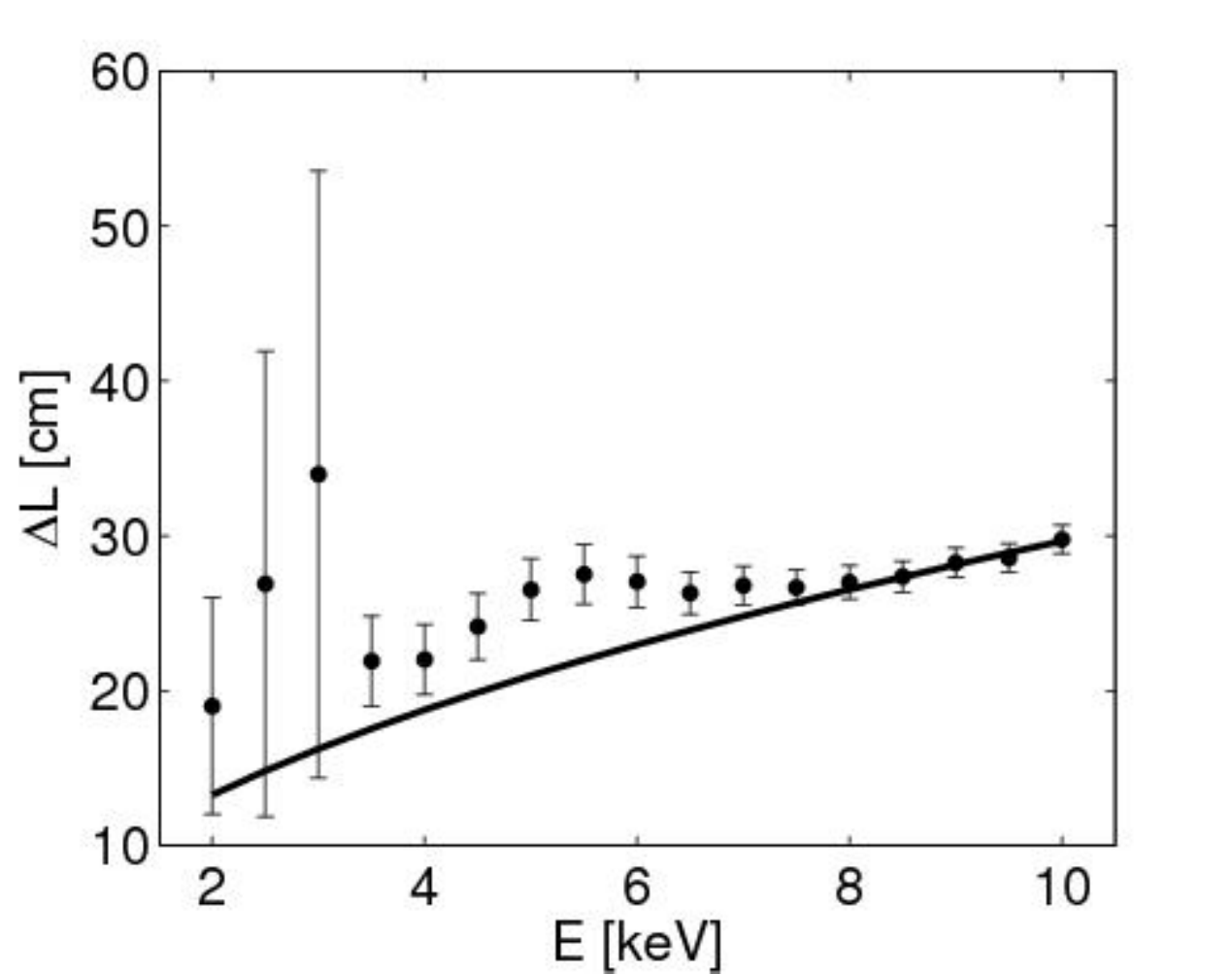}
\end{center}
\caption{\label{length}Electron pulse length estimated by the product $\Delta L=v_b FWHM$. The continuous curve is the theoretical $E^{1/2}$ trend for a constant FWHM.}
\end{figure}

\begin{figure}
\begin{center}
\includegraphics[scale=0.7]{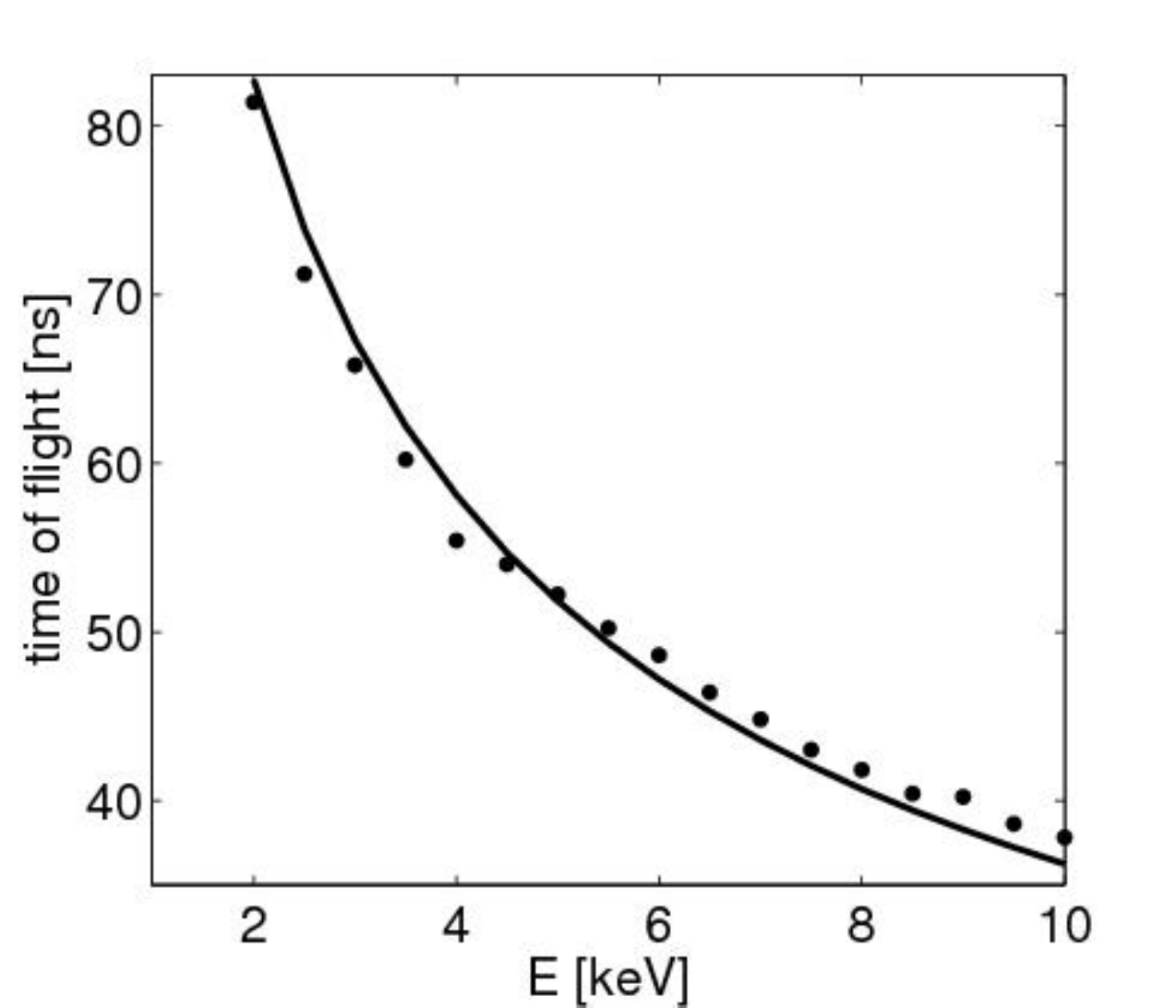}
\end{center}
\caption{\label{timeoff}Estimated time of flight of the electron bunch as a function of extraction energy $E$. The continuous curve represent a $E^{-1/2}$ fit. The delay between bunch emission and data acquisition is included.}
\end{figure}

The characteristic time width of the bunch after the flight through the trap is measured with the electrostatic diagnostics in low impedance mode $Z_L=50$ $\Omega$. This information is extracted in terms of FWHM using the curves obtained numerically (see fig. \ref{numerical} (b)), assuming a Gaussian axial density distribution of the bunch $g(z)$ with $\sigma_t=2.55$ ns and extracting the ratio $\zeta =V_{min}/V_{max}$ from the data. The measurements were obtained averaging on 50 signals for every extraction potential in the range 2 - 10 kV with a magnetic field of 330 G and an antenna's capacitance $C_A=300$ pF. In figure \ref{signals} we report the characteristic signals for 3, 5, 8 and 10 kV. The time width of the deconvoluted signals, rappresented in  figure \ref{timewidth}, are constant for higher energy due to the constancy of the laser pulse time width, but for lower energy the time width increase, highlighting the presence of a longitudinal spread of the bunch due to space charge effects. The error bars are calculated considering the electronic noise (4 m$V_{pp}$) and the statistical error, obtained averaging the signals on 50 samples. Note that error increase for lower energy i.e. for higher FWHM because an error in the known of $\zeta$ means an error in the time width measurement mainly for higher FWHM. The estimate of the bunch length (see fig. \ref{length}) was obtained considering the product $\Delta L=FWHM\, v_b$ with $v_b=\sqrt{2\, E/m_e}$ and where $E$ is the bunch energy. The assumption of motion with constant velocity $v_b\propto E^{1/2}$ used in the previous analysis can roughly be confirmed measuring the time of fligth $t_F$ versus the energy of the bunch and fitting the data with the function $t_F=a_0\, E^{1/2}+\, t_0$, as represented in figure \ref{timeoff}, where $a_0$, $t_0$ are free parametes. The acquisition was triggered on the optical trigger out of the UV laser with a jitter of $\approx 500$ ps. The measured $t_F$ is the time position of the principal maximum $V_{max}$ of the acquired signal. The measured lengths of the bunch are in the range 20 - 30 cm for energies between 4 - 10 keV. 

\subsection{One-dimensional fluid model}

In order to explain the longitudinal spread of the bunch for lower energy we analize its time width as a function of its energy with the analytic solution of a 1-dimensional fluid model. This approximation is assumed considering that the radial expansion is strongly prevented by the axial magnetic field. This method was used in a similar way \cite{Faltens} to study the longitudinal expansion of a $C_s^+$ beam with current in mA range and with an energy of $\approx120$ keV. The one-dimensional model consists of the equation of continuity, 

\begin{equation}\label{hydro1}
\frac{\partial \lambda}{\partial t}+\frac{\partial \lambda \, v}{\partial z}=0
 \end{equation}

the force equation,

\begin{equation}\label{hydro2}
 \frac{\partial v}{\partial t}+v\frac{\partial v}{\partial z}-\frac{e}{m_e}\frac{\partial \Phi}{\partial z}+\frac{1}{m_e \lambda}\frac{\partial P}{\partial z}=0
 \end{equation}

and the "long wavelength" field equation,

\begin{equation}\label{hydro5}
 \frac{\partial \Phi}{\partial z}= g \frac{\partial \lambda}{\partial z}
 \end{equation}

where $\lambda$ is the line charge density of a cylindrical electron beam of radius $R_p$ sorrounded by a cylindrical conductor of radius $R_W$, $v$ the electrons velocity and $\Phi$ is the electric potential. In equation \ref{hydro5} g is defined as

\begin{equation}\label{gfactor}
 g=\frac{1}{4\pi\epsilon _0}\left(\frac{1}{2}+2\ln\frac{R_W}{R_p}\right) \, .
 \end{equation}

\begin{figure}
\begin{center}
\includegraphics[scale=0.8]{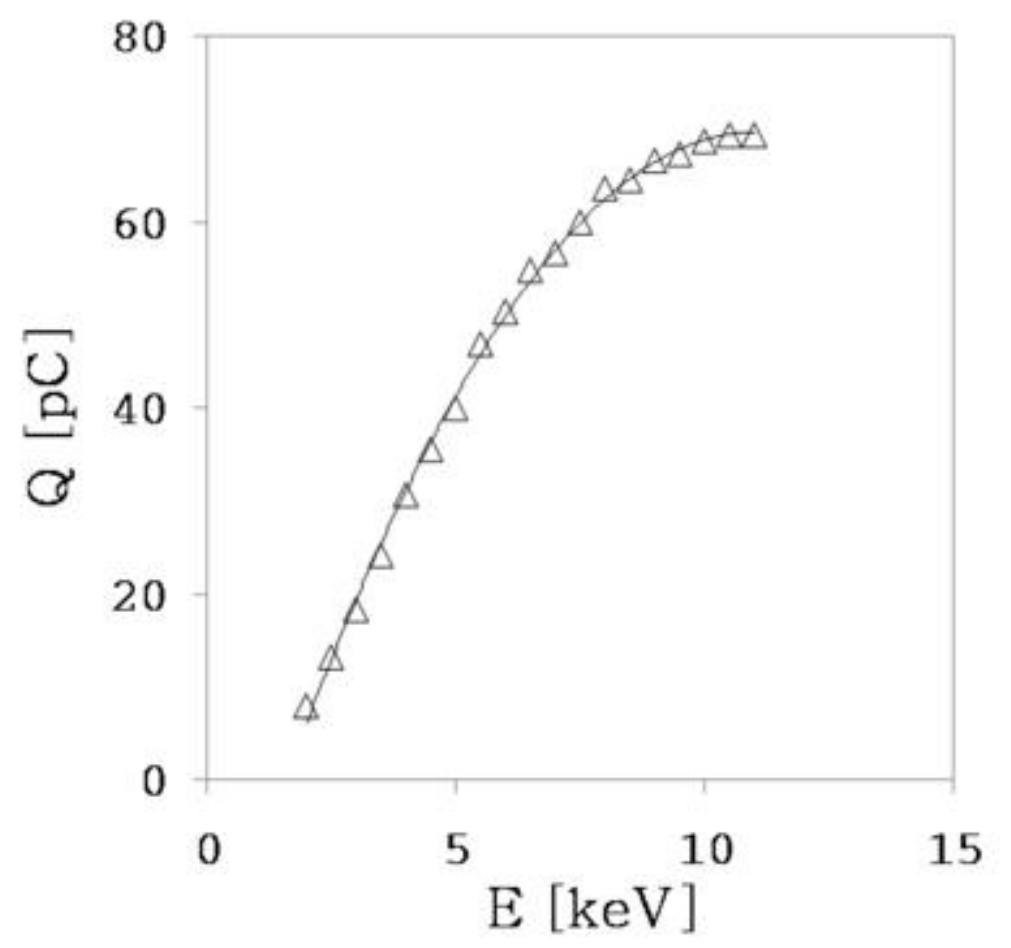}
\end{center}
\caption{\label{chargebunch}Charge measurements versus energy. The data are fitted with a polynomial of second order (full line). This polynomial is used to estimate the initial linear density $\lambda_0$ in the fluid model varying the bunch energy.}
\end{figure}

\begin{figure}
\begin{center}
\includegraphics[scale=0.8]{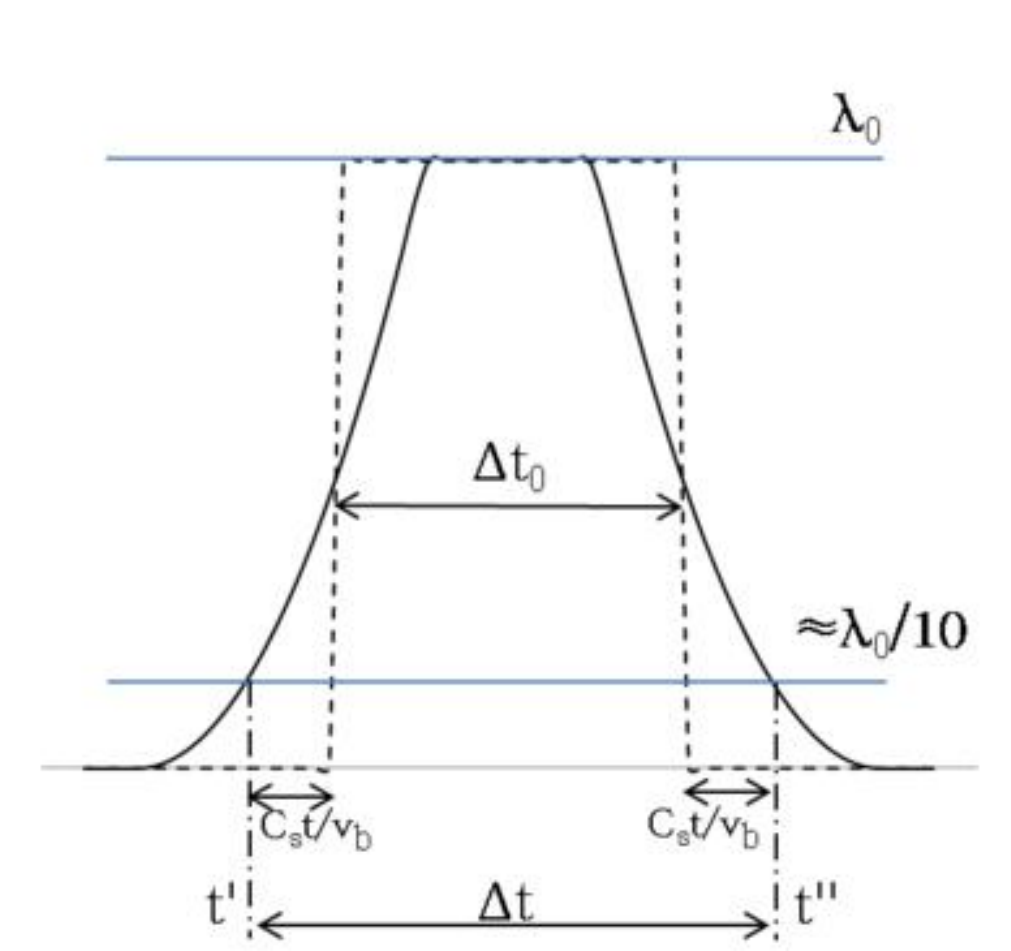}
\end{center}
\caption{\label{fluidprofile}Tipical shape of the linear density profile (full line) obtained with the solution (\ref{fluidsoldens}) on both sides of the initial square profile (dashed line) after the evolution $t>0$. The parameter used as time width is $\Delta t=\Delta t_0+2 c_s t/ v_b$ obtained measuring in a fixed point the linear density and rigidly traslating the profile through this point with velocity $v_b$. The points $t'$, $t''$ correspond with a density value of $\approx\lambda_0/10$.}
\end{figure}

\begin{figure}
\begin{center}
\includegraphics[scale=0.8]{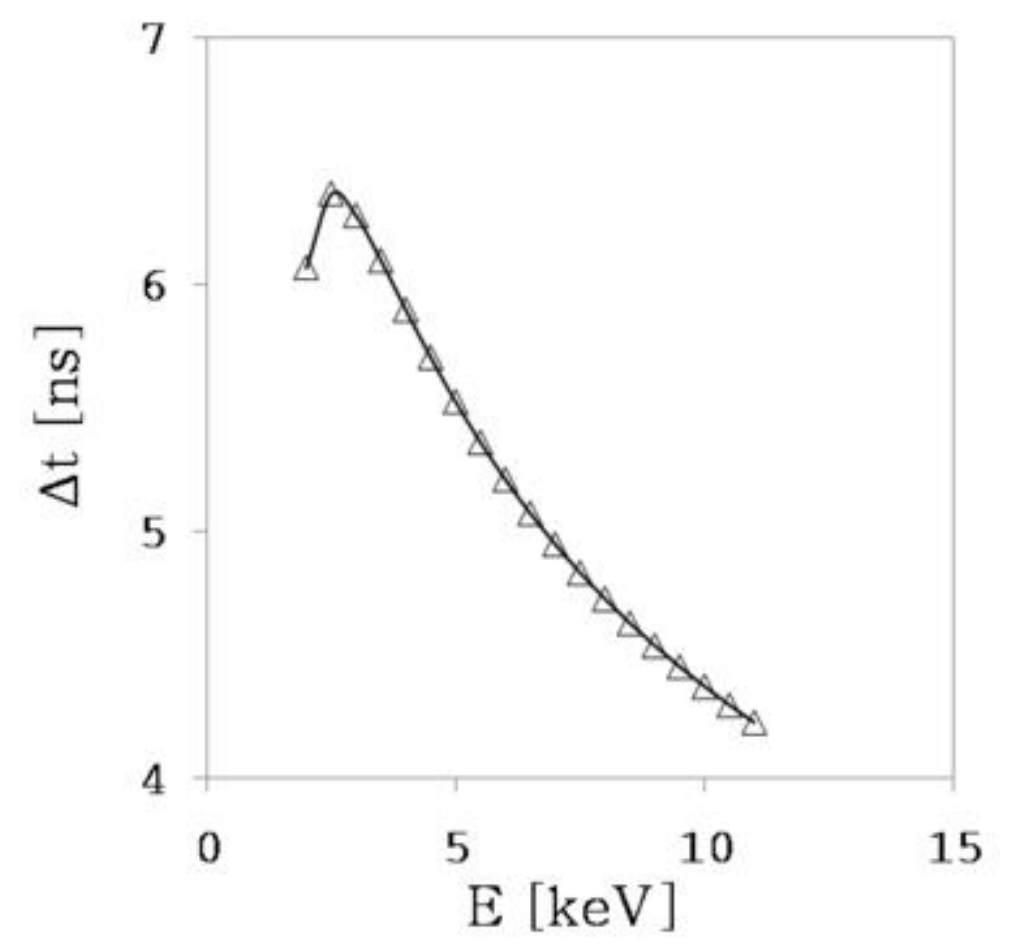}
\end{center}
\caption{\label{fitfluid}One-dimension fluid solution vs the bunch energy. The parameters are $R_p=0.5$ mm, $\Delta t_0=3$ ns, $R_W=4.5$ cm and $L_F=1$ m.}
\end{figure}

For a cold plasma the evolution of a initial uniform linear density profile $\lambda_0$ in the reference frame of the beam is 

\begin{equation}\label{fluidsoldens}
\frac{\lambda (z,t)}{\lambda_0}=
\left\{
\begin{array}{rl}
1 &\mbox{if} \: z<-c_st\\
\left(\frac{2}{3}-\frac{z}{3c_st}\right)^2 &\mbox{if} \: -c_st\leq z\leq 2c_st\\
0 &\mbox{if} \: z>2c_st
\end{array}
\right.
\end{equation}

where $c_s$ is the sound speed defined as $c_s=\sqrt{eg\lambda_0/m_e}$. The initial density profile depending on the energy is obtained measuring in high impedance mode the total charge $Q$ of the bunch (see fig. \ref{chargebunch}) and assuming that the radius of the bunch remains unchanged  for a constant magnetic field so $\lambda_0(E)=Q(E)/L_b(E)$, where $L_b(E)=\Delta t_0\sqrt{2E/m_e}$ and $\Delta t_0$ is the initial time width of the bunch.
The function $Q(E)$ is well approximated with a polynomial of second order expressed as
$Q(E)=aE^2+bE+c$ with the coefficients $a=-0.7781$ pC/keV$^2$ $b=17.176$ pC/keV $c=-25.054$ pC. For the time width variation of the bunch as a function of the energy $\Delta t=\Delta t(E)$ we use the assumption

\begin{equation}\label{deltat}
 \Delta t=\Delta t_0+2 c_s t/v_b
 \end{equation}
 
 with this definition $\Delta t$  is the time width of the bunch measured between the times $t'$, $t''$ such that the linear density profile is $\lambda(t')=\lambda(t'')\approx 0.1 \lambda_0$ (see fig. \ref{fluidprofile}) and neglecting the bunch evolution during the measurement. Substituting the sound speed $c_s$ and using the initial profile $\lambda_0$ in equation \ref{deltat} we obtain the solution

\begin{equation}\label{fluidsolution}
 \Delta t (E)=\Delta t_0+k\left(aE^{-1/2}+bE^{-3/2}+cE^{-5/2}\right)^{1/2}
 \end{equation}

where $k$ is a constant defined as

\begin{equation}\label{kappaconst}
 \left[\frac{L_F^2\: e\:m_e^{3/2}}{\sqrt{2}\:\Delta t_0\:4\pi \epsilon _0} \left(\frac{1}{2}+2\ln\frac{R_W}{R_p}\right)\right]^{1/2}
 \end{equation}

 where $L_F$ is the length of flight, $R_W$ is the radius of the drift tube, $R_p$ is the radius of the bunch and $\Delta t_0$ is the initial time width of the bunch. For typical experimental parameters $L_F=1$ m, $R_W=4.5$ cm, $R_p=0.5$ mm and $\Delta t_0=3$ ns we obtain the solution (\ref{fluidsolution}) as represented in figure \ref{fitfluid}. The time width $\Delta t$ of the bunch decreases of $\approx 2$ ns varing the bunch energy from 3 keV to 11 keV. So the bunch spread due to free expansion induced by space charge effects is qualitatively in agreement with the experimental measurements (see fig. \ref{timewidth}), i.e. few nanoseconds in a range of energy of 3 - 11 keV. We have to considered that a one-dimensional model is a great simplification of the real system and that $\Delta t$ is chosen arbitrarily, so this is a rough analysis of the longitudinal dynamic and more information about the longitudinal and transversal dynamic are needed.  

\subsection{Spread measurement versus the magnetic field}

The axial magnetic field of the trap acts focusing the bunch and limiting the transversal spread and distortion. The radial compression has as a consequence an increment of the brightness of the beam and therefore we expect a strong dependence of the axial spread on the magnetic field. 
The measured time width was obtained experimentally using the electrostatic diagnostics in low impedance mode (see fig. \ref{diagnostics}). The magnetic field was varied from 30 to 1000 G, for a constant extraction energy of 7 keV . The corresponding  total charge  was measured with the electrostatic diagnostics in high impedance mode (see fig. \ref{chargemagn}). The measurements show that for higher values of the magnetic field the longitudinal spread increases even if there is a reduction of the total charge from 35 pC at 200 G to 18 pC at 1000 G. This is probably due to the fact that for lower magnetic fields, a portion of the total charge is not efficiently focused around the center of the electron distribution so the resulting density of the bunch is lower and the bunch spread remain low.    

\begin{figure}
\begin{center}
\includegraphics[scale=0.8]{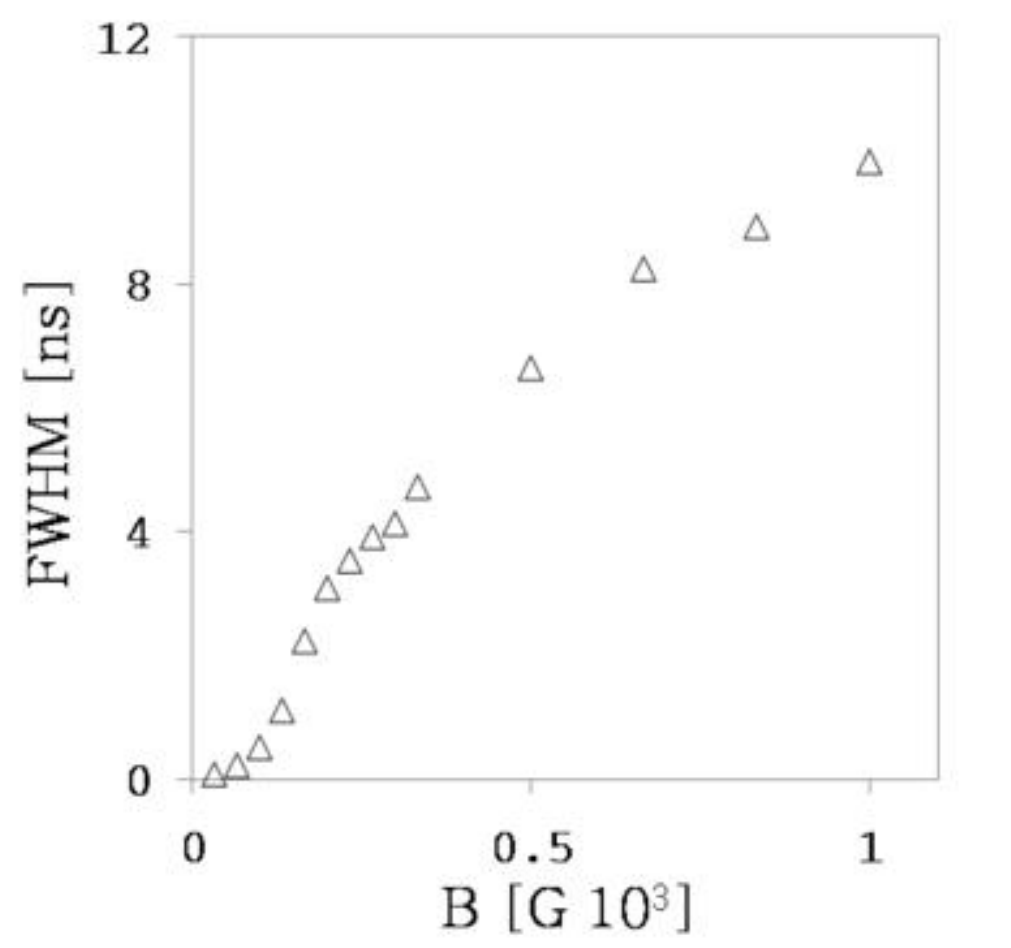}
\end{center}
\caption{\label{timemagn}Time width measurements vs magnetic field from 30 to 1000 G. Higher values of the FWHM for higher magnetic field are due to the radial focusing of the bunch.}
\end{figure}

\begin{figure}
\begin{center}
\includegraphics[scale=0.8]{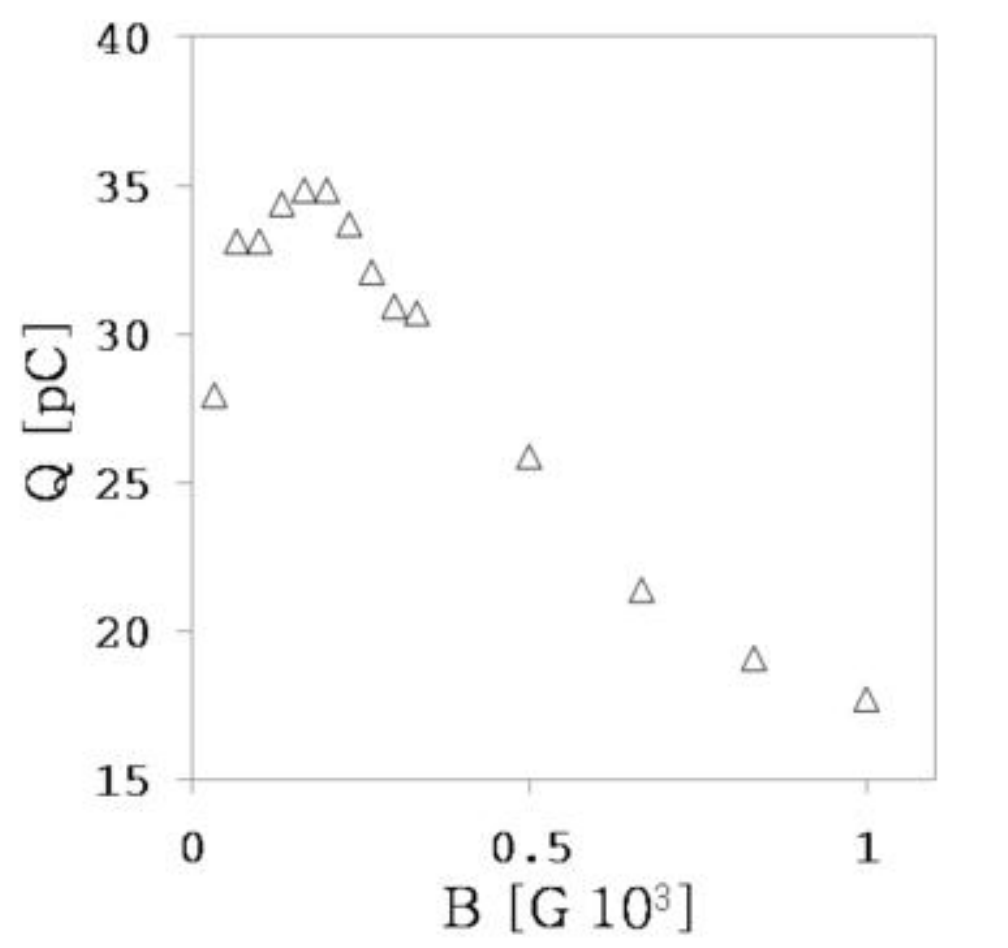}
\end{center}
\caption{\label{chargemagn}Total charge measurement versus the magnetic field obtained in the same experimental conditions as in figure \ref{timemagn}.}
\end{figure}

\section{Non-destructive electrostatic diagnostics}

Non-destructive electrostatic diagnostics is a useful tool to know the bunch properties, without perturbing the motion, reading the induced current produced by the passage of the bunch near to one or more electrodes.  The electrodes used for this diagnostics are the cylinders C1 - C8  and the cylinders  S2 and S4, sectored respectively in two and four patches.  The bunch passing  through these cylinders induce a measurable current without perturbing its dynamic appreciably. The information that can be obtained are the bunch length, spread velocity and time of flight, consistently with the main limitations of this diagnostics. A first limitation is the electrode's capacity $C_p$ that limits the bandwidth. In fact the produced current pulse is read as a voltage drop on a 50 $\Omega$ impedance and so the cut-off frequency is in general lower than $f_c=1/(2\pi\,50\Omega\,C_p)$. The minimum parasitic capacity of C1 - C8 can be estimated calculating the capacity between the base plane of the cylinders and the ground plane were they are mounted, $C_{min}=\epsilon_0 (S_e/d_e)$, where $d_e$ is the distance from the base to the ground plane, $S_e$ is the base surface and $\epsilon_0$ is the vacuum permittivity. Assuming $d_e=4.4$ mm, $S_e=72$ cm$^2$ we obtain $C_{min}\approx 14$ pF and the resulting cut-off frequency is 230 MHz, while the minimum required frequency, for a bunch time width of $\approx$ 4 - 5 ns, should be 200 - 250 MHz. The sectored electrodes have a greater bandwidth because the surface is smaller, but the amplitude of the induced signal is reduced by a factor $1/2$ for S2 and by a factor $1/4$ for S4. So the choice of the geometry is a compromise between sensitivity and response time. A second limitation is the electrode length that should be lower than the bunch length, otherwise small variations in the bunch axial density distribution are not measurable, because the induce charge depends on the charge contained in the cylindrical electrode and not on how it is spatially distributed. If the length of the electrode is comparable with the bunch length only an estimate of the rough properties of the bunch is possible.

\begin{figure}
\begin{center}
\includegraphics[scale=0.8]{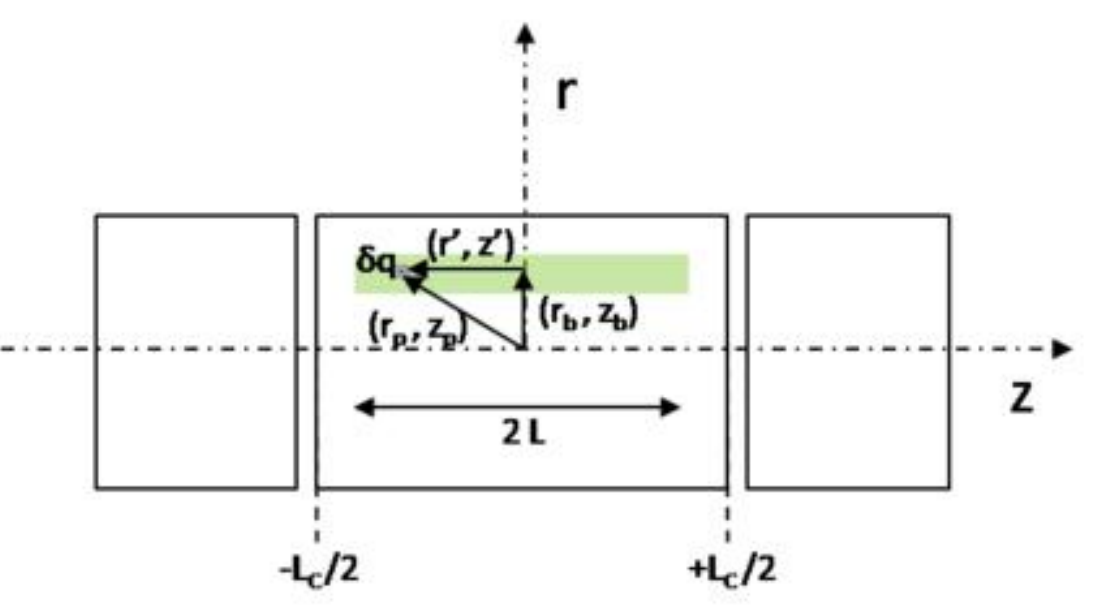}
\end{center}
\caption{\label{inducescheme}Scheme of the geometry used for the analysis of the induced charge. The charge element position is identified by the coordinates $(r_p,z_p)$ with respect to the axis of symmetry and by $(r',z')$ with respect to the point $(r_b,z_b)$ (center of the bunch). The angular coordinates are considered equal $\theta_p=\theta_b=\theta'$.}
\end{figure}
\subsection{Signal formation}
We assume a cylindrical electron  bunch of radius $R_b$ and length $L_b = 2 L$ with a uniform charge distribution $Q_b$ that moves rigidly  in a conducting cylinder of radius $R_W$ and length $L_C$, positioned at the center of other two coaxial  grounded cylinders. Defining $r_b$ the distance between the trajectory of the bunch and the cylindrical axis (considered parallel) and fixing a reference frame at the center of the floating cylinder,  as represented in figure \ref{inducescheme}, we can estimate the induced charge $\delta Q_{ind}$ for a small charge element $\delta q$ of the charge distribution, positioned  in $(\theta_p,r_p,z_p)$  with the Ramo theorem \cite{Ramo1}, that in simpler form is expressed as \cite{Ramo2} :

\begin{equation}\label{Ramo}
 \delta Q_{ind}=-\delta q \: \phi_p(\theta_p,r_p,z_p)
 \end{equation}

where $\phi_p$ is the weight electric potential that would exist at $\delta q$'s instantaneous position $(\theta_p(t),r_p(t),z_p(t))$ under the following  circumstances: the selected electrode is at unit potential, all other electrodes at zero potential, and all charges are removed. The electric potential is determined, considering that $\phi_p$ is independent by $\theta$ and solving the Laplace problem:

\begin{equation}\label{Laplace}
\left\{
\begin{array}{rl}
\nabla^2\phi_p=0 \\
\phi_p(R_W,z)=1 &\mbox{if} \: |z|\leq L_C/2\\
\phi_p(R_W,z)=0 &\mbox{if} \: |z|>L_C/2
\end{array}
\right.
\end{equation}

With the Fourier transform of the potential in the $z$ coordinate, $\widetilde{\phi}(r,k)=\int{e^{-i2\pi k z }\phi_p(r,z)dz}$ we obtain the modified Bessel equation:
 
\begin{equation}\label{besseleq}
 \frac{1}{r}\frac{\partial}{\partial r} r \frac{\partial}{\partial r}\widetilde{\phi}-4\pi^2k^2\widetilde{\phi}=0
 \end{equation}

with general solution:

\begin{equation}\label{solbessel}
 \widetilde{\phi}(r,k)=A(k)I_0(2\pi k r)+B(k) K_0(2\pi k r)
 \end{equation}

where $B(k)=0$ for the convergence of the solution, while $A(k)$ is determined by the boundary conditions espressed as $\phi_p(R_W,z)=\int{e^{i2\pi k z }c(k)dk}=1$ with:

\begin{equation}\label{coeffc}
c(k)=\int{e^{-i2\pi k z }\phi_p(Rw,z)dz}=\frac{1}{\pi k}\sin(\pi k L_c)
 \end{equation}

and from equation:

\begin{equation}\label{equalcoeffc}
c(k)=\widetilde{\phi}(R_W,k)=A(k)I_0(2\pi k R_W)
 \end{equation}

we obtain

\begin{equation}\label{coeffa}
A(k)=\frac{\sin(\pi k L_C)}{\pi k I_0(2\pi k R_W)} \: .
 \end{equation}

The solution for the weight electric potential is

\begin{equation}\label{potsol}
\phi_p(r,z)=\int{e^{i2\pi k z }\frac{\sin(\pi k L_C)}{\pi k I_0(2\pi k R_W)}I_0(2\pi k r)dk}
 \end{equation}

and from equation (\ref{Ramo}) we have:

\begin{equation}\label{qsolution}
\delta Q_{ind}=-\delta q\int{e^{i2\pi k z_p }\frac{\sin(\pi k L_C)}{\pi k I_0(2\pi k R_W)}I_0(2\pi k r_p)dk}
 \end{equation}

that can be rewritten as

\begin{equation}\label{qsolution2}
\delta Q_{ind}=-2\delta q\int_0^\infty{\frac{\sin(\pi k L_C)}{\pi k I_0(2\pi k R_W)}I_0(2\pi k r_p)\cos (2\pi k z_p)dk} \: .
 \end{equation}

The total induced charge of the entire bunch is given by the linear contribution of all elements of the bunch volume $\int{\delta Q_{ind}\:dV}$. The position of the charge elements can be expressed with respect to the reference frame of the bunch's center $r_b,z_b$ as:

\begin{equation}\label{Laplace}
\left\{
\begin{array}{l}
z_p=z_b+z' \\
r_p=\sqrt{r'^2+r_b^2-2r'r_b\cos(\pi-\theta')}
\end{array}
\right.
\end{equation}

where $r'$ $z'$ are the coordinate of $\delta q$ in the reference frame of the bunch's center. The total induced charge depending on the new coordinates $r_b$, $z_b$ is:

\begin{eqnarray}\label{inducedcharge}
Q_{ind}(r_b,z_b)=4\pi n e\int_{-L}^{L}{dz'} \int_0^{R_b}{r'dr'}\int_0^\infty{\frac{\sin(\pi k L_C)}{\pi k I_0(2\pi k R_W)}}\cdot\nonumber \\
\nonumber\\
I_0\left(k \sqrt{r'^2+r_b^2-2r'r_b\cos(\pi-\theta')}\right)\cos (2\pi k (z_b+z'))\,dk\quad\quad.
\end{eqnarray}

where $\delta q=n\,e\,dV$ is assumed, with a corresponding induced current $I_{ind}=dQ_{ind}/dt$. 


\subsection{Bunch length measurement}

Starting from equation (\ref{inducedcharge}) and considering $R_b << R_W$, $r_b=0$ we can rewrite the induced charge as:

\begin{equation}\label{simpleinduce}
 Q_{ind}(z_b)=2\pi R_b^2 n e\int_{-L}^{L}{dz'}\int_0^\infty{\frac{\sin(\pi k L_C)}{\pi k I_0(2\pi k R_W)}\cos (2\pi k (z_b+z'))dk}
 \end{equation}

where $I_0(2\pi k r')\approx 1$ is considered. Note that the dependence of the integral by $\theta'$ and $r'$ is removed. Defining the function 

\begin{equation}\label{gfunction}
 g(z';z_b,R_W,L_C)=2\int_0^\infty{\frac{\sin(\pi k L_C)}{\pi k I_0(2\pi k R_W)}\cos (2\pi k (z_b+z'))dk}
 \end{equation}

dependent on $z'$, the geometrical and bunch parameters $z_b$, $R_W$, $L_C$. Defining the linear density $\lambda_0=\pi R_b^2 n e$ the induced charge is expressed by the formula:

\begin{equation}\label{simplesol}
 Q_{ind}(z_b)=\lambda_0\int_{-L}^{L}{g(z';z_b,R_W,L_C) dz'}\quad.
 \end{equation}

\begin{figure}
\centering
\subfigure[]
{\includegraphics[scale=0.7]{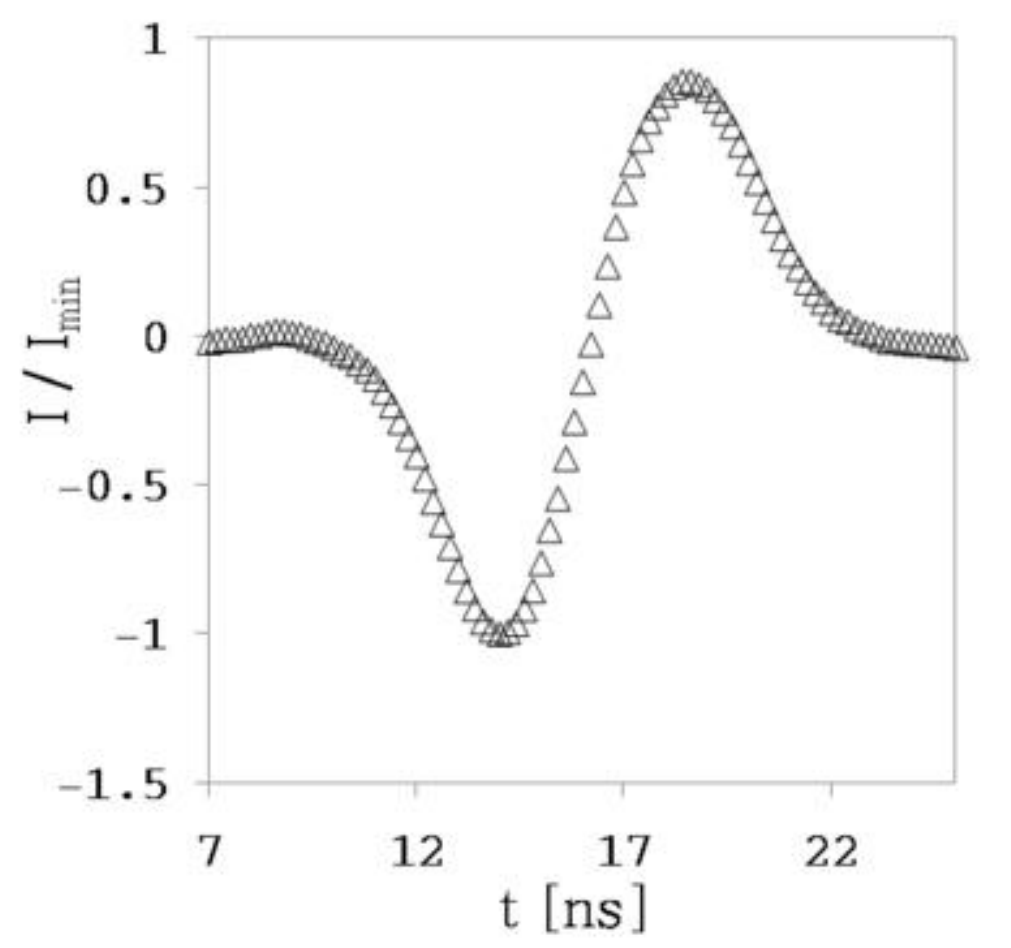}}
\hspace{5mm}
\subfigure[]
{\includegraphics[scale=0.7]{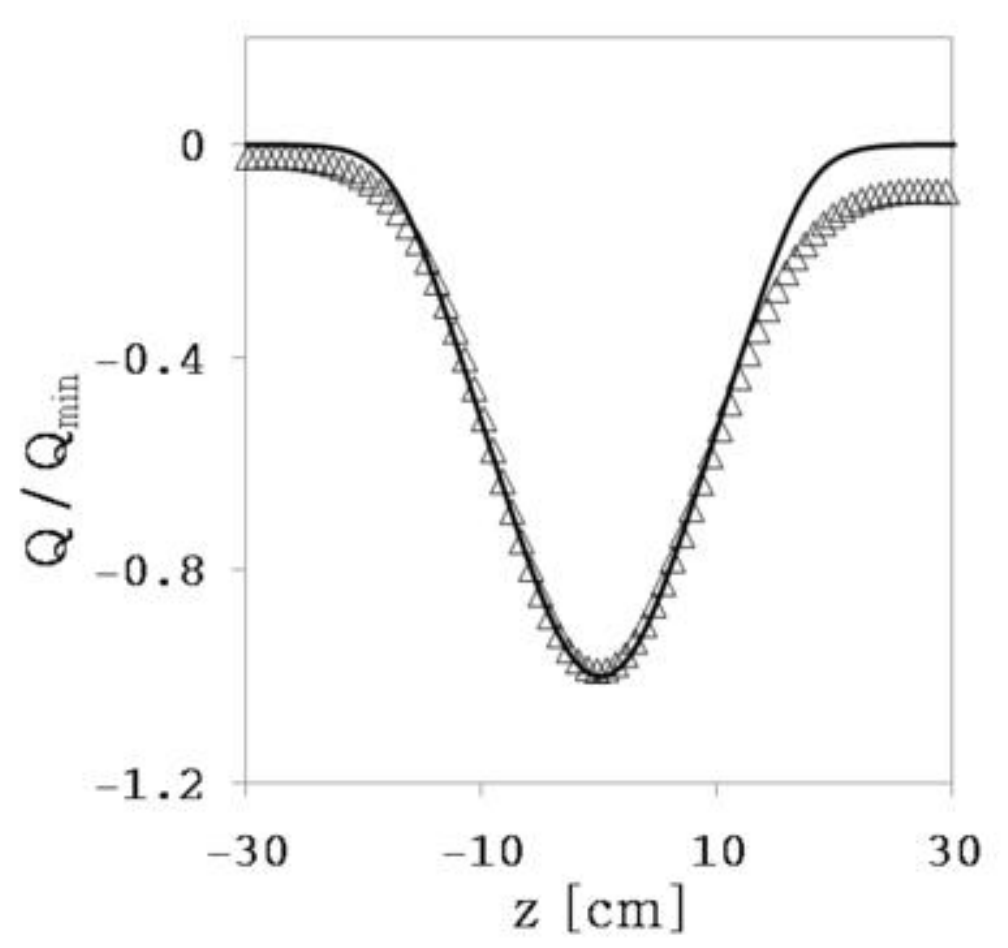}}
\caption{\label{inducedsignal}(a) Induced current signal measured on S4R for a bunch with an energy of 15 keV. The integral of this signal (b) is fitted with the function \ref{simplesol} to extract the bunch length $L_b=2L=20$ cm.}
\end{figure}

The parameters $R_W$ and $L_C$ are known (4.5 cm and 15 cm respectively), so we can fit a measured signal with the equation \ref{simplesol} leaving as free parameters the linear density $\lambda_0$ and the bunch length 2$L$. The estimate of $L_b$ with this technique was performed  for a bunch in a magnetic field of 330 G and with an energy of 15 keV in order to reduce the spread effects measured in the previous diagnostics and to get as close as possible to the assumption of rigid motion. The induced current  signal, represented in figure \ref{inducedsignal} (a) was measured with the electrode S4R in order to minimize the signal distortion introduced by the parasitic capacity. Note that the signal is simmetric i.e. the negative part is very similar to the positive part as predicted by equations \ref{inducedcharge} and \ref{simplesol} in ideal conditions. The integral of the measured induced current (see fig. \ref{inducedsignal} (b)) is fitted by the equation \ref{simplesol} and the extracted parameter is $L=10$ cm for a resulting bunch length of $\approx 20$ cm.  

\subsection{Asymmetries introduced by the electrode's capacitance}

Comparing the experimental signals of the same bunch measured  with different electrodes e.g. S4 and C5  we obtain two different shapes of the induced current integrated over time. The time-integrated signal measured with the electrode S4 is approximately  symmetric with respect the minimum, but this symmetry is broken if measured on C5, the signal is generally more smooth and the amplitude is lower. This effect is a distortion of the voltage signal $V_{out}$ measured on a load resistor $R$ introduced by the electrode capacity $C_e$ of the electrode C5. The proposed model of the equivalent circuit to study this effect is a current generator $i(t)$ and a parallel between the capacitance $C_e$ and the resistor $R$. From the Kirchhoff equation for the node we have:

\begin{equation}\label{nodi}
\frac{R}{\tau} i(t)=\frac{d}{dt}V_{out}(t)+\frac{1}{\tau}V_{out}(t)
 \end{equation}

where $\tau=R\,C_e$. Integrating equation (\ref{nodi}) we obtain

\begin{equation}\label{intnodi}
\int_{-\infty}^t{i(t')\,dt'}=\tau\frac{d}{dt}\int_{-\infty}^t{\frac{1}{R}V_{out}(t')\,dt'}\int_{-\infty}^t{\frac{1}{R}V_{out}(t')\,dt'}
 \end{equation}

This equation is written in terms of $Q_{out}=\int_{-\infty}^t{\frac{1}{R}V_{out}(t')\,dt'}$ and $Q_{in}=\int_{-\infty}^t{i(t')\,dt'}$ as:

\begin{equation}\label{diffnodi}
Q_{in}(t)=\tau \frac{d}{dt}Q_{out}(t)+Q_{out}(t)\quad.
 \end{equation}

Approximating the undistorted integrated current pulse as a Gaussian $Q_{in}(t)=\frac{1}{\sqrt{2\pi\sigma}}\,e^{\frac{-t^2}{2\sigma^2}}$ the output signal calculated from eq. \ref{diffnodi} is:

\begin{equation}\label{funcnofit}
Q_{out}(t)=\frac{1}{2\tau}\exp\left(-\frac{t}{\tau}-\frac{\sigma^2}{2\tau^2}\right)\left[1+Erf\left(\frac{t-\frac{\sigma^2}{\tau}}{\sigma\sqrt{2}}\right)\right]
 \end{equation}

This function can be used as a fit function adding an amplitude parameter A and the translation parameter $t_0$:  

\begin{equation}\label{fitfunc}
Q_{out}(t)=\frac{A}{2\tau}\exp\left(-\frac{t+t_0}{\tau}-\frac{\sigma^2}{2\tau^2}\right)\left[1+Erf\left(\frac{t+t_0-\frac{\sigma^2}{\tau}}{\sigma\sqrt{2}}\right)\right]
 \end{equation}

\begin{figure}
\begin{center}
\includegraphics[scale=0.8]{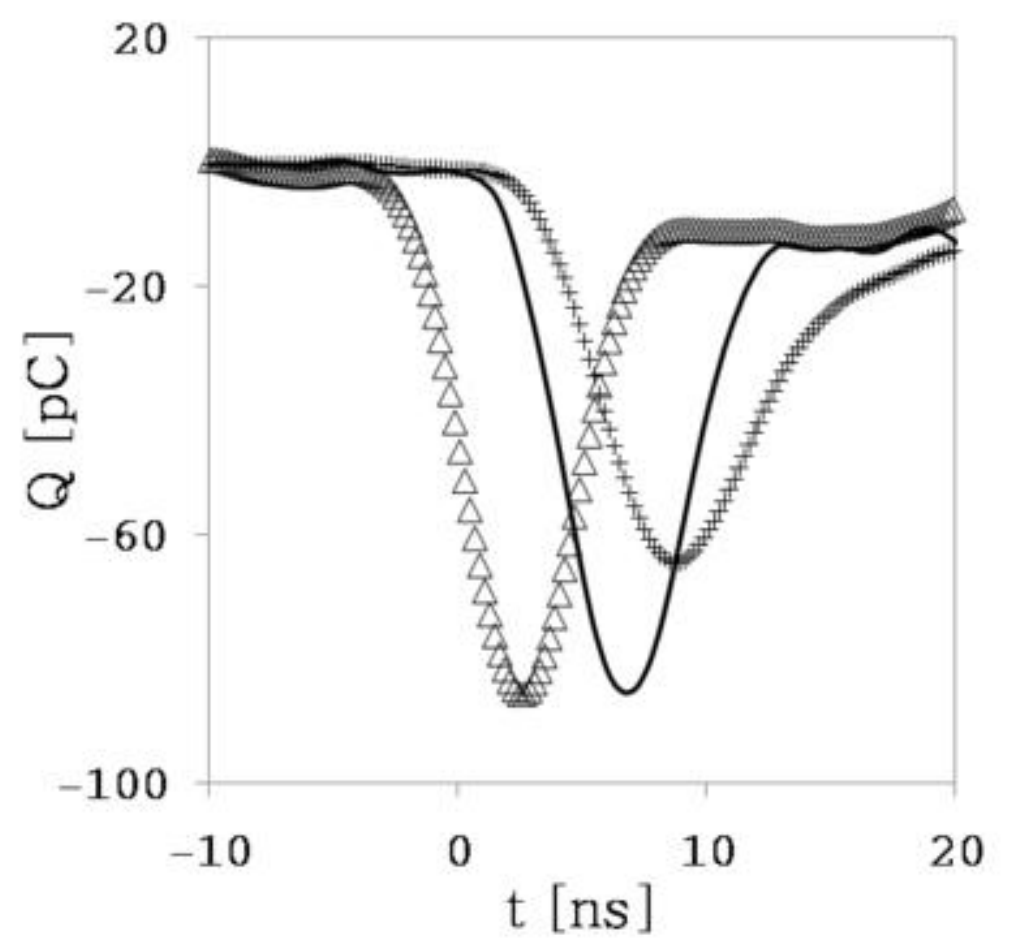}
\end{center}
\caption{\label{indcurrcap}Signals of the induced current measured on S4R (triangles) and C5 (crosses) integrated over time for a bunch with an energy of 15 keV. The full line is the signal obtained with the described deconvolution technique of the signal measured on C5. The resulting capacity of the cylinder C5 is $\approx$ 61 pF.}
\end{figure}

The deconvolution can be obtained with equation (\ref{diffnodi}) where $Q_{out}$ is the best fit with the function \ref{fitfunc} of the integrated measured signal and leaving as free parameters $A$, $t_0$, $\tau$. The calculated value $Q_{in}$ is the integral of the induced current pulse approximated as a Gaussian pulse. The integrated measured signals were obtained for a bunch in a magnetic field of 330 G and with an energy of 15 keV using the electrodes S4 and C5. With this bunch energy we minimize the energy spread due to the space charge, furthermore we assume that the distortion indroduced by the capacity of the sector S4 is negligible. The signal acquired with S4 is multiplied by a factor 4 to be compared with the signal measured on C5. In fact the induced charge on a sector of S4 is a quarter of that measured from an entire cylinder. The resulting signal obtained using the described deconvolution technique (full line in fig. \ref{indcurrcap}) on the distorted signal measured on C5 (signal represented with crosses in fig. \ref{indcurrcap}) is well compared with the undistorted signal measured on S4R (signal represented with triangles in fig. \ref{indcurrcap}). Because the signals measure on S4R and C5 are obtained from the same bunch and the only difference between this two measurements is the electrode capacity, (that is negligible for S4) we conclude that the distorted signal on C5 is due to its capacity. The resulting $\tau$ parameter is 3.04 ns. Considering a load resistor of $R=50$ $\Omega$ we obtain a capacity $C_e \approx 61$ pF of the electrode C5.

\subsection{Spread velocity measurements}

\begin{figure}
\begin{center}
\includegraphics[scale=0.8]{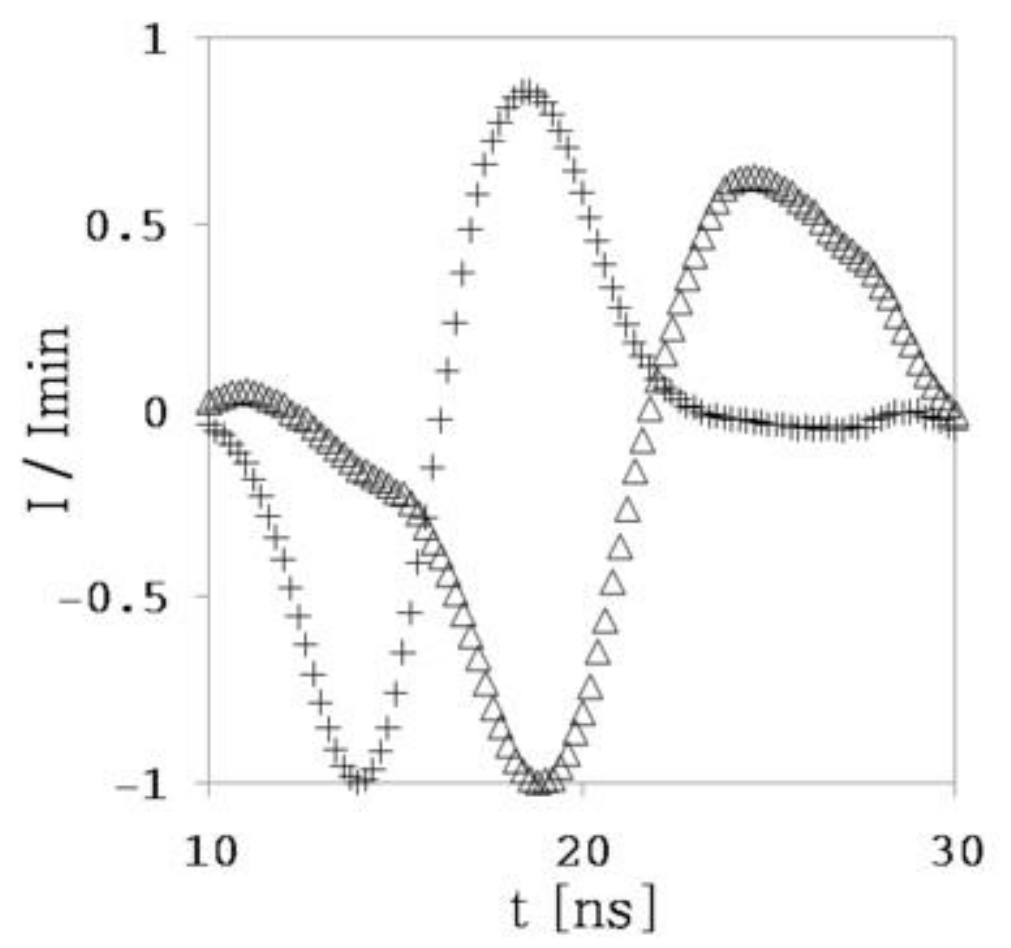}
\end{center}
\caption{\label{asyspacecharge}Induced current signals for bunch energies of 15 keV (crosses) and 10 keV (triangles) normalized to the minimum peak. The symmetry of the current signal is broken for lower energies.}
\end{figure}

Another effect of asymmetry in the signals measured on S4 occurs when the energy decrese from 15 keV to 8 keV as represented in figure \ref{asyspacecharge}. Since the expected time width of the bunch is higher decreasing the energy  (see fig. \ref{timewidth}), the cause of these asymmetries can not be attributed to the signal distortion introduced by the electrode's capacity (the 15 keV signal should be mainly distorted) but they are related to a real bunch distortion introduced by space charge effects. The relation between the spread velocity and the measured asymmetry was obtained starting from equation (\ref{qsolution2}). In the approximation $R_b << R_W$ and in analogy to eq. (\ref{simpleinduce}) the induced charge can be rewritten as:

\begin{equation}\label{qsolution2delta}
\delta Q_{ind}=-2\delta q\int_0^\infty{\frac{\sin(\pi k L_C)}{\pi k I_0(2\pi k R_W)}\cos (2\pi k z_p)dk}\:.
 \end{equation}
 
The density of the bunch is approximated with a cylindrical charge distribution with an axial density profile $g(z-v_bt,t)$ i.e. $n(r,z,t)=n_0 [1-H(r-R_b)]g(z-v_bt,t)$. This charge distribution moves with velocity $v_b$ along the z axis and the dependence of $g$ on $t$ shows that this motion is not generally rigid. The conservation of the initial total charge $Q_0$ implies that $\int{n(r,z,t)\,dV}=Q_0$ for every time. If the function $g$ satisfies $\int{g(z-v_bt,t)dz}=1$ the charge conservation require $n_0=Q_0/(\pi R_b^2)$. We introduce now an axial distortion of the bunch rappresented by a Gaussian that expandes in time

\begin{equation}\label{Gaussianexp}
g(z-v_bt,t)=\frac{1}{\sigma(t)\sqrt{2\pi}}\exp(-\frac{(z-v_bt)^2}{2\sigma(t)^2}) 
 \end{equation}

where $v_b$ is the bunch translation velocity and $\sigma(t)=\sigma_0+v't$, with $\sigma_0$ the bunch length parameter and $v'$ the spread velocity. Defining $\delta q=n(r,z,t)\,dV$ the total induced current at the time $t$ is $\frac{d}{dt}\int{\delta Q_{ind}\,dV}$ and so:

\begin{equation}\label{fitinducedfunc}
 I_{ind}(t)=2 Q_0\:\frac{d}{dt}\left[\int_{-\infty}^{\infty}{dz'\:g(z'-v_b t,t)}\int_0^\infty{\frac{\sin(\pi k L_C)}{\pi k I_0(2\pi k R_W)}\cos (2\pi k z')dk}\right]\: .
 \end{equation}

\begin{figure}
\begin{center}
\includegraphics[scale=0.8]{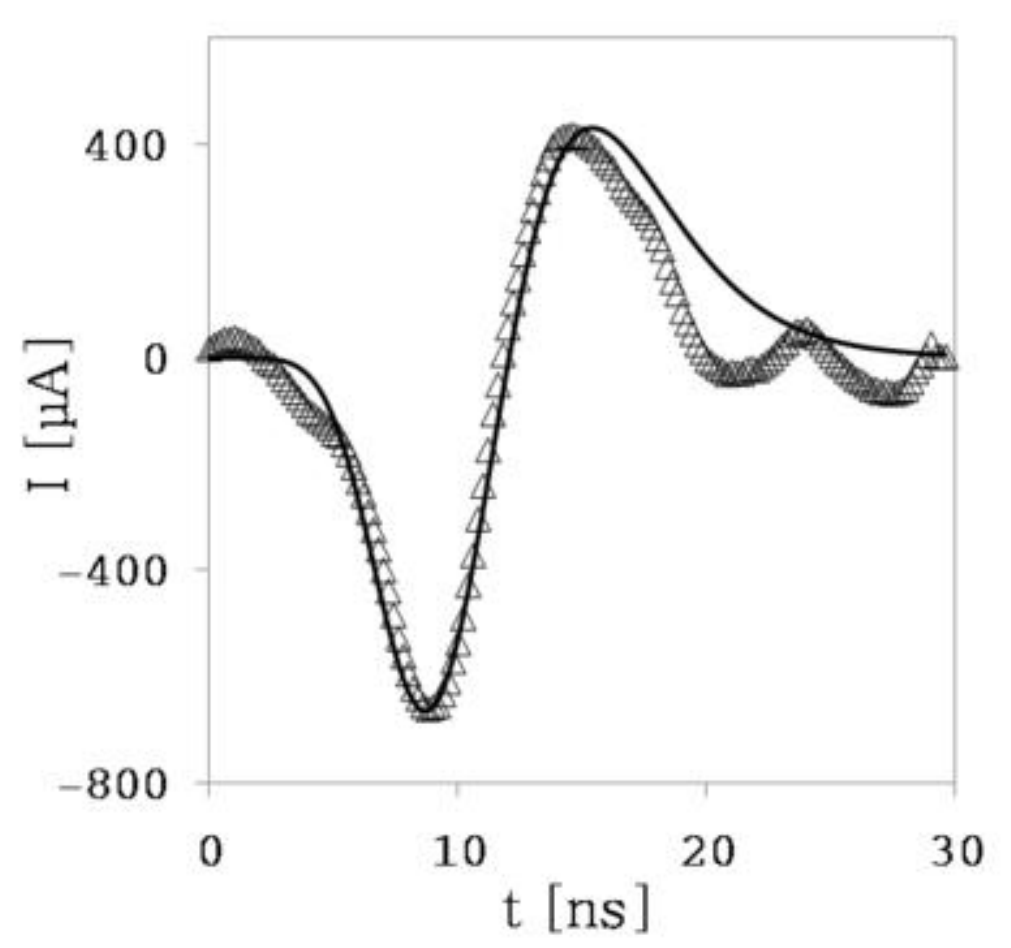}
\end{center}
\caption{\label{fitinducedris}Induced current signal measured for an energy bunch of 10 keV. The signal is fitted with the function \ref{fitinducedfunc}. Note that the asymmetry is well reproduced for a spread velocity of $v'=0.9\cdot 10^7$ m/s.}
\end{figure}

\begin{figure}
\begin{center}
\includegraphics[scale=0.8]{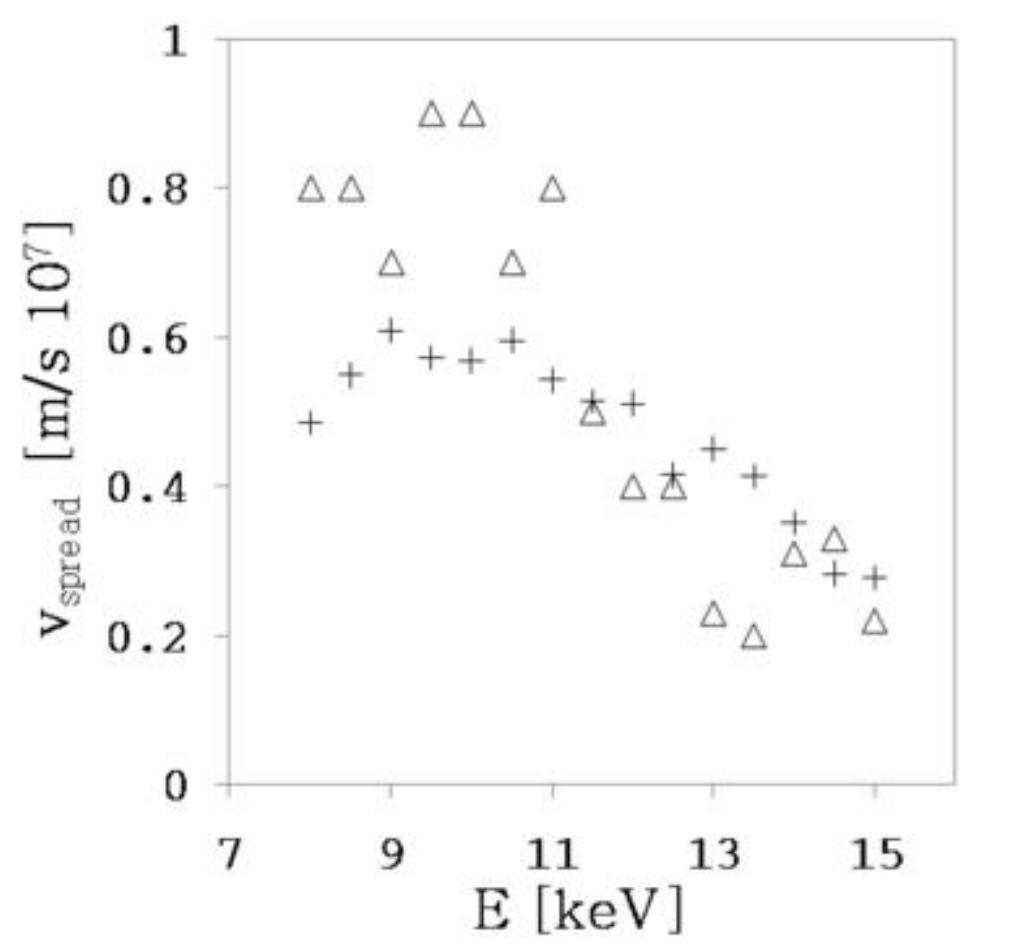}
\end{center}
\caption{\label{vspread}Spread velocity versus the bunch energies for an expanding Gaussian profile (triangles) and an expanding cylindrical uniform charge distribution (crosses). The velocities differ of $\approx$ 30\% in average.}
\end{figure}

This function can be used as a fit function fixing the parameters $R_W$, $L_C$, $v_b$ and leaving as free parameters $Q_0$, $\sigma_0$, $v'$. The measured signals were produced with bunches travelling in a magnetic field of 330 G and for energies of 8 - 15 keV. An example of the signal fitted with the function \ref{fitinducedfunc} is reported in figure \ref{fitinducedris}, note that the asymmetry is well reproduced. The resulting spread velocities from the fits were compared with the spread velocities obtained assuming a different density profile i.e. an expanding axially cylindrical uniform charge distribution $n_l(r,z,t)=n_0' [1-H(r-R_b)] g_l(z-v_bt,t)$  (see fig. \ref{vspread}) with velocity $v_l$ and initial length $L_0$. The function $g_l$ is defined as:

\begin{equation}\label{profi2}
g_l(z-v_bt,t)=H(z-v_bt+\frac{L_0+v_lt}{2})-H(z-v_bt-\frac{L_0+v_lt}{2})
 \end{equation}

The average variations of the spread velocities ($v'$, $v_l$) between these two different assumed density profiles are of $\approx$ 30\% and the order of magnitude of the spread velocity is $10^6$ - $10^7$ m/s for bunches with energies between 8 and 15 keV.

\subsection{Time of flight measurement}

\begin{figure}
\begin{center}
\includegraphics[scale=1]{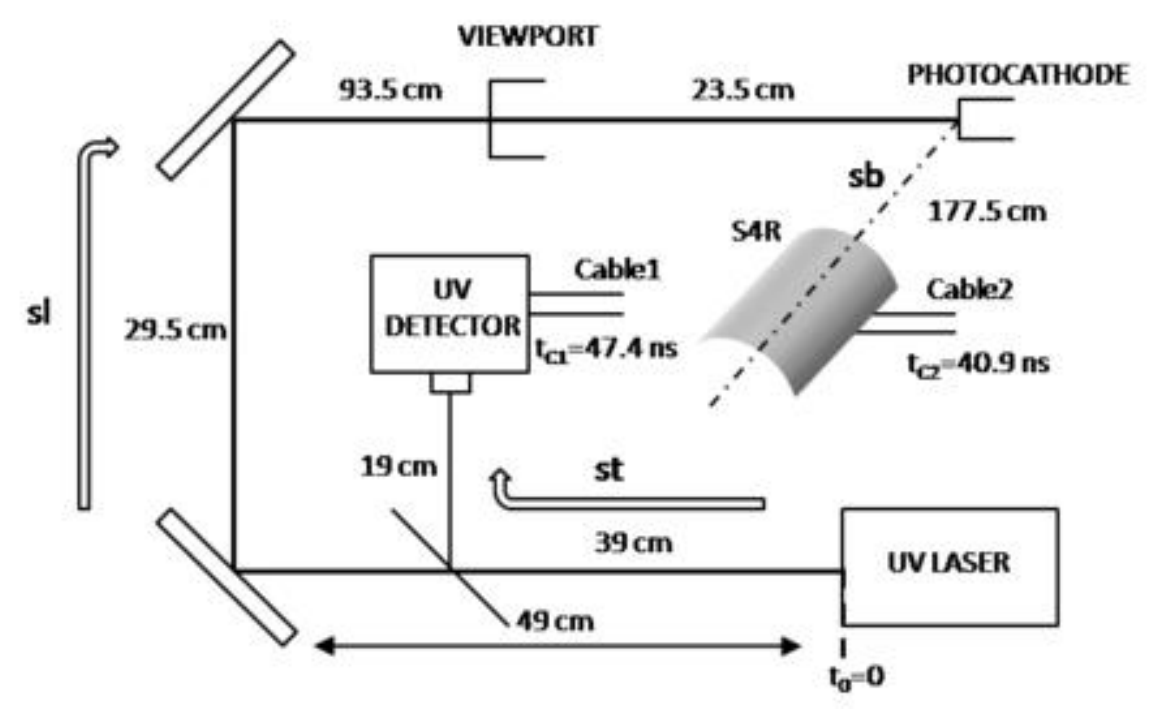}
\end{center}
\caption{\label{flightscheme}Scheme of the time of flight measurements. The bunch position is estimated knowing all delay times and distances. The main paths are $sl$ from the laser to the photocathode, $st$ from the laser to the detector and $sb$ from the photocathode to the center of the sector S4R.}
\end{figure}

\begin{figure}
\begin{center}
\includegraphics[scale=0.8]{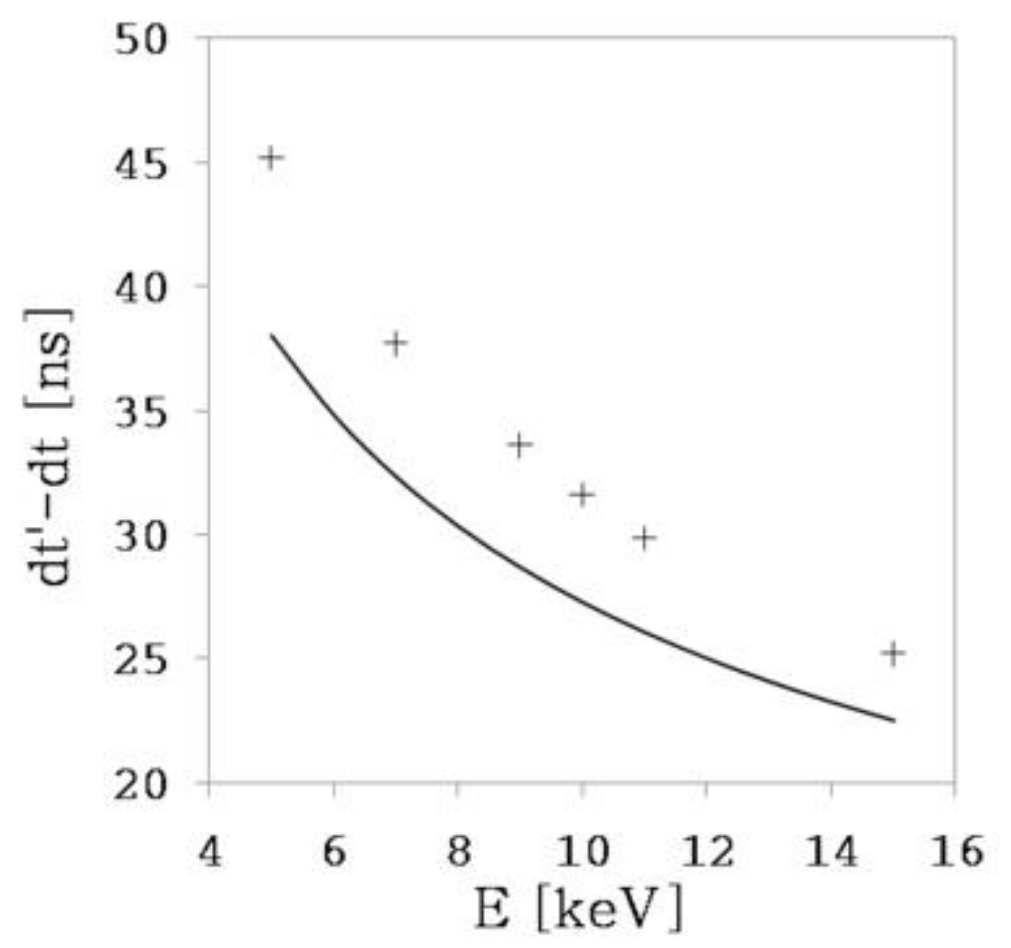}
\end{center}
\caption{\label{timeofflightabs}Time of flight measurements expressed as difference $dt'-dt$ in order to estimate the systematic error considering a uniform motion of the bunch. The error is $\approx$ 2.5 ns at 15 keV and increase for lower energies.}
\end{figure}

The uniform linear rigid motion of the bunch was verified in section 3.2.2, comparing the time of flight measurement with the fit formula $t_F=a_0 E^{-1/2}+t_0$. The initial time $t_0$ and the length of flight $a_0$ in this function were left as free parameters and relativistic effects are neglected. We want to verify the energy dependence of the time of flight including the experimental measurement of $t_0$ and $a_0$. This analysis is important to quantify the systematic error in the knowledge of the bunch position, approximating the bunch dynamics as a uniform rigid motion. The relativistic effect for a bunch with a maximum kinetic energy of 15 keV is negligible, in fact the time of flight differs about 2$\%$ from the non-relativistically computed time, i.e. $\approx 0.6$ ns for the characteristic drift time of the bunch in the trap. The measuring scheme is sketched in figure \ref{flightscheme}. A portion of the laser beam is split toward a UV detector with a low jitter, the other portion illuminates the photocathode after an optical path $sl=1.955$ m. The produced bunch is detected after a distance $sb=1.54$ m from the photocathode to the center of S4R. The delay times introduced by the cables are measured with the reflectometry technique describe in appendix A. Now we define the initial time $t_i=0$ as the time when the laser beam is at the output of the laser, the intervall $dt$ as the time between $t_i$ and the measured signal coming from the UV detector and the interval $dt'$ as the time between $t_i$ and the detected pulse coming from the electrode S4R. The difference $dt'-dt$ is calculated as:

\begin{equation}\label{profi2}
dt'-dt=\frac{sl}{c}+ \frac{sb}{\sqrt{2E/m_e}}+\frac{t_{C2}}{2}-\left(\frac{st}{c}+\frac{t_{C1}}{2}\right)   
 \end{equation}

where $st=58$ cm is the optical path from the laser output to the detector, $t_{C1}=47.4$ ns is the reflectometry measurement of the detector's cable and $t_{C2}=40.9$ ns is the reflectometry measurement of the S4R's cable. The theoretical $dt'-dt$ is compared with the measured value (see fig. \ref{timeofflightabs}) of a bunch travelling in a magnetic field of 330 Gauss and with energies between 5 -15 keV. A systematic error of about 2.5 ns was obtained for higher energies and become 4 ns at 8 keV. This discrepancy is reasonable considering that the space charge effects limit the efficiency of extraction, and the bunch injection is more complex of a single particle acceleration. Moreover some systematic errors are introduced in the reflectometry measurements mainly in the measurement of $t_{C2}$, because the electrode's capacity introduce a distortion in the measured signal.  

\section{Optical diagnostics}
An optical diagnostics based on a phosphor screen coupled with a CCD digital camera was used to characterize the transversal charge distribution of the bunch. The phosphors (P43 type) are deposited on a glass disc with a diameter of 11 cm covered by alluminium and the screen is positioned at  the end of the cylindrical stack. The CCD digital camera is a Hamamatzu C8484-05G with a resolution of 1.3 million pixels a wide dynamical range of 12 bit and a readout noise of 10 electrons r.m.s. The screen is polarized with a positive high voltage 1 - 15 keV used to additional accelerate bunches with lower energies. The image is formed by the impact of the electron bunch on the screen and the persistence of the phosphors $\approx 1$ ms permits the capture of the image controlled by an external trigger. The result is a charge distribution of the bunch integrated along the z axis. The corresponding data were elaborated with a C code in order to extract information about the shape and the size of the bunch spot. In particular an algorithm to search the maximum profile $n(i,j_{max})$ of the charge distribution $n(x,y)$ was used, where $j_{max}$ is the value of $j$ by which is maximum the function  $M_p(j)=\sum_i n(i,j)$ and $i$, $j$ are the quantization of $x$ and $y$ respectively. A complementary maximum profile $n(i_{max},j)$ can be extracted with the function $ M_p(i)=\sum_j n(i,j)$. For a symmetric charge distribution i.e. depending only on $r=\sqrt{(x-x_0)^2+(y-y_0)^2}$ and not by $\theta=\arctan \frac{y-y_0}{x-x_0}$, where $(x_0,y_0)$ is the centre of symmetry, we can define the density profile $\rho_c(r)$ as  

\begin{equation}\label{profconserv}
\rho_c(r)=-r\int_0^{2\pi}{e\,n(r,\theta) d\theta}\:
 \end{equation}

This profile is the charge per unit radius of the transversal charge distribution and satisfies the property $\int {\rho_c(r)\,dr}=Q $, where $Q$ is the total charge of the distribution. The graphic representation of this profile is useful because the fraction of the total charge in a region $r_a<r<r_b$ is simply the area under the graph in that region. Numerically the $\rho_c$ profile is calculated using

\begin{equation}\label{profconservn}
\rho_c(k)=\sum_{i,j\,:\,k<\,\sqrt{(i-i_0)^2+(j-j_0)^2}\leq\,k+1} {-e\,n(i,j)}
 \end{equation}

where $i_0$ and $j_0$ are the center of the symmetry of the charge distribution. The conservation property is given by 

\begin{equation}\label{conservprop}
\sum_k {\rho_c(k)}=\sum_{i,j}{-e\,n(i,j)}\quad .
 \end{equation}

\begin{figure}
\begin{center}\includegraphics[scale=1]{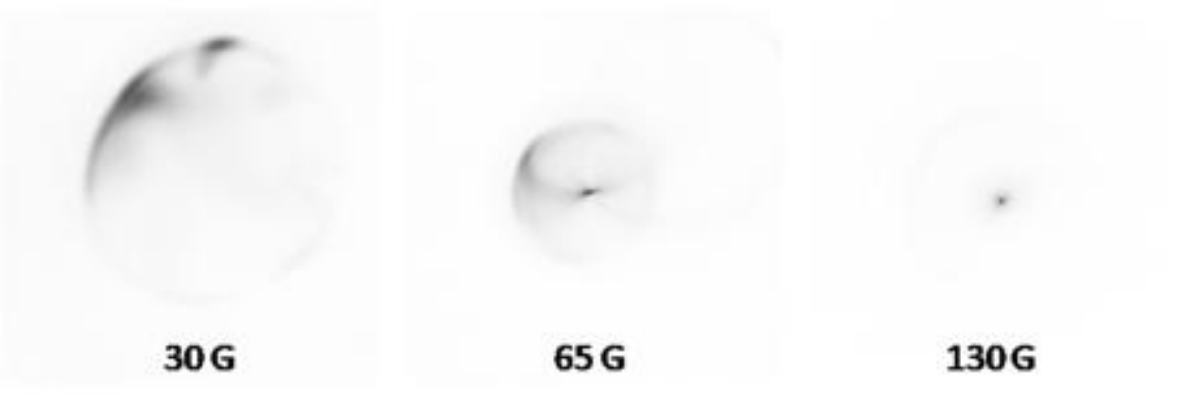}
\end{center}
\caption{\label{spotbrutti}Images of the bunch charge distribution obtained with the optical diagnostics for lower values of the axial magnetic field. The focus occurs efficiently for magnetic field strengths greater than $\approx 100$ G.}
\end{figure}

\begin{figure}
\begin{center}
\includegraphics[scale=0.6]{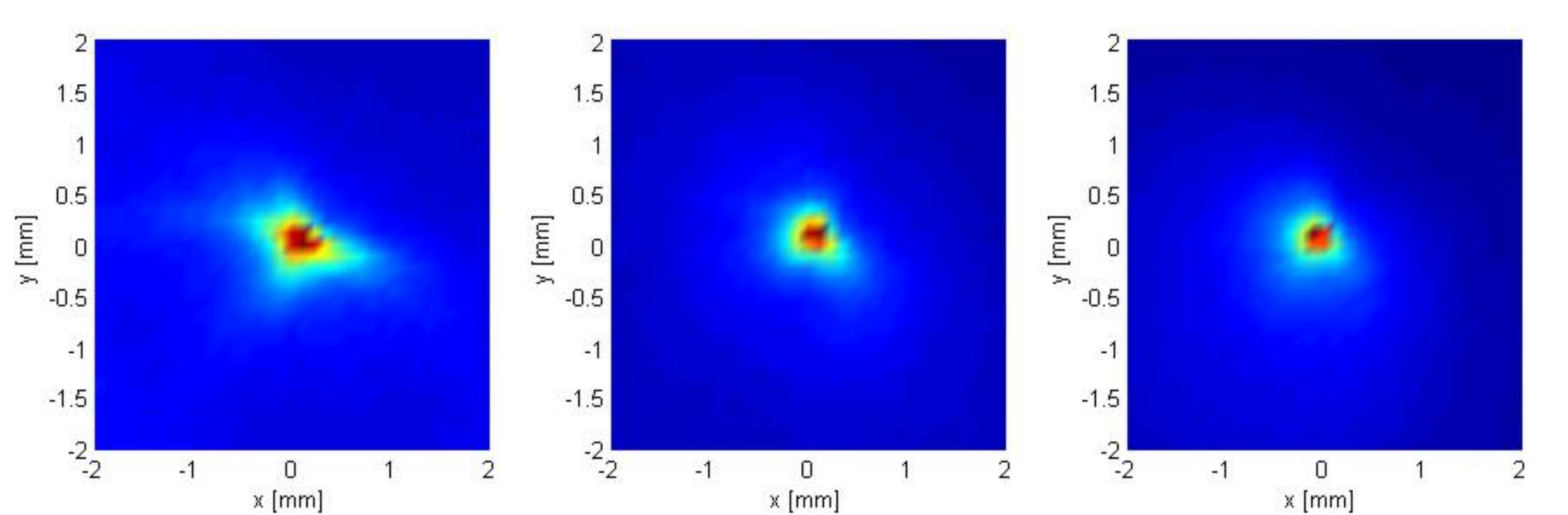}
\end{center}
\caption{\label{spotbelli}Spots of a 6 keV electron bunch at B = 90 G (left), B = 300 G (center), B = 900 G (right).}
\end{figure}

\begin{figure}
\begin{center}
\includegraphics[scale=0.7]{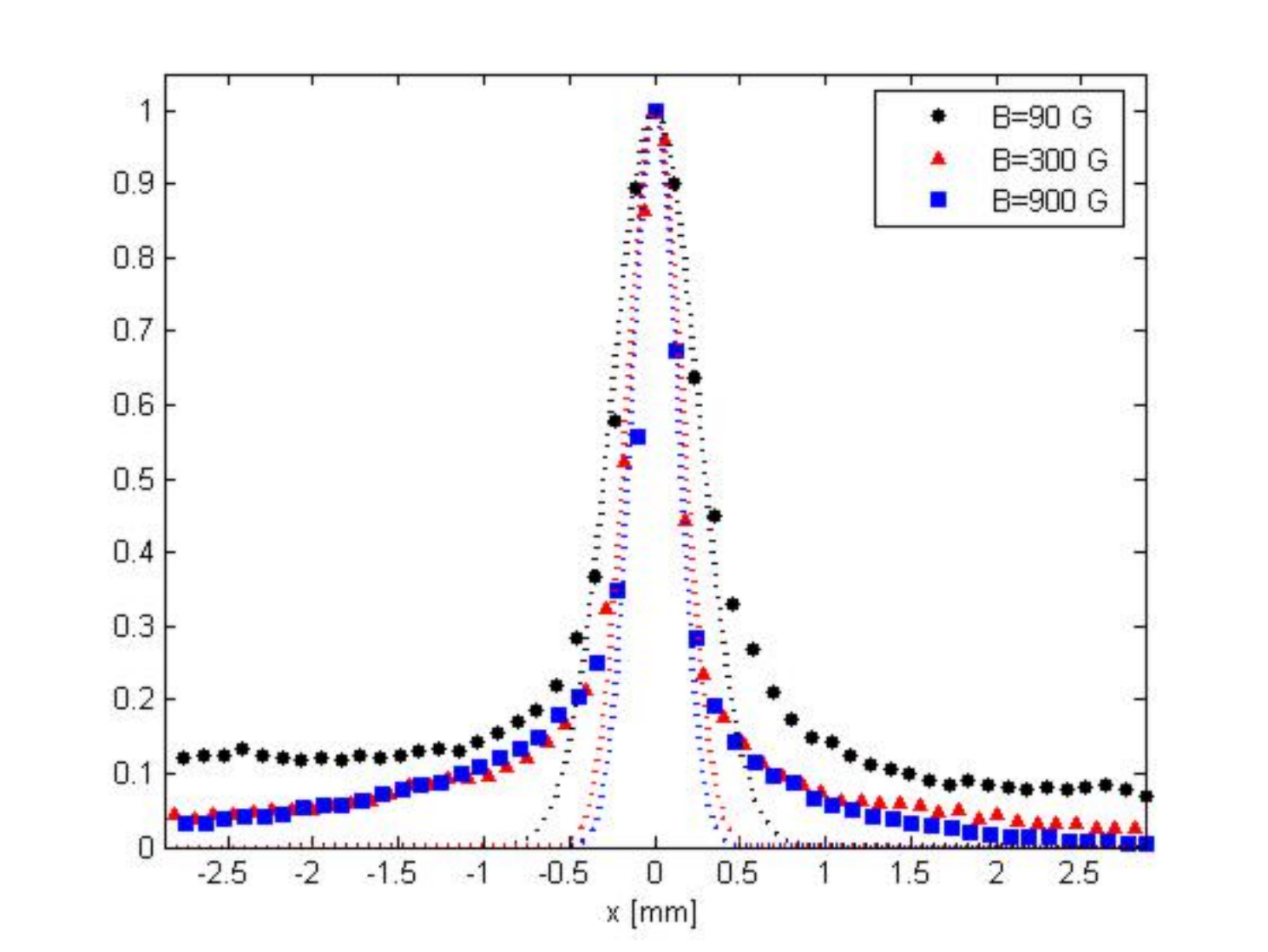}
\end{center}
\caption{\label{fitprofiles}Maximum profiles $n(i,j_{max})$ of the beam. The FWHM are 0.6 mm (B = 90 G), 0.35 mm (B = 300 G), 0.3 mm (B = 900 G).The dotted lines are Gaussian fits with the same FWHM.}
\end{figure}

\begin{figure}
\begin{center}
\includegraphics[scale=0.6]{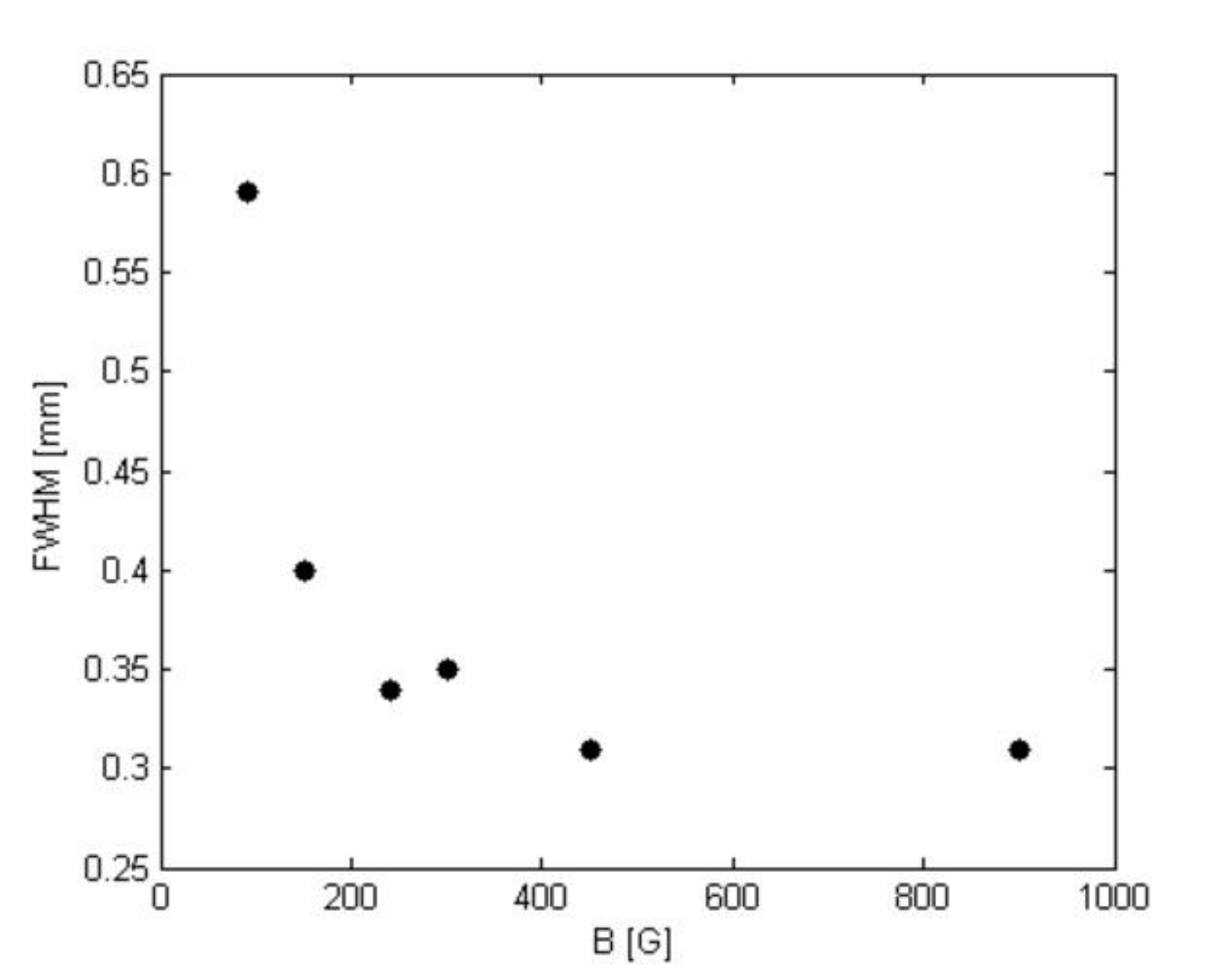}
\end{center}
\caption{\label{fwhmmagn}Spots-size of a 6 keV bunch versus the magnetic field (B = 90 - 900 G). The FWHM of the transversal profile does not change aprreciably for magnetic field larger than 400 G.}
\end{figure}

\begin{figure}
\centering
\subfigure[]
{\includegraphics[scale=0.7]{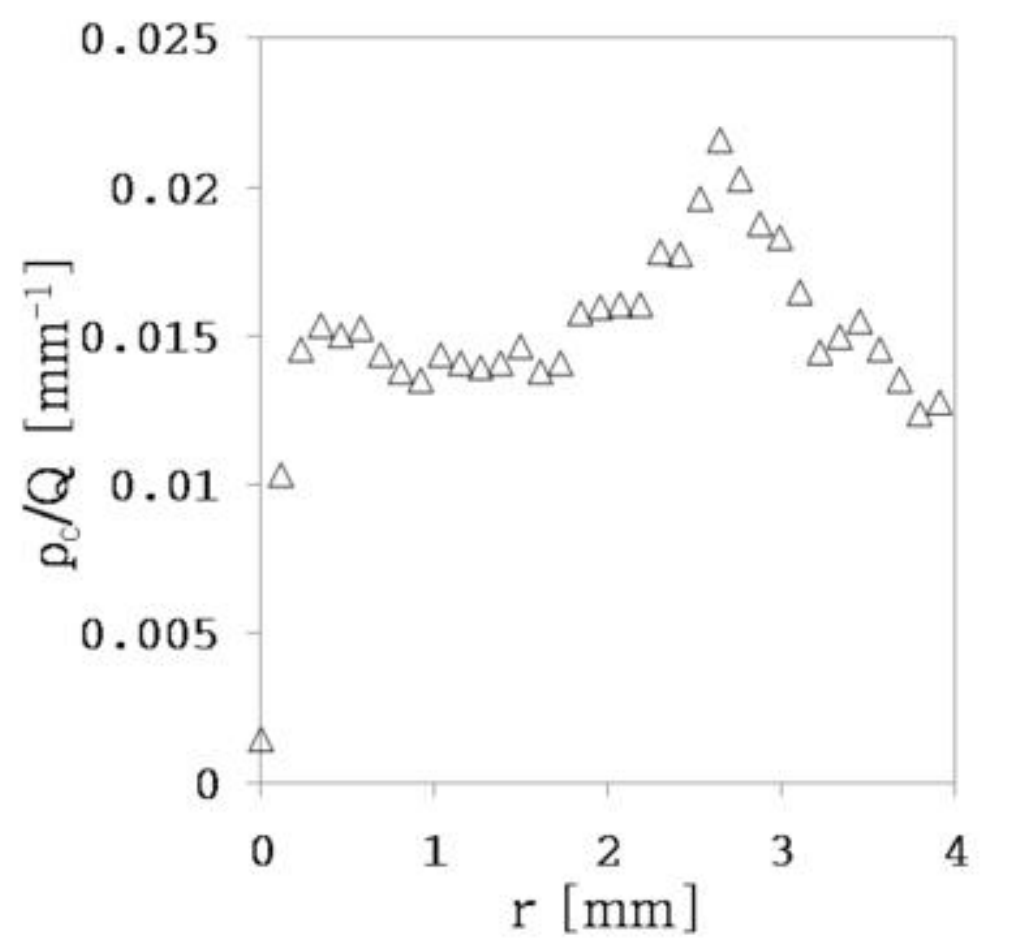}}
\hspace{5mm}
\subfigure[]
{\includegraphics[scale=0.7]{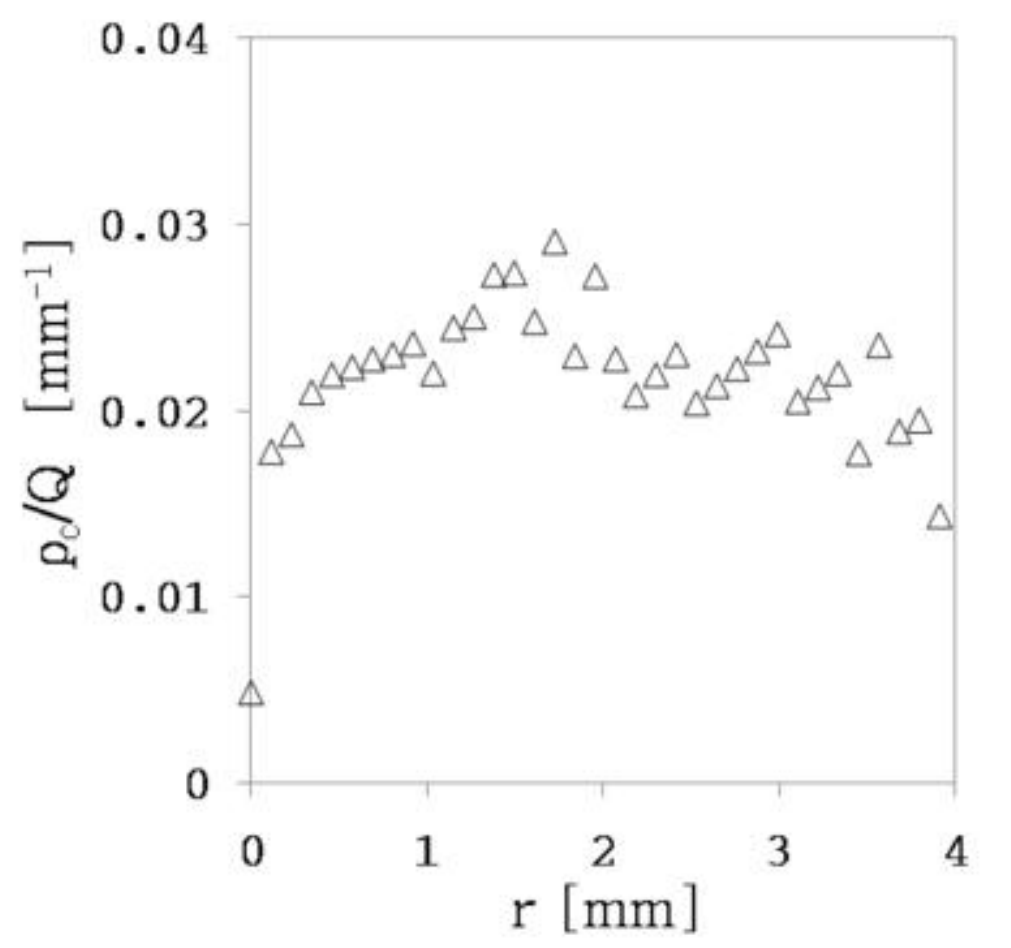}}
\caption{\label{conservprof}Charge per unit radius profiles $\rho_c(r)$ normalized to the total charge $Q$ for a magnetic field of 100 G (a) and 300 G (b). The fraction of the total charge in the dense region around the center is $\approx 6\%$ and $\approx 4\%$ for 100 and 300 G respectively.}
\end{figure}

\begin{figure}
\begin{center}
\includegraphics[scale=0.7]{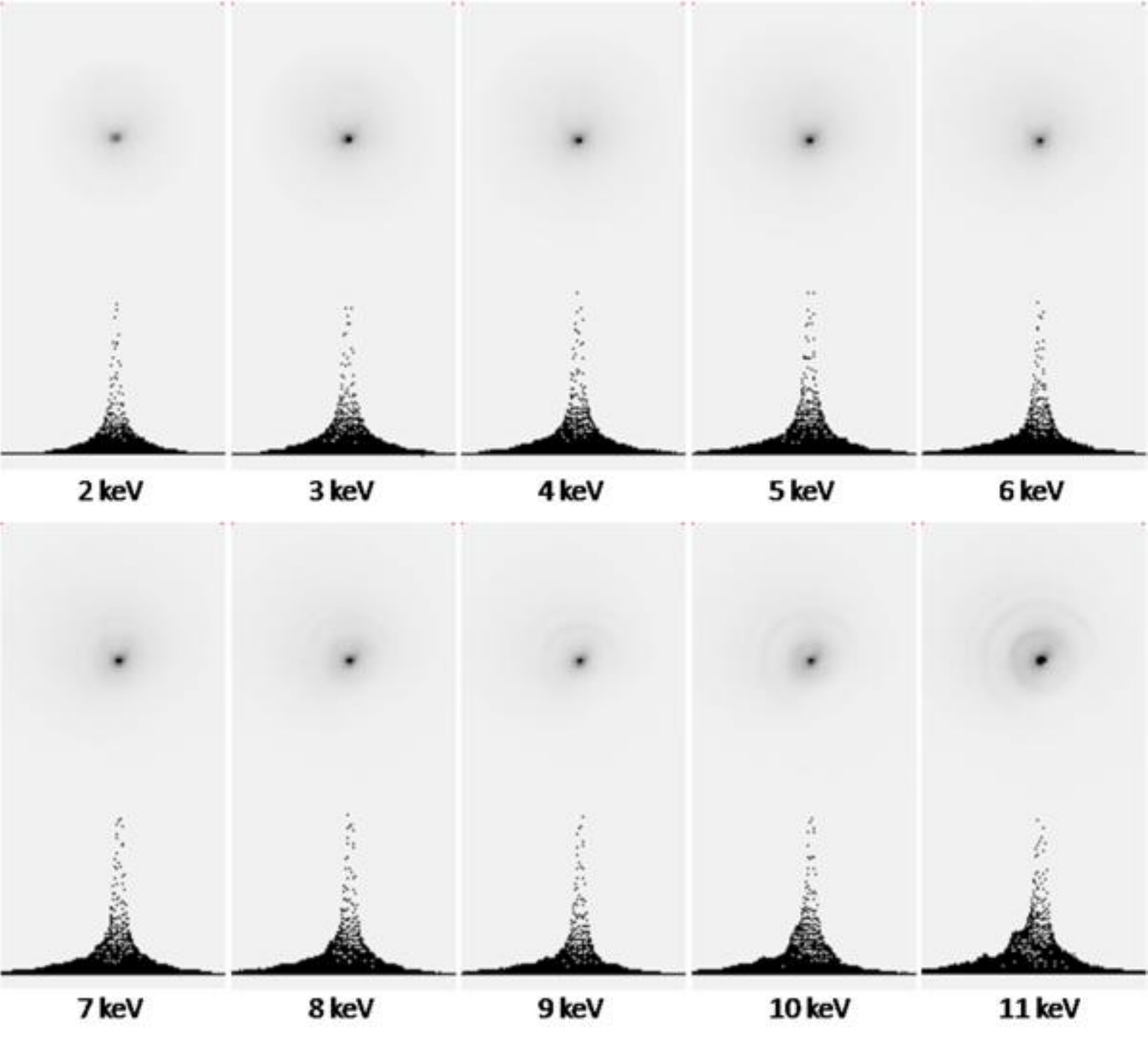}
\end{center}
\caption{\label{profenergy}Images of the bunch charge distribution and relatives profiles obtained with the optical diagnostics varying the bunch energy from 2 to 11 keV. Note the formation of radial dense coherent structures for higher energies.}
\end{figure}

\begin{figure}
\begin{center}
\includegraphics[scale=0.7]{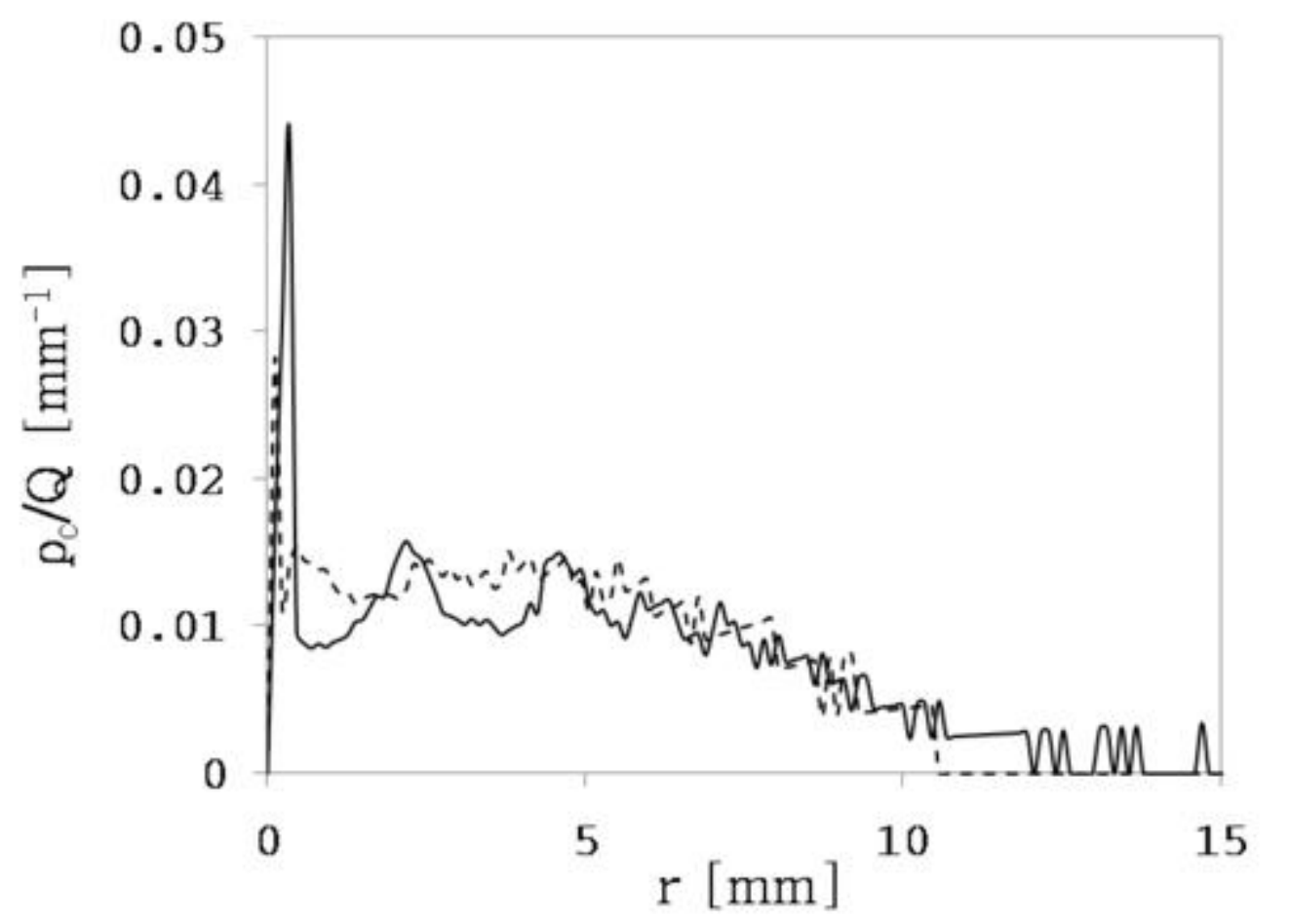}
\end{center}
\caption{\label{consvprofcomp}Comparison of the $\rho_c$ profiles at 6 keV (dotted line) and 11 keV (full line). Charge depressions in the 11 keV profile are redistributed on the tail and on the main peak.}
\end{figure}

A first set of measurements using the optical diagnostics was performed for  bunches with energy of 6 keV and varying the magnetic field in the range 30 - 900 G. The intensity of the magnetic field to obtain a localized spot, start from values of $\approx 90$ G as represented in figure \ref{spotbrutti}. For lower magnetic field the charge is asymmetrically distributed in a large region. When the focus is reached efficiently we obtain localized spots (see fig. \ref{spotbelli}). The maximum profiles $n(i,j_{max})$ was obtained for these spots and fitted with Gaussian functions to estimate the FWHM. The resulting fits are represented in figure \ref{fitprofiles} for magnetic field of 90, 300, 900 G. The profiles are well represented by Gaussian functions near to the center but then deviate from it. The FWHM versus the intensity of the magnetic field (see fig. \ref{fwhmmagn}) shows that a significant focus occurs mainly from 100 G and then the bunch doesn't reduce its transversal size appreciable for magnetic fields greater then 400 G. The fraction of the total charge around the center of the bunch's spot is well represented by the $\rho_c$ profile figure \ref{conservprof}. In the range $0\leq r \leq r_a$ we have a fraction of $\approx 6\%$ ($r_a=0.46$ mm) and $\approx4\%$ ($r_a=0.23$ mm) for 90 and 300 G respectively. Where $r_a$ is the radius that contains the main peak of the profile $\rho_c$. A second set of measurements were obtained for a constant magnetic field of 330 G and varying the bunch energy from 2 to 11 keV.  The respective images and profiles (see fig. \ref{profenergy}) show that the bunches are efficiently focused and symmetric, but increasing the energy a small asymmetry occurs moving the center of charge on the left of the main peak of density and  some dense coherent radial structures are formed. A substantial amount of the total charge is distributed in these regions. The formation of these structures doesn't limit the bunch brightness in the main peak, comparing the normalized $\rho_c$ profiles at 6 keV and 11 keV i.e. with  and without the rings (see fig. \ref{consvprofcomp}) we observe that the main peak contain a bigger fraction of the total charge when the rings are present $\approx 10\%$ with respect to $\approx 4\%$ when the rings are absent. In fact the rings are formed like a depression of charge that is redistributed mainly on the tail of the profile and on the main peak.

\section{Conclusions}

With the developed electrostatic diagnostic we have characterized the global properties of the bunch varying the extraction potential and the magnetic field focusing. The total charge obtained experimentally is of the order of 100 pC for lower magnetic field. A small fraction $4\% - 10\%$ of this charge  is distributed transversally in good approximation like a Gaussian function near to the main peak with a FWHM ranging from 0.6 mm at 90 G to 0.3 mm at 900 G. The other fraction of the total charge is dispersed around the main peak.  An estimate of the bunch density reachable in the region of uniformity for the magnetic field is now possible. The length of the bunch can be computed by the formula $\Delta L=\Delta t \sqrt{2 E/m_e} $, where $\Delta t$ is the characteristic time width of the bunch, but not for lower energy because the space charge effects introduce a substantial deviation from the ideal condition of rigid motion. Considering for example a bunch diameter of $\approx 0.6$ mm at 90 Gauss and $\Delta L=29$ cm at 15 keV, we have $n_b=Q/(\pi r_b^2 \Delta L)=4.3\cdot10^8$ cm$^{-3}$, where the assumed charge $Q$ is $\approx 6$ pC. Greater densities should be reached in the optimal experimental conditions. We have to consider that these estimates are made starting from length measurements that require post-processing techniques of an assumed initial undistorted shape signal and the result change if the initial assumption is changed. For example in the spread velocity measurements the average error was 30$\%$ for two different shapes of the initial signal. These limitations can be overcome with the development of a Thomson backscattering diagnostics described in the next chapter. 

\chapter{Thomson backscattering diagnostics}
\section{Introduction}

In the previous chapter we characterized the electron beam with an experimental estimate of its density, length and radius and we pointed out the main limitations of the electrostatics diagnostics used in this characterization. We describe now the Thomson backscattering diagnostics implemented in this apparatus and based on the laser-electron interaction. To this purpose we start considering the elementary interaction between a photon and an electron, described in the rest frame of the electron by the famous Klein-Nishina formula \cite{Quantum} 

\begin{equation}\label{KleinNishina}
\frac{d\sigma}{d\Omega}=\frac{r_0^2}{2} \left(\frac{\omega'}{\omega}\right)^2 \left(\frac{\omega'}{\omega}+\frac{\omega}{\omega'}-\sin^2\theta\right)
\end{equation}

where $\frac{d\sigma}{d\Omega}$ is the differential cross section of the scattered photon with energy $\omega'$, $r_0$ is the classical radius of the electron, $\theta$ is the angle between the incoming and  outgoing photon and $\omega$ is the energy of the incoming photon. Explicitly, introducing the dimensionless parameter $\eta=\frac{h\nu}{m_e c^2}$, we get

\begin{equation}\label{energyrel}
\omega'=\frac{\omega}{1+\eta\cos\theta}
\end{equation}

For visible light we have $\eta\approx10^{-5}$ so $\omega'\approx\omega$ is a good approximation and we get the following differential cross section 

\begin{equation}\label{diffcrossection}
\frac{d\sigma}{d\Omega}=\frac{r_0^2}{2} \left(1+\cos^2{\theta}\right)
 \end{equation}

that is the classical Thomson scattering formula. The total cross section $\sigma=\frac{8}{3}\pi r_0^2$ is of the order of the square of the classical radius $r_0=2.8 \cdot 10^{-15}$ m. For an electron sample interacting with a monochromatic radiation, the intensity of the scattered radiation is proportional to the density of the electron sample and the energy distribution of the scattered photons (proportional to the photon frequency) depends on the velocity distribution of the electron ensemble. Measuring the intensity and the spectrum of the scattered radiation it is possible to estimate the temperature and the density of the electron distribution. In the case of incoherent scattering $\frac{1}{\Delta k\,\lambda_D}<<1$ (where $\Delta k$ is the difference between the wave vectors of the incident and scattered radiation and $\lambda_D$ the Debye length) for an electron velocity distribution $f(v)=\frac{n_e}{\sqrt{\pi}v_{th}}\exp{\left(-\frac{v^2}{v_{th}^2}\right)}$, the intensity of the spectrum $I(\Delta\omega)$ (as a function of the difference between the frequencies of the incident and scattered radiation) is

$$
\frac{I(\Delta\omega)}{I_0} \propto \frac{n_e}{\Delta k v_{th}} \exp{\left(-\frac{\Delta\omega^2}{\Delta k^2 v_{th^2} }\right)}
$$

where $v_{th}=\left(\frac{2kT_e}{m_e}\right)^2$ is the thermal velocity. 

The very low cross section $\sigma_T=6.65\cdot 10^{-29}$ m$^2$ implies that the number of incident photons must be relatively high. More precisely taking the typical values \cite{Soltwisch} $L_S=1$ cm (scattering length), $n_e=10^{11}$ cm$^{-3}$ (electron density), $\Delta\Omega=0.2$ sr (solid angle  of the detected photons) the number of scattered photons can be estimated by

$$
N_T= \frac{3\sigma_T}{8\pi}\cdot L_S \cdot \Delta\Omega \cdot n_e \cdot N_0 \: .
$$

And we find $N_T/N_0=1.6\cdot 10^{-15}$. The use of a high energy pulse laser in the order of 1J (ruby or Nd:Yag) would produce $N_T\approx 8\cdot10^3$ photons/pulse further reduced by the optical and quantum efficiencies of the photon detector. On the other hand the use of a high power light source means that an appreciable portion of the stray light, coming from the inner walls of the vacuum chamber, falls in the solid angle of the optics collection. The reduction of the stray light is so one of the main problem of the diagnostics to increase its sensitivity. Thomson scattering is used for the detection and the diagnostics of pulsed electron beams solving the problem of the interaction (in space and in time) between the laser and the electron bunches. When the electron beams have an energy of some tens of keV two effects should be considered. The intensity of the scattered radiation increase appreciably along the direction of the beam propagation and the back scattered radiation results in a wavelength shift $\lambda_s=\lambda_i\frac{1-\beta\cos\theta}{1+\beta}$ where $\beta=v/c$, $v$ is the electron velocity, $\theta$ is the angle between the incident and scattered radiation and $\lambda_i$, $\lambda_s$ are the wavelengths of the incident and scattered radiation, respectively. When $\theta=0$ we are in the back scattering condition, i.e. the wavelength shift and the differential cross section are maximum. The use of this diagnostics for charged particle beams can in principle provide information about beam density and density profile from the intensity of the scattered radiation and about beam energy and energy spread from the radiation spectrum. In this chapter after a theoretical estimate of the scattered photons number, we describe two different set-up of the Thomson back scattering diagnostics implemented in Eltrap. In the first set-up the laser was injected collinearly with the beam propagation. The interaction point can be varied during the experiment along the laser trajectory. In the second set-up the laser is focused in a particular interaction point optimized for the photons collection. In both cases we describe the solutions for the laser injection in the vacuum chamber, photons detection, space and time coincidence of electron and laser pulses. We summarize the results with an estimate of the minimum sensitivity obtained.

\section{Theoretical estimate of the scattered photons number}
A theoretical analysis on the scattered photon number is necessary for both the optimization of the geometry and the estimate of the minimum sensitivity. The analysis starts from the equation (\ref{diffcrossection}) in the rest frame and considering the Lorentz trasformation of the quantities, that are not invariant, to obtain the number of scattered photons in the lab frame. We note that for a bunch energy of some tens of keV we are near to the classical case, e.g. for a bunch energy of 15 keV the usual relativistic factors $\beta$ and $\gamma$ are 0.24 and 1.03, respectively. As pointed out in section 3.3.5 the relativistic effects are negligible for the time of flight measurements as well as for length and spread velocity measurements. The discrepancy is about $2\%$ from the non-relativistically computed time. However in back scattering condition and for a collinear interaction the number of scattered photons increase of about $60\%$ passing from $\beta=0$ to $\beta=0.24$. For this reason we will compute the scattered photons relativistically. In the following treatment we consider two reference frames. The frame where the electrons are at rest is called $K'$ and all quantities referred to this frame are primate. The lab frame called $K$ is chosen with the axis parallel to $K'$ and with the origins of the axes coinciding at $t=t'=0$. All quantities referred to $K$ are not primate. Finally the electron velocity $\vec{v_e}$ is the relative velocity between $K$ and $K'$. 

\subsection{Integral of the Thomson cross section in the lab frame}

For an unpolarized electromagnetic wave interacting with an electron at rest, the differential cross section of the scattered radiation in the limit $h\nu<<m_e\,c^2$, is due to the Thomson formula:

\begin{equation}\label{diffcrossection2}
\frac{d\sigma'}{d\Omega'}=\frac{r_0^2}{2} \left(1+\cos^2{\theta'}\right)
 \end{equation}

Where $\theta'$ is the angle between the wave vector of the incident radiation $\vec{k_i'}$ and the scattered radiation $\vec{k_s'}$ in the reference frame K' of the electron at rest. Using the unit vectors $\vec{e_i'}=\vec{k_i'}/|\vec{k_i'}|$, $\vec{e_s'}=\vec{k_s'}/|\vec{k_s'}|$ of the  wave vectors the formula (\ref{diffcrossection2}) becomes

\begin{equation}\label{diffcrossectione}
\frac{d\sigma'}{d\Omega'}=\frac{r_0^2}{2} \left[1+\left(\vec{e_i'}\cdot\vec{e_s'}\right)^2\right]=\tilde{\sigma'}(\vec{e_i}',\vec{e_s}')
 \end{equation}

The unit vectors $\vec{e_i'}$, $\vec{e_s'}$ are represented in spherical coordinates as
\begin{equation}\label{versspherical}
\vec{e_i'}=(\sin\theta_i'\cos\phi_i',\sin\theta_i'\sin\phi_i',\cos\theta_i'),\:\vec{e_s'}=(\sin\theta'\cos\phi',\sin\theta'\sin\phi',\cos\theta')\, .
 \end{equation}

With this representation the function $\tilde{\sigma'}$ is a function of the angles $\theta_i'$, $\phi_i'$, $\theta'$, $\phi'$:

\begin{equation}\label{diffcrossectiondepend}
\tilde{\sigma'}=\tilde{\sigma'}(\theta_i',\phi_i',\theta',\phi') \, .
 \end{equation}

With the trasformation $(\theta_i',\phi_i',\theta',\phi')\mapsto(\theta_i,\phi_i,\theta,\phi)$ the integral of the differential cross section, considering the infinitesimal solid angle $d\Omega'=\sin\theta' d\theta'd\phi'$, change as

\begin{equation}\label{trasfintegr}
\int_{\Delta\Omega'}{\tilde{\sigma'}(\theta_i',\phi_i',\theta',\phi')}\,d\Omega'=\int_{\Delta\Omega}{\tilde{\sigma'}(\theta_i,\phi_i,\theta,\phi)}\left|\frac{\partial(\theta',\phi')}{\partial(\theta,\phi)}\right|\frac{\sin\left[\theta'(\theta,\phi)\right]}{\sin\theta}\,d\Omega
 \end{equation}

where $\theta_i'=\theta_i'(\theta_i,\phi_i)$, $\phi_i'=\phi_i'(\theta_i,\phi_i)$, $\theta'=\theta'(\theta,\phi)$, $\phi'=\phi'(\theta,\phi)$ and $\theta_i$, $\phi_i$, $\theta$, $\phi$  are the spherical coordinates of the unit vectors of the wave vectors of the incident and scattered radiation in an arbitrary inertial reference frame K and the quantitity 

$$
\left|\frac{\partial(\theta',\phi')}{\partial(\theta,\phi)}\right|
$$

is the determinant of the Jacobian represented by the matrix 

$$
{\frac{\partial(\theta',\phi')}{\partial(\theta,\phi)}}=\left(
\begin{array}{cc}
\frac{\partial{\theta'}}{\partial{\theta}} & \frac{\partial{\theta'}}{\partial{\phi}}\\
\frac{\partial{\phi'}}{\partial{\theta}} & \frac{\partial{\phi'}}{\partial{\phi}} \, .
\end{array}
\right)
$$ 

Note that the Jacobian of the trasformation is 

$$
{J}=\left(
\begin{array}{cc}
\frac{\partial{\cos\theta'}}{\partial{\cos\theta}} & \frac{\partial{\cos\theta'}}{\partial{\phi}}\\
\frac{\partial{\phi'}}{\partial{\cos\theta}} & \frac{\partial{\phi'}}{\partial{\phi}} 
\end{array}
\right)
$$ 

thus

$$
\det J = \left|\frac{\partial(\theta',\phi')}{\partial(\theta,\phi)}\right|\frac{\sin\theta'}{\sin\theta} \, .
$$ 

In order to estimate the derivatives in the Jacobian, we introduce the quadrivectors $k_i'^\mu$, $k_s'^\mu$ of the incident and scattered radiation in $K'$ as

\begin{equation}\label{tetravector}
k_i'^\mu=(\frac{\omega_i'}{c},\vec{k_i'}),\:k_s'^\mu=(\frac{\omega_s'}{c},\vec{k_s'})
 \end{equation}

where $\omega_i'$ and $\omega_s'$ are the angular frequencies of the incident and scattered electromagnetic waves in $K'$. The Lorentz transformations of the quadrivectors  $k_i'^\mu$, $k_s'^\mu$ from $K'$ to $K$, ($K$ is a frame in standard configuration with the axis $x$, $y$, $z$ parallel to $x'$, $y'$, $z'$) for a boost in any arbitrary direction $\vec{\beta}=(\beta_x, \beta_y, \beta_z)$, are

\begin{equation}\label{transformations}
k_i'^\mu=\Lambda_\nu^\mu\,k_i^\nu,\:\: k_s'^\mu=\Lambda_\nu^\mu\,k_s^\nu
 \end{equation}

where $k_i^\nu=(\frac{\omega_i}{c},\vec{k_i})$, $k_s^\nu=(\frac{\omega_s}{c},\vec{k_s})$ are the quadrivectors of the incident and scattered radiation in the frame K and $\Lambda_\nu^\mu$ is the matrix of the Lorentz transformation:

$$
{\Lambda_\nu^\mu}=\left(
\begin{array}{cccc}
\gamma & -\beta_x\gamma & -\beta_y\gamma & -\beta_z\gamma\\
-\beta_x\gamma & 1+(\gamma-1)\frac{\beta_x^2}{|\beta|^2} & (\gamma-1)\frac{\beta_x\beta_y}{|\beta|^2} & (\gamma-1)\frac{\beta_x\beta_z}{|\beta|^2}\\
-\beta_y\gamma & (\gamma-1)\frac{\beta_y\beta_x}{|\beta|^2} & 1+(\gamma-1)\frac{\beta_y^2}{|\beta|^2} & (\gamma-1)\frac{\beta_y\beta_z}{|\beta|^2}\\
-\beta_z\gamma & (\gamma-1)\frac{\beta_z\beta_x}{|\beta|^2} & (\gamma-1)\frac{\beta_z\beta_y}{|\beta|^2} & 1+(\gamma-1)\frac{\beta_z^2}{|\beta|^2}
\end{array}
\right)
$$

The variables $\theta'$, $\phi'$ can be written as functions of $\theta$, $\phi$ considering the components of the unit vector $\vec{e_s}$ in spherical coordinates after the transformations (\ref{transformations}):

\begin{equation}\label{componentx}
\sin\theta'\cos\phi'=\frac{\Lambda_\nu^1 k^\nu (\theta,\phi)}{\sqrt{\sum_{\mu=1}^{3}(\Lambda_\nu^\mu k^\nu)^2} (\theta,\phi)}=\frac{\Lambda_\nu^1 k^\nu (\theta,\phi)}{\Lambda_\nu^0 k^\nu (\theta,\phi)}
 \end{equation}

\begin{equation}\label{componenty}
\sin\theta'\sin\phi'=\frac{\Lambda_\nu^2 k^\nu (\theta,\phi)}{\sqrt{\sum_{\mu=1}^{3}(\Lambda_\nu^\mu k^\nu)^2} (\theta,\phi)}=\frac{\Lambda_\nu^2 k^\nu (\theta,\phi)}{\Lambda_\nu^0 k^\nu (\theta,\phi)}
 \end{equation}

\begin{equation}\label{componentz}
\cos\theta'=\frac{\Lambda_\nu^3 k^\nu (\theta,\phi)}{\sqrt{\sum_{\mu=1}^{3}(\Lambda_\nu^\mu k^\nu)^2} (\theta,\phi)}=\frac{\Lambda_\nu^3 k^\nu (\theta,\phi)}{\Lambda_\nu^0 k^\nu (\theta,\phi)}
 \end{equation}

written in explicit form as

\begin{equation}\label{thetaprimo}
\theta'=\arccos\left(\frac{\Lambda_\nu^3 k^\nu (\theta,\phi)}{\Lambda_\nu^0 k^\nu (\theta,\phi)}\right)
 \end{equation}

\begin{equation}\label{phiprimo}
\phi'=\arctan\left(\frac{\Lambda_\nu^2 k^\nu (\theta,\phi)}{\Lambda_\nu^1 k^\nu (\theta,\phi)}\right)
 \end{equation}

The determinant of the Jacobian in the integral (\ref{trasfintegr}) is so

\begin{equation}\label{determinante}
\left|\frac{\partial(\theta',\phi')}{\partial(\theta,\phi)}\right|=D'\cdot\left[\frac{\partial}{\partial\theta}\left(\frac{\Lambda_\nu^3 k^\nu}{\Lambda_\nu^0 k^\nu }\right) \frac{\partial}{\partial\phi}\left(\frac{\Lambda_\nu^2 k^\nu }{\Lambda_\nu^1 k^\nu }\right)- \frac{\partial}{\partial\phi}\left(\frac{\Lambda_\nu^3 k^\nu}{\Lambda_\nu^0 k^\nu}\right)\frac{\partial}{\partial\theta}\left(\frac{\Lambda_\nu^2 k^\nu}{\Lambda_\nu^1 k^\nu }\right)  \right]
 \end{equation}

where $D'$ is 

$$
D'=\frac{-1}{\left[1+\left(\frac{\Lambda_\nu^2 k^\nu }{\Lambda_\nu^1 k^\nu }\right)^2\right]\sqrt{1-\left(\frac{\Lambda_\nu^3 k^\nu }{\Lambda_\nu^0 k^\nu }\right)^2}}
$$ 

Defining now the quantities $\partial k_{\theta i}$, $\partial k_{\phi i}$, ($i=0,1,2,3$) as

\begin{equation}\label{tetravectorderiv}
\partial k_{\theta i}=(0, \cos\theta\cos\phi, \cos\theta\sin\phi, -\sin\theta), \: \partial k_{\phi i}=(0, -\sin\theta\sin\phi, \sin\theta\cos\phi, 0)
 \end{equation}
 
The equation (\ref{determinante}) is written as 

\begin{eqnarray}\label{determinantesimpl}
\left|\frac{\partial(\theta',\phi')}{\partial(\theta,\phi)}\right|=D\cdot\left[\left(\Lambda_\nu^0 k^\nu\,\sum_{i=0}^{3}\Lambda_i^3\partial k_{\theta i}-\Lambda_\nu^3 k^\nu\,\sum_{i=0}^{3}\Lambda_i^0\partial k_{\theta i}\right)\right.\nonumber\\
 \left(\Lambda_\nu^1 k^\nu\,\sum_{i=0}^{3}\Lambda_i^2\partial k_{\phi i}-\Lambda_\nu^2 k^\nu\,\sum_{i=0}^{3}\Lambda_i^1\partial k_{\phi i}\right)-\nonumber\\
\left(\Lambda_\nu^0 k^\nu\,\sum_{i=0}^{3}\Lambda_i^3\partial k_{\phi i}-\Lambda_\nu^3 k^\nu\,\sum_{i=0}^{3}\Lambda_i^0\partial k_{\phi i}\right)\nonumber\\
\left.\left(\Lambda_\nu^1 k^\nu\,\sum_{i=0}^{3}\Lambda_i^2\partial k_{\theta i}-\Lambda_\nu^2 k^\nu\,\sum_{i=0}^{3}\Lambda_i^1\partial k_{\theta i}\right)\right]
 \end{eqnarray}
with the coefficient

$$
D=\frac{-1}{\left[1+\left(\frac{\Lambda_\nu^2 k^\nu }{\Lambda_\nu^1 k^\nu }\right)^2\right]\sqrt{1-\left(\frac{\Lambda_\nu^3 k^\nu }{\Lambda_\nu^0 k^\nu }\right)^2}\left(\Lambda_\nu^0 k^\nu\,\Lambda_\nu^1 k^\nu\right)^2}
$$

or by

\begin{equation}\label{compact}
\left|\frac{\partial(\theta',\phi')}{\partial(\theta,\phi)}\right|=\frac{P_\phi^{0,3}P_\theta^{1,2}-P_\theta^{0,3}P_\phi^{1,2}}{\left[1+\left(\frac{\Lambda_\nu^2 k^\nu }{\Lambda_\nu^1 k^\nu }\right)^2\right]\sqrt{1-\left(\frac{\Lambda_\nu^3 k^\nu }{\Lambda_\nu^0 k^\nu }\right)^2}\left(\Lambda_\nu^0 k^\nu\,\Lambda_\nu^1 k^\nu\right)^2}
 \end{equation}

with the definitions $P_\theta^{i,j}=\Lambda_\nu^i k^\nu\,\sum_{l=0}^{3}\Lambda_l^j\partial k_{\theta l}-\Lambda_\nu^j k^\nu\,\sum_{l=0}^{3}\Lambda_l^i\partial k_{\theta l}$, 
$P_\phi^{i,j}=\Lambda_\nu^i k^\nu\,\sum_{l=0}^{3}\Lambda_l^j\partial k_{\phi l}-\Lambda_\nu^j k^\nu\,\sum_{l=0}^{3}\Lambda_l^i\partial k_{\phi l}$ ($i,j=0,1,2,3$).

Substituting now the equation (\ref{compact}) in the second integral of equation (\ref{trasfintegr}) and considering $\theta'$ as in (\ref{thetaprimo}) we obtain 

\begin{equation}\label{finalintegr}
\int_{\Delta\Omega'}{\tilde{\sigma'}(\theta_i',\phi_i',\theta',\phi')}\,d\Omega'=\int_{\Delta\Omega}{\frac{\tilde{\sigma'}(\theta_i,\phi_i,\theta,\phi)\left(P_\phi^{0,3}P_\theta^{1,2}-P_\theta^{0,3}P_\phi^{1,2}\right)}{\sin{\theta}\left[1+\left(\frac{\Lambda_\nu^2 k^\nu }{\Lambda_\nu^1 k^\nu }\right)^2\right]\left(\Lambda_\nu^0 k^\nu\,\Lambda_\nu^1 k^\nu\right)^2}}\,d\Omega
\end{equation}

and the determinant of the Jacobian of the trasformation is given by

\begin{equation}\label{detjacobian}
\det J=\frac{\left(P_\phi^{0,3}P_\theta^{1,2}-P_\theta^{0,3}P_\phi^{1,2}\right)}{\sin{\theta}\left[1+\left(\frac{\Lambda_\nu^2 k^\nu }{\Lambda_\nu^1 k^\nu }\right)^2\right]\left(\Lambda_\nu^0 k^\nu\,\Lambda_\nu^1 k^\nu\right)^2} \, .
\end{equation} 

The second integral in (\ref{finalintegr}) is the integral of the differential cross section with all variables in the lab frame and for a boost in an arbitrary direction.
\subsection{Differential cross section in the collinear scattering}
Let's consider now a collinear scattering $\vec{e_i}\times\vec{v_e}=0$ with the direction of the incident radiation and the electron velocity collinear along the Z axis and counterpropagating. For semplicity we write $\beta_z=\beta$. The matrix of the Lorentz trasformation is given by
$$
{\Lambda_\nu^\mu}=\left(
\begin{array}{cccc}
\gamma & 0 & 0 & -\beta\gamma\\
0& 1& 0 & 0\\
0 & 0 & 1 & 0\\
-\beta\gamma & 0 & 0 & \gamma
\end{array}
\right)
$$ 

used to compute the quantities $P_\phi^{0,3}$, $P_\theta^{1,2}$ $P_\theta^{0,3}$, $P_\phi^{1,2}$, $\Lambda_\nu^0 k^\nu$, $\Lambda_\nu^1 k^\nu$, $\Lambda_\nu^2 k^\nu$, $\Lambda_\nu^3 k^\nu$  in (\ref{detjacobian}), explicitly written as

\begin{eqnarray}
P_\phi^{0,3}=0\label{esp1}\\
P_\theta^{1,2}=0\\
P_\theta^{0,3}=-\sin\theta\,\frac{\omega^2}{c^2}\\
P_\phi^{1,2}=\sin^2\theta\,\frac{\omega^2}{c^2}\\
\Lambda_\nu^0 k^\nu=\gamma\frac{\omega}{c}(1-\beta\cos\theta)\\
\Lambda_\nu^1 k^\nu=\frac{\omega}{c}\sin\theta\cos\phi\\
\Lambda_\nu^2 k^\nu=\frac{\omega}{c}\sin\theta\sin\phi\\
\Lambda_\nu^3 k^\nu=\gamma\frac{\omega}{c}(\cos\theta-\beta)\label{esp2}
 \end{eqnarray}
 
Substituting the equations (\ref{esp1}) $\div$ (\ref{esp2}) in (\ref{detjacobian}) and using 

\begin{equation}
\tilde{\sigma}=\frac{d\sigma}{d\Omega}=\frac{d\sigma'}{d\Omega'} \det J 
\end{equation}

we obtain

\begin{equation}\label{crosssectioncoll}
\tilde{\sigma}=\frac{r_0^2}{2}\left[1+\left(\frac{\cos\theta-\beta}{1-\beta\cos\theta}\right)^2\right]\frac{1}{\gamma^2(1-\beta\cos\theta)^2} \, .
\end{equation}

\begin{figure}
\centering
\subfigure[]
{\includegraphics[scale=0.55]{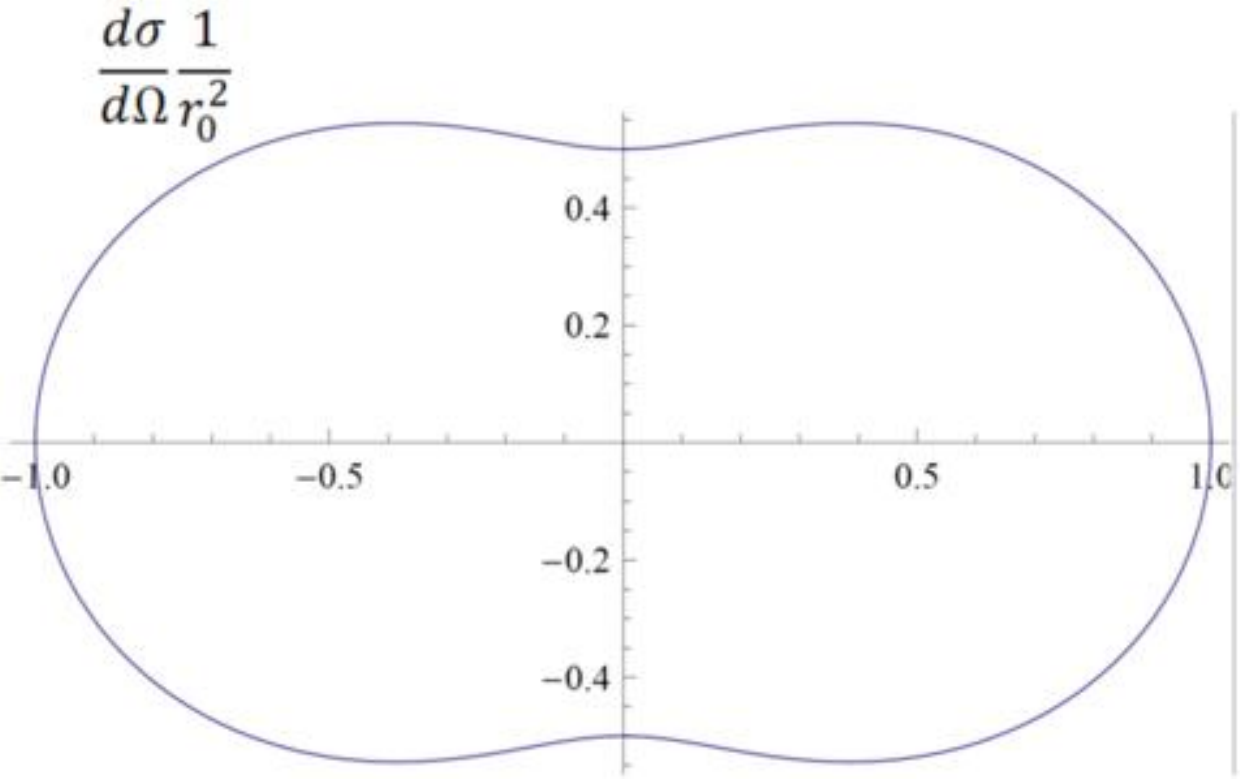}}
\hspace{5mm}
\subfigure[]
{\includegraphics[scale=0.5]{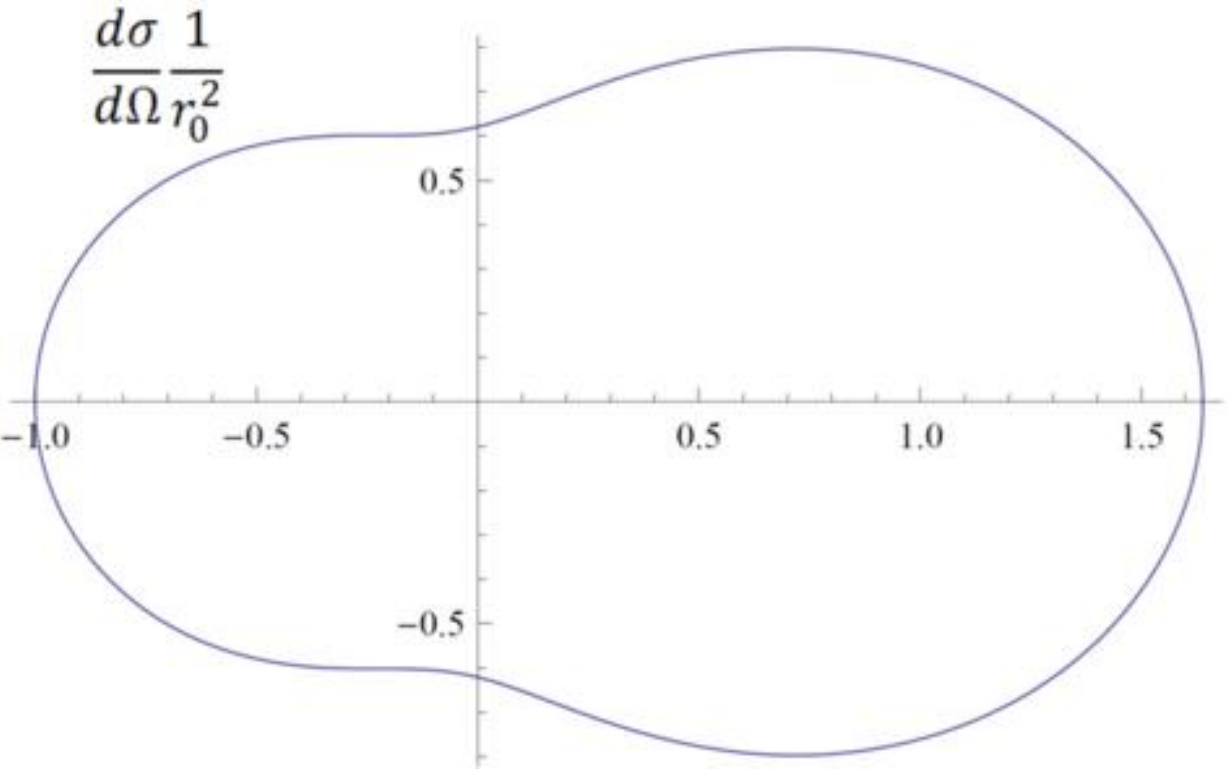}}
\caption{\label{polarcross}Polar diagrams of the differential cross sections normalized to the classical electron radius $r_0$ in a collinear scattering $\vec{e_i}\times\vec{v_e}=0$, for an electron at rest (a) and with an energy of 15 keV (b). The maximum is for $\theta=0$ in back scattering condition.}
\end{figure}

The differential cross section, normalized to $r_0^2$ is represented in figure \ref{polarcross} in a polar diagram with the angular coordinate $\theta$ and for two different values of the bunch energy $E=0$, $E=15$ keV. The cross section increases along the electron velocity direction and decreases in the opposite direction (with respect to the cross section with the electron at rest $\beta=0$). The maximum is at $\theta=0$, i.e. in the back scattering condition, where (\ref{crosssectioncoll}) takes the value

\begin{equation}\label{backscattering}
\tilde{\sigma}=\frac{r_0^2}{\gamma^2(1-\beta)^2} \, .
\end{equation}

In this particular condition the cross section increase of about $60\%$ at $E=15$ keV with respect to $E=0$. 

\subsection{Relativistic invariants and compressional photon flux}
Let's consider the number of scattered photons from a small "volume" in the momentum space $d^3x'd^3p'$ in the time interval $dt'$ and in the reference frame $K'$ in which the particles with momentum $\vec{p'}$ are at rest 

\begin{equation}\label{deltaennep}
dN'= n_p' f'(\vec{x'},\vec{p'}) d^3x'\,d^3p'\, v_p'\, dt'\, \tilde{\sigma'} d\Omega' 
 \end{equation}

where $f'(\vec{x'},\vec{p'}) d^3x'\,d^3p'$ is the number of scattering centers at the position $\vec{x'}$ for particles with momentum $\vec{p'}$, $n_p'$ is the density of the incident particles, $v_p'$ the velocity of the incident particles and $\tilde{\sigma'}, d\Omega'$ the differential cross section and the element of solid angle, respectively.

Since the number $dN'$ is by its very nature an invariant quantity, we try to express it in a form which is applicable in the reference frame K

 \begin{equation}\label{deltaennep}
dN'= dN = n_p f(\vec{x},\vec{p}) d^3x\,d^3p\, \tilde{v}\, dt\, \tilde{\sigma} d\Omega 
 \end{equation}

where these quantities are now relative to the reference frame K and $\tilde{v}$ is the quantity to be determined according to the assumptions $dN=dN'$ and $\tilde{\sigma'} d\Omega' =\tilde{\sigma} d\Omega$.

So we can write

\begin{equation}\label{denneeqdnp}
n_p' f'(\vec{x'},\vec{p'}) d^3x'\,d^3p'\, v_p'\, dt'\, \tilde{\sigma'} d\Omega' = n_p f(\vec{x},\vec{p}) d^3x\,d^3p\, \tilde{v}\, dt\, \tilde{\sigma} d\Omega 
 \end{equation}
 
from the invariance of $f d^3x d^3p$ and considering the time dilation from $K'$ to K $dt'=dt/\gamma$ the equality (\ref{denneeqdnp}) is written as

\begin{equation}\label{denneeqdnprid}
n_p'v_p'\frac{dt}{\gamma}=n_p \tilde{v}\, dt
 \end{equation}

The density $n_p'$ in the reference frame at rest can be expressed using the Lorentz trasformation $J'^\mu=\Lambda^\mu_\nu J^\nu$ of the quadrivector that describe the photon flux $J^\mu=(n_p,n_p\vec{\beta_p})$ where $\vec{\beta_p}=\vec{v_p}/c$. We have $n_p'=J'^0=n_p\gamma(1-\vec{\beta}\cdot\vec{\beta_p})$. Replacing in equation (\ref{denneeqdnprid}) we obtain

\begin{equation}\label{compress1}
\tilde{v}= v_p' (1-\vec{\beta}\cdot\vec{\beta_p})
 \end{equation}

For the photons $v_p'=c$, $\vec{\beta_p}=|\vec{\beta_p}| \vec{e_i}$, $|\vec{\beta_p}|=1$ and so

\begin{equation}\label{compress2}
\tilde{v}= c (1-\vec{\beta}\cdot\vec{e_i}) .
 \end{equation}
 For a collinear scattering the equation (\ref{compress1}) gives 

$$
\tilde{v}=|\vec{v_p}-\vec{v_e}| 
$$ 

and the compressional photon flux takes the physical meaning of relative velocity between the incident particles and the target particles. 
\subsection{Number of scattered photons}
Whit the results of the section 4.2.2, 4.2.3 we can now write the integral of the scattered phothon number $N_s$ for an electron distribution function $f(\vec{x},\vec{p},t)$ and an incident photon flux of density $n_p(\vec{x},t)$

\begin{equation}\label{numberscattered}
N_s=\int{[1-\vec{\beta}(\vec{p})\cdot\vec{e_i}]\,n_p(\vec{x},t)f(\vec{x},\vec{p},t)}\int_{\Delta\Omega}{\frac{\tilde{\sigma'}\left(P_\phi^{0,3}P_\theta^{1,2}-P_\theta^{0,3}P_\phi^{1,2}\right)}{\sin{\theta}\left[1+\left(\frac{\Lambda_\nu^2 k^\nu }{\Lambda_\nu^1 k^\nu }\right)^2\right]\left(\Lambda_\nu^0 k^\nu\,\Lambda_\nu^1 k^\nu\right)^2}}\,d\Omega\,cdt\,d^3x\,d^3p
\end{equation} 

where $\vec{\beta}(\vec{p})=\frac{\vec{p}}{\sqrt{m^2c^2+p^2}}$ and $\Delta\Omega$ is a portion of solid angle of the scattered radiation. In the simpler case of an electron distribution with a momentum function

\begin{equation}
f(\vec{x},\vec{p},t)= g(\vec{x},t)\,\delta(\vec{p}-\vec{p_0})
\end{equation}  

where all electrons of the distribution propagate with momentum $\vec{p_0}$. Integrating on the momentum space, eq. (\ref{numberscattered}) gives 

\begin{equation}\label{integral}
N_s=\int{[1-\vec{\beta}(\vec{p}_0)\cdot\vec{e_i}]\,n_p(\vec{x},t)n_e(\vec{x},t)}\int_{\Delta\Omega}{\frac{\tilde{\sigma'}\left(P_\phi^{0,3}P_\theta^{1,2}-P_\theta^{0,3}P_\phi^{1,2}\right)}{\sin{\theta}\left[1+\left(\frac{\Lambda_\nu^2 k^\nu }{\Lambda_\nu^1 k^\nu }\right)^2\right]\left(\Lambda_\nu^0 k^\nu\,\Lambda_\nu^1 k^\nu\right)^2}}\,d\Omega\,cdt\,d^3x \,.
\end{equation} 

In the matrix $\Lambda_\nu^\mu=\Lambda_\nu^\mu(\beta)$, $\beta$ must be considered as $\beta=\frac{\vec{p_0}}{\sqrt{m^2c^2+p_0^2}}$. The quantity $n_e(\vec{x},t)=g(\vec{x},t)$ is the density of the electron distribution. The integral (\ref{integral}) will be used in section 4.4 to choose the optimal interaction point of the set-up and to estimate the number of scattered photons.

\section{Experimental set-up with collimated laser injection}

\begin{figure}
\begin{center}
\includegraphics[scale=0.55]{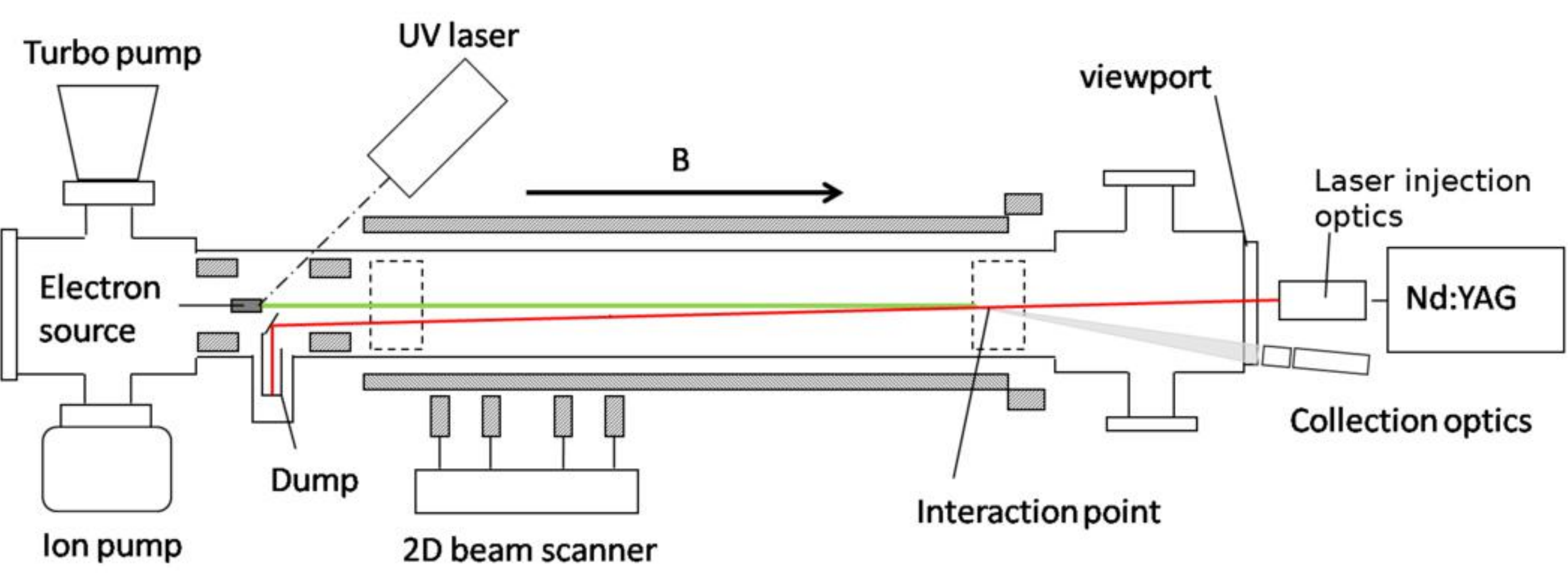}
\end{center}
\caption{\label{basicconf}Schematics of Thomson scattering diagnostics set-up implemented in Eltrap . A pulsed ($<4$ ns) electron bunch is produced by a UV laser with a photocathode. The beam is focused by a magnetic field B with a maximum intensity of 0.2 T. A 2D beam scanner moves the bunch trajectory in the transverse plane passing through the interaction point and interacts with a collimated laser pulse (Nd:Yag at 1064 nm). The scattered radiation is collected by a photomultiplier based optics.}
\end{figure}

Thomson backscattering diagnostics was set up on the Eltrap apparatus described in Chapter 2. This diagnostics is characterize by four main systems that provide a solution to the problem of the laser injection, the collection of the scattered radiation, and the interaction of the laser with the bunch. In the basic experiment configuration (see fig. \ref{basicconf}) a bunch produced by the photocathode source illuminated with a UV radiation (337 nm) is accelerated by a potential 1-20 kV and interacts with  a high power Nd:Yag laser ($\approx1$ J energy and 1064 nm ) in the vacuum chamber. In this set-up the laser was injected maintaining collinear the laser beam. With this choice the interaction point can be changed rapidly along the drift tube (changing the timing of the IR laser), but the spatial resolution is lost (we can't measure the density profile because the laser interacts with all the electrons of the beam) and the cross sections of the electron and laser beams are not matched. The interaction of the laser with the bunch occurs by means a 2D  beam scanner that moves the bunch trajectory in both directions (X and Y) in the transverse plane. The interaction in the Z direction is optimized to have the maximum probability with an appropriate sinchronization between the IR and UV lasers. The scattered light is collected by an optical system optimized to reduce the stray light produced with the laser dump. 

\subsection{Laser injection}

\begin{figure}
\centering
\subfigure[]
{\includegraphics[scale=0.55]{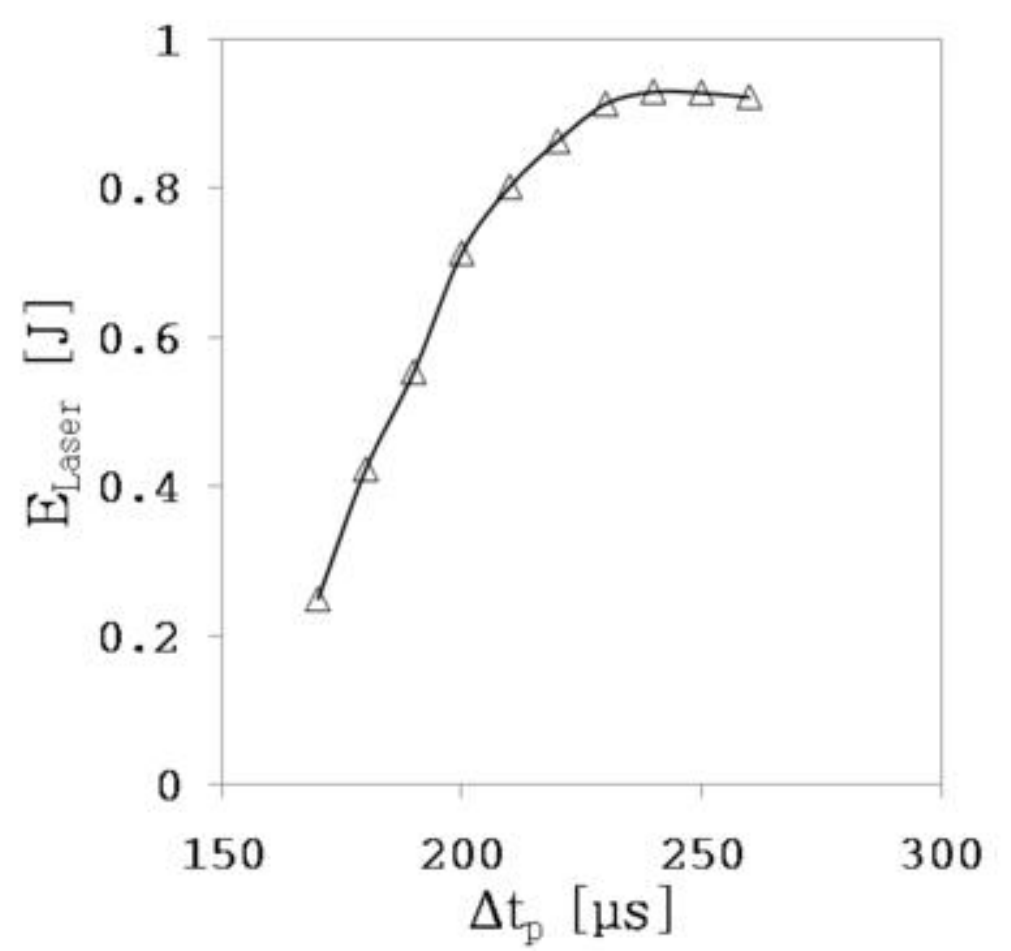}}
\hspace{5mm}
\subfigure[]
{\includegraphics[scale=0.5]{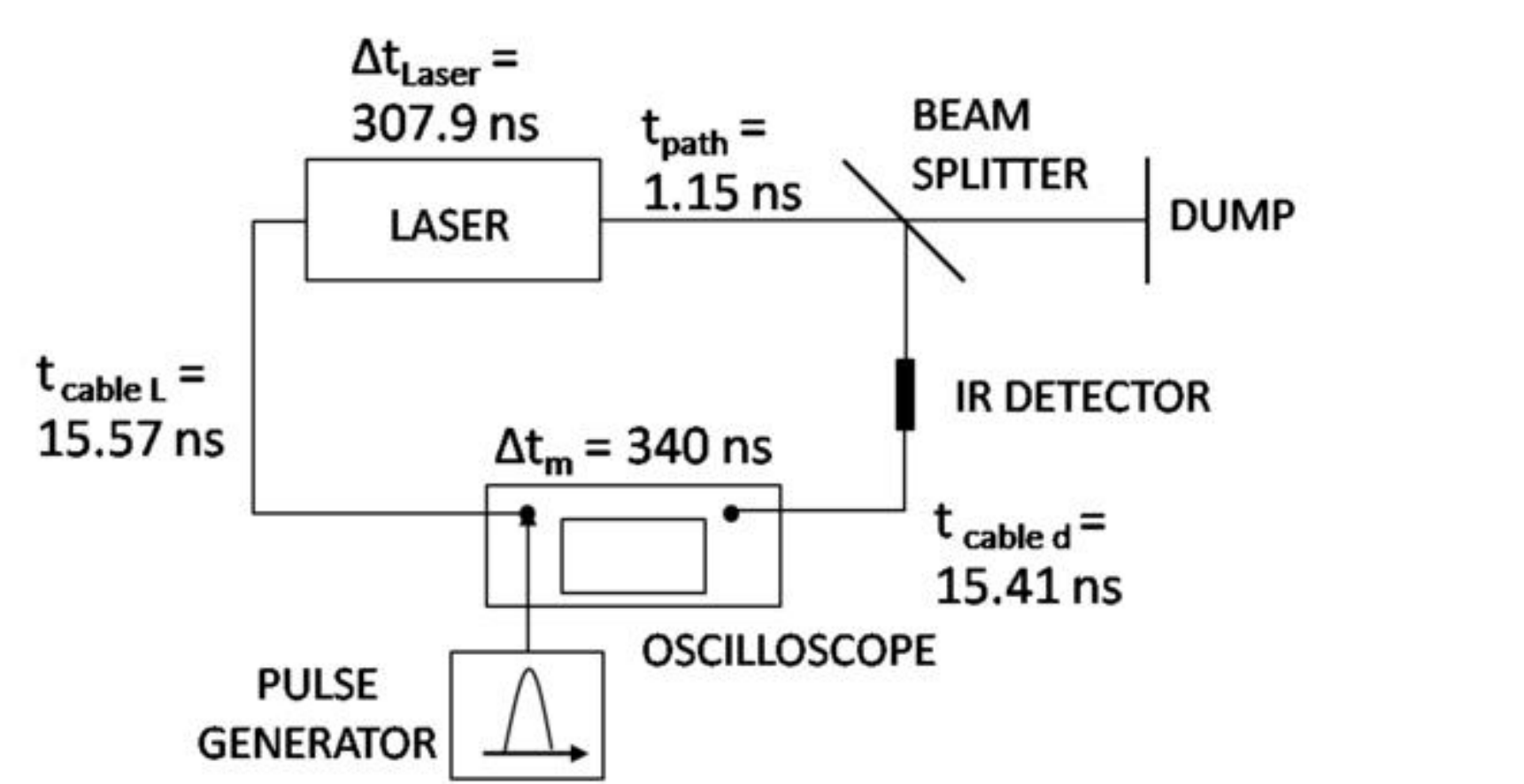}}
\caption{\label{laserchar}(a) Laser energy vs pumping time measured with pyroelectric sensor. (b) Schematics of the laser delay characterization set-up. The delay is evaluated measuring the time $\Delta t_m$ and compensating the delays introduced by the cables ($t_{cable L}$, $t_{cable d}$) and by the optical path $t_{path}$.}
\end{figure}

\begin{figure}
\begin{center}
\includegraphics[scale=0.6]{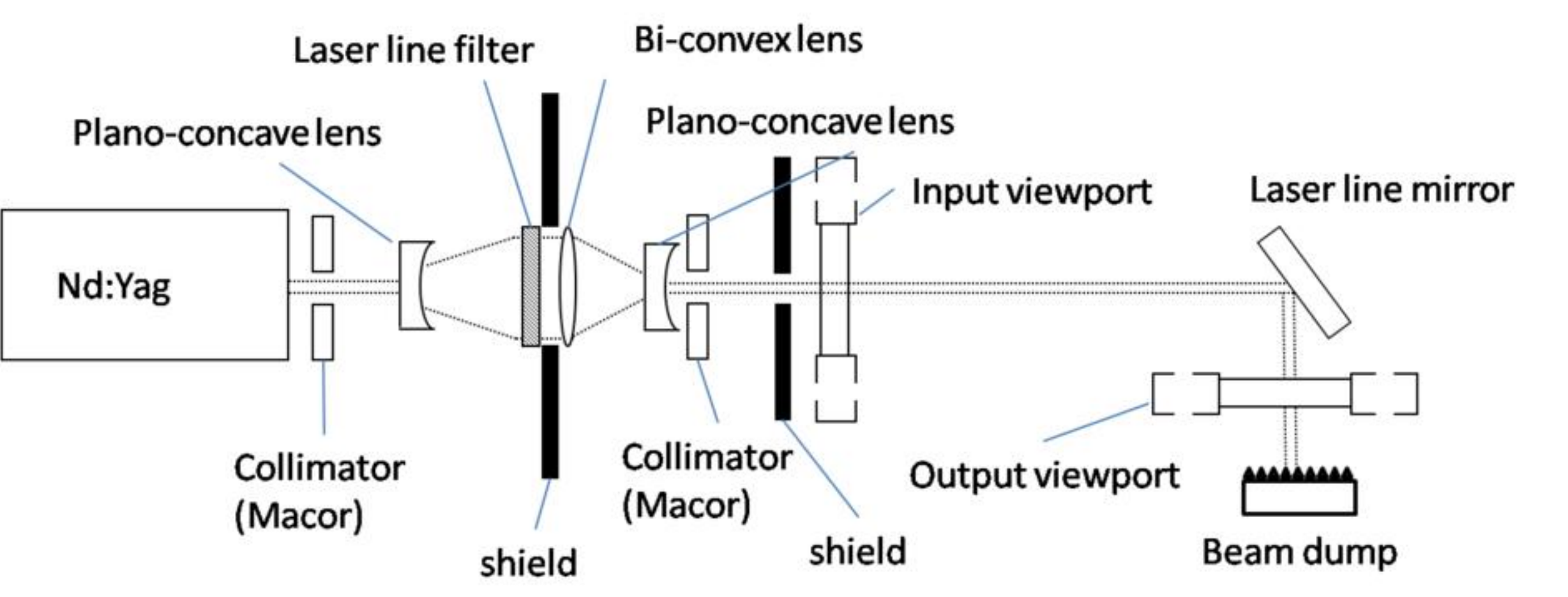}
\end{center}
\caption{\label{opticscheme}Optical system for the IR laser injection in the vacuum chamber. A power filter (composed by three lenses and a laser line filter) reduce the stray-light of the pumping flash-lamp. The injected laser is then dumped by a laser line mirror.}
\end{figure}

\begin{figure}
\begin{center}
\includegraphics[scale=0.7]{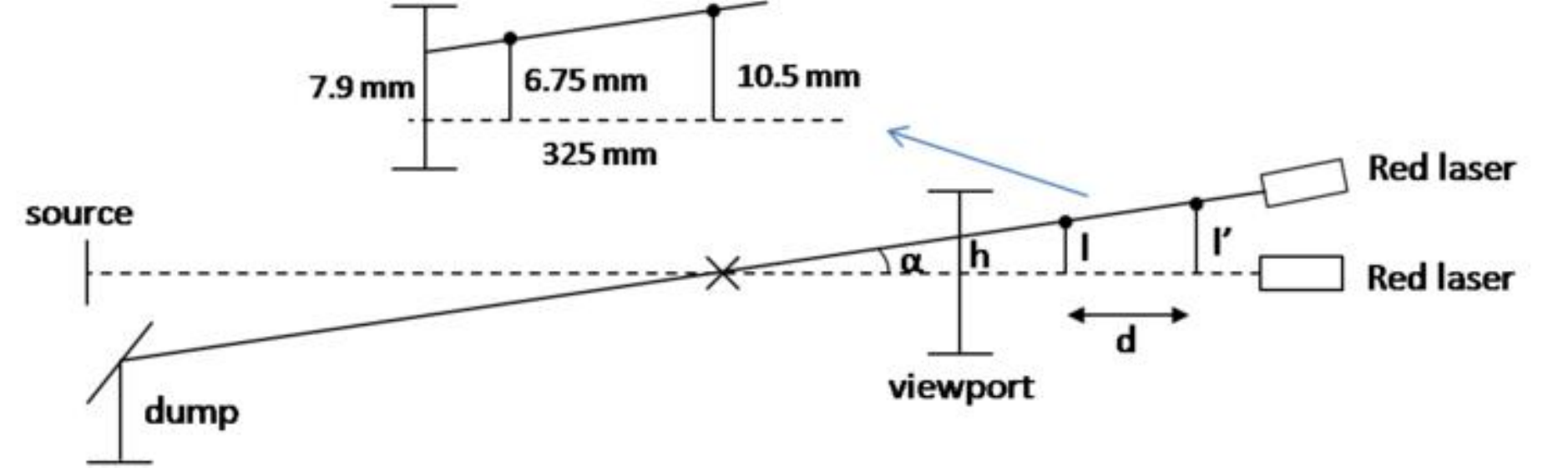}
\end{center}
\caption{\label{alignment}Geometry of the laser trajectory in the vacuum chamber.The intersection between the laser trajectory and the geometrical axis is at about 68 cm from the viewport along the axis. The collimators are at a distance from the axis of $l$, $l'$. The laser interacts with the viewport at $h =7.9$ mm from the axis. The distance between the collimators is $d=325$ mm.}
\end{figure}

\begin{figure}
\centering
\subfigure[]
{\includegraphics[scale=0.7]{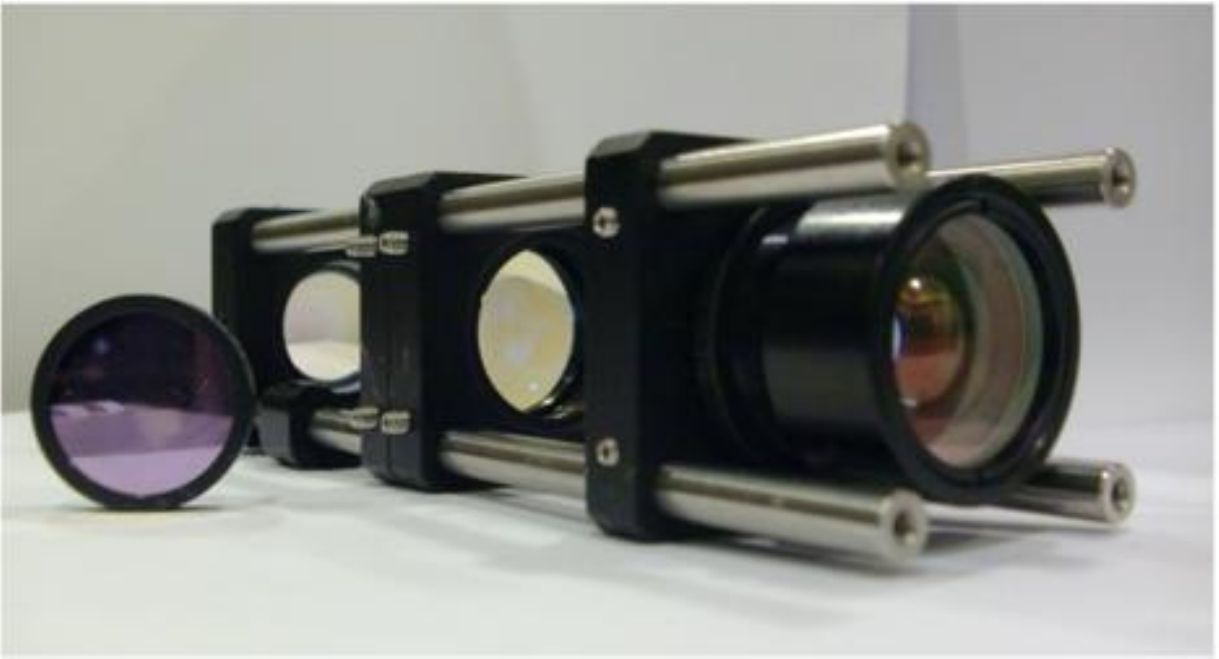}}
\hspace{5mm}
\subfigure[]
{\includegraphics[scale=0.5]{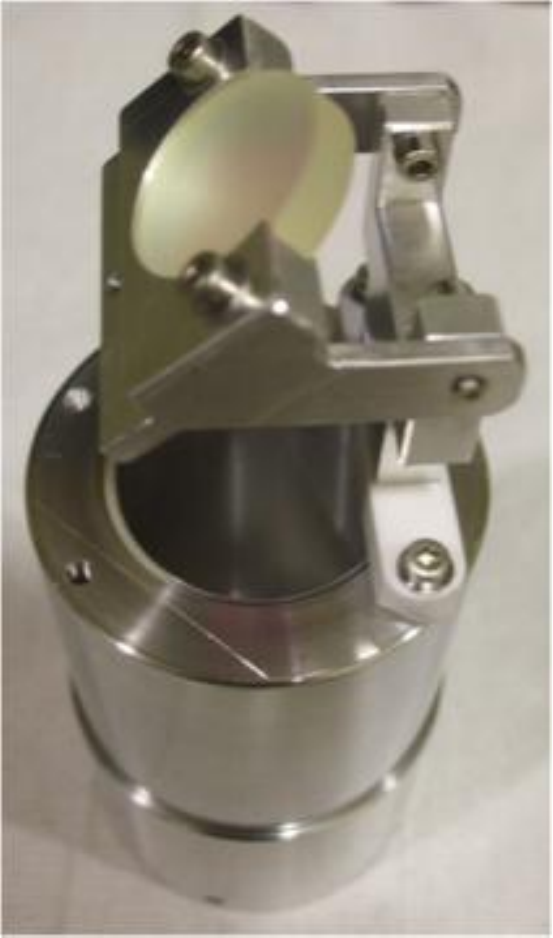}}
\caption{\label{filterdump}(a) Picture of the high power filter to reduce the stray-light produced during the flash of the laser pumping lamp. On the left the dichroic laser line filter centered at 1064 nm.   (b) Picture of the high power laser line mirror used to deviate the laser beam out of the chamber. The support is optimized to free up the top edge of the lens.}
\end{figure}

The source of the incident radiation used in this diagnostics is a Nd:Yag laser with a wavelength of 1064 nm an energy pulse of $\approx 1$ J and a repetition rate of 10 Hz. The jitter of the laser $\approx 1$ ns is less than the characteristic time of both the laser and bunch pulses. The pumping time $\Delta t_p$   can be changed to optimize the laser energy and the maximum value of 0.929 J is obtained for $\Delta t_p=240$ $\mu$s. The energy measurements (see fig. \ref{laserchar} (a)) was performed with a high energy pyroelectric sensor with a measurement uncertainty of $5\%$. In order to synchronize the laser with the bunch we characterized the delay $\Delta t_{laser}$ between the IR laser trigger and the laser beam emission with a precise pulse generator (jitter $< 500$ ps). The delay was computed measuring the time between the detection of the IR radiation (with a fast PIN photodiode) and the sending of the trigger and compensating the delay of the cables and the laser path (see fig. \ref{laserchar} (b)). The time $\Delta t_p$ is so given by  
$$
\Delta t_{laser}= \Delta t_m - (t_{cable L}+t_{path}+t_{cable d})
$$
where $\Delta t_m=340$ ns is the time interval measured on the oscilloscope, $t_{path}$ in the laser path from the output port to the detector, $t_{cable L}$ and $t_{cable d}$ are the delays of the cables of the laser trigger and of the IR detector, respectively. The resulting time delay is $\Delta t_{laser}=307.9$ ns. This time is not a constant and in general change varying the pumping time $\Delta t_p$, in this case the measurement was performed at $\Delta t_p=170$ $\mu$s corresponding to a laser pulse energy of $\approx$ 0.25 J. The optics for the laser injection (see fig. \ref{opticscheme}) is needed to reduce the visible light  generated by the flash lamp of the laser, and the stray light produced during the laser beam dump. The alignment of the laser is done by means of a red laser pointer, tracing the laser trajectory from a point on the viewport to the centre of the final beam dump (see fig. \ref{alignment}) and marking the trajectory with two collimators. To check if this trajectory is in agreement with the geometry of the system, we measure the angle between the laser trajectory and the geometrical axis. The reference of the geometrical axis is taken aligning the red laser between the viewport center and the center of the electron source. The measured angle is $\alpha=\arctan (\frac{l'-l}{d})=0.661^\circ$, where d is the distance between the two collimators measured along the geometrical axis and $l'$, $l$ are the distances between the geometrical reference and the first and second collimators, respectively. The measurement is in agreement with the technical drawing $\alpha=0.662^\circ$, corresponding to an interaction point at a distance $\approx 68$ cm from the viewport. The high power filter (see fig. \ref{filterdump} (a)) is positioned between the two collimators. It is formed by a first lens that diverges the laser beam in order to reduce the power per unit area. The light is filtered by a band-pass  dichroic filter centered at a wavelength of 1064 nm and with a band width of 10 nm. This filter with an optical density (OD ) $< 4$ strongly reduce the light with frequencies different from the laser line frequency. The filtered radiation is refocused and collimated by a convergent-divergent optics. Two shields positioned after the dichroric filter and close to the viewport are of fundamental importance to reduce the stray light coming from the laser port and from the viewport  when the laser is injected. The maximum working energy of the filter is 0.25 J for a laser pulse of 5 ns. After the injection the laser is diverted out from the vacuum chamber by means a dichroic high power mirror designed for Nd:Yag laser line. The mirror was mounted on a particular support (see fig. \ref{filterdump} (b)), in order to bring the mirror as close as possible to the source. The characterization of the stray-light, produced by the input viewport interacting with the laser, is needed because the optics collection to detect the scattered light is positioned close to it. A reflection test was performed in order to estimate the number of collected photons. A photomultiplier (PMT) with a gain of $10^5$ was positioned as close as possible to the laser beam $\approx 3.4$ cm and oriented toward the point of intersection between the laser trajectory and the second face of the viewport. To reduce the intensity of the collected light two filters with a total attenuation of $10^8$ (at 1064 nm)  were positioned in front of the PMT. The number of collected photoelectrons was of the order of $\approx 10^3$. This means that the stray-light produced in the laser-viewport interaction is minimized with a filter of optical density $\geq 11$.  

\subsection{Optical collection}
\begin{figure}
\begin{center}
\includegraphics[scale=0.8]{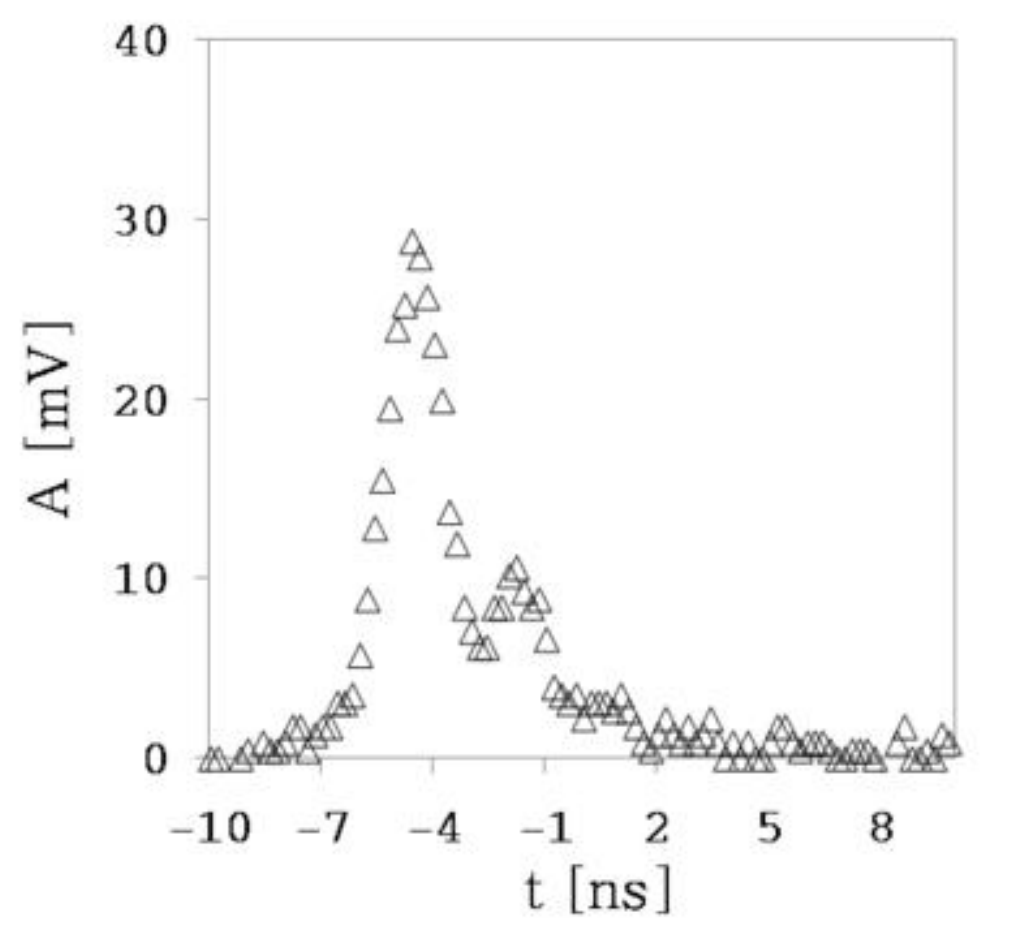}
\end{center}
\caption{\label{singlephoton}Single-photon counting pulse measured closing the PMT in a black box and with a PMT gain of $10^6$. The FWHM of the characteristic pulse is $\approx 3$ ns.}
\end{figure}

The scattered photons were detected by a PMT filtering the radiation out of the spectrum of interest in order to increase the signal to noise ratio. The filters were taken to leave a pass band in the visible range around 650 nm and to reduce drastically the main source of the stray light, i.e. the incident radiation in the IR region and the UV radiation of the electron source (at 337 nm). For this purpose, two filters were placed in front of the PMT. The first filter is a composed colored filter with an optical density OD$\geq 7$ in the intervals 190 - 534 nm, 960 - 1064 nm, an optical density OD$\geq 5$ in the range 850 - 925 nm and an optical density OD $\geq 6$ in the range 925 - 1070 nm, while the transmittance is 35$\%$. The second filter is a dichroic short-pass filter with a cut-off wavelength at 750 nm, an optical density OD$\geq 4$ and a transmission coefficient of 0.85. With these two filters is guaranteed a total optical density $\geq 11$ in the range 960 - 1064 nm for the attenuation of the stray-light produced by the IR laser and an optical density $\geq 7$ to attenuate the stray-light produced by the UV laser (at 337 nm). The total transmission coefficient, out of the rejection region, is $\approx 30\%$. The detector is a high gain, high stability  photomultiplier designed for fast time response. The active area is 3.8 cm$^2$ and the quantum efficiency at peak (400 nm) is 21$\%$. The quantum efficiency decreases by the wavelength (starting from 400 nm) and reaches a value of 5$\%$ at about 650 nm. The gain at nominal anode sensitivity (50 A/lm) is $3\cdot10^5$ with a corresponding dark count rate of 3000 s$^{-1}$ (at 20$^\circ$). The timing of a single photon pulse was characterized putting the PMT in a closed box and illuminating it with a red light by a current regulated light emitting diode, with a PMT gain of $\approx 10^6$ and measuring the voltage signal with a 1 GHz band width oscilloscope on a impedance load of 50 $\Omega$. The single-photon counting pulses have a FWHM of $\approx3$ ns (see fig. \ref{singlephoton}). The transit time of the PMT, plus the delay introduced by its cable $\Delta t_{PMT}$, was characterized in order to know the effective arrival time of the photons on the PMT windows with respect to the signal measured on the oscilloscope. To this purpose we used an UV laser with a characteristic  pulse time of $\leq 4$ ns. The light was splitted in two different directions. The main beam after an optical path of about 4.12 m reaches an optical diffuser and the diffused light is detected by the PMT at a distance of 1.32 m (from the optical diffuser). A second fraction of the laser beam, deviated by the beam-splitter, is collected by a fast UV detector. The time $\Delta t_{PMT}$   is computed knowing the optical path $L_{main}$ of the laser beam from the exit port of the laser to the PMT, the optical path $L_{split}$  of the laser from the exit port of the laser to the UV detector and the delay introduced by the cable $t_{cableUV}$, connecting the detector UV to the oscilloscope (see fig. \ref{laserchar} (b)).  The time $\Delta t_{PMT}$ is given by

\begin{equation}\label{pmtdelay}
\Delta t_{PMT} = \Delta t_m + \frac{L_{split}}{c} + t_{cableUV}-\frac{L_{main}}{c}= 36\,ns  
\end{equation}

where $\Delta t_m$ is the interval time measured on the oscilloscope between the two signals. The PMT with the filters was mounted at a distance of 3.4 cm with respect to the geometrical axis of the trap. 

\begin{figure}
\begin{center}
\includegraphics[scale=0.7]{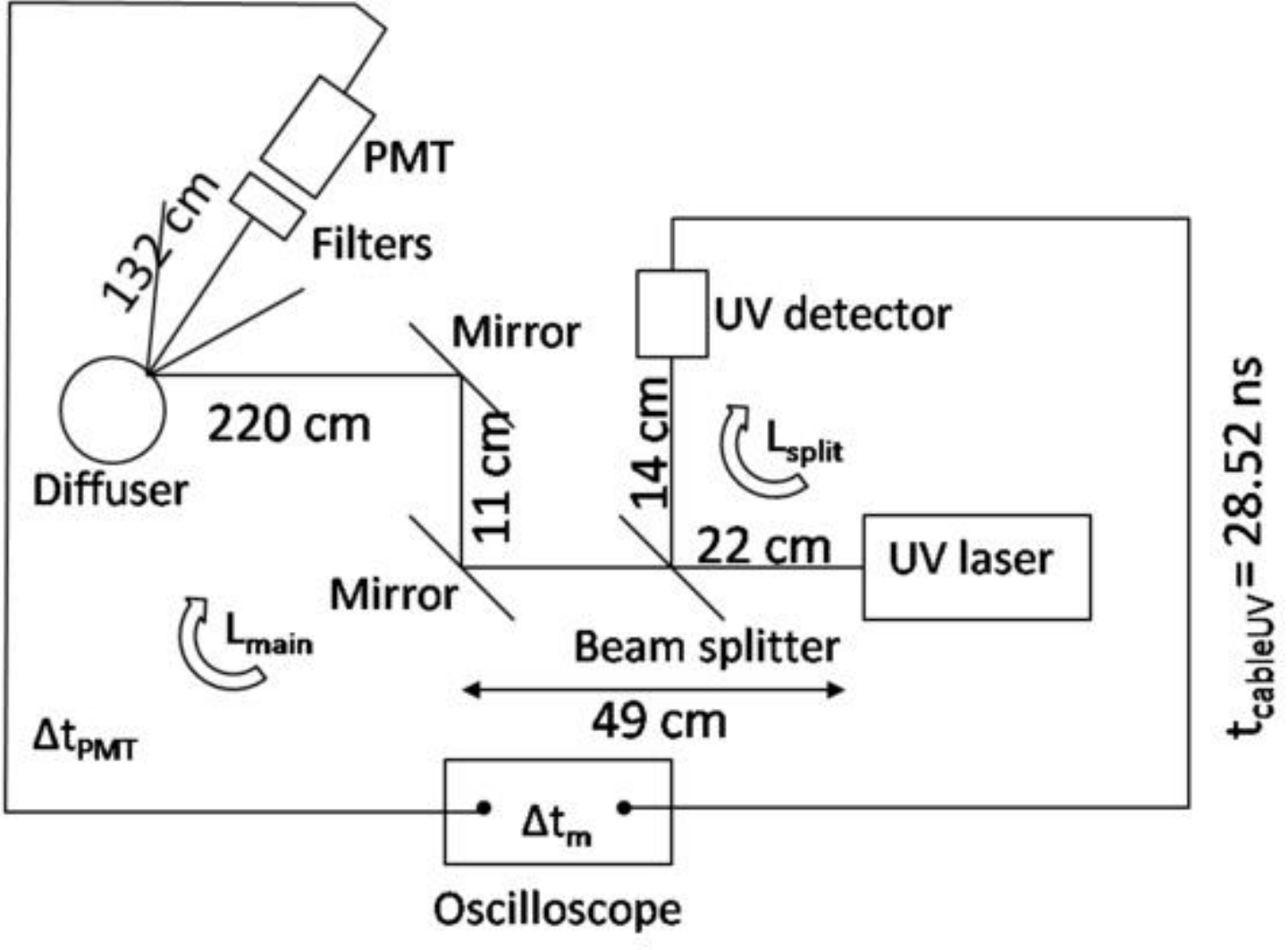}
\end{center}
\caption{\label{singlritardopmt}Experimental set-up to characterize the transit time of the PMT plus the delay introduced by its cable $\Delta t_{PMT}$. The time is computed knowing all time delays and the optical paths of the main beam $L_{main}$ and of the splitted beam $L_{split}$.}
\end{figure}

\subsection{Space coincidence}

The bunch trajectory that is aligned to the main magnetic field has a small misalignment with respect to the geometrical axis, i.e. the electron bunch reaches the end of the trap electrode stack with an appreciable offset. The correction of the bunch trajectory to obtain the spatial coincidence of the bunch with the laser in the transversal plane passing through the interaction point is realized with a two dimensional beam scanner that steers the bunch trajectory in both directions (X and Y). The scanner acts using two pairs of dipole coils that introduce small corrections of the direction of the main magnetic field. The maximum deflection of the bunch was characterized at the end of the trap with the optical diagnostics. The intensity of the currents needed to correct the offset of the bunch with respect to the geometrical axis was characterized moving the center of the bunch spot (that is acquired with the optical diagnostics) with respect to the center of the phosphor screen (that pass through the geometrical axis), varying both the magnetic field and the bunch energy. The data are reported in tables \ref{deflect50},  \ref{deflect100} and \ref{deflect150} for three values of the magnetic field strength 170 G, 330 G and 500 G and for bunch energies from 5 to 16 keV. The two currents called $I_A$ and $I_D$ are the currents corresponding to the coil A, that acts moving the bunch in the X direction, and the coil D that acts moving the bunch in the Y direction. In the tables we report the values of the currents $I_{A\,center}$, $I_{B\,center}$ needed to center the bunch spot on the phosphor screen, the currents $I_{A\,max}$, $I_{A\,min}$ to move the bunch spot in the X direction at a distance from the center of 5 cm and -5 cm respectively and the currents $I_{D\,max}$, $I_{D\,min}$  to move the bunch spot in the Y direction at a distance from the center of 5 cm and -5 cm respectively. 


\begin{table}[htbp]
\begin{center}
\begin{tabular}{|l||l||l||l||l||l||l|}
\hline
 
Energy $[keV]$ & $I_{A\,center} [A]$ & $I_{D\,center} [A]$ & $I_{A\,max} [A]$ & $I_{A\,min} [A]$ & $I_{D\,max} [A]$ & $I_{D\,min} [A]$ \\
\hline
5 & 0.13 & 0.80 & -0.49 & 0.71 & 0.07 & 1.48 \\
\hline
6 & 0.14 & 0.93 & -0.34 & 0.61 & 0.24 & 1.43 \\
\hline
7 & 0.03 & 0.96 & -0.37 & 0.44 & 0.37 & 1.42 \\
\hline
8 & 0.01 & 0.91 & -0.37 & 0.39 & 0.37 & 1.37 \\
\hline
9 & 0.00 & 0.86 & -0.33 & 0.37 & 0.37 & 1.32 \\
\hline
10 & 0.02 &0.86 & -0.30 & 0.37 & 0.39 & 1.28 \\
\hline
12 & 0.03 & 0.83 & -0.29 & 0.36 & 0.36 & 1.22 \\
\hline
14 & 0.04 & 0.79 & -0.29 & 0.36 & 0.34 & 1.22 \\
\hline
16 & 0.04 & 0.79 & -0.28 & 0.36 & 0.32 & 1.22 \\
\hline
\end{tabular}
\end{center}
\caption{Values of the currents of the coils A and D needed to correct the initial offset of the bunch to the center ($I_{A\,center}$, $I_{D\,center}$), to move the bunch along the X direction at a distance of $\pm 5$ cm from the center ($I_{A\,max}$, $I_{A\,min}$) and to move the bunch along the Y direction at a distance of $\pm 5$ cm from the center ($I_{D\,max}$, $I_{D\,min}$). The measurements was taken for a magnetic field of 170 G and for bunch energies in the range 5 - 16 keV.}
\label{deflect50}
\end{table}


\begin{table}[htbp]
\begin{center}
\begin{tabular}{|l||l||l||l||l||l||l|}
\hline
 
Energy $[keV]$ & $I_{A\,center} [A]$ & $I_{D\,center} [A]$ & $I_{A\,max} [A]$ & $I_{A\,min} [A]$ & $I_{D\,max} [A]$ & $I_{D\,min} [A]$ \\
\hline
5 & 0.04 & 1.33 & -0.12 & 0.43 & 0.77 & 1.79 \\
\hline
6 & 0.29 & 1.30 & -0.16 & 0.70 & 0.68 & 1.86 \\
\hline
7 & 0.30 & 1.33 & -0.18 & 0.80 & 0.59 & 1.97 \\
\hline
8 & 0.32 & 1.33 & -0.24 & 0.92 & 0.46 & 2.09 \\
\hline
9 & 0.34 & 1.30 & -0.29 & 1.04 & 0.28 & 2.26 \\
\hline
10 & 0.37 &1.30 & -0.42 & 1.19 & 0.10 & 2.38 \\
\hline
12 & 0.40 & 1.27 & -0.62 & 1.48 & -0.26 & 2.66 \\
\hline
14 & 0.42 & 1.27 & -0.81 & 1.66 & -0.54 & 2.86 \\
\hline
16 & 0.39 & 1.33 & -0.84 & 1.65 & -0.56 & - \\
\hline
\end{tabular}
\end{center}
\caption{Values of the currents of the coils A and D needed to correct the initial offset of the bunch to the center ($I_{A\,center}$, $I_{D\,center}$), to move the bunch along the X direction at a distance of $\pm 5$ cm from the center ($I_{A\,max}$, $I_{A\,min}$) and to move the bunch along the Y direction at a distance of $\pm 5$ cm from the center ($I_{D\,max}$, $I_{D\,min}$). The measurements was taken for a magnetic field of 330 G and for bunch energies in the range 5 - 16 keV.}
\label{deflect100}
\end{table}


\begin{table}[htbp]
\begin{center}
\begin{tabular}{|l||l||l||l||l||l||l|}
\hline
 
Energy $[keV]$ & $I_{A\,center} [A]$ & $I_{D\,center} [A]$ & $I_{A\,max} [A]$ & $I_{A\,min} [A]$ & $I_{D\,max} [A]$ & $I_{D\,min} [A]$ \\
\hline

8 & 0.01 & 1.86 & -0.53 & 0.55 & 1.05 & 2.51 \\
\hline
9 & 0.11 & 1.86 & -0.42 & 0.63 & 1.11 & 2.52 \\
\hline
10 & 0.15 &1.86 & -0.38 & 0.68 & 1.07 & 2.53 \\
\hline
12 & 0.19 & 1.80 & -0.32 & 0.78 & 1.02 & 2.57 \\
\hline
14 & 0.22 & 1.86 & -0.35 & 0.90 & 0.91 & 2.68 \\
\hline
16 & 0.25 & 1.86 & -0.41 & 1.02 & 0.80 & - \\
\hline
\end{tabular}
\end{center}
\caption{Values of the currents of the coils A and D needed to correct the initial offset of the bunch to the center ($I_{A\,center}$, $I_{D\,center}$), to move the bunch along the X direction at a distance of $\pm 5$ cm from the center ($I_{A\,max}$, $I_{A\,min}$) and to move the bunch along the Y direction at a distance of $\pm 5$ cm from the center ($I_{D\,max}$, $I_{D\,min}$). The measurements was taken for a magnetic field of 500 G and for bunch energies in the range 8 - 16 keV.}
\label{deflect150}
\end{table}

\begin{figure}
\begin{center}
\includegraphics[scale=0.45]{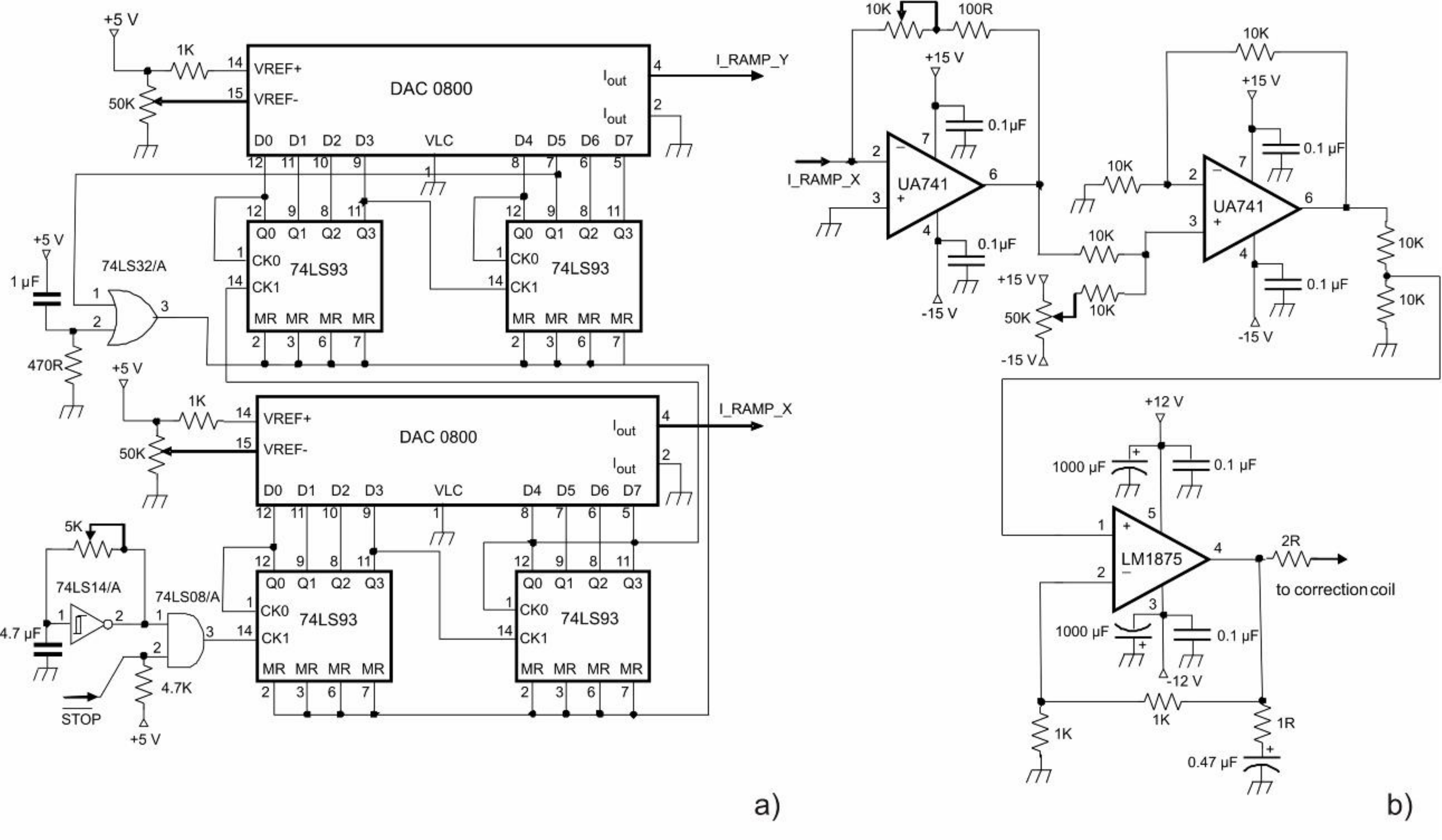}
\end{center}
\caption{\label{electroniccircuit}Electronic circuit of the 2D beam scanner. a) Two digital ramps are formed by four 74LS93 counters connected with two DAC0800 digital-to-analog converters and synchronized by a 74LS14 internal RC astable multivibrator. b) Each ramp is conditioned by two UA741 differential amplifiers ant the output current is buffered with a LM1875 amplifier. The current is then fed to the dipole coil through a 2 $\Omega$ resistor connected in series.}
\end{figure}

\begin{figure}
\begin{center}
\includegraphics[scale=0.8]{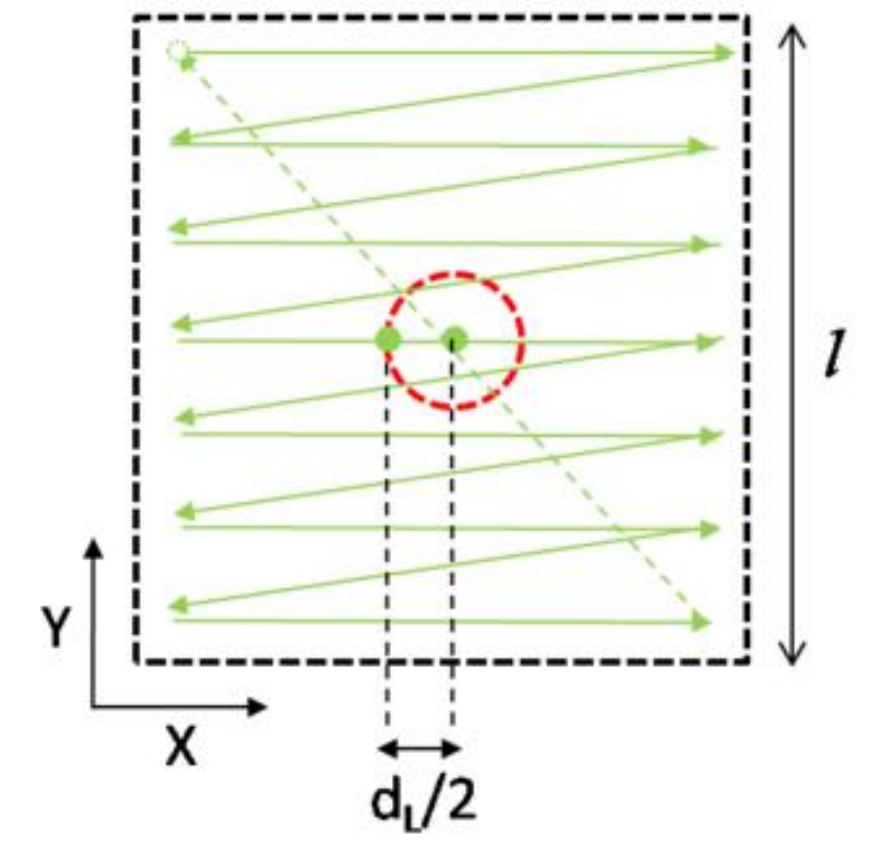}
\end{center}
\caption{\label{scanmethode}Scan mode of the bunch trajectory on a tranverse plane (X,Y). During the scan in a square window of side $l$, the bunch spot (green circle) overlap the IR laser spot (red circle). The position change of a quantity $d_L/2$ between two laser shots.}
\end{figure}

Using the maximum and minimum currents values we move the bunch trajectory on a square window with side 5 cm in the transverse plane. Because the alignment depends on the energy and the magnetic field, to prevent a continuous adjustment of the currents when the experimental parameters and the interaction point are changed, a continuous scan of the bunch trajectory in the windows is realized by  an automatic digitally controlled electronic circuit (see fig. \ref{electroniccircuit}) designed specifically for this purpose and realized with a photolithography technique. The currents $I_A$ and $I_D$ are generated starting from four circuit integrated digital counters 74LS93 synchronized by an internal oscillator realized with a 74LS14 astable multivibrator. The output signals of the DAC0800 digital-to-analog converters are conditioned by two UA741 operational amplifiers and are then amplified by a LM1875 linear current amplifier (for each stage). Each output current is monitored reading the potential on two resistors (with a value of 2 $\Omega$) connected in series with the correcting coils. Using a 74LS08 logic AND gate the internal oscillator can be stopped by a feedback signal and the digital outputs are stored in the counter outputs, i.e. the scan is stopped and the bunch trajectory remains fixed in the direction at the time of the stop action. In the set-up presented here this last function is not used and the system work in open loop configuration, i.e. when the scan of the windows is finished the system starts automatically another scan. With this system is possible to choose the scan rate, acting on the feedback resistor of the internal oscillator, as well as the window size and position.  Because the spot size $d_L$ of the IR laser beam has a diameter of $\approx 6$ mm and the repetition rate $f_L$ is 10 Hz we estimate from these parameters the scan time needed to cover the whole area of the scan windows. The distance traveled by the bunch between two IR laser shots should be exactly $d_L/2$. With this choice there is at least an overlap between the bunch and the laser spots during the scan as represented in figure \ref{scanmethode}. So the time $t_W$ to cover the whole square windows of side $l$ will be

\begin{equation}\label{scanwindow}
t_W= \left(\frac{l}{d_L/2}\right)^2 \frac{1}{f_L}
\end{equation}

Assuming $l=5$ cm we obtain $t_W\approx 28$ s. Note that this time is computed considering only the coincidence in the X, Y plane and it increases considering that the interaction obviously involves also the Z component along the geometrical axis. This problem is discussed in the next section.

\subsection{Temporal coincidence}

\begin{figure}
\begin{center}
\includegraphics[scale=0.8]{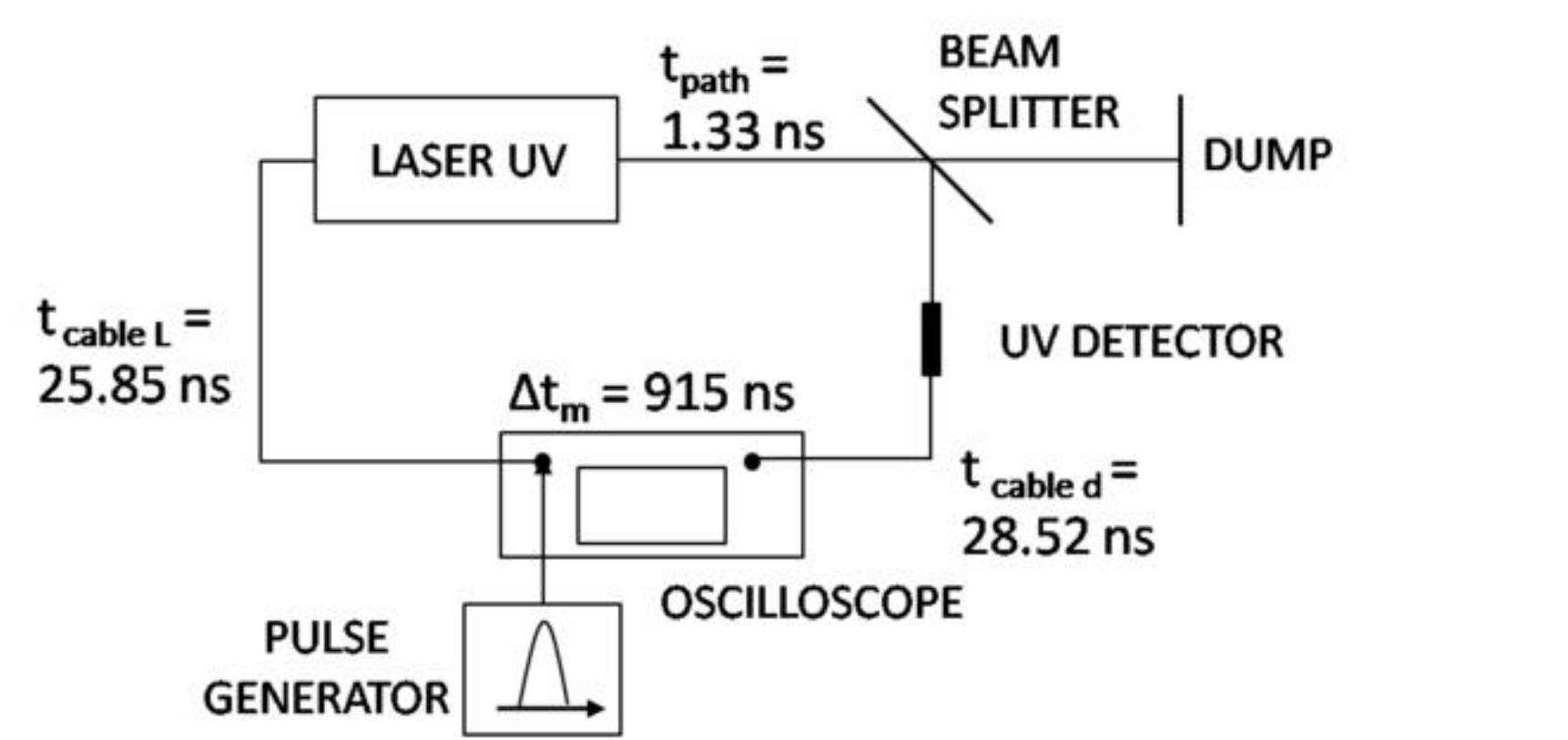}
\end{center}
\caption{\label{UVdelay}Sketch of the set-up for the jitter measurement of the UV laser. The delay is computed measuring the time $\Delta t_m$ between the signal from the UV detector and the signal from the pulse generator. The delays in the cables ($t_{cable L}$, $t_{cabled}$) and the delays in the optical path ($t_{path}$) must be considered.}
\end{figure}

\begin{figure}
\begin{center}
\includegraphics[scale=0.6]{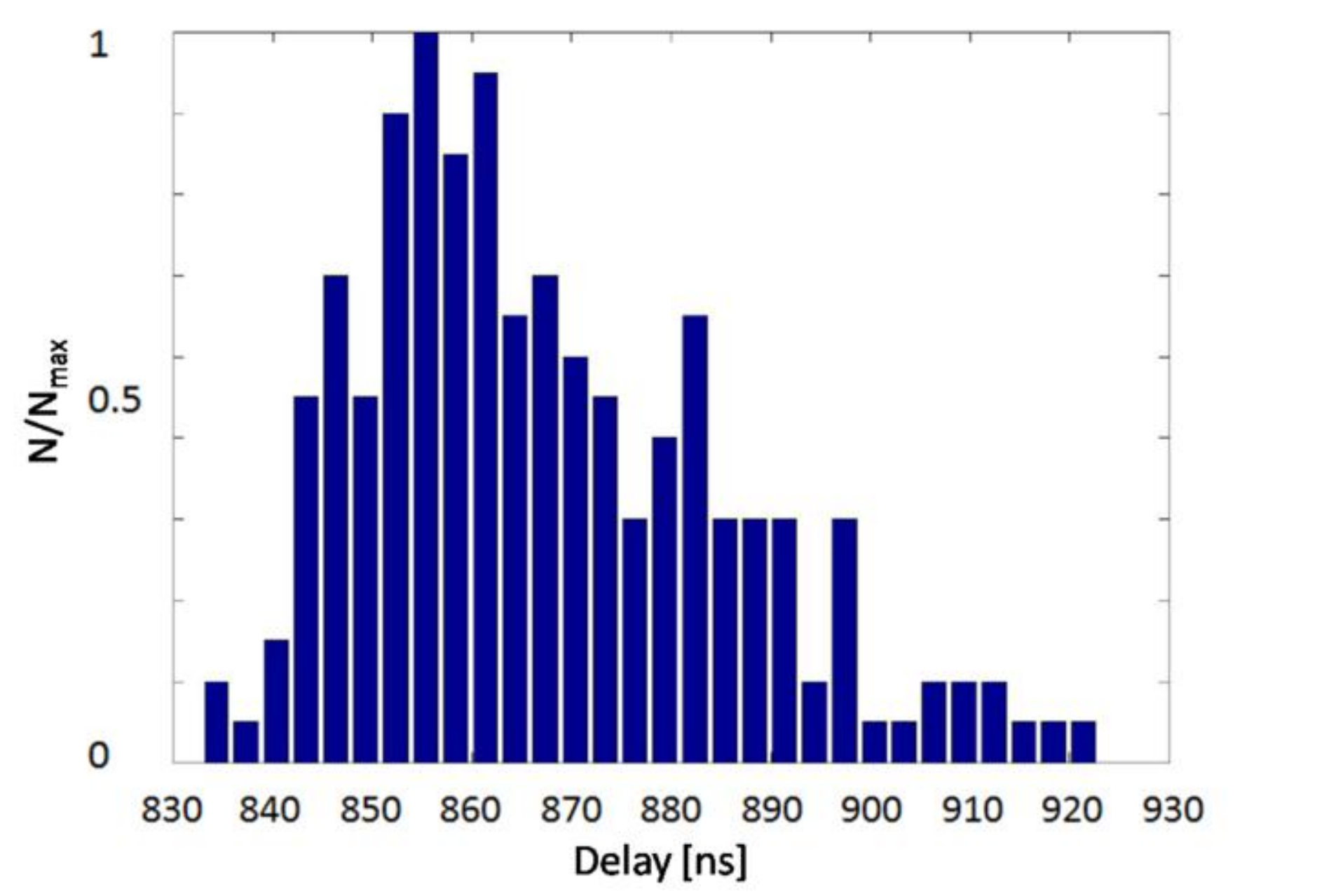}
\end{center}
\caption{\label{delays}Distribution of the delays of the UV laser normalized to the maximum peak (at $\approx859$ ns). The root mean square of the distribution is 20 ns.}
\end{figure}

The interaction along the geometrical axis of the system is determined trivially in ideal condition (i.e. without errors in the synchronization of the bunch and laser triggers and for a uniform motion of the bunch) searching a solution 
\begin{equation}\label{system}
Z_b(t-t')=Z_L(t-t'') 
\end{equation}

for $t'$, $t''$, where $Z_b(t-t')$ and $Z_L(t-t'')$ are the bunch and laser beam evolution in the space along the z axis. However the bunch emission and the laser fire are controlled by electrical signals propagating in cables and subject to delays in the lasers and in the cables themselves. These delays are generally affected by systematic and random errors. The sources of random behavior in the laser-bunch synchronization is due to the jitter of the device that produce the triggers $\sigma_D$ and the jitters of the IR and UV lasers ($\sigma_{IR}$, $\sigma_{UV}$). When these errors are negligible with respect to the characteristic time of the interaction, i.e. the pulse durations of both laser and electron beams, we can use the result of the system (\ref{system}) to synchronize the laser IR with the laser UV. In our case the IR laser and the delay generator used to generate the trigger signals have jitters of $\approx 1$ ns and $\approx 0.5$ ns respectively, but the UV jitter is $\approx20$ ns. The delays distribution of the UV laser was characterized with a fast UV detector specially designed to reach a fast response ($<1$ ns) and a small delay ($<1$ ns). The laser delays are estimated with the same configuration described in section 4.3.1 splitting a portion of the main beam toward the UV detector and measuring the delay between the signal of the detector and the trigger (generated by a pulse generator) and knowing the delays introduced by the cables and the optical path (see fig. \ref{UVdelay}).  The distribution of the delays (see fig. \ref{delays}) has a peak at 859 ns and a root mean square of 20 ns. Knowing the maximum of the distribution we synchronize the lasers to have the maximum probability for the laser-bunch interaction in a particular point.  For example considering an interaction at 68 cm from the viewport the time between the signal generated by the delay generator and the arrival of the laser beam to the interaction point is $\approx332$ ns (see fig. \ref{timecoincidence}), while the time between the signal generated by the delay generator and the arrival of the bunch at the interaction point is $\approx927$ ns (for a bunch energy of 10 keV), so the trigger must be generated with a delay of (927 - 332) ns in order, for the laser beam and the bunch, to arrive simultaneously in the interaction point.  We can now estimate the time $t_{scan}$ needed to obtain at least one interaction with a probability $p$ and with an error $\epsilon$. When the IR laser beam arrives at the interaction point the bunch arrives at a distance from the interaction point within $\pm \epsilon\, v_b$ ($v_b$ is the bunch velocity). Is obvious that if $\epsilon$ decrease is more improbable that  the delay of the UV laser is so that the bunch will arrive exactly between $z_0-v_b\epsilon$ and $z_0+v_b\epsilon$, i.e. in a intervall $2v_b \epsilon$ around the interaction point $z_0$. Considering a Gaussian distribution of the delays introduced by the UV laser $\frac{1}{\sigma\sqrt{2\pi}}\exp{-\frac{(t-t_0)^2}{2\sigma^2}}$ (where $t_0 =859$ ns and $\sigma=20$ ns) assuming that the synchronization is optimized so that when the laser beam arrives to the interaction point the bunch arrives simultaneously at the interaction point  with the maximum probability, the probability that the bunch arrives before or after a time less than $\epsilon$, with respect to the arrival time of the laser beam is 

$$
\frac{1}{\sigma\sqrt{2\pi}}\int_{-\epsilon}^{+\epsilon}{e^{-\frac{t^2}{2\sigma^2}}}\, .
$$

The probability that in $n$ shots the bunch arrive at a distance from the interaction point between $z_0-v_b\epsilon$ and $z_0+v_b\epsilon$ one or more times is

$$
p=\sum_{k=1}^{n} \left(^n_k\right) \left(\frac{1}{\sigma\sqrt{2\pi}}\int_{-\epsilon}^{+\epsilon}{e^{-\frac{t^2}{2\sigma^2}}}\right)^k \left(1-\frac{1}{\sigma\sqrt{2\pi}}\int_{-\epsilon}^{+\epsilon}{e^{-\frac{t^2}{2\sigma^2}}}\right)^{n-k}
$$

where $(^n_k)=\frac{n!}{k!(n-k)!}$ is the usual binomial coefficient. This probability is a function of $n$ for a fixed value of $\epsilon$. With the definition $F(n;\epsilon)\equiv p$ the inverse function $F^{-1}(p;\epsilon)$ is a function of $p$ for the same fixed value $\epsilon$. The meaning of this function is that for a probability $p$ and an error $\epsilon$ we need $n=F^{-1}(p;\epsilon)$ shots to have at least one (i.e. one or more) interactions in the interval $( z_0-v_b\epsilon, z_0+v_b\epsilon)$. Considering this result and the time needed to cover a windows of side $l$ computed in equation (\ref{scanwindow}) we get
$$
t_{scan}(p,\epsilon)=\left(\frac{l}{d_L/2}\right)^2 \frac{1}{f_L} F^{-1}(p;\epsilon) 
$$
Considering the parameters $l=5$ cm, $d=6$ mm, $f_L=10$ Hz, $p\approx70\%$, $\epsilon=1$ ns and $\sigma=20$ ns the resulting scan time is 14 min. We have to consider that this is a maximum value because in general a smaller scan window is needed.  

\begin{figure}
\begin{center}
\includegraphics[scale=0.7]{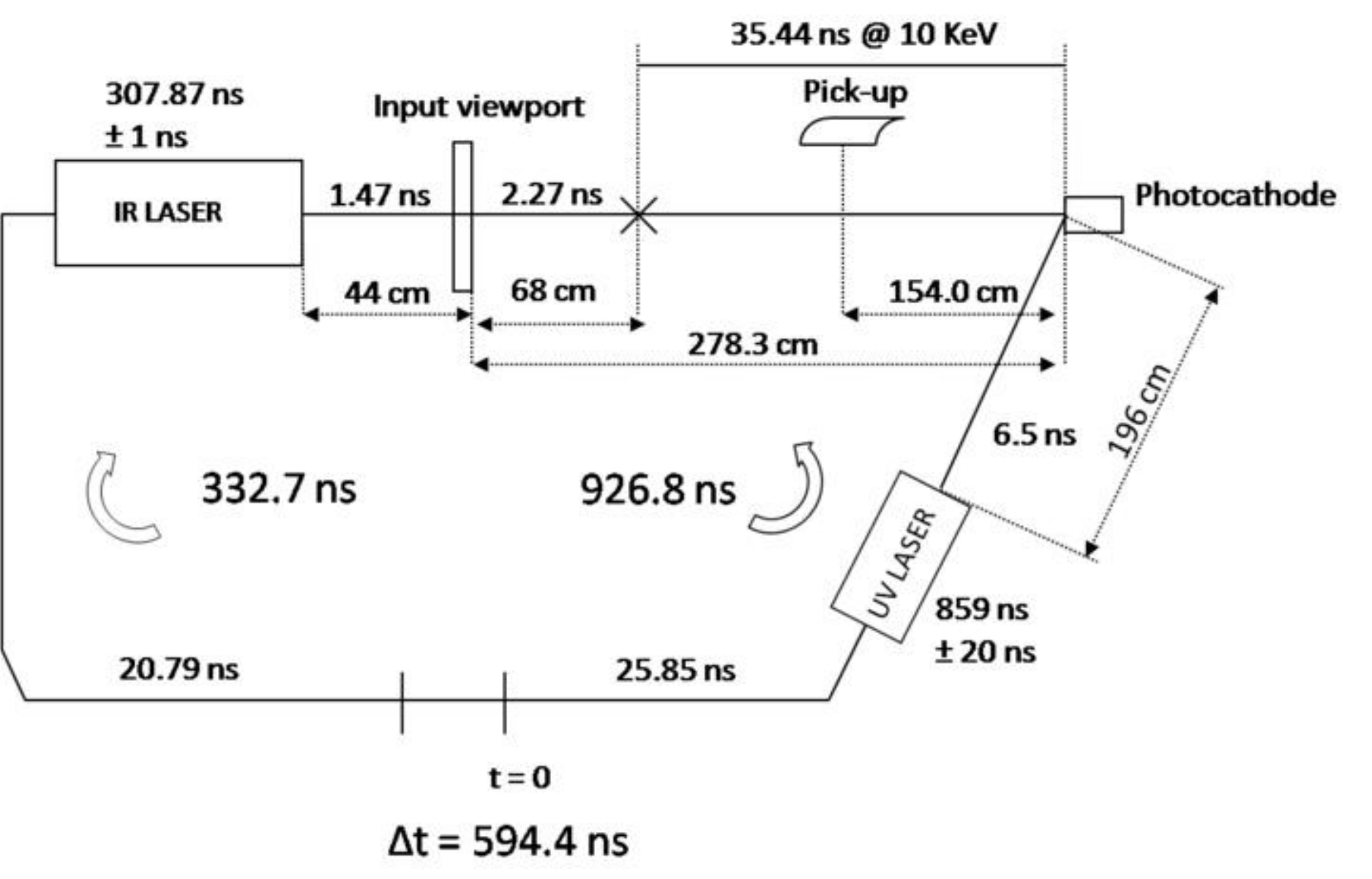}
\end{center}
\caption{\label{timecoincidence}Schematic of the laser-bunch synchronization. The total times are $926.8$ ns from $t=0$ to the interaction point (right branch) and $332.7$ ns from $t=0$ to the interaction point (left branch). The difference $926.8-332.7$ ns is the delay between the trigger of the UV laser and the trigger of the IR laser for the synchronization.}
\end{figure}


\subsection{Timing and noise measurements}

\begin{figure}
\begin{center}
\includegraphics[scale=0.7]{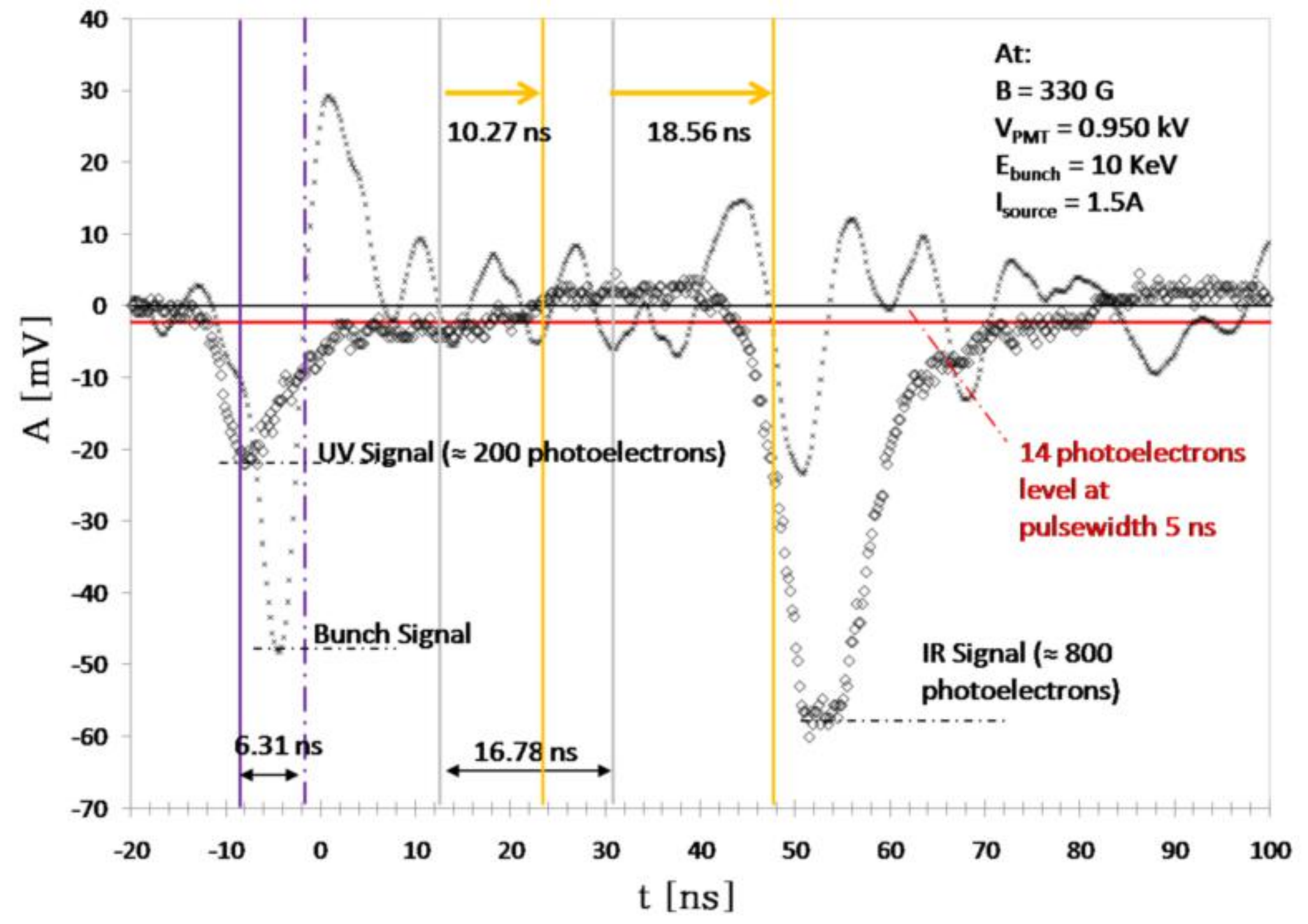}
\end{center}
\caption{\label{primoeltest}Timing and noise measurements. The signal measured with S4 (crosses) is taken as reference (violet dasched line) to check the syncronization. The peak of the UV light acquired with the PMT (diamonds) is expected  at 6.31 ns before the reference. The scattered radiation is expected between the gray lines for an interaction between the viewport and S4. The orange lines are the times at which is expected the  edge of the IR laser signal for an interaction between S4 (first line) and the viewport (second line). The noise level (RMS) between the gray lines is $\approx2.3$ mV (i.e. 14 photoelectrons for an interaction of 5 ns and a PMT gain of $10^5$).}
\end{figure}

The time $t_{S4}$ measured with the non destructive electrostatic diagnostics described in section 3.3 is used as a reference to check the synchronization in the laser-bunch interaction. Both the UV and the IR laser beams are visible in the PMT signal (produced by the impact of the UV laser on the photocathode and the impact of the IR laser on the beam dump) because the stray-light noise is not totally filtered and so we can compare the instants corresponding to the UV and laser pulses with the time $t_{S4}$. We expect that the UV signal is temporally fixed with respect to $t_{S4}$ because the bunch emission is a consequence of the impact of the UV laser beam on the photocathode, while the IR signal change continuously its time with respect to $t_{S4}$ because (as described in the previous section) the interaction occur with an error given by the UV jitter of 20 ns. Timing and noise measurements was taken with the present geometrical configuration, for a magnetic field of 330 G and a bunch energy of 10 keV. With a delay time $\Delta t$ of 594.4 ns between the IR laser trigger and the UV laser trigger the maximum probability of interaction occur at 68 cm from the viewport. Increasing this time the interaction point (i.e. the point where the interaction occur with maximum probability) moves toward the viewport while reducing it the point moves towards the source (along the direction of the laser beam trajectory). In figure \ref{primoeltest} we report an example of measurement with both the S4R and PMT signals. The S4R signal is the signal produced by the induced current due to the passage of the bunch and in particular we are interested to the zero crossing of the signal between the minimum and the maximum peaks, that is the instant when the bunch is centered in S4. In the PMT signals (obtained with a PMT gain of $10^5$) we have two pulses temporally separated. The first pulse is the stray-light noise of the UV laser beam, and it is always fixed in time, the second pulse is the stray-light noise of the IR laser beam and change its time position at every shot.   The violet dashed line is the reference time for all following evaluations and pass through the zero of the induced current signal measured on S4R $t_{ref}$=-1.7 ns. The vertical violet line is the computed instant (considering all delay times of the light propagation and of the signals in the cables) when is expected the maximum peak of the UV signal with the following assumptions:
i)	The UV light comes from the electron source during the bunch emission.
ii)	The bunch propagates with a uniform motion with energy $E=10$ keV and the bunch spread due to the space charge effects is negligible.
The vertical gray lines are at the ends of a temporal windows of width 17 ns, where we expect the observation of the signal of the scattered light produced in the laser-bunch interaction. Assuming that the interaction can occur between the center of S4 (first line) and the viewport (second line). The vertical orange lines are the instant when we expect the edge of the IR signal detected by the PMT. The first line when the interaction occur at the center of S4, the second line when the interaction occur near to the viewport. Note that the first line is shifted of 10.27 ns from the first gray line and the second orange line is shifted of 18.56 ns from the second gray line, this means that the scattered light is never overlapped with the IR pulse (with an interaction between the viewport and S4) assuming that the origin of the stray-light noise of the incident radiation comes from the beam-dump is very close to the electron source. The scattered radiation is not measureable with the electron source used in this test because the density is not sufficiently high to be observable with the actual noise (i.e. electronic noise, stray-light noise). The number of scattered photons in ideal condition is estimated starting from equation (\ref{backscattering}) with $\theta=0$ and $\vec{e_i}\times\vec{\beta}=0$ so that 

$$
Ns=r_0^2 \frac{(1+\beta)^2 }{(1-\beta)}\frac{N_iN_e }{A }\Delta\Omega
$$

where $\Delta \Omega$ is the solid angle of the detected scattered radiation $\Delta \Omega\approx\frac{ S}{ d^2}$ i.e. the ratio between the active area of the PMT and the square of the distance between the interaction point and the PMT, while $A$ is the laser beam cross section. 
With the typical bunch parameters $Q=100$ pC, $E=10$ keV, $S=3.8$ cm$^2$, $A=0.28$ cm$^2$ and $d=68$ cm the scattered photons are less than one, $N_{ph}=0.34$, and the signal corresponding to a single photon is not observable because the electronic noise level (between the gray lines) has a RMS of $\approx2.3$ mV, while the level of the single photoelectrons is 267 $\mu V$. The minimum electron beam density (matched with the laser beam) required for the observation of the scattered radiation is of the order of $3\cdot10^{11}$ cm$^{-3}$ corresponding to $\approx14$ photoelectrons per shot, a signal to noise ratio equal to 1, a bunch length of 24 cm and an interaction time of 5 ns.  

\section{Experimental set-up with focused laser injection}

\begin{figure}
\begin{center}
\includegraphics[scale=0.6]{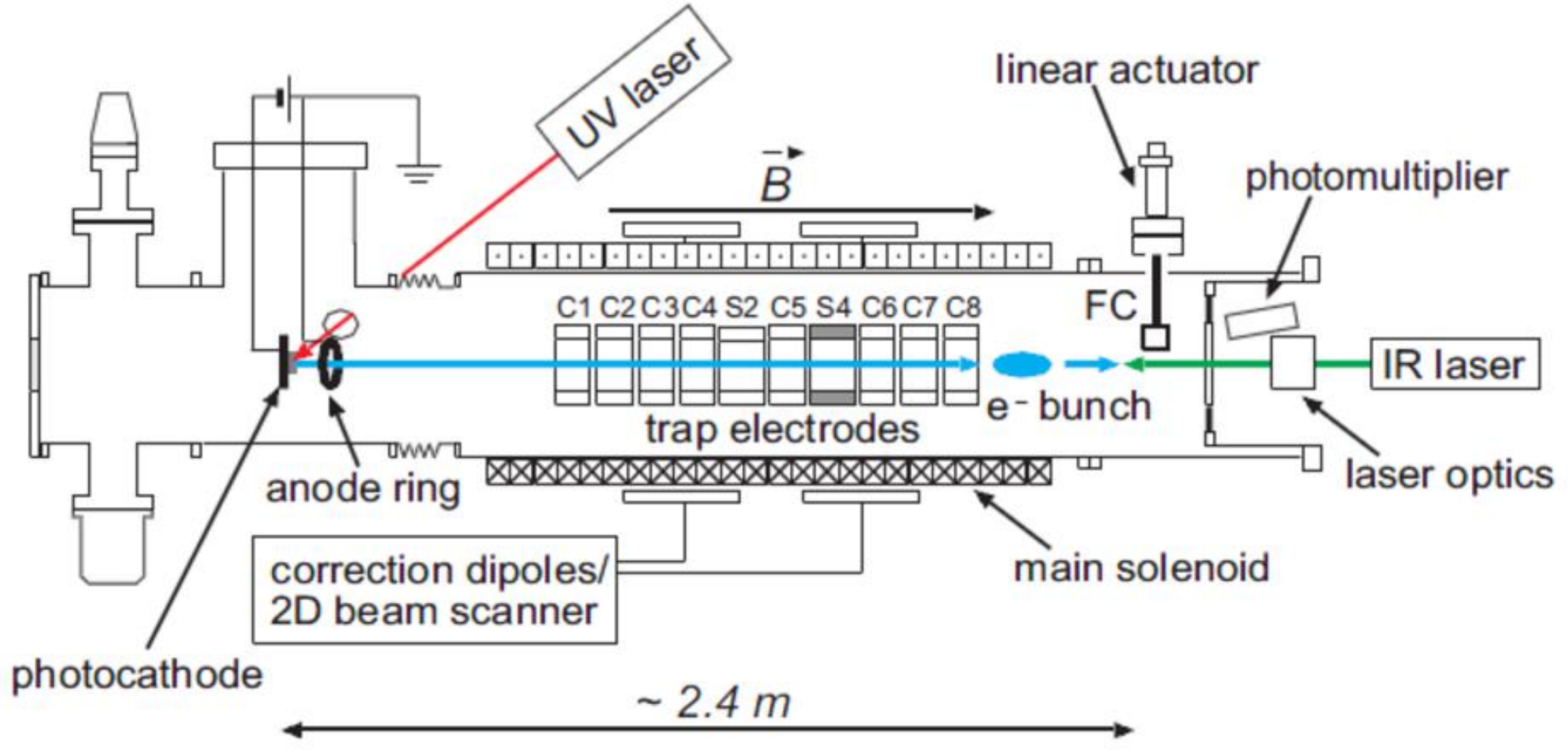}
\end{center}
\caption{\label{setup2}Sketch of the Thomson backscattering diagnostic set-up. A pulsed electron bunch is produced by an ultraviolet (UV) laser impinging on a photocathode set at an extraction voltage of 1-20~keV. The bunch is focused by the axial magnetic field $B\leq0.2$~T of the Penning-Malmberg trap. The trap electrode S4 can be used to detect the bunch crossing via induced current. A 2D beam scanner exploiting two pairs of correction dipoles, combined with the charge readout from a Faraday cup, automatically deflects the bunch transversally until the bunch transverse position reaches the interaction point. The IR radiation is filtered and focused onto the same point by an optical system. The scattered radiation is collected by a photomultiplier.}
\end{figure}

The main advantages of the previous set-up are that the interaction point can move along the laser trajectory simply changing the time between the triggers of the UV and IR lasers and that the main noise source is the electronic noise (the stray-light is temporally separated by the scattered radiation)  while the main limitations are:  a) increasing the distance $d$ from the PMT the sensitivity reduces as $ 1/d^2$,  b) The bunch and laser cross sections are different (i.e. non matched) reducing the efficiency in the laser-bunch interaction,  c) the interaction does not occur at every laser shot, d) the diagnostics has no spatial resolution. To solve partially these limitations a different set-up was designed and implemented on the same apparatus.  The focusing  of the laser in a point along the geometrical axis is optimized for the maximum collection (considering the geometrical  and optical limits), the matching between the bunch and the laser cross sections is obtained at least in a characteristic length defined as scattering length, that is also a measurement of the spatial resolution of the diagnostics. With this set-up the laser beam of the incident radiation is not efficiently dumped (i.e. after the focus the beam diverges and impacts the vacuum chamber) and the stray-light can be overlapped with the scattered radiation during the measurement. The sensitivity can be reduced introducing a monitoring system that  helps to check if the interaction occur along the Z axis and in the transverse plane (X, Y), i.e. to discriminate the shots resulting in the simultaneous arrival of the electron and laser bunch in the interaction point.  The advantage is that these shots can be averaged and the sensitivity increased, because the uncorrelated noise is reduced in principle by a factor $1/\sqrt{n }$ (where $n$ are the shots number). The basic configuration of the experimental set-up is sketched in figure \ref{setup2}. The electron bunch emitted by the photocathode source travels through the grounded trap cylinders. The S4R sector is used to monitor the bunch passage via induced current. The bunch interacts with the incident radiation produced by the Nd:Yag laser focused onto the interaction point and the backscattered photons are collected by a PMT. Most of the stray-light is discriminated by a set of filters placed in front of the PMT in order to enhance the signal-to-noise ratio. In order to compensate the mismatches in the relative transverse position of the electron and laser pulses the 2D beam scanner (described in the previous section) is used in closed loop configuration to steer the bunch trajectory and to check the position with a removable Faraday cup. The Faraday cup and the electrostatic signal from the S4 electrode are also used for the laser-bunch time coincidence.

\subsection{Laser injection and optical collection}

\begin{figure}
\begin{center}
\includegraphics[scale=1]{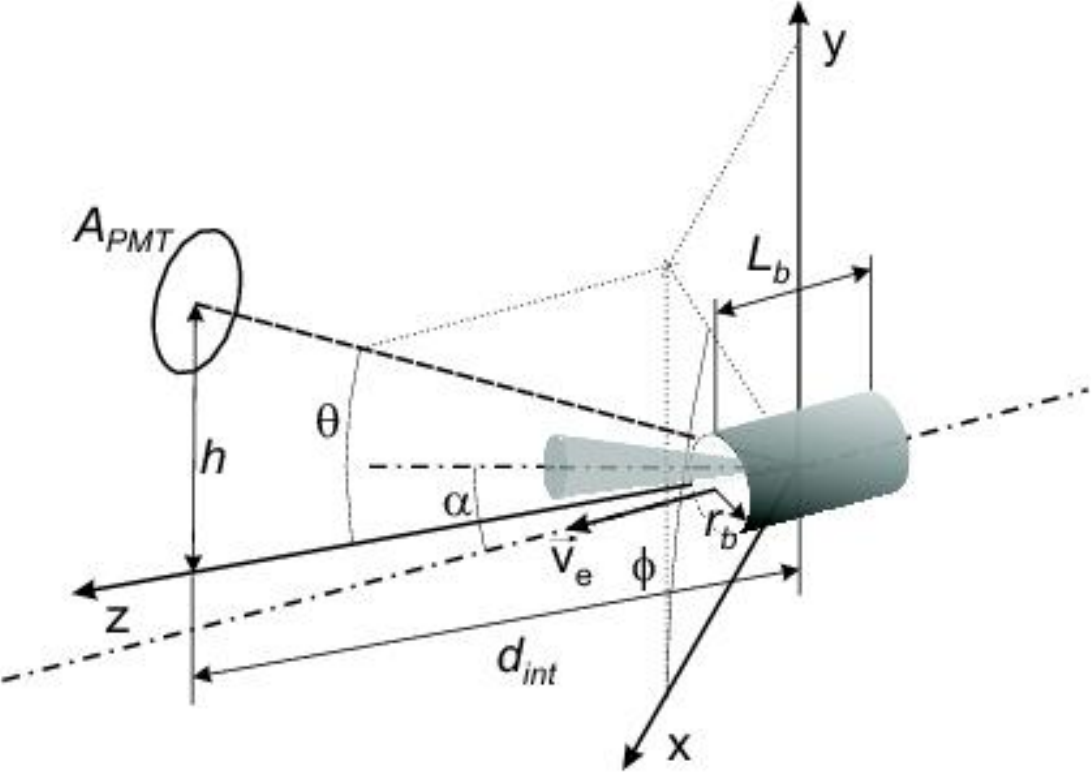}
\end{center}
\caption{\label{geometry}Geometry of the interaction between a laser pulse and a cylindrical electron bunch of radius $r_b$ and length $L_b$. The axis of the gray cone is the direction of the incident light, forming an angle $\alpha$ with the longitudinal axis of the system. The center of the active cross-section $A_{PMT}$ of the photomultiplier is at a distance $d_{int}$ from the interaction point along the longitudinal axis and at a distance $h$ from the axis. The direction of the scattered light is defined by the spherical coordinates $(\theta,\phi)$.}
\end{figure}

\begin{figure}
\begin{center}
\includegraphics[scale=1]{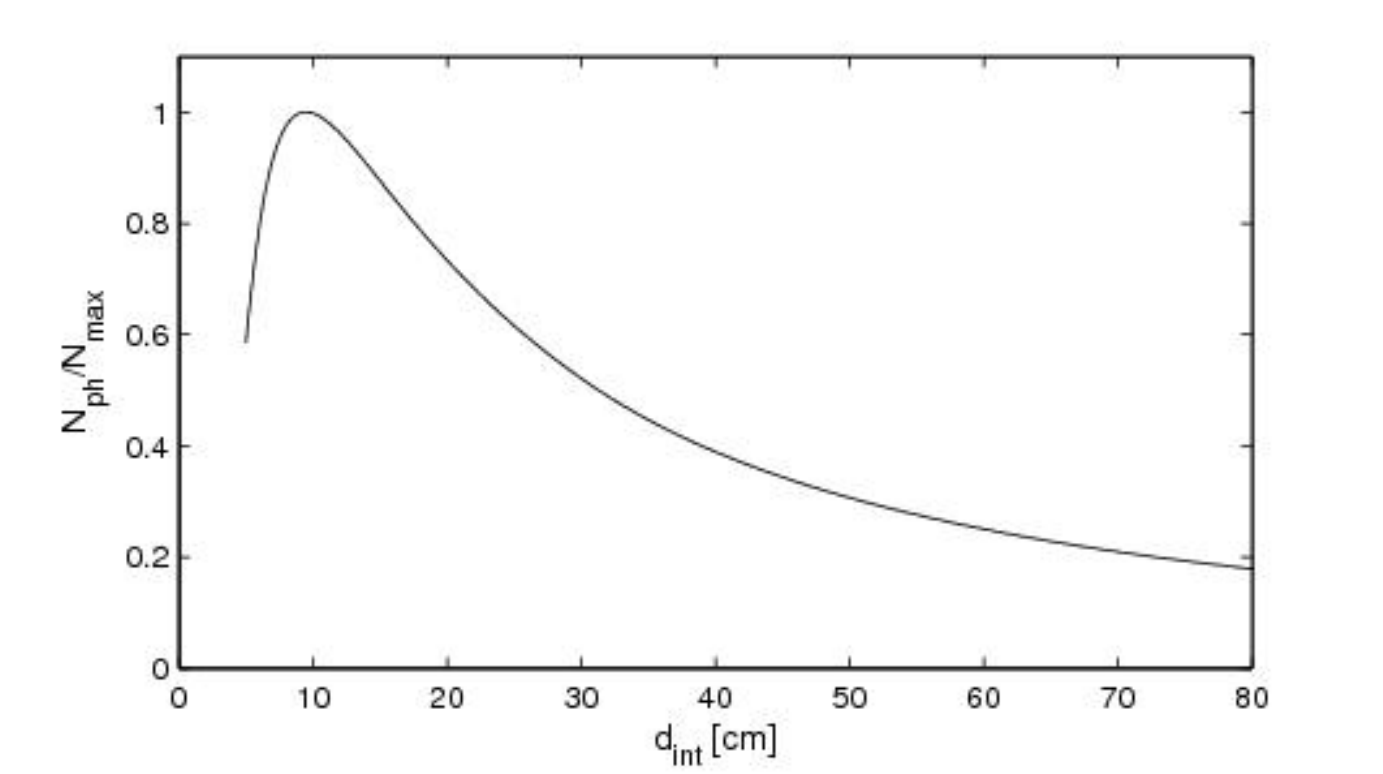}
\end{center}
\caption{\label{distance}Number of detectable scattered photons $N_{ph}$ as a function of the distance of the interaction point from the viewport, calculated using Eq.~(\protect\ref{integral}) and normalized to the maximum. Taking into account the physical and technical constraints of our experimental apparatus, the maximum is obtained for $d_{int}\approx 10$~cm.}
\end{figure}

\begin{figure}
\begin{center}
\includegraphics[scale=0.8]{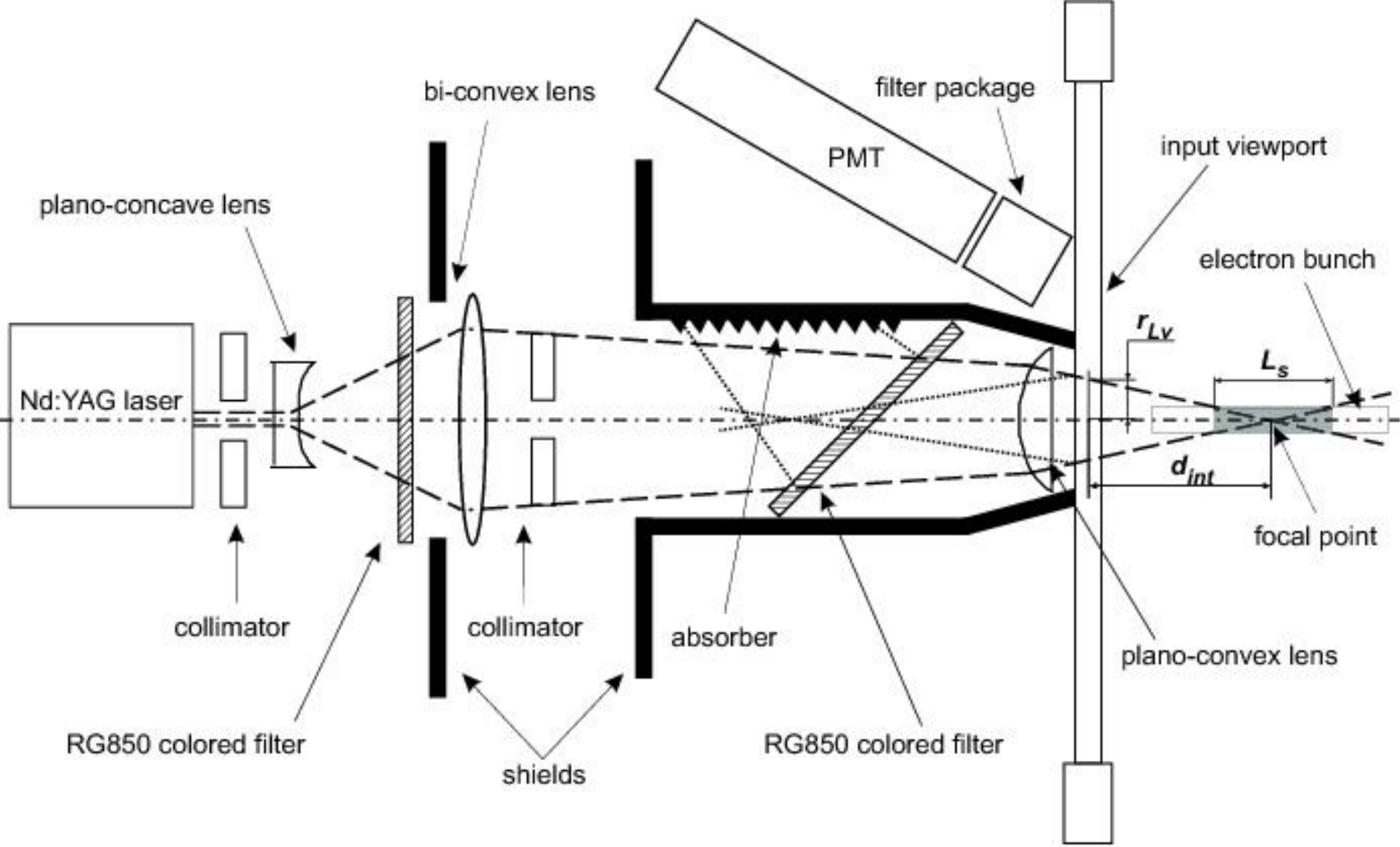}
\end{center}
\caption{\label{opticinjection}Schematic of the optical systems for the infrared laser injection and scattered radiation detection. A series of three lenses and a longpass RG850 Schott colored glass filter focus the $1064$~nm beam into the interaction point at a distance $d_{int}$ from the viewport while filtering out the unwanted radiation. Two removable collimators are used to align the laser and thus allow the matching between focal and interaction point. The scattering length $L_s$ is a measure of the matching between laser and electron pulses. The backscattered radiation is collected by a photomultiplier whose bandwidth is limited to part of the visible range by a package of eight filters. Thanks to a second, tilted RG850 filter housed in a black box, the fraction of incident radiation reflected by the viewport is deviated onto an absorber.}
\end{figure}

A fundamental issue in the design of the optics for the focused laser injection is the choice of the focus point (i.e. the interaction point) to optimize the collection of the scattered photons. We consider a geometry (see fig. \ref{geometry}) with a uniform cylindrical electron bunch propagating with velocity $V_e$ interacting with a focused laser beam. The direction of the incident radiation form an angle $\alpha$ with respect to the longitudinal axis of the system and the PMT is positioned so that its active area $A_{PMT}$ is oriented toward the focal point of the incident radiation, at a distance  $h$ from the axis and at a distance $d_{int}$ from the interaction point. The direction of the scattered radiation, oriented towards the PMT cross section, is represented in the frame (X, Y, Z) by the spherical coordinates ($\theta,\phi$). This notation is consistent with that of the section 4.2.1 with the unit vector $\vec{e_s}=\vec{e_s}(\theta,\phi)$. Note that the distance h must be reduced as much as possible because both the differential cross section and the frequency shift increase for smaller $\theta$. This distance is limited by the radius of the PMT and by the injection optics.  Given the distance $h=30$ mm, a PMT active area $A_{PMT}=3.8$ cm$^2$, a bunch radius $r_b\approx0.3$ mm, a time duration of the bunch $\Delta t_b=4$ ns, a bunch energy of 15 keV and a square laser pulse of $\Delta t_L=5$ ns, we can compute the integral of (eq. \ref{integral}) leaving as free parameter the distance $d_{int}$ in a collinear interaction $\alpha=0$. As shown in fig. \ref{distance} the maximum number of scattered photons occurs at $d_{int}\approx10$ cm. The distance $d_{int}$ fixes the spatial resolution of the diagnostics  i.e. the characteristic length $L_s$ (scattering length) where the laser-bunch interaction is strong. In ray optics approximation and with a uniform cylindrical bunch propagating collinearly with the laser beam this length can be defined as two times the length between the focal point and the point along the laser trajectory where the laser beam and the bunch cross sections are matched (see fig \ref{opticinjection}).   Out of the region represented by $L_s$ a portion of laser beam does not interact with the electron beam. The scattering length depends on the geometry of the interaction $L_s\approx 2 (r_b/r_{Lv}) d_{int}$ where $r_{Lv}$ is the laser spot radius on the viewport. As $L_s$ increases moving the focal point away from the viewport, the spatial resolution decreases. The distance $h$ is minimized reducing the radius $r_{Lv}$ so that the PMT is not along the path of the laser beam. The minimum value of $r_{Lv}$ depends on the maximum power that can pass from the viewport. For a  BK7 glass viewport of the vacuum vessel and a laser energy of 0.92 J we have chosen $r_{Lv}\approx6$ mm. The injection optics was designed to filter the stray-light (the second and third laser harmonic at 532 nm and 355 nm and the light emitted by the flash lamp ), to focus the laser beam at a distance $d_{int}$ from the viewport and to minimize $r_{Lv}$.  The laser beam is defocused by a plano concave lens so that the power per unit area is reduced when it passes with a radius of 20 mm through a longpass RG850 Schott colored glass filter. A bi-convex lens immediately follows and provides a first refocusing. A third plano-convex lens is placed right in front of the viewport. We found out that the laser reflection on the viewport glass, which is about $10\%$ for a BK7 glass, is sufficient to create a small spark in air, adding to sources of stray light. In order to eliminate this effect we have introduced another $45^\circ$ tilted longpass RG850 filter. The $10\%$ of the incident light is collected on the absorbing wall of a black box. Two shields  after the first filter and before the black box overshadow the viewport and the PMT. Given a transmission coefficient 0.99 for all coated lenses and 0.9 for the filters and viewport we obtain a total transmission for the optical system $\eta_L\approx0.7$. The total attenuation of the stray-light is $\approx 10^8$ for a wavelength $<700$ nm. The alignment of the optics for the laser injection is obtained manually using two ceramic collimators.  A He-Ne red laser trace the trajectory passing between the center of the viewport and the center of the electron source. The collimators are positioned along the red laser beam trajectory and are used as reference to the alignment of the others optical elements (lens and filters). For the detection of the scattered radiation a PMT with a set of filters was used. The composition of the filter package was modified in this set-up in order to increase both the efficiency of the optics and the total optical density in the rejection region. The bunch energy used to test this set-up was 15 keV. The scattered radiation is expected to be centered around 650 nm, and the bandpass of the detected scattered radiation between 610 nm to 700 nm. In particular the package is composed by three dichroic shortpass filters with a cut-off wavelength of 850 nm and a total optical density (OD)$<12$ that attenuate the IR laser radiation. Two dichroic shortpass filter with cut-off wavelength of 700 and 750 nm reduce the fluorescence induced by the IR laser radiation hitting the inner structures of the interaction chamber. Two dichroic longpass filters with cut-off at a wavelength of 450 nm reduce the UV radiation of the nitrogen laser by a factor $10^8$. Finally a multi-coated longpass RG610 Schott colored glass filter attenuates  the fluorescence in the visible range induced by the UV laser. The total trasmission coefficient of this filter package is 0.5. Taking into account a PMT quantum efficiency of $5\%$ at 650 nm we obtain an optical efficiency of $\rho=0.025$. The PMT was aligned manually in a position as close as possible to the axis of the system and near to the viewport.

\subsection{Space coincidence}

\begin{figure}
\begin{center}
\includegraphics[scale=0.65]{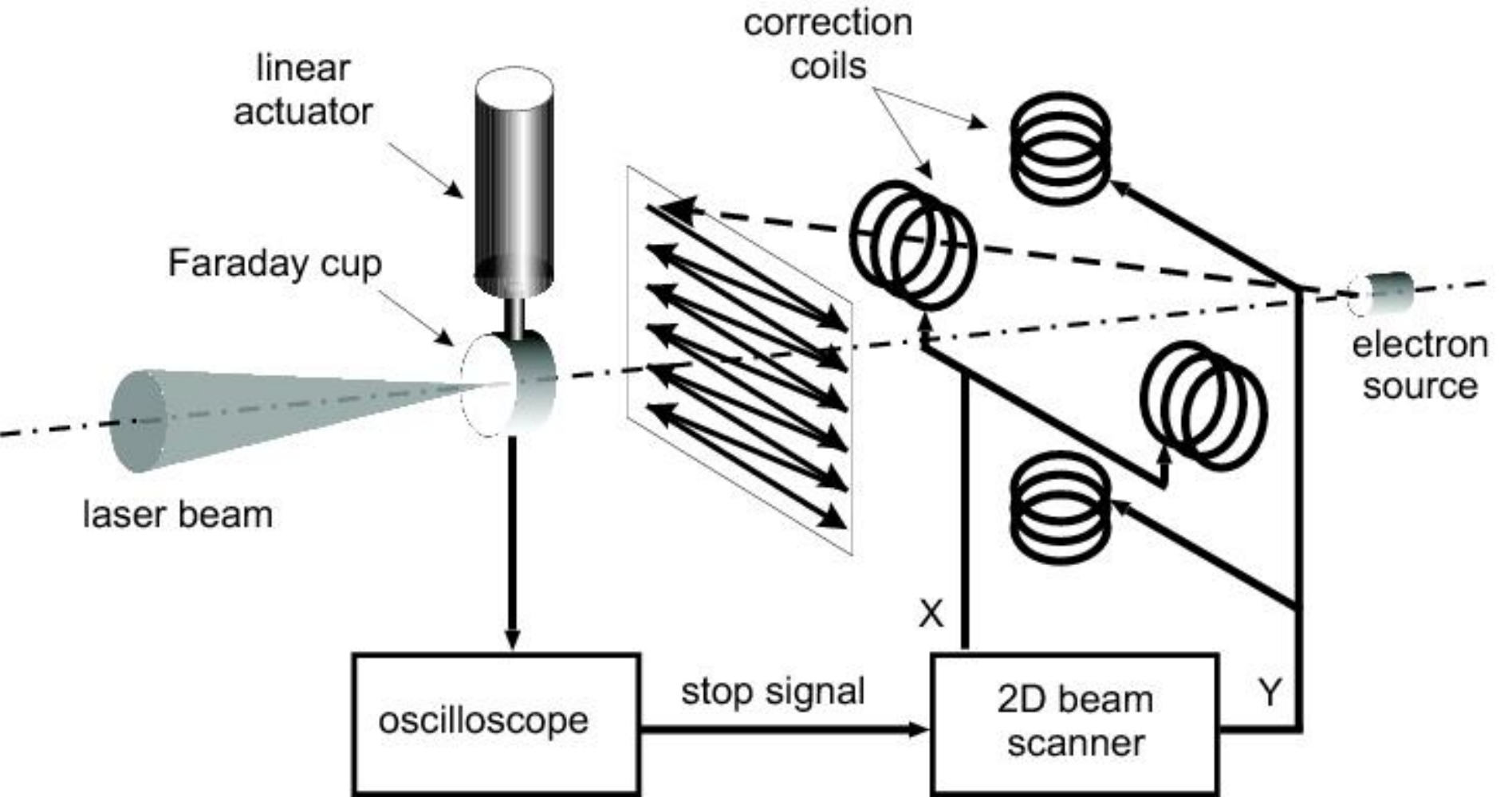}
\end{center}
\caption{\label{faradaycup}Sketch of the space-coincidence system. The infrared laser is aligned with the Faraday cup. The bunch trajectory is steered using two orthogonal pairs of correction dipoles. A 2D beam scanner automatically scans a square region in the transverse plane ramping the currents in the dipole coils. When the bunch is detected by a digital oscilloscope connected to the Faraday cup the scan is stopped by a feedback signal.}
\end{figure}

\begin{figure}
\begin{center}
\includegraphics[scale=1]{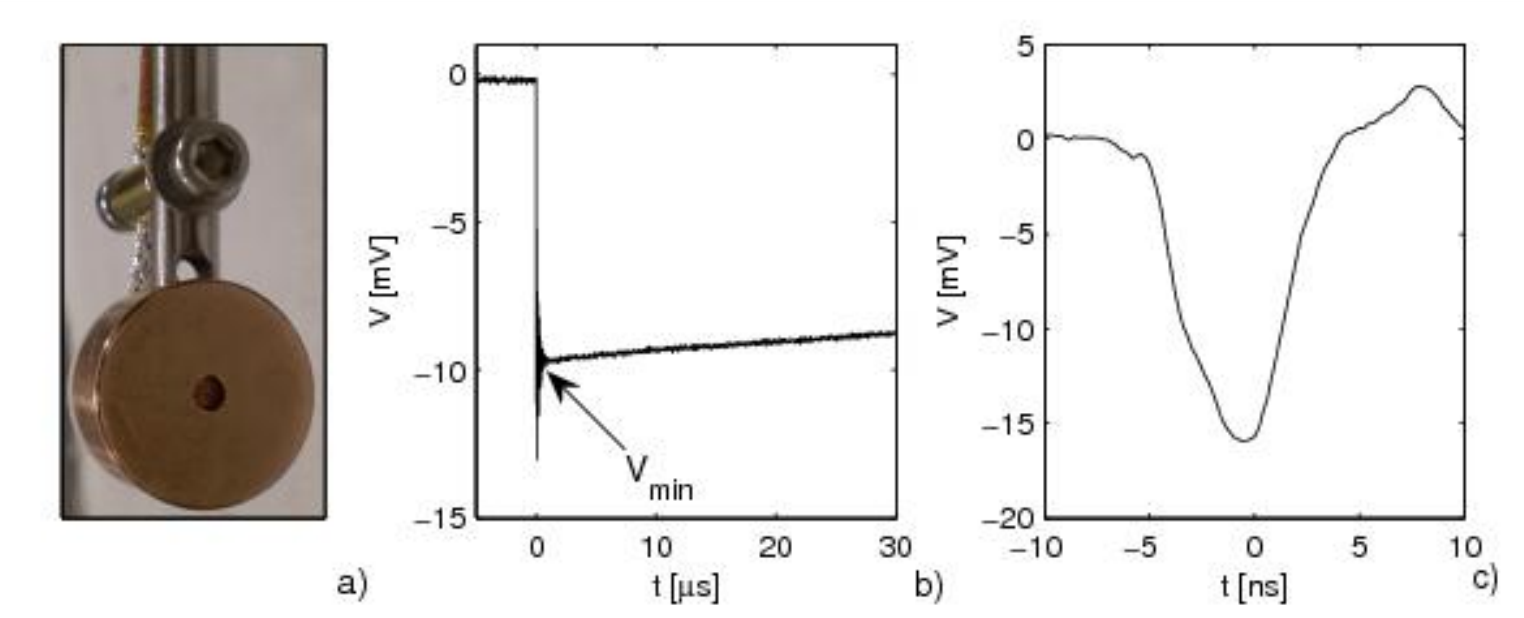}
\end{center}
\caption{\label{selfalignment}a) Photograph of the Faraday-cup. The central cylindrical conductor (active area) has a diameter of $3$~mm. The grounded shield has a diameter of $\approx 20$~mm. The two OFHC copper conductors are connected by a macor insulator. The shield is screwed to the shaft of a linear actuator. b) Charge measurement obtained from the Faraday cup signal with a load $R_L=1$~M$\Omega$ on the oscilloscope. The characteristic time of the discharge signal is about $659$~$\mu$s. The signal minimum $V_{min}$ is clearly visible even in the presence of an overshoot. c) Faraday cup signal read on the oscilloscope in low-impedance mode ($R_L = 50$~$\Omega$). This measurement is used to calculate the laser-bunch time coincidence.}
\end{figure}

In the previous set-up the spatial coincidence was obtained with a scan of the bunch trajectory using an 2D beam-scanner in open loop configuration, this method introduce a wait time ($\approx28$ s) for the coincidence in transverse plane. The advantage is that moving the interaction point along the laser trajectory the interaction is however guaranteed in a transverse square window of side 5 cm. Because in this set-up the interaction point is fixed, it is advantageous to introduce a method to fix the laser bunch  coincidence in the transverse plane passing through the interaction point. In this way the coincidence is spatially guaranteed at every laser shot. For this purpose a 2D beam scanner was configured in a closed loop introducing a Faraday cup (see fig. \ref{faradaycup} (a)) to check the alignment of the bunch with the IR laser beam trajectory. The Faraday cup was designed with coaxial geometry. The inner electrodes realized in copper OFHC has a diameter of 3 mm and is shielded by a copper cylinder of $\approx 2$ cm diameter. These parts are mounted on a ceramic disc that is electrically insulated.  The dieletric between the electrodes is vacuum to reach an impedance of about 50 $\Omega$, matched with the coaxial line. The cross section of the central charge collector was chosen as a compromise between sensitivity and resolution. The Faraday cup is coupled with the 2D beam scanner to realize in practice a self-alignment system (see fig. \ref{selfalignment}). First the Faraday cup is positioned by means a linear actuator on the IR laser beam trajectory, than the electron bunch trajectory is steered with the  beam scanner. When the bunch hits the active area of the Faraday cup, the signal of the collected charge triggers the acquisition of the oscilloscope that stops the scan. The feedback signal acts on the And gate to stop the internal oscillator of the beam scanner. The bunch and the laser trajectories remain so aligned. Than some manual adjustments are needed to optimize the final alignment.  An example of the measured characteristic signal produced by the FC during the charge collection of the bunch is represented in fig. \ref{faradaycup} (b). The signal is acquired setting an input impedance of the oscilloscope at 1 M$\Omega$. The negative fast edge triggers the acquisition to generate the feedback signal. Note that the signal rise exponentially with a characteristic time $\tau= R C$ of 659 $\mu$s, where R is the oscilloscope load impedance and C is the Faraday cup plus cable capacitances. Because $\tau<1/f_L$ and the propagation time needed to generate the feedback signal is less than $\tau$, the time response of the system is adequate to capture the bunch during the scan at every shot. The maximum time to obtain the self-alignment is 28 s but the difference with the previous set-up is that we have to wait only once for the alignment. 

\subsection{Time coincidence}

\begin{figure}
\begin{center}
\includegraphics[scale=0.9]{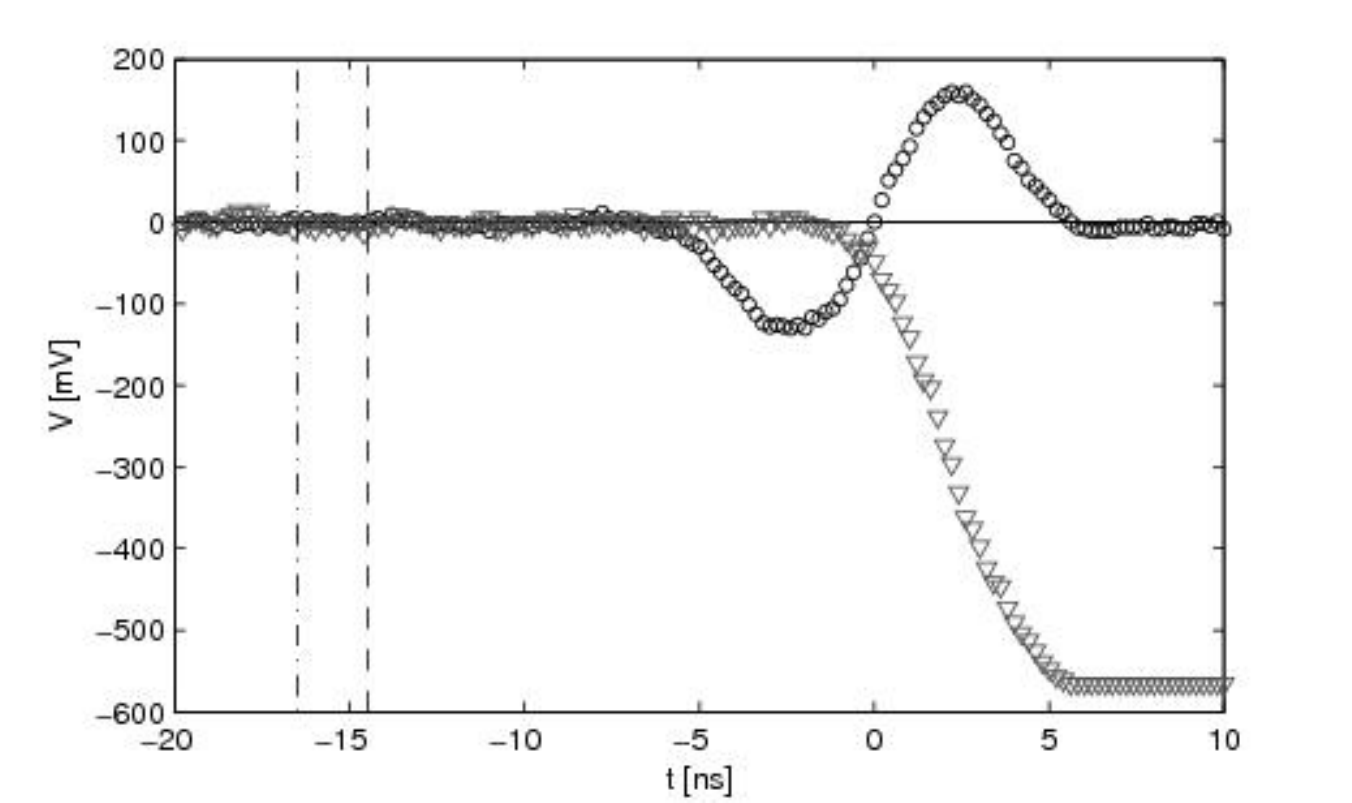}
\end{center}
\caption{\label{timecoincmeas}Measurement of the laser-bunch coincidence. The zero time instant is set by the S4R signal (circles). The vertical dashed line is the actual arrival time of the bunch in the interaction point., i.e. $14.5$~ns before. From the signal of the infrared detector (triangles) and the calculated delays the arrival time of the IR pulse in the interaction point is determined (vertical dash-dotted line). In this case, the difference in the two arrival times is $\approx 2$ ns.}
\end{figure}

Because in this set-up the space coincidence  is provided by the self-alignment system the interaction along the Z coordinates must be managed by a system that controls the synchronization between the UV and IR lasers. Similarly with the previous set-up the main problem is that the UV laser has a jitter of 20 ns and the synchronization can be computed, in the sense of maximum probability of interaction knowing all delay times due to the propagation  of the signals in cables and the laser beams optical paths. The difference with respect to the previous set-up is that the indetermination is only in the Z direction at every laser shot. The design of a monitoring system that measures the position of the bunch and the laser beam along the Z axis is useful to discriminate the shots resulting in the simultaneous arrival of the electron and laser bunch in the interaction point. In analogy with the consideration of the section 4.3.4 we can solve the problem of the lasers synchronization with a delay generator that triggers the UV laser in advance of the IR laser of a time $\Delta t$ that is the difference between the following two times. The first is the sum of all delay times from the delay generator to the interaction point.  For a uniform bunch propagation with an energy of 15 keV the total time (right branch of the schematic in  figure \ref{timecoincidence}) is 926.3 ns. The second time is the sum of all time from the delay generator to the arrival of the laser beam at the interaction point (left branch of the schematic in figure \ref{timecoincidence}). This time is computed to be 294.7 ns. Note that the delay time of 271 ns due to the IR laser differs from that of the previous set-up of 307.87 ns. This discrepancy occurs because these times depend on the laser pumping time $\Delta t_p$ that determinates the output energy of the beam. In the previous set-up $\Delta t_p$ is 170 $\mu$s  for a laser energy of 0.25 J, while in this set-up we use $\Delta t_p=240$ $\mu$s for a laser energy of 1 J. The monitoring system was implemented with three basic elements: 1) A fast silicon Pin photodiode with a rise time less than 5 ns to detect the infrared radiation, 2) the Faraday cup connected with the linear actuator to detect the bunch, 3) the sector S4R to detect the passage of the bunch when the Faraday cup is removed for the laser-bunch interaction.  The arrival time of the pulsed laser beam at the interaction point is known from the rising edge of the signal produced by the photodiode, placed off-centered between the laser output port and  the viewport. The photodiode detects the passage of the IR radiation reading the light scattered by the optical elements during the laser injection. The Faraday cup measure the arrival time of the bunch at the interaction point. It is connected to the oscilloscope(1 GHz bandwidth ) setting the input impedance at 50 $\Omega$ to match the load with both the Faraday cup and the cable impedances. This matching reduce the distortion of the signal and increase the response time. As shown in fig. \ref{faradaycup} (c), the characteristic time of the pulse is in agreement with the pulse duration of the UV laser considering both the smoothing effects introduced by the 1 GHz limited band of the oscilloscope and the longitudinal spread of the bunch. Because the Faraday cup must be removed from the laser and bunch trajectories during the interaction we use the sector S4R during the experiment  as a reference. A calibration is needed to know the time between the arrival time of the electron bunch at the Faraday cup and the time when the bunch passes through S4R (i.e. the time of flight of the bunch from the electrode to the Faraday cup). The zero crossing of the induced current signal  is used as reference. Knowing these times we can acquire the scattered radiation signal from the PMT only if the two arrival times coincide, i.e. if they fall within a given time interval. This interval is set by the delay generator and has a width of 3 ns, which is consistent with the minimum width of measurable signal of  FWHM $\approx3.1$ ns (width of the signal of a single photon count). An example of the measurement of the laser-bunch time coincidence is represented in fig. \ref{timecoincmeas}. The zero time instant is set by the S4R signal (circles). Considering all delay times in the electron bunch production branch of the system the arrival time of the bunch, indicated by the vertical dashed line, is shifted to the left by 14.5 ns. Considering the delays in the IR laser branch the arrival time of the IR pulse, indicated by the vertical dash-dotted line, is shifted by 14 ns with respect to the signal of the IR photodiode detector (triangles). The maximum time interval between the two vertical lines must then be less than 3 ns in order to consider the signals as coincident and therefore trigger the signal acquisition. 

\begin{figure}
\begin{center}
\includegraphics[scale=0.6]{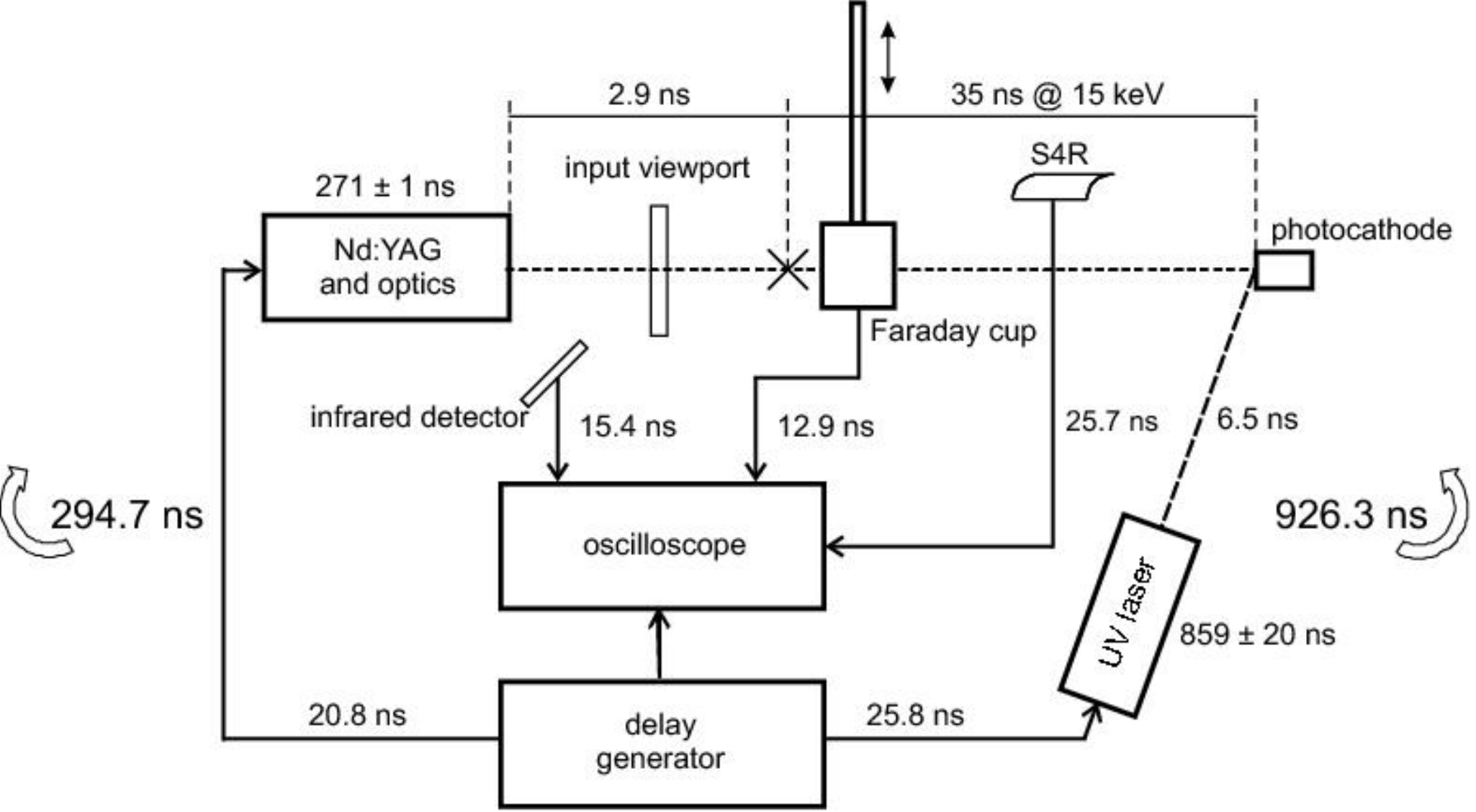}
\end{center}
\caption{\label{timecoincidence}Schematic of the time-coincidence system. All laser, time of flight and signal propagation delays are indicated. A delay generator triggers the IR and the UV lasers with a delay of (926.3 - 294.7) ns such that the IR pulse and electron bunch arrive simultaneously in the interaction point. To account for the large $20$~ns jitter of the UV laser, the arrival of electron bunch and IR pulse are monitored with the Faraday cup and the infrared detector, respectively. The sector S4R gives the time reference signal when the Faraday cup is removed from the axis during the laser-bunch interaction.}
\end{figure}

\subsection{Estimate of the set-up minimum sensitivity}

\begin{figure}
\begin{center}
\includegraphics[scale=1]{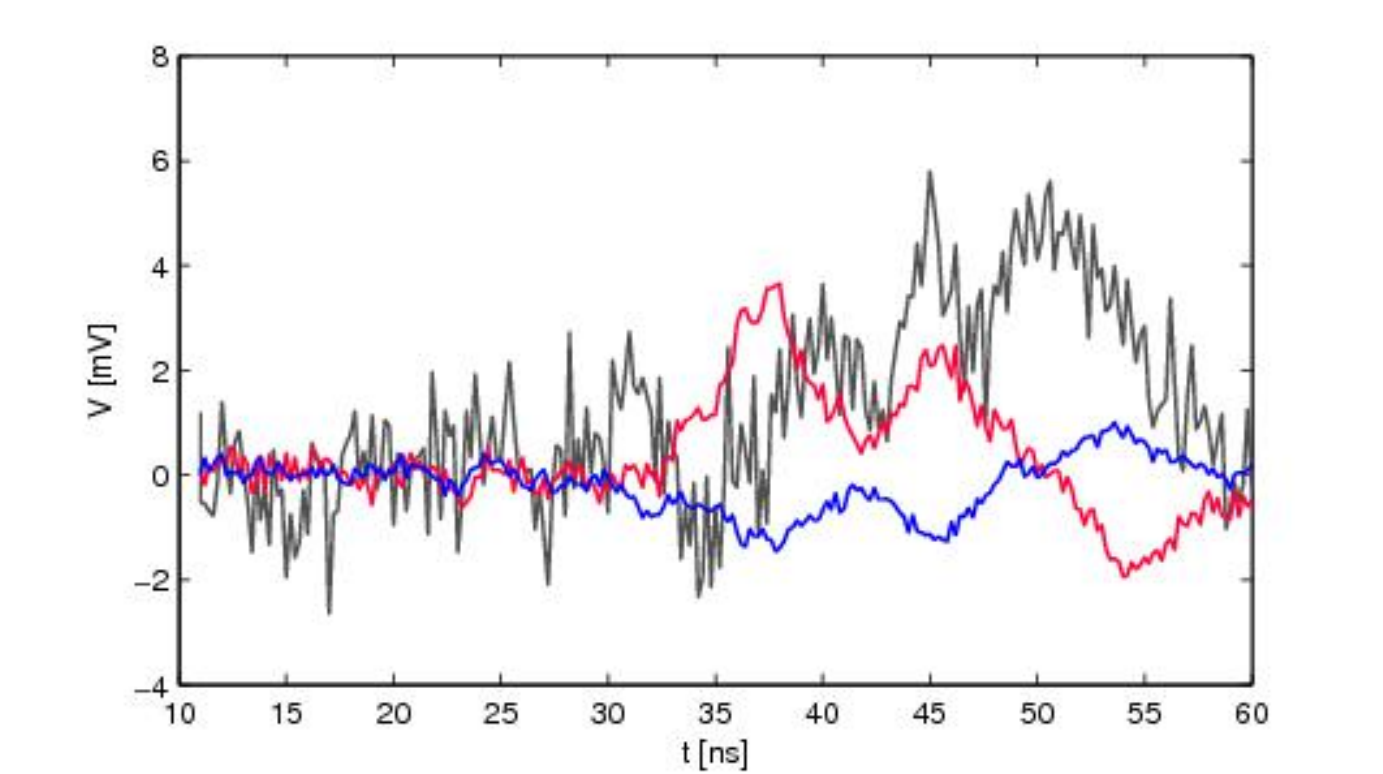}
\end{center}
\caption{\label{noise measurements}Residual background noise measured by PMT after subtracting the coherent noise for integration times of $0.1$~s (gray), $1$~s (red) and $5$~s (blue), respectively. After $t=30$~ns the noise level increases due to the stray light produced by the laser hitting the internal structures of the vacuum chamber.}
\end{figure}

\begin{figure}
\begin{center}
\includegraphics[scale=1]{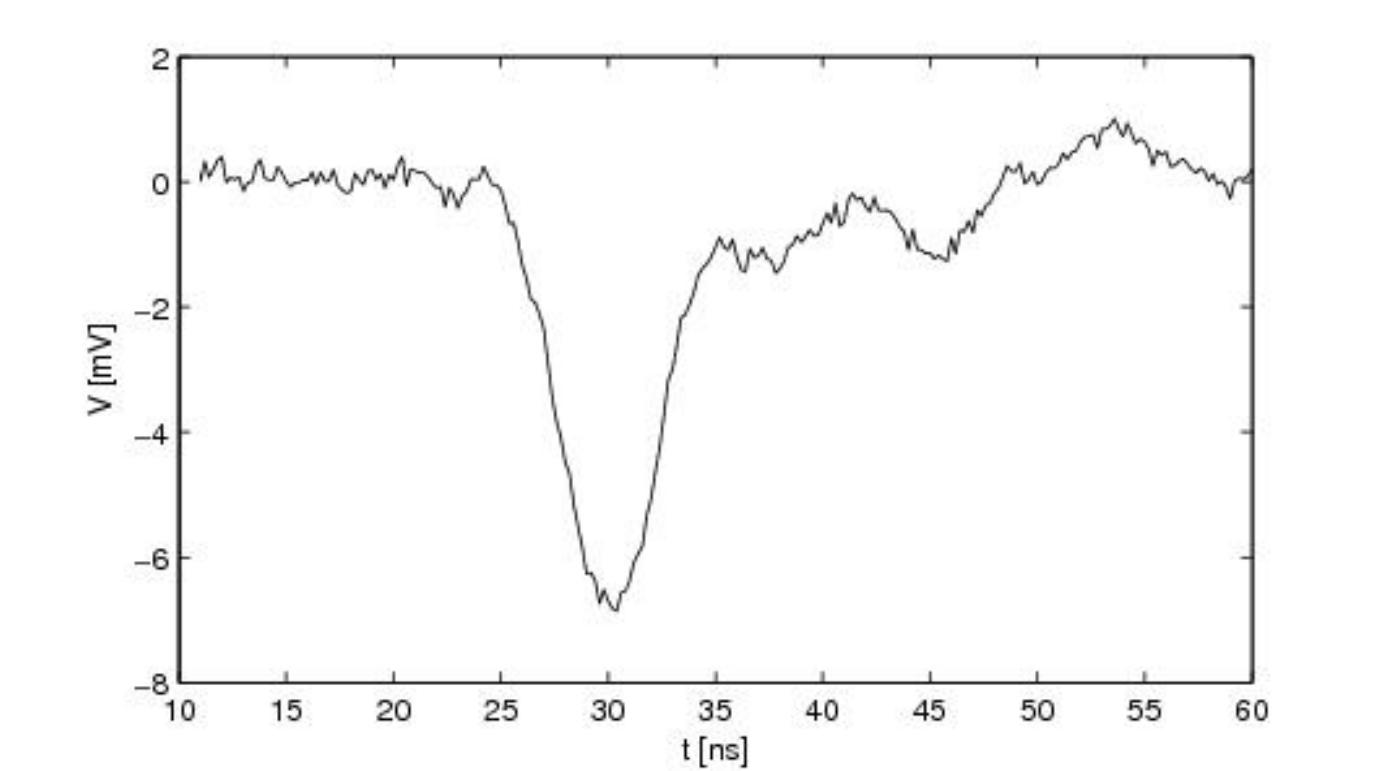}
\end{center}
\caption{\label{prevision}Expected scattered photons signal for a bunch of density $\approx 3.6\cdot 10^{11}$~cm$^{-3}$ and radius $r_b = 0.3$~mm in the present set-up configuration. We assume a Gaussian signal profile with a FWHM of $5$~ns. Part of the signal is overlapped with the stray light noise (see also Fig.~\protect\ref{noise measurements}).}
\end{figure}

In order to estimate the minimum density of the set-up we define the sensitivity as the amplitude of the signal detected by the PMT equivalent to the noise i.e. the signal amplitude which is necessary to obtain a signal-to-noise ratio ($S/N$) equal to 1. The noise level is experimentally measured and includes the stray-light of the UV and IR laser beam, the electronic noise and the coherent noise of the laser lamp discharges. The measurements were performed averaging the signals of 1, 10, 50 shots respectively after subtracting the coherent noise components averaged on 100 shots (see fig. \ref{noise measurements}). The computed root-mean-square (RMS) values were taken for ten different measurements and the maximum values are 1028 $\mu$V averaging on 10 shots and 736 $\mu$V averaging on 50 shots. Note that the noise level increase starting from 30 ns due to the stray light of the IR laser beam. To measure the expected time of the scattered light a time-resolved technique was used. The Faraday cup is positioned near to the laser trajectory until a portion of the laser beam hits the ceramic back face of the Faraday cup. The diffused light is acquired by the PMT and occur at a time of $\approx 2$ ns before the stray light of the IR laser beam. This means that the scattered radiation signal would be expected to start at 28 ns. For an interaction time of 5 ns a portion of the scattered radiation signal will be overlapped with the stray-light signal. Considering that the noise signal also limits the maximum gain of the PMT at $G=6\cdot10^4$ we can estimate the signal level as $S=eGR_L N_{phe}/\Delta t_{int}=100$ $\mu$V $N_{phe}$, where $N_{phe}$ is the number of produced photoelectrons, $R_L$ is the impedance load, $e$ is the elementary electron charge and $\Delta t_{int}=5$ ns is the characteristic laser-bunch interaction time. Considering the condition  $S/N=1$ we obtain $N_{phe}\approx10$ averaging on 10 shots and $N_{phe}\approx7$ averaging on 50 shots. The number of detectable photons is $N_{ph}=N_{phe}/\eta$. The minimum density can now be estimated computing the integral of the equation (\ref{integral}). The values $\alpha=\vec{e_i}\cdot\vec{\beta}$ and $\Delta \Omega$ are fixed taking into account the present set-up geometry. The photon density  is taken considering an energy laser beam of $0.92 \eta_L$ J and the geometry of the laser injection optics. The bunch density is estimated for a cylindrical uniform electron charge distribution of radius $r_b=0.3$ mm propagating rigidly with a velocity $V_e=\sqrt{2E/m_e}$ ($E=15$ keV) and for a collinear interaction. The minimum electron density is $n_e=5.1 \cdot 10^{10}$ cm$^{-3}$ and $n_e=3.6 \cdot 10^{10}$ cm$^{-3}$ averaging on 10 and 50 shots, respectively. As we have previously shown at the end of chapter 3 we have achieved densities up to some $10^8$ cm$^{-3}$  in the high magnetic field region and therefore the electron bunch is not detectable with the present set-up configuration. In figure \ref{prevision} we report the expected signal for a density $n_e=3.6 \cdot 10^{11}$ cm$^{-3}$, i.e. ten times the value corresponding to $S/N=1$, assuming a Gaussian profile for the detected pulse with FWHM $= 5$ ns and averaging on 50 shots.

\section{Conclusions}

A Thomson back scattering diagnostics was designed and implemented for the diagnostics of electron bunches in nanosecond regime. Two experimental configurations were set-up and discussed. In the first set-up the laser was injected collinearly  and dumped out of the vacuum chamber with the advantage that the position of the interaction point can be fastly changed along the laser trajectory during the experiment . The space coincidence in this set-up requires a continuous scan of a 2D beam scanner that steer the trajectory in the transverse plane. The maximum time required for a laser-bunch coincidence is about 14 min. and the detection of the scattered radiation needs a good sensitivity to detect the scattered photons in a single shot. The scattered photons are temporally separated by the main source of noise i.e. the stray light produced during the laser dumps and the sensitivity reached is of the order of $3\cdot10^{11}$ cm$^{-3}$. A second set-up was designed to focus the injected beam in a particular point in the chamber . The optimal position of the interaction point was estimated with a theoretical analysis of the scattered photons. The main problem in the design of the laser injection optics is the reduction of the stray-light  that is produced by the interaction of the laser with the elements of the optics itself . This problems must be solved considering that the high power density per unit area of the injected laser limits the use of materials and optical technologies. The problem of the space and time coincidence was solved with a 2D beam scanner configured in closed loop. The laser-bunch coincidence is always guaranteed at every shot in the transverse plane while in the Z direction the coincidence was tested by a monitoring system. Whit this set-up the laser and bunch cross-sections are matched, the position of the interaction point is optimized to increase the solid angle of the detection as well as the spatial resolution. The dump of the laser beam into the vacuum chamber and the overlap in time between the scattered radiation and the stray-light during the detection increase in principle the noise with respect  to the set-up with collimated injection. The combined use of a high quality dichroic and colored filters in a filter package allows one to reduce the stray light near to the electronic noise level and a better sensitivity was reached with respect to the first set-up,  because the residual noise was further reduced averaging the measurements on 50 shots. The minimum measured sensitivity was 7 photoelectrons  corresponding with a density of $\approx3.6\cdot10^{10}$ cm$^{-3}$. This sensitivity is obtained advantageously exploiting the blue-shift of the infrared radiation in the visible range. The most stringent limit to the signal-to-noise ratio is presently the stray-light noise, which is now $\approx3.5$ times larger than the electronic noise.

\chapter{Conclusions}
A Thomson backscatterig diagnostics was designed and implemented in the Eltrap apparatus for the study of the dynamics of electron beams and nanosecond bunches in high space charge regime. This apparatus was originally a trap for nonneutral plasmas but it was used here in open configuration. The bunch travels in a cylindrical drift tube of 9 cm diameter and $\approx 1$ m length, with a uniform magnetic field $< 0.2 $ T and in UHV condition $10^{-9}$ mbar. A low density bunch produced with a photocathode source and radially focused by the axial magnetic field was characterized  with two suitably developed electrostatic diagnostics. The first electrostatic diagnostics uses a phosphor screen as a charge collector for charge and length measurements. A deconvolution technique  was used to characterize the time duration of the bunch and length. The length is well approximated (at higher energies) by $\Delta_L=\sqrt{2E/m_e}\Delta t$. At lower energy the density and the time of flight of the bunch increase and space charge effects are experimentally observed . At higher energy or lower density the bunch motion is well described as a uniform motion. This results are also confirmed by a second diagnostics developed specifically as non-destructive electrostatic diagnostics in order to know the bunch position, bunch velocity spread and length. Transversally the bunch was characterized with an optical diagnostics acquiring with a CCD camera the image formed by the electrons impact on the phosphor screen. A systematic analysis varying the magnetic field and the bunch energy to characterize the beam profile and its spot size was done. This analysis shows that $\approx10\%$ of the total charge is contained in the most dense region and that the bunch density is of the order of $10^8$ cm$^{-3}$ in the region of uniformity of the magnetic field. This bunch was used to validate the space and time coincidence system of the Thomson backscattering diagnostics. To increase the accuracy in time of flight measurement for the laser-bunch synchronization the bunch energy was manteined  $>10$ keV in order to reduce the longitudinal spread. 

To increase the measureable beam parameters a Thomson backscattering diagnostics was designed and tested with two different set-ups. In the first set-up (with a collinear laser injection) a sensitivity of  14 photoelectrons is estimated (for a bunch-laser matched interaction), corresponding to a density of $10^{11}$ cm$^{-3}$, while in a second set-up (with a focused laser injection) the measured sensitivity was 7 photoelectrons for a density of $3.6\cdot 10^9$ cm$^{-3}$. We briefly list the advantages and disadvantages of the presented set-ups. The advantages of the collinear set-up are the following: 1) the interaction point can be moved along the laser trajectory, 2) the stray-light of the IR laser is adequately dumped out of the vacuum chamber, 3) the IR stray-light noise component is temporally separated by the scattered radiation (in a wide region for the interaction).
While the disadvantages are: 1) no spatial resolution, 2) the bunch and the laser are not in coincidence at every shot and a wait time of $\approx 14$ min. is needed for the interaction (this time can be appreciably reduced by decreasing the initial error between the laser and bunch trajectories or decreasing the UV laser jitter). In the second focused set-up the benefits that emerge are: 1) spatial resolution $\approx1$ cm for density profile measurements, 2) optimisation of the laser bunch interaction (e.g. matching between laser-bunch cross sections), 3) ability to implement a monitoring system to obtain an acquisition at every interaction, 4) optimisation of the optical collection (increasing the number of collected photons). With the following disadvantages: 1) the interaction point is fixed along the laser trajectory, 2) the dumping of the beam is not controlled and it hits the internal structures of the chamber, 3) the stray light is temporally overlapped with the scattered light. The previous advantages in both set-ups are obtained thanks to a space and time coincidence systems. The space coincidence is used in the collinear set-up to search continuously the laser-bunch interaction point in the transversal plane, that changes moving the interaction point along the laser trajectory or changing the experimental parameters (e.g. magnetic field, bunch energy). In the set-up with the focused laser the space coincidence system, configured in closed loop, realizes a self-alignment system to search automatically the focal point of the incident radiation. This system was realized with an electronic digital controller, a coaxial Faraday cup working in the nanosecond regime and a fast IR detector. The time coincidence system have the same goal in both the set-ups, i.e. the interaction along the geometrical axis (Z). Time of flight measurements and reflectometry technique were used to test the laser-beam synchronization.  An optical system based on photomultipliers and a set of high quality dichroic and colored filters was optimized to reduce the stray-light produced by the IR and UV lasers and the relatives induced fluorescence. 

With this work we have proposed an alternative diagnostics showing experimentally that densities $\geq 10^{10}$ cm$^{-3}$ can be measured also at low beams energies (ten of keV) and that the problem of the misalignment introduced, for example changing the experimental parameters, can be solved with a self-alignment system. This diagnostics is suitable for the study of the dynamic of bunch and beams in high space charge regime thanks to the quantities that are simultaneously measureable, i.e. density, density profile, energy, energy spread, and to its non-perturbative nature. The main sources of uncertainty in the energy and energy spread measurements are the PMT finite sensitive area and the finite scattering length. The PMT radius of  1.1 cm determines a maximum error in the angle $\theta$  (between the incident and scattered radiation) of $\pm 0.1$ rad. As a consequence of the 1 cm scattering length there is a spread in the interaction distance $d_{int}$ of  $\pm 0.5$ cm and in turn an error in $\theta$ of 28 mrad. Altogether, the two factors give a relative energy uncertainty of  $8-9\%$ in the whole range from 1 to 20 keV. These considerations are valid in the simpler case of an incident plane wave and for a linear scattering. Non-linear scattering, gaussian laser and electron beams transversal profiles require a more complete analysis also for the spectrum of the scattered radiation emitted by a single electron \cite{Wang}, \cite{Krafft}. A spectrum could be obtained using for instance a monochromator, whose typical resolution is higher than our experimental accuracy. 

Some improvements are possible to increase the sensitivity of the diagnostics. The actual stray-light noise can be reduced because the laser beam dump is presently not efficiently controlled. We are currently exploring a solution, namely the design of a suitable light shield allowing the interaction and radiation collection to take place while preventing the light coming from the chamber walls to reach the photomultiplier. The reduction of the stray-light allows to increase both the gain and the dynamical range of the PMT. The electronic noise can be reduced inserting an amplifier between the photomultiplier and the acquisition system. A low noise figure of a commercial amplifier $\approx 1.4$ dB should reduce the present electronic noise by a factor of 4. The replacement of the present Nitrogen UV laser with a jitter of 20 ns with a Nd:Yag laser (working at the third harmonic $\approx 355$ nm) with a lower jitter ($\leq 1$ ns) could reduce the uncertainty of the interaction along the Z axis. The waiting time in the interaction is advantageously reduced to $\approx 28$ s in the first set-up and to $1/10$ s in the second set-up. The signal-to-noise ratio can be increased by increasing the current emitted by the source. A better extraction geometry will be evaluated. As an alternative, quasi-continuous or pulsed electron sources reaching densities up to $10^{11}$ cm$^{-3}$ are already available \cite{Bugaev}, \cite{Boscolo}. Finally we consider that this diagnostics can be extended at higher energies (some hundreds of keV), where the Thomson backscattering becomes more efficient and the scattered radiation is detected with higher quantum efficiency. A similar method is recently under development as a non-destructive diagnostics of low energy electron beams in cooling devices \cite{Weilbach}.

\appendix
\chapter{}

\section{Discontinuities in cables}

The voltage wave propagating in ideal TEM (Transverse ElectroMagnetic) transmission lines are described by the wave equation:

\begin{equation}\label{waveeq}
\frac{\partial^2 v}{\partial z^2}-L C \frac{\partial^2v}{\partial t^2}=0
\end{equation}

where $L$, $C$ are the inductance and capacitance per unit length of the line, respectively. The general solution $v(z,t)=v^+(z-v_f t)+v^-(z+v_f t)$ is expressed in term of a propagating $v^+$ and counter-propagating $v^-$ voltage waves with velocity $v_f$.  With the Fourier transformation $V(z,\omega)=F\{v(z,t)\}$ of the waves $v^+(z,t)$ and $v^-(z,t)$  the solution of (\ref{waveeq})  takes the form (in the new variable $\omega$): 

\begin{equation}\label{solomega}
V(z,\omega)=V_0^+(\omega) e^{-jkz}+ V_0^-(\omega) e^{jkz}
\end{equation}

where $V_0^\pm(\omega)$ are arbitrary functions independent by $z$. 
The ratio $\Gamma=\frac{V_0^-}{V_0^+}$ in a point $z$ along the transmission line is defined as voltage reflection coefficient. In the particular case in which a line with characteristic impedance $Z_\infty$ is terminated with an impedance load $Z_L$, the  reflection coefficient takes the form:

\begin{equation}\label{reflect}
\Gamma=\frac{Z_L-Z_\infty}{Z_L+Z_\infty} \,.
\end{equation}
  
When  $Z_L=\infty$ and $Z_L=0$ we obtain $\Gamma=1$, $\Gamma=-1$, respectively and the reflected wave is equal and opposite to the incident wave. 
If a transmission line with impedance $Z_{1\infty}$  is connected with a second transmission line of impedance $Z_{2\infty}$ the reflection coefficient  is:

\begin{equation}\label{reflect1}
\Gamma=\frac{ Z_{2\infty}-Z_{1\infty}}{ Z_{2\infty}+Z_{1\infty}} \, .
\end{equation}

The equations (\ref{solomega} $\div$ \ref{reflect1}) suggest that measuring the reflected wave $V_0^-(\omega)$ or $v^-(z,t)$, information about the discontinuities along the transmission line can be obtained. The reflectometry technique uses this method to characterize the presence of discontinuities, short or open circuits.    

\section{Experimental test on the transmission lines of ELTRAP by means of time-domain reflectometry technique}

\begin{figure}[h]
\begin{center}
\includegraphics[scale=0.7]{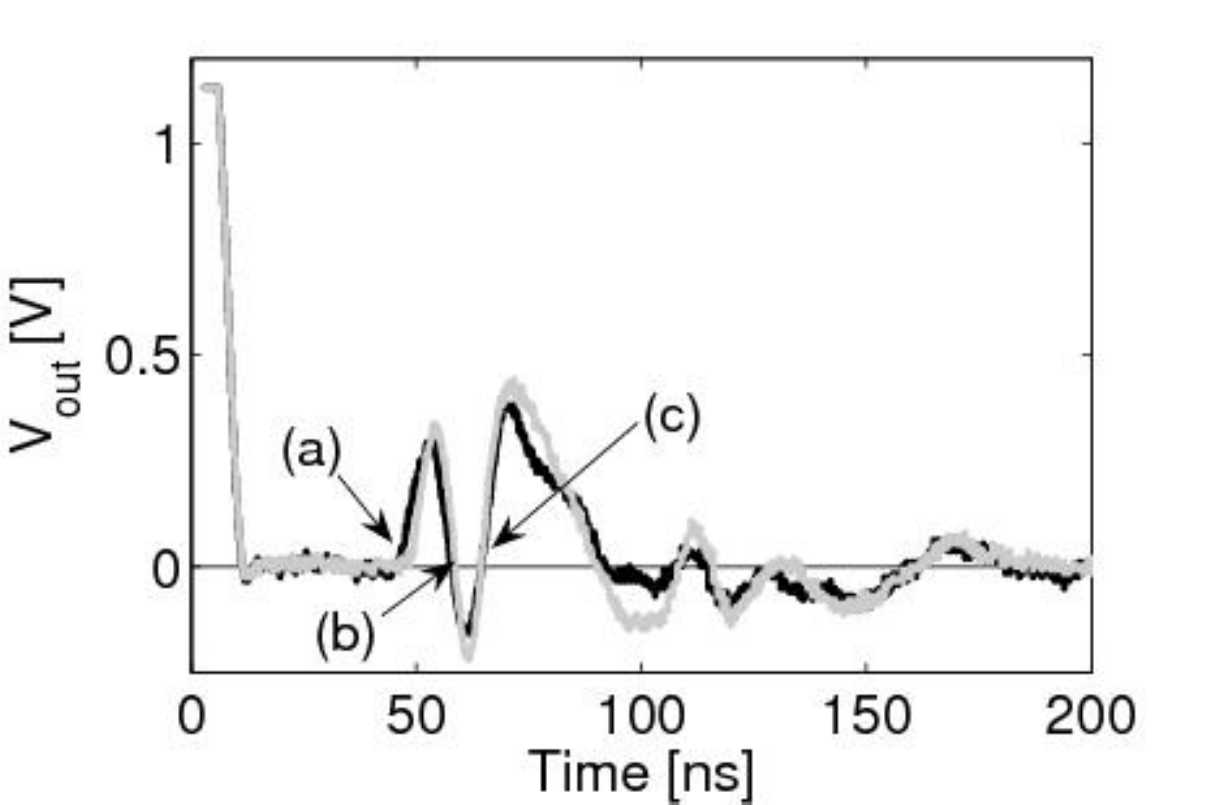}
\end{center}
\caption{Comparison between experimental (black line) and test bench (gray line) signals. The latter is the reflection signal obtained on a test bench with the reconstructed line and optimized parameters. The zero crossings (a), (b), (c) indicate the presence of discontinuities.}
\label{sign_c}
\end{figure}

As described before the measurement of a voltage reflected wave (e.g. a pulse ) provides informations about discontinuities in the trasmission line. This technique was used to characterize the cables connected with the electrodes of the experimental apparatus, i.e. generating an impulse of amplitude 1 V and duration 8 ns with a function generator and recording the line response.
The measured signals show some oscillations after the  first trasmitted pulse. These oscillations are attributed to impedance discontinuities in the transmission line, caused by the fact that the antenna is connected to the oscilloscope by a series of different conductors: the $18$ $\Omega$ kapton-insulated wire from the electrode to the vacuum feedthrough, enclosed within the ultra-high vacuum vessel of the Malmberg-Penning trap, and the $50$ $\Omega$ coaxial cable from the trap flange up to the measuring device. The connection between the two parts is not matched and multiple reflections at the discontinuities occur. In order to characterize the transmission line a model has been built on a test bench and the results have been compared with the original signal to verify the initial assumption.
\begin{figure}
\begin{center}
\includegraphics[scale=0.8]{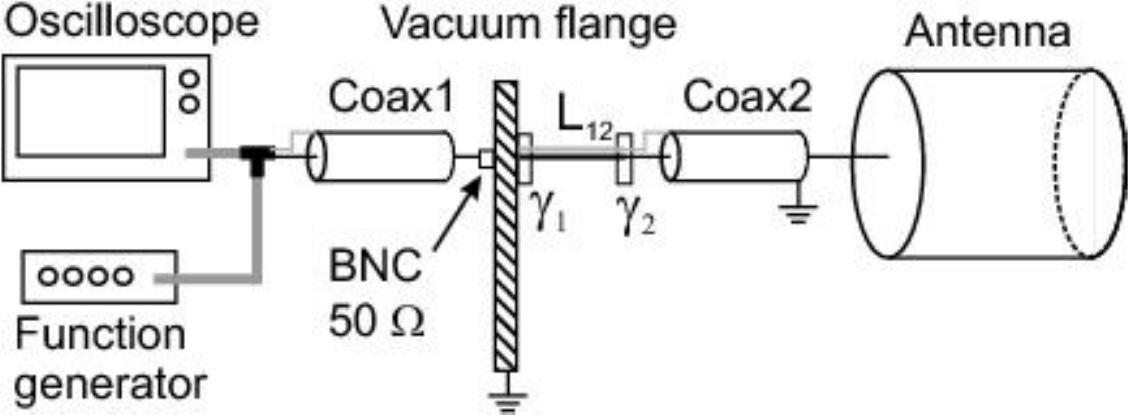}
\end{center}
\caption{Scheme of the signal acquisition line from the antenna reconstructed on a test bench. The discontinuities are indicated by $\gamma_1$ and $\gamma_2$.}
\label{testbench}
\end{figure}
The test system, shown in Fig.~\ref{testbench}, consisted in a 50 $\Omega$ coaxial cable (Coax1), connected by non-coaxial wires of length $L_{12}$ to a kapton-insulated coaxial cable (Coax2) of impedance 18 $\Omega$ and length $L_2$. $L_{12}$ and $L_2$ could be varied to match the experimentally detected signal. The antenna has been simulated with an open circuit of infinite impedance (the capacity of the antenna being included into that of the cable). An example of the output is shown in Fig.~\ref{sign_c}, where we compare an experimental signal to the best-fitting results of the test bench system, obtained with the following parameters: $L_{12}=0.16$ m, $Z_{12}=300$ $\Omega$, $L_2=1.10$ m, $Z_2=18$ $\Omega$. The comparison shows that the main features of the signal are well reproduced. The three points (a), (b), (c) indicate the presence of two discontinuities $\gamma_1$ and $\gamma_2$ and a third one due to the open termination of the line at the antenna. Successive zero crossings appear as a consequence of multiple reflections.

\begin{figure}
\begin{center}
\includegraphics[scale=0.8]{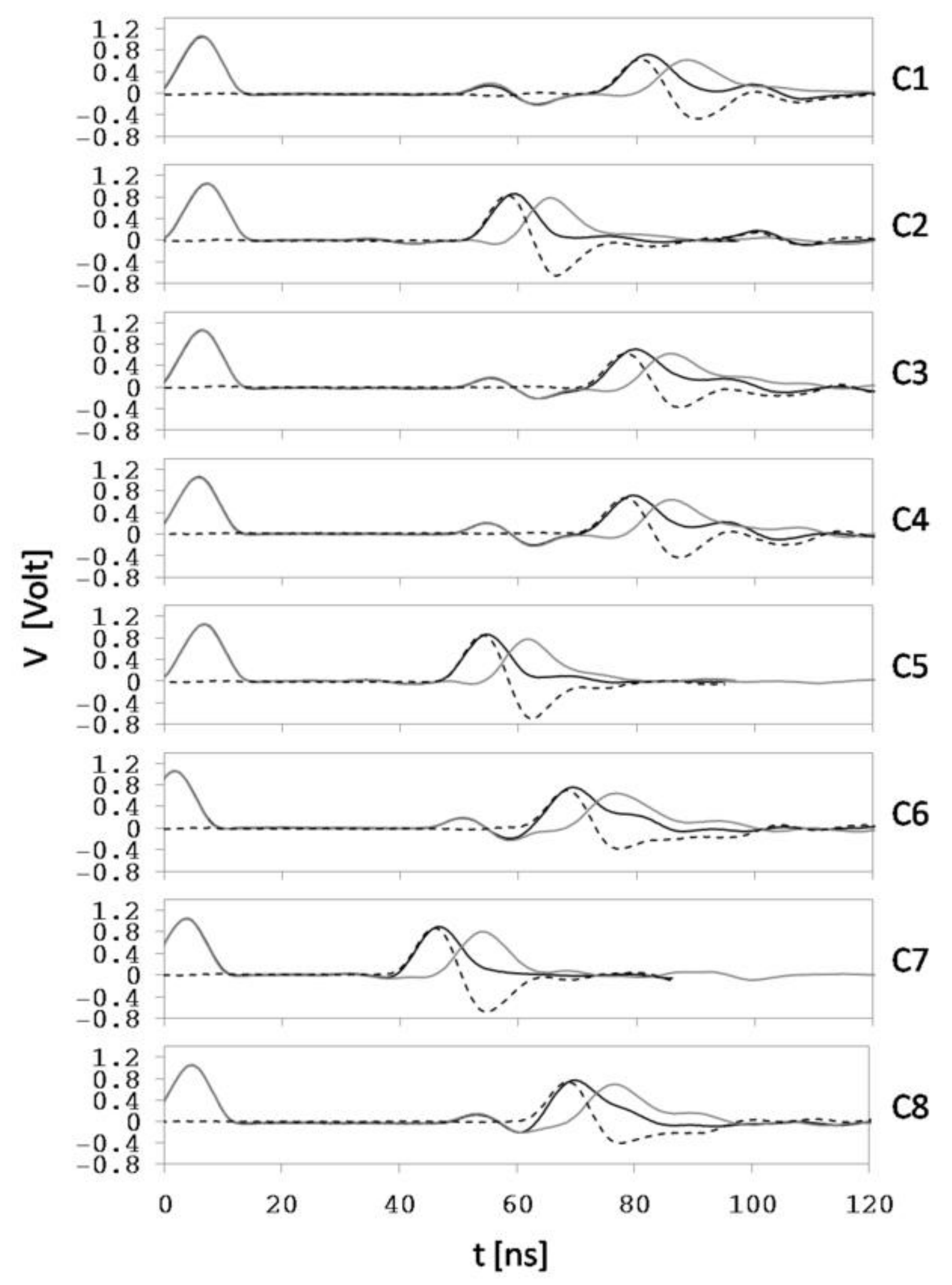}
\end{center}
\caption{\label{coaxiallines1}Reflectometry measurements of C1 - C8 cylinders with the trasmission lines in high impedance (black lines) or with the electrode connected at the line end (grey lines). The effect of the electrode is shown by the difference between the two reflected signals (dotted line).}
\end{figure}

\begin{figure}
\begin{center}
\includegraphics[scale=0.8]{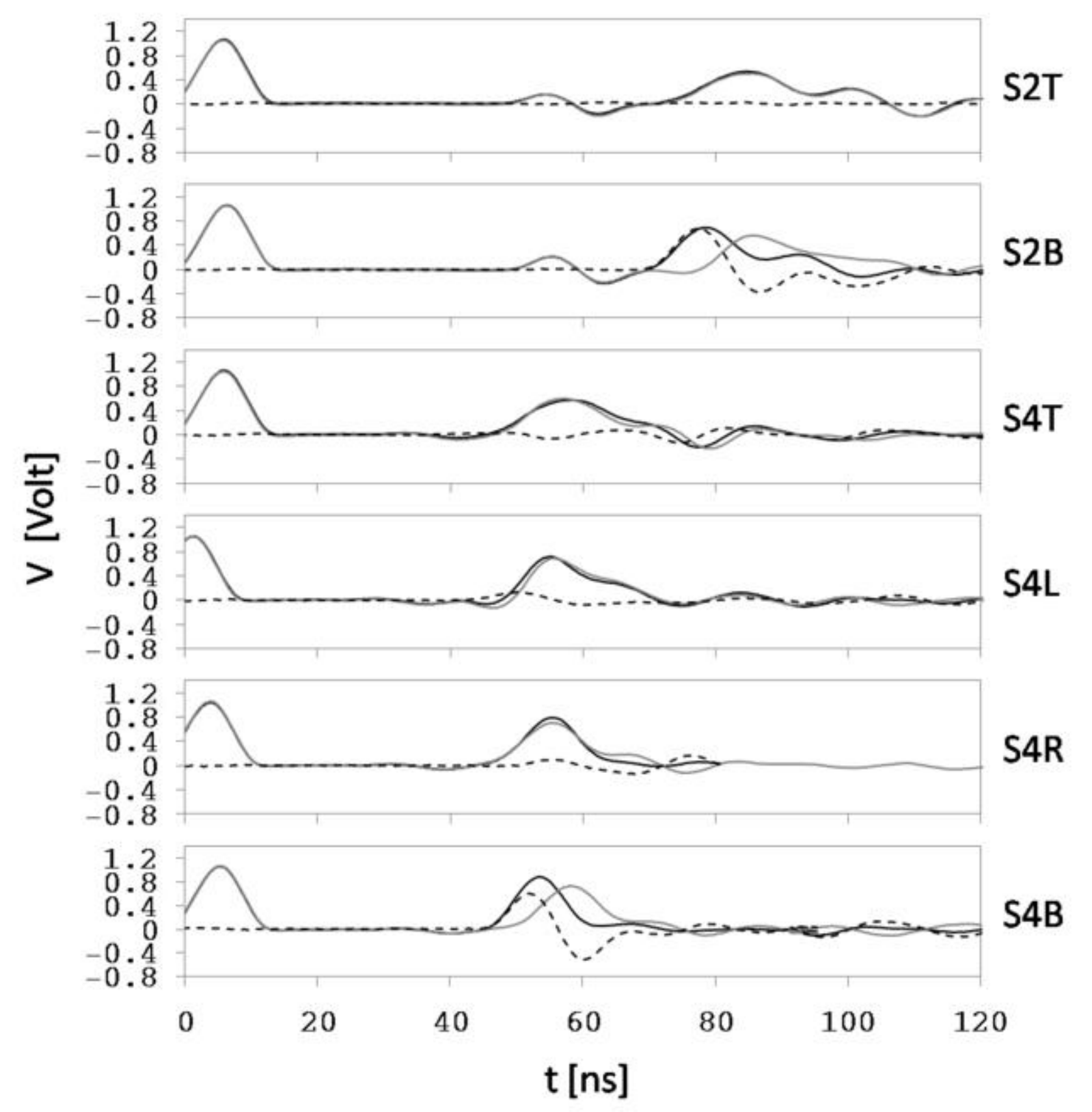}
\end{center}
\caption{\label{coaxiallines2}Reflectometry measurements of S2, S4 cylinders with the trasmission lines in high impedance (black lines) or with the electrode connected at the line end (grey lines). The effect of the electrode is shown by the difference between the two reflected signals (dotted line).}
\end{figure}

In order to reduce the multiple discontinuities the not-matched transmission lines was substituted with 50 $\Omega$ coaxial transmission line. Six lines were connected with the electrodes C1, C3, C4, C6, C8, S4 and with high power feedthrough (non-coaxial) while four lines were connected with the electrodes C2, C5, C7, S2 and with coaxial feedthrough.  The line were then characterized as described before. In order to distinguish the effect introduced by the electrode itself, the measurements were performed connecting the electrode at the end of the transmission line or leaving the electrode at high impedance (see fig. \ref{coaxiallines1}, \ref{coaxiallines2}). The measurements show a first distortion in the line-feedthrough transitions that is evidently greater for the cylinders where the feedthrough are not coaxial. A second distortion is observed subtracting the reflected signal, measured with the lines terminated at high impedance, with those measured with the lines connected to the electrodes. This distortion is introduced by the impedance of the electrode and it is smaller for S4 and S2 due to their lower capacity.  These results show that the multiple discontinuities are efficiently removed and the distortion in the reflected signal is appreciably reduced. We choose the sector S4R as the better matched electrode to be used in the electrostatic diagnostics described in chapter 3.
  
\chapter{}

\section{Broadband radio frequency plasma generation in a Malmberg-Penning  trap}

\begin{figure}
\begin{center}
\includegraphics[scale=0.5]{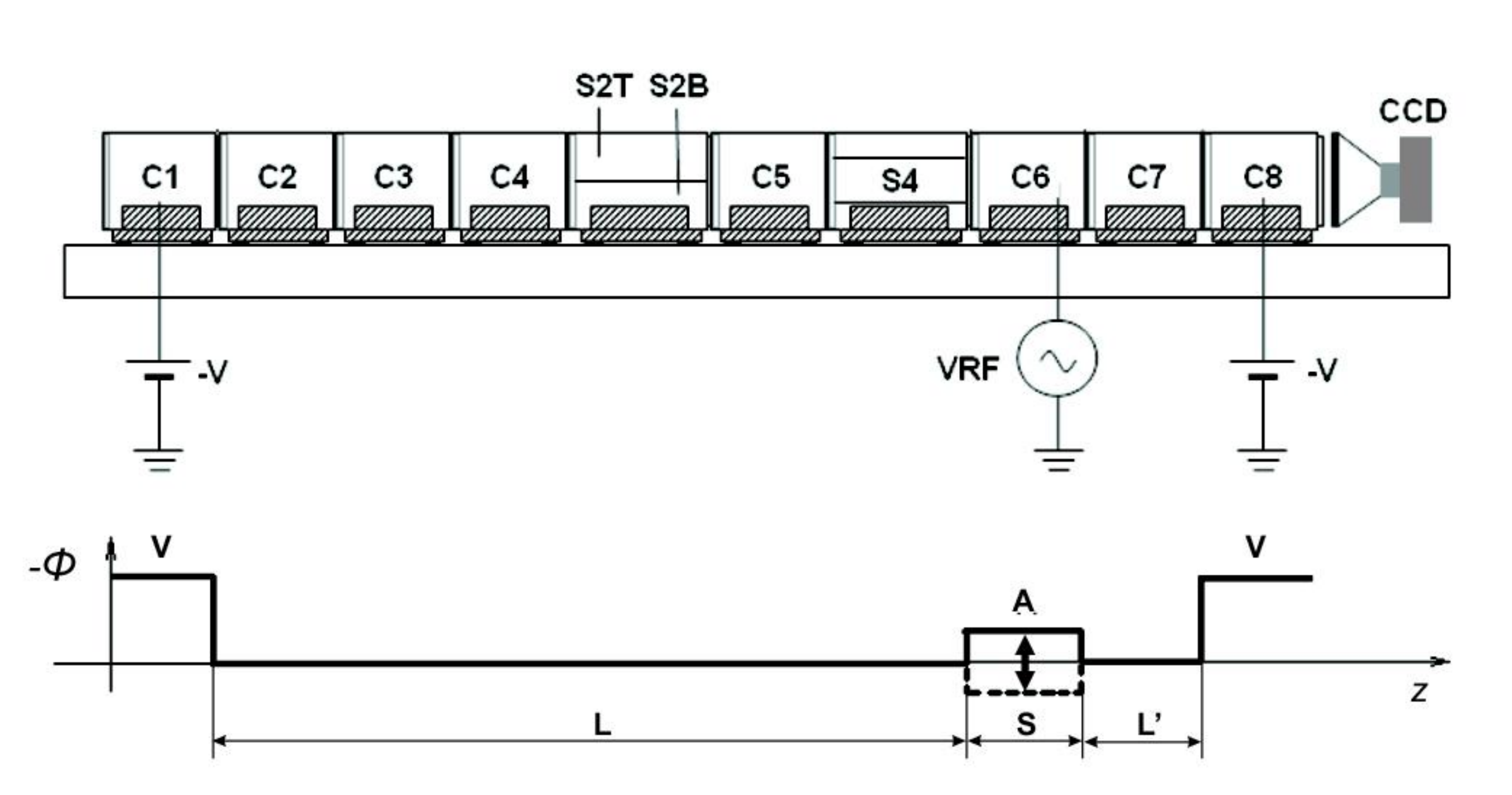}
\end{center}
\caption{\label{fig:setup}ELTRAP setup as used for RF discharge and plasma confinement. Top: sketch of the electrode stack, mounted on a holding bar and followed by a CCD camera for optical diagnostics. Bottom: indicative scheme of the electrode potentials, with confinement between electrodes C1 and C8 biased at a negative voltage $V$ and RF drive of amplitude $A$ on C6.}
\end{figure}

In this appendix we describe the formation of non-neutral electron plasmas in UHV condition by means a Radiofrequency (RF) heating in the ELTRAP apparatus \cite{ParoliRF}. These plasmas could be used to study their interaction with the electron beam produced by the photocathode source with important applications in charged particles acceleration. The goal of this work is limited to the experimental analysis of the plasma formation and the systematic study on the plasma characteristics at the dynamical equilibrium. Furthermore a simple one-dimensional model is also discussed to demonstrate the feasibility of sufficient electron heating and plasma generation via the proposed mechanism. As shown below this production scheme appears as a valid alternative to conventional sources of low-energy non-neutral plasmas. In the ELTRAP apparatus (see fig.\ref{fig:setup}) two negative potentials of -80 V were applied on two electrodes of the trap to confine axially the electrons produced by the ionization of the residual gas and to promote the self-sustainment of the discharge necessary to obtain appreciable charge densities. The radial confinement is provided by the axial magnetic field with typical values of the order of 0.1 T.  The RF power for plasma generation and heating is given by a power supply capable of producing sinusoidal waveforms of amplitude up to $10$ V and $80$ MHz. The power absorbed is of few hundreds mW and therefore the RF signal is directly imposed on a trap electrode suitably chosen as antenna through a $50$ $\Omega$ impedance coaxial cable without any matching network. The detection of the confined plasma is performed by lowering the trapping potential and dumping the sample onto the P43 three-layer aluminium-coated phosphor screen of the optical diagnostics describe in chapter 2. The plate is biased at $\approx 15$ kV so that the energy is sufficient to produce an axially-integrated image of the plasma. The image is acquired by the Charge-Coupled Device (CCD) camera of the diagnostics. The screen can also be used without high-voltage bias as a charge collector to yield the value of the total trapped charge .

\section{Plasma formation and Fermi-like heating}

\begin{figure}
\begin{center}
\includegraphics[scale=0.65]{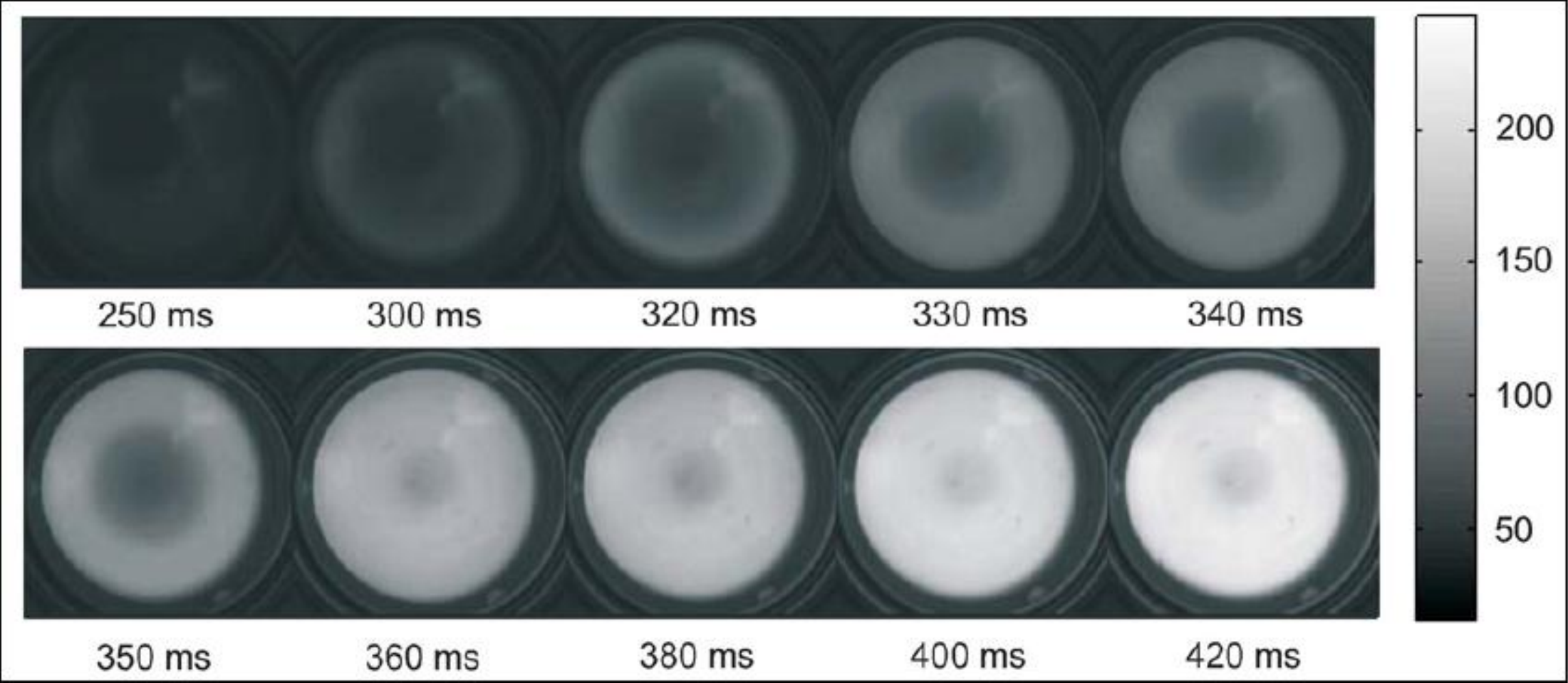}
\end{center}
\caption{\label{fig:formation}Optical measurement of the transverse density profile during the discharge. The plasma can be observed after $\approx300$ ms. The generation takes place mostly in the periphery in its early stage and successively fills the whole space.}
\end{figure}

\begin{figure}
\begin{center}
\includegraphics[scale=1]{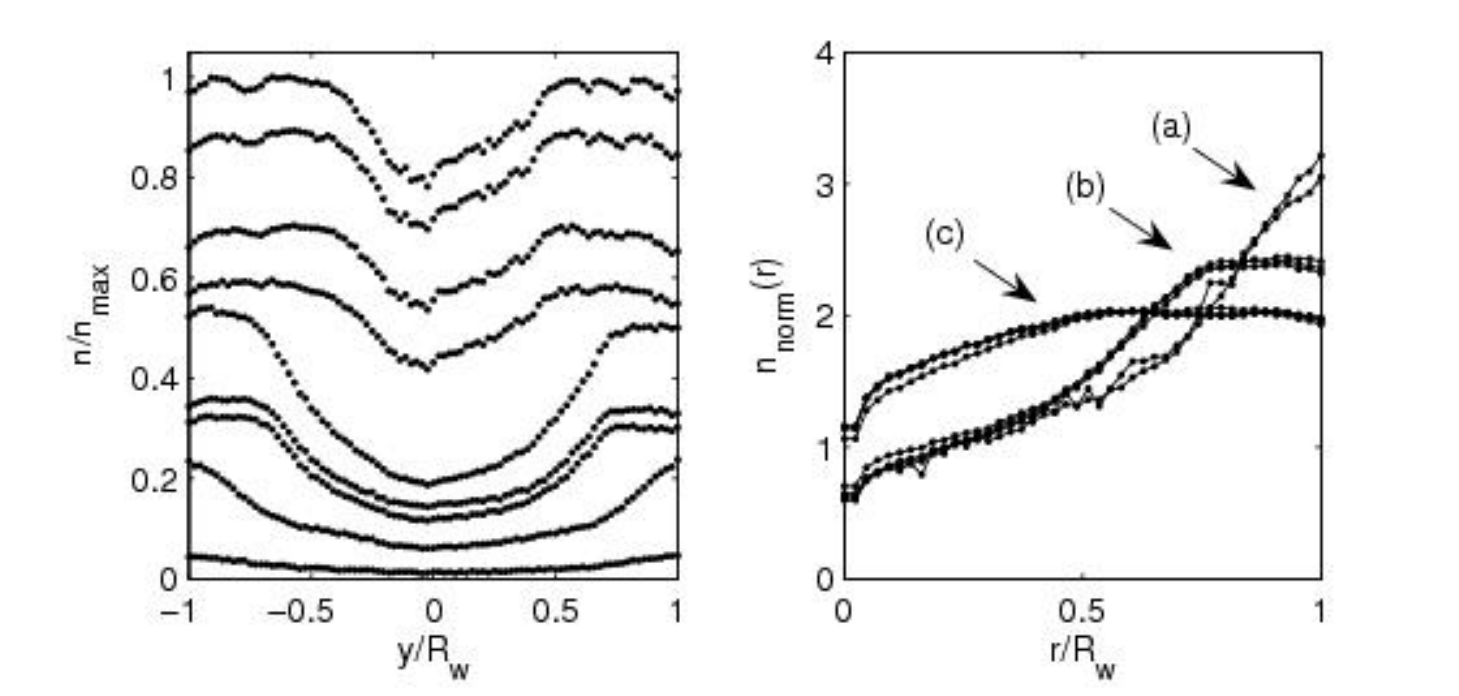}
\end{center}
\caption{\label{fig:radialprofile}Axially-integrated density profiles during plasma formation, for times between $300$ and $420$ ms. On the left, profiles along the vertical axis $y$, normalized to the maximum measured value. On the right, azimuthally-integrated profiles, normalized to the total charge. The profile evolution is not continuous but follows three successive shape groups: (a) for $300-320$ ms, (b) for $330-350$ ms, (c) for $360-420$ ms.}
\end{figure}

The formation of electrons in the chamber by ionization processes induced by RF heating was experimentally observed in a characteristic time of hundreds of ms for a RF amplitude of 3.8 V, a frequency of 8 MHZ and a magnetic field of 0.1 T. The electrons were confined between the electrodes C1 and C8, while the RF was applied on C7. The axially integrated charge distribution, measured via our optical diagnostics (see fig. \ref{fig:formation}) shows that an annular distribution forms close to the trap wall, followed by an increasing occupation of the central region in a time of about 300 ms. These observations are also confirmed by the radial profiles (a vertical cut passing through the symmetry center) of the distributions (see fig. \ref{fig:radialprofile} left). Integrating azimuthally the profiles, normalized to the total charge (see fig. \ref{fig:radialprofile} right) we obtain a grouping in three different shapes: (a) for $300-320$ ms, (b) for $330-350$ ms, (c) for $360-420$ ms. This suggests the presence of complex collective phenomena beyond the basic, continuous diffusion process which deserve further investigation.
The density growth at the trap wall is compared with a theoretical estimate of the ionization rate where no secondary ionization mechanism take place:   

\begin{equation}
\frac{\textrm{d}n\left(t\right)}{\textrm{d}t}=N_n\left\langle \sigma\left(v\right) v \right\rangle n\left(t\right)
\label{eqn:ionizrate}
\end{equation}

\begin{figure}
\begin{center}
\includegraphics[scale=0.9]{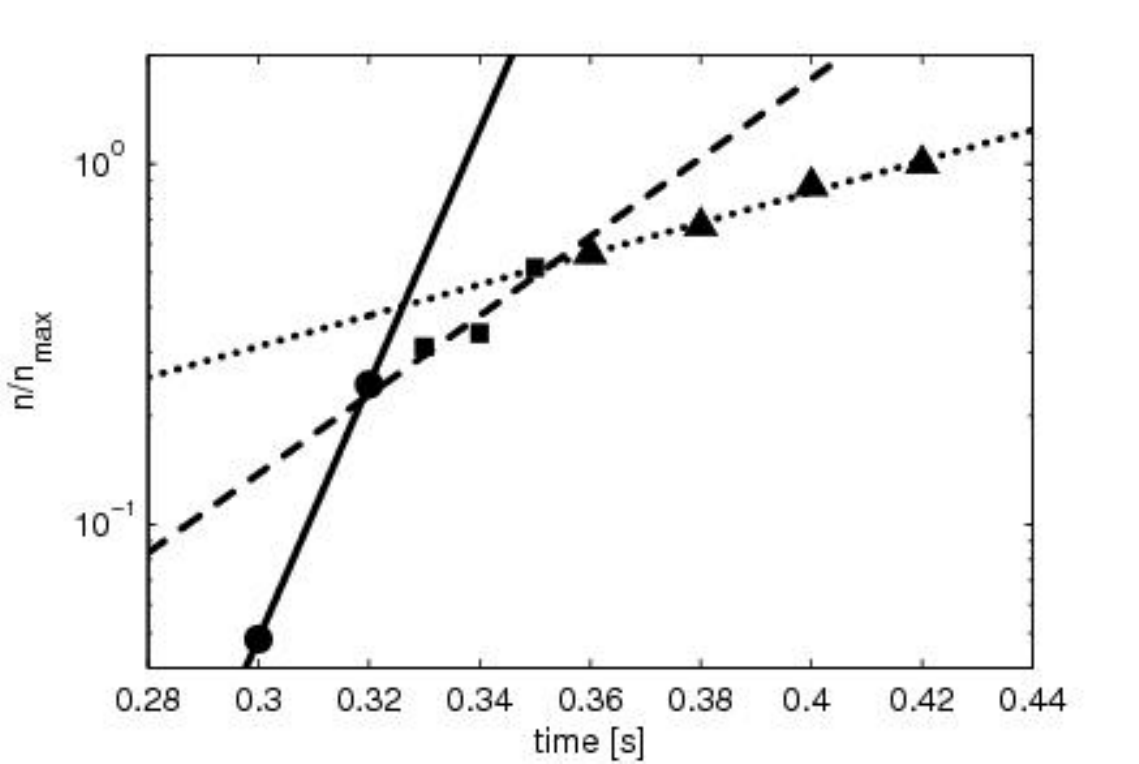}
\end{center}
\caption{\label{fig:ionizationrate}Ionization rate measured in terms of density growth at the trap wall, normalized to the maximum measured value. The data are grouped according to Fig.~\ref{fig:radialprofile} (circles correspond to group (a), squares to (b) and triangles to (c)) and are fitted with exponential laws of inverse time constants $81.3$, $25.2$ and $9.9$ s$^{-1}$, respectively.}
\end{figure}

with $N_n$ the density of the considered residual gas, i.e. H$_2$ in our case, at the working pressure of $\approx4\cdot 10^{-9}$ mbar and $\sigma$ its first ionization cross section at the electron velocity $v$.
The rates rappresented in figure \ref{fig:ionizationrate} are grouped consistently with the three groups found before. Partial fits yield $1/\tau=81.3$, $1/\tau=25.2$ and $1/\tau=9.9$ s$^{-1}$ for groups (a), (b) and (c), respectively. In every case the experimental ionization appears much higher than expected, therefore we can conclude that secondary production mechanisms (which we do not investigate here) must play a dominant role in the plasma density growth.

In order to explain some typical behavior of the experimentally observed plasma formation a Fermi-like one dimensional model was studied. This model explains at least qualitatively that a plasma can be created and brought beyond the energy threshold of the first ionization cross section for light gases ($\approx 10-20$ eV) with a low power RF drive of the like of our experiment. Fig.~\ref{fig:setup}-bottom sketches the model. An electron of charge $-e$ and mass $m$ is confined in a square potential well of depth $V$ and interacts with a square barrier of amplitude $A \sin{(\omega t)}$, where $\omega$ is the frequency of the sinusoidal oscillation. When the electron interacts with the edges of the barrier its energy  changes instantaneously  of a quantity  $\tilde{E}=E_i- e A \sin{(\omega t)}$. This variation occurs only when the electron energy exceeds the amplitude of the barrier otherwise the electron is reflected. Taking into account many interactions, the electron energy state $E_i$  at the interaction $i$ is written in term of an iterative map: 

\begin{equation}\label{eqn:map}
\left\{
\begin{array}{l l l}
\widetilde{E} & = & E_i+\left(-1\right)^{k_i}eA\sin \left[\omega \left(\sum_{j=0}^{i}\frac{l_{k_j}}{\sqrt{2E_j/m}}\right)+ \varphi \right] \\
E_{i+1} & = & \widetilde{E} \qquad \qquad \, \textrm{if}\;\widetilde{E}>0 \\
E_{i+1} & = & E_i \qquad \qquad \textrm{if}\;\widetilde{E}<0
\end{array}
\right.
\end{equation}

where $l_k=(2L,S,2L',S)$ is a vector indicating the lengths of the regions that the particle would go through over a complete bounce period without being reflected at the oscillating barriers. The region $S$ has been replicated for a practical reason, namely that if no reflection takes place ($\widetilde{E}>0$), the sequence of the indexes $k=\left(0,1,2,3\right)$, i.e. the sequence of regions travelled by the particle, is repeated always in the same order. On the contrary, every time that the particle is reflected by an oscillating barrier ($\widetilde{E}<0$), it crosses again the last region and the order of the sequence $k$ is inverted. Therefore, for convenience in the implementation of the numerical algorithm, we define a flag $\sigma$ that changes sign when a particle reflection occurs, i.e. $\sigma_{i+1}=\sigma_i \textrm{ for } \widetilde{E}>0 \textrm{, }\sigma_{i+1}=-\sigma_i \textrm{ for } \widetilde{E}<0 \textrm{ and }\sigma_0=1$. The iteration rules for $k_i$ will then be:

\begin{itemize}

\item
$\textrm{if } \widetilde{E}>0 \textrm{ and } \sigma_i>0 \rightarrow k_{i+1}=\textrm{mod}\left(k_i+1,3\right)$

\item
$\textrm{ if } \widetilde{E}>0 \textrm{ and } \sigma_i<0 \rightarrow k_{i+1}=\textrm{mod}\left(k_i-1,3\right)$

\item
$\textrm{ if } \widetilde{E}<0  \rightarrow k_{i+1}=k_i$.

\end{itemize}

\begin{figure}
\begin{center}
\includegraphics[scale=1.0]{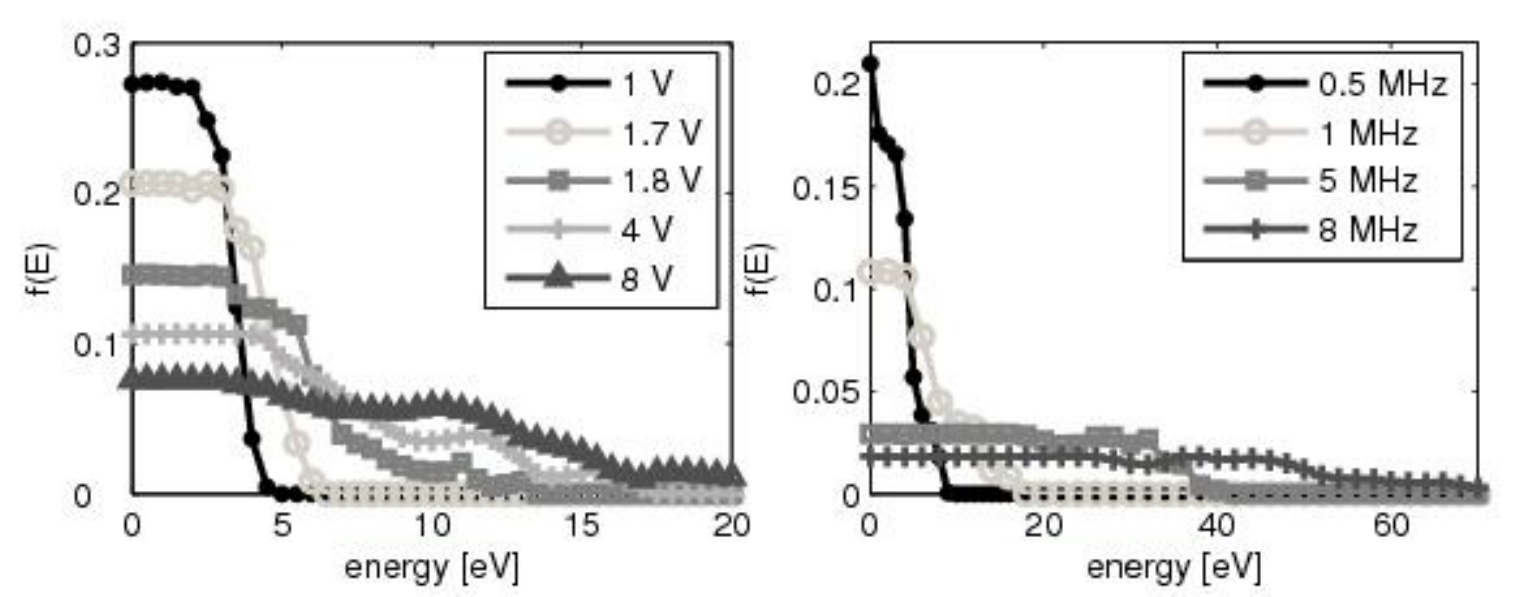}
\end{center}
\caption{\label{distributions}Limit energy distributions $f(E)$ of a trapped electron, after $10^7$ interactions with an 
oscillating barrier. Geometrical parameters of the ELTRAP device have been used: confinement between electrodes C1 and C8, RF drive on electrode C7. In the left panel, the amplitude of RF drive is varied while keeping the frequency at $1$ MHz. In the right panel, the frequency is varied at a constant amplitude of $3.8$ V.}
\end{figure}

the equation \ref{eqn:map} can be solved recursively changing the amplitude $A$ and for different values of $\omega$. For a number of interactions of the order of $10^7-10^8$, corresponding to few seconds, the energy distribution $f\left( E \right )$, i.e. the count of energy values $E_i$ recorded at the interaction instants $i$, tends to a limit that is independent of the initial phase $\varphi$ of the RF drive. Figure \ref{distributions} shows the distribution function $f(E)$ varying the amplitude and the frequency of the RF drive and for a geometry corresponding with a potential well between the electrodes C1 and C8 and with the RF applied on C7. An appreciable number of electron energies exceeds the value of 10 eV for RF amplitude greater than 1.8 V (for a drive frequency of 1 MHz), while for an  amplitude of 3.8 V the electron energies are distributed to higher values increasing the drive frequency between 1 to 8 MHz. In both cases with a maximum RF amplitude of 3.8 V the electron reach values exceeding the first ionization energy of the residual gas (molecular hydrogen). This model is a strong simplification of the real system because the electron motion is forced to be one-dimensional, the electrons formed in the trap are non-interacting and the potential square well is an ideal case. The real potential gets smoother towards the symmetry axis, so that the interaction between the particle and the oscillating field takes place along a finite length. To obtain more realistic results we can extend this model to the case of the actual, azimuthally-symmetric potential $\Phi\left(r,z\right)$ of a cylindrical trap. Let us consider a case where a RF potential is applied to a cylinder far from the trapping electrodes and all other cylinders are grounded. Then $\Phi$ (which is easily evaluated analytically or computationally) is solved using the boundary condition $\Phi=A\sin(\omega t+\varphi)$ on the electrode chosen for RF input. An electron of initial velocity $v_o$ undergoes a variation of energy $\delta E$ after the interaction with a single edge of the potential barrier, reaching a velocity $v_f$. To evaluate quantitatively the effect of the electron-potential interaction we can calculate the mean square of $\delta E$, i.e.

\begin{equation}
\overline{\delta E^2}(r,v_0)=\frac{1}{2\pi}\int_0^{2\pi}\left[\frac {m}{2}\left( v_f^2(r,\varphi)-v_0^2\right)\right]^2 \textrm{d}\varphi\,.
\label{eqn:dEsquare}
\end{equation} 

Hence we can conveniently define the equivalent potential amplitude $V_{eq}$, i.e. the amplitude of a square barrier that would have the same $\overline{\delta E^2}$ of the real potential $\Phi$, as

\begin{equation}
V_{eq}(r,v_0)\equiv \frac{\sqrt{2\overline{\delta E^2}}}{e}\,.
\label{eqn:Veff}
\end{equation} 

\begin{figure}
\begin{center}
\includegraphics[scale=0.8]{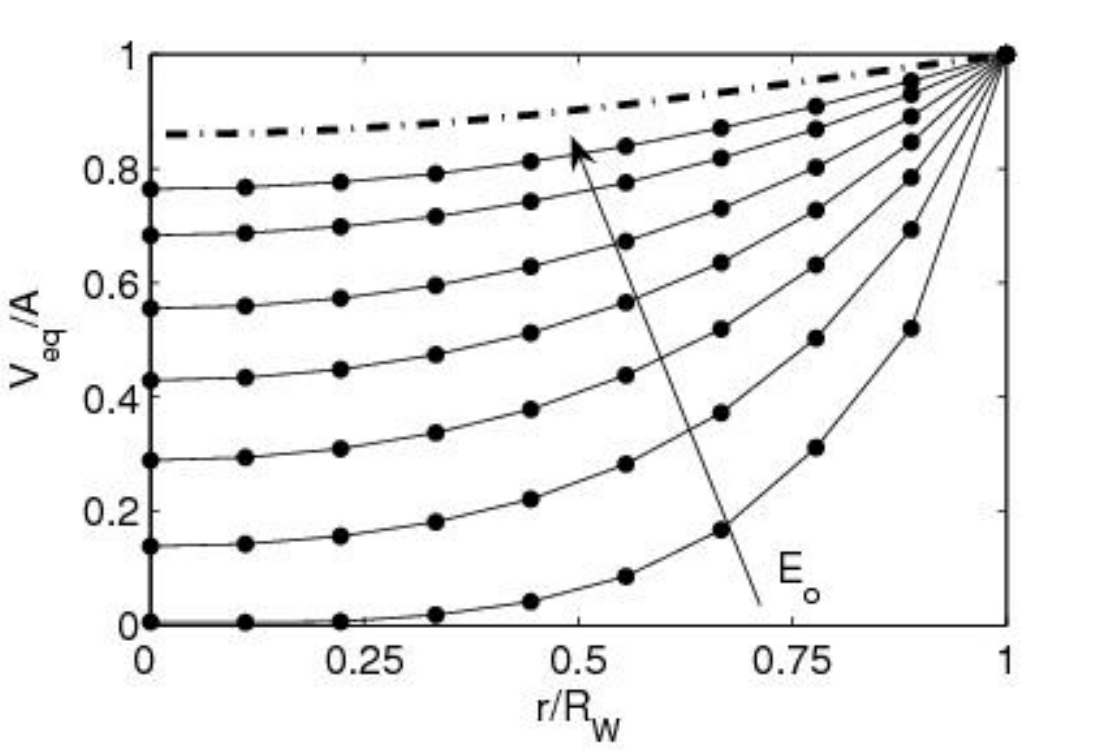}
\end{center}
\caption{\label{fig:Veff} Equivalent potential $V_{eq}$ of an oscillating barrier versus radial position, for different values of the initial electron energy $E_o$. $V_{eq}$ curves correspond to $E_o=0.4,\,1,\,1.8,\,3,\,5,\,10,\,20$ eV. The dash-dotted line is the potential $\Phi(r)$.}
\end{figure}

This definition is consistent with the fact that $V_{eq}=A$ when $r=R_W$, i.e. at the electrode surface. Fig.~\ref{fig:Veff} shows the trend of $V_{eq}$ as a function of the radial position with the initial electron energy as parameter. The energy gain is smaller for lower initial energy and it also decreases towards the center, so that we can expect that most of the ionization takes place close to the trap wall. This is indeed what we have observed in the experiments.

\begin{figure}
\begin{center}
\includegraphics[width=\textwidth]{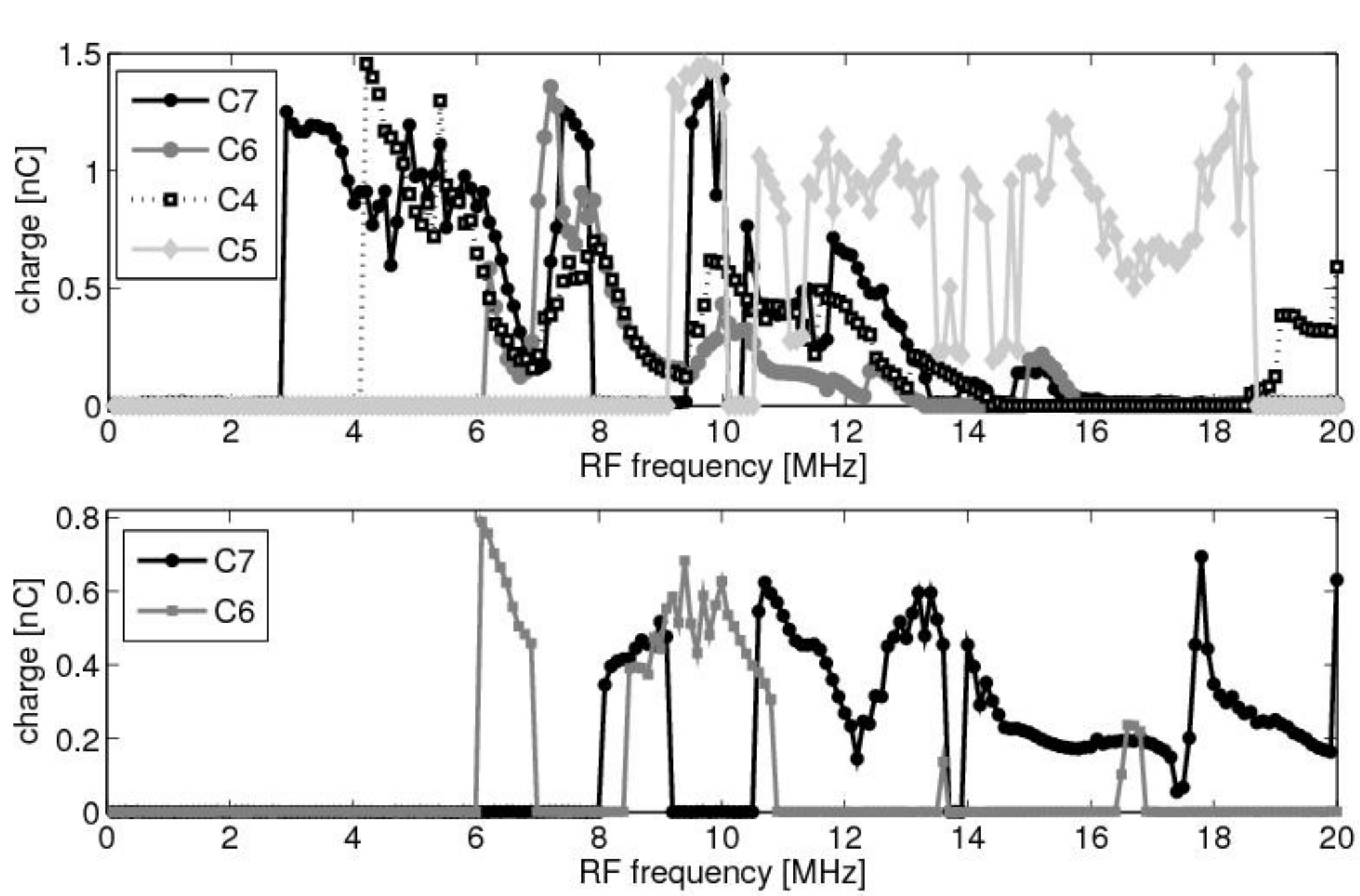}
\end{center}
\caption{\label{longandshort}Total charge confined for $4.5$ s in a condition of dynamical equilibrium versus the frequency of the RF drive. The plasma is formed and confined between C1 and C8 (top, \textit{long trap}) or between S2 and C8 (bottom, \textit{short trap}). The legend specifies the electrode used as antenna for the RF excitation.}
\end{figure} 

\section{Plasma at dynamical equilibrium}

After the formation, if the RF is continuously applied, the electrons that are lost radially and axially are replaced by those produced by collisions .  The total confined charge was experimental measured in this dynamical equilibrium changing the drive frequency and  the geometry of the set-up (length of the trap, position and length of the cylinder where the RF is applied ). In particular two trapping lengths were considered. In the first case the trapping region was between the electrode C1 and C8 (long trap) while in the second case was between S2 and C8 (short trap). For both configurations different electrodes were used for the RF application. For each frequency value, the excitation has been applied for $4.5$ s, after which the trapping voltage on the C8 electrode has been lowered and the plasma dumped on the phosphor screen, used as charge collector. The discharge signal has been filtered from the random noise (typically $\leq90$ mV rms) with a digital low-pass filter of the third order with cutoff frequency $500$ kHz. The collected charge is then calculated as $Q=-V_{min}C$. Here $V_{min}$ is the minimum of the voltage discharge signal and $C$ the capacity of the measurement system, i.e. essentially the capacity of the coaxial cable. The latter is obtained directly from the time constant $1/RC$ of the discharge, with a resistance $R=1$ M$\Omega$ given essentially by the load of the oscilloscope. The experimental results are shown in figure \ref{longandshort}. The phenomena appear to be non-resonant because is observed in a wide range of frequencies . The results are qualitatively in agreement with the model described before in that the lowermost threshold for plasma creation is lower for the choice of the C7 electrode as RF antenna, while higher frequencies are needed with electrodes closer to the center of the trap. When the confinement length is reduced this argument is apparently no longer valid and in general we can say that the creation of the plasma is more difficult. In both geometries the total charge is of the order of 1 nC corresponding with a density of $\approx 10^{-6}$ cm$^{-3}$ that is comparable to the thermocathode sources used in our past experiments as well as in similar set-ups. These densities are not suitable to promote a strong beam-plasma interaction, nevertheless the combined effects of plasma formation and the compressional phenomena induced by the same RF used for the electron heating can increase the plasma density of some order of magnitude. These issue is not the goal of this work and more studies are needed, yet compressional effects are observed in our experiment aimed to the stabilization of the plasma column in this regime of dynamical equilibrium \cite{gmaerostab}.

\addcontentsline{toc}{chapter}{Bibliography}
\bibliography{refs}        
\bibliographystyle{unsrt}  

\end{document}